COSTAS J. PAPACHRISTOU

INTRODUCTION TO

# ELECTROMAGNETIC THEORY

AND THE PHYSICS OF CONDUCTING SOLIDS

Introduction to

# ELECTROMAGNETIC THEORY

## and the Physics of Conducting Solids

**Costas J. Papachristou**

*Department of Physical Sciences*
*Hellenic Naval Academy*

*To Loulou*

# PREFACE

This textbook is a revised, expanded and translated version of the author's lecture notes (originally in Greek) for his sophomore-level Physics course at the Hellenic Naval Academy (HNA). It consists of two parts. Part A is an introduction to the physics of conducting solids (Chapters 1-3) while Part B is an introduction to the theory of electromagnetic fields and waves (Chap. 4-10). Both subjects are prerequisites for the junior- and senior-level courses in electronics at HNA.

Besides covering specific educational requirements, a unified presentation of the above two subjects serves a certain pedagogical purpose: it helps the student realize that classical and quantum physics are not necessarily rivals but may supplement each other, depending on the physical situation. Indeed, whereas the study of conducting crystalline solids at the microscopic level necessitates the use of quantum concepts such as quantum states, energy bands, the Pauli exclusion principle, the Fermi-Dirac distribution, etc., for the study of electromagnetic phenomena at a more or less macroscopic level the classical Maxwell theory suffices. On the other hand, the student learns from the outset that this latter theory cannot explain things such as the stability or the emission spectra of atoms and molecules.

The basic goal of the first two chapters of Part A is an introduction to crystalline solids and an understanding of the mechanism by which they conduct electricity, with special emphasis on the differences between metals and semiconductors. In the last chapter of Part A (Chap. 3) the conducting solids are studied from the point of view of quantum statistical physics. In particular, the distribution of energy to the charge carriers in metals and semiconductors is examined and the important concept of the Fermi energy is introduced.

The beginning chapter of Part B (Chap. 4) is, so to speak, a mathematical interlude in which certain concepts and theorems on vector fields, to be used subsequently, are summarized. Chapters 5-8 are then devoted to the study of static (time-independent) electric and magnetic fields, while in Chap. 9 the full Maxwell theory of time-dependent electromagnetic fields is presented. Finally, in Chap. 10 it is shown that the Maxwell equations lead, in a rather straightforward manner, to the prediction of the wave behavior of the electromagnetic field; the propagation of electromagnetic waves in both conducting and non-conducting media is examined, and the concept of electromagnetic radiation is introduced. Several important theoretical issues are separately discussed in the Problems at the end of each chapter. Most problems are accompanied by detailed solutions, while in other cases guiding hints for solution are given.

I am indebted to my colleague and friend, Dr. Aristidis N. Magoulas, for many fruitful discussions on Electromagnetism (despite the fact that, being an electrical engineer, he often disagrees with me on issues of terminology!). I also thank the Hellenic Naval Academy for publishing the original, Greek version of the textbook.

Costas J. Papachristou
Piraeus, July 2019

# CONTENTS

## PART A: THE PHYSICS OF CONDUCTING SOLIDS

### CHAPTER 1: ATOMS, MOLECULES, AND CRYSTALS

### CHAPTER 2: ELECTRICAL CONDUCTIVITY OF SOLIDS

### CHAPTER 3: DISTRIBUTION OF ENERGY

## PART B: ELECTROMAGNETIC FIELDS AND WAVES

**CHAPTER 9: TIME-DEPENDENT ELECTROMAGNETIC FIELDS**

**CHAPTER 10: ELECTROMAGNETIC WAVES**

## CHAPTER 1

# ATOMS, MOLECULES, AND CRYSTALS

### 1.1 States of Matter

Most physical substances can exist in each of the three *states of matter* – solid, liquid, or gas – depending on external factors such as temperature and pressure. Let us examine the physical processes involved in each case.

The natural state of a substance is the result of a "contest" between two effects acting in opposition to each other: (*a*) an *attractive* (or *cohesive*) force, of electromagnetic origin, between the atoms (or molecules, or ions) of the substance, which force tends to bring the atoms close to one another; this is done most effectively if the atoms form some sort of regular arrangement, as in a *crystal lattice*; (*b*) thermal energy, which causes a *random* motion of the atoms; this motion becomes more intense as the temperature increases.

Let us consider a substance that is initially in the *solid* state. In this state the attractive interatomic forces predominate, tending to bring the atoms together into a regular arrangement called *crystal* [1]. Thermal energy cannot compete with the strong forces that hold the atoms at fixed positions relative to one another within the crystal structure; the only thing it can achieve is to set the atoms into *vibration* about these fixed positions.

As the temperature increases, the amplitude of vibration of the atoms increases accordingly, until the moment when the attractive forces between the atoms are no longer sufficient to hold the atoms at fixed positions in the crystal. At this temperature (*melting point*) the crystal structure breaks up and the solid *melts*, becoming a *liquid*.[1] At this stage there is equilibrium, so to speak, between electromagnetic and thermal effects.

With the further increase of temperature, atomic thermal motion begins to gain the upper hand in the contest. When the temperature reaches the *boiling point*, the atoms have enough energy to finally "escape" from the liquid and form a *gas*. The attractive interatomic forces are now very small (in the case of an *ideal* gas they are considered negligible). Note that the melting and boiling points of a substance are closely related to the strength of the electromagnetic bonds between the atoms of that substance.

### 1.2 Amorphous and Crystalline Solids

We mentioned earlier that the crystalline structure, which is characterized by a regular arrangement of atoms, provides maximum stability to the solid state of a substance. In Nature, however, we also see materials that *look like* solids (e.g., they are rigid and have fixed shape) without, however, possessing an actual crystalline structure. Such solids are called *amorphous* [1] and are in many respects similar to liquids. That is, their atoms are arranged at random, not exhibiting the regular arrangement of a crystalline solid. We may thus consider an amorphous solid as a liquid with an extremely high viscosity. A material belonging to this category is ordinary glass (don't be deceived by the everyday use of expressions containing the word *"crystal"*, e.g., "crystal bowl"!).

[1] Diamond is a notable exception. Theoretically, its melting point is at 5000 K; it is never reached in practice, however, since diamond transforms to graphite at about 3400 K ! (See [1], p. 406.)

Let us now get back to real crystals. What accounts for their greatest stability? Let us consider a much simpler system, that of a pendulum. Stable equilibrium of the pendulum bob is achieved when the bob is at rest at the lowest point of its path (i.e., when the string is vertical). At this position, the gravitational potential energy of the bob is minimum. By analogy, the internal potential energy of certain solids is minimized when their atoms form a regular crystalline structure. This arrangement provides maximum stability to these solids. On the other hand, some other solids are amorphous due to the fact that, because of increased viscosity in the liquid state, their atoms or molecules are not able to move relative to one another in order to form a crystal as the temperature is decreased. We also note that the transition of a crystalline solid to the liquid state takes place *abruptly* when the temperature reaches the melting point. On the contrary, the transition from amorphous solid to liquid takes place *gradually*, so that no definite melting point can be specified in this case.

Crystalline solids such as *metals* exhibit significant *electrical conductivity*. This is due to the fact that they possess *free* (or *mobile*) charges (electrons) that can move in an oriented way under the action of an electric field. At the other extreme, some solids do not have such free charges and therefore behave as electrical *insulators*. Finally, there exist certain "double agents", the *semiconductors*, which carry characteristics from both the above categories and which, under normal conditions, have conductivity that is smaller than that of metals.

Another interesting property of solids[2] is *thermal conductivity*. Heat transfer in these substances takes place by means of two processes: (*a*) vibrations of the crystal lattice (in all solids) and (*b*) motion of free electrons (in metals). The excellent thermal conductivity of metals owes itself to the contribution of both these mechanisms.

Solids can be classified according to the type of bonding of atoms (or molecules, or ions) in the crystal lattice. The most important types of solids are the following:

1. *Covalent solids:* Their atoms are bound together by covalent bonds. Such solids are the crystals of diamond, silicon and germanium. Due to their stable electronic structure, these solids exhibit certain common characteristics. For example, they are hard and difficult to deform. Also, they are poor conductors of heat[3] and electricity since they do not possess a significant quantity of free charges that would transfer energy and electric charge through the crystal.

2. *Ionic solids:* They are built as a regular array of positive and negative ions. A characteristic example is the crystal of sodium chloride (NaCl), consisting of $Na^+$ and $Cl^-$ ions. Because of the absence of free electrons, these solids are poor conductors of heat and electricity. Also, they are hard and they have a high melting point due to the strong electrostatic forces between the ions.

---

[2] By *"solid"* we will henceforth always mean *crystalline* solid.

[3] Again, diamond is a notable exception given that its thermal conductivity exceeds that of metals at room temperature [1]. This conductivity is, of course, exclusively due to lattice vibrations.

3. *Hydrogen-bond solids:* They are characterized by the presence of polar molecules (see Sec. 8.2) containing one or more hydrogen atoms. Ice is an example of such a solid.

4. *Molecular solids:* They consist of non-polar molecules (see Sec. 8.2). An example is $CO_2$ in its solid state.

5. *Metals:* They consist of atoms with small ionization energies, having a small number of electrons in their outermost shells. These electrons are easily set free from the atoms to which they belong by using part of the energy released during the formation of the crystal. The crystal lattice, therefore, consists of *positive ions* through which a multitude of electrons move more or less freely. These mobile electrons, which were originally the outer or *valence* electrons of the atoms of the metal, are called *free electrons*. To these electrons the metals owe their electrical conductivity as well as a significant part of their thermal conductivity (another part is due to vibrations of the ions that form the lattice). Free electrons also provide the coherence necessary for the stability of the crystal structure, since the repulsive forces between positive ions would otherwise decompose the crystal! It may be said that the free electrons are the "glue" that holds the ions together within the crystal lattice [1].

Before we continue our study of crystalline solids it would be useful to familiarize ourselves with some basic notions from Quantum Physics and to examine the structure of simpler quantum systems such as atoms and molecules.

## 1.3 Rutherford's Atomic Model

The first modern atomic model was proposed by Rutherford in 1911. Let us consider the simplest case, that of the hydrogen atom. According to Rutherford's model, the single electron in the atom is moving on a circular orbit around the nucleus (proton) with constant speed $v$ (Fig. 1.1). The centripetal force necessary for this uniform circular motion is provided by the attractive Coulomb force between the proton and the electron. We call $m$ the mass of the electron and $q$ the *absolute* value of the electronic charge, equal to $1.6\times10^{-19}C$. Therefore, the proton has charge $+q$ while the charge of the electron is $-q$.

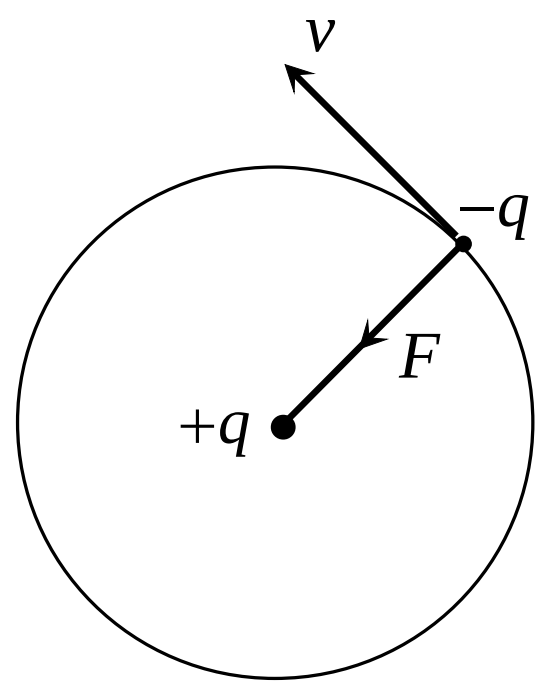

Fig. 1.1. Rutherford's picture of the hydrogen atom.

The total force on the electron is

$$F = \frac{mv^2}{r} = \frac{q^2}{4\pi\varepsilon_0 r^2} \quad \Rightarrow \quad v = \left( \frac{q^2}{4\pi\varepsilon_0 mr} \right)^{1/2} \tag{1.1}$$

where $r$ is the radius of the circular orbit. The kinetic energy of the electron is

$$E_k = \frac{1}{2} mv^2 = \frac{q^2}{8\pi\varepsilon_0 r}$$

while the electron's potential energy in the Coulomb field of the proton is

$$E_p = -\frac{q^2}{4\pi\varepsilon_0 r}$$

where we have arbitrarily assigned a zero value to the potential energy at an infinite distance from the nucleus ($r = \infty$). The total mechanical energy of the electron is

$$E = E_k + E_p = -\frac{q^2}{8\pi\varepsilon_0 r} \tag{1.2}$$

Notice that $E \to 0$ as $r \to \infty$. The negative sign on the right-hand side of (1.2) is due to our choice of the zero level of $E_p$ at infinity and has no special physical significance. Indeed, what is physically significant in Atomic Physics is the energy *difference* $\Delta E$ between two states of motion, *not* the energy $E$ itself. The quantity $\Delta E$ is well defined, independent of the choice of zero level for the potential energy.

The angular velocity $\omega$ of the electron is a function of the total energy $E$. Indeed, by combining the expression (1.1) for $v$ with the relation $v = \omega r$, we find that

$$r = \left( \frac{q^2}{4\pi\varepsilon_0 m\omega^2} \right)^{1/3}$$

That is, the radius $r$ is proportional to $\omega^{-2/3}$. Hence, according to (1.2), $E$ is proportional to $\omega^{2/3}$, or, $\omega$ is proportional to $|E|^{3/2}$.

Although mechanically sound, this "planetary" model of the hydrogen atom presents some serious problems if one takes into account the laws of classical electromagnetism. Indeed, as we will learn in Chapter 10, every *accelerating* electric charge emits energy in the form of electromagnetic radiation. If the charge performs periodic motion of angular frequency $\omega = 2\pi f$ (where $f$ is the frequency of this motion) the emitted radiation will also be of angular frequency $\omega$. In our example, the angular frequency of the radiation emitted by the electron will be equal to the angular velocity $\omega$ of the electron's uniform circular motion (note that $\omega = 2\pi/T = 2\pi f$, where $T$ is the period of this motion). But, if the electron emits energy, its total energy $E$ must constantly decrease, with a parallel decrease of the radius $r$ of the orbit, according to (1.2). If this is to happen, the electron will spiral into the nucleus and the atom will collapse (as has been estimated, the time for this process to take place would be of the order of only $10^{-8}s$ !). Fortunately this doesn't happen in reality, since the atoms are stable.

Another problem of the model is the following: As the energy $E$ changes continuously, the frequency $\omega$ of the emitted radiation must also change in a continuous fashion since, as we have shown, $E$ is a continuous function of $\omega$. Thus the *emission spectrum* of hydrogen must span a continuous range of frequencies. In reality, however, hydrogen (and, in fact, all atoms) exhibits a *line spectrum*, consisting of a discrete set of frequencies characteristic of the emitting atom.[4]

The Rutherford model was thus a bold first step in atomic theory but suffered from serious theoretical problems. The main reason for its failure was that it treated a particle of the microcosm – the electron – as an ordinary classical particle obeying Newton's laws; the model thus ignored the quantum nature of the electron. Did this mean that Rutherford's model had to be completely abandoned, or was there still a possibility of "curing" its problems? In 1913, a young physicist working temporarily at Rutherford's lab decided to explore that possibility...

## 1.4 Bohr's Model for the Hydrogen Atom

To overcome the theoretical difficulties inherent in Rutherford's model, Bohr proposed the following *quantum conditions* for the hydrogen atom:

1. The electron may only move on specific circular orbits around the nucleus, of radii $r_1$, $r_2$, $r_3$, ..., and with corresponding energies $E_1$, $E_2$, $E_3$, ... When moving on these orbits, the electron does *not* emit electromagnetic radiation.

2. When the electron falls from an orbit of energy $E$ to another orbit of lesser energy $E'$, the atom emits radiation in the form of *a single photon* of frequency

$$f = \frac{E - E'}{h} \qquad (1.3)$$

where $h$ is Planck's constant, equal to $6.63 \times 10^{-34}\, J \cdot s$.

3. The allowed orbits and associated energies are determined by the condition that the angular momentum of the electron may assume an infinite set of discrete values given by the relation

$$mvr = n\frac{h}{2\pi} \ , \quad n = 1, 2, 3, \cdots \qquad (1.4)$$

The property of the energy and the angular momentum to take on specific values only, instead of the arbitrary values allowed by classical mechanics, is called *quantization* of energy and angular momentum, respectively.

---

[4] The name "line spectrum" is related to the fact that each frequency appears as a line in a spectroscope.

We will now calculate the allowable orbits $r_n$ and corresponding energies $E_n$ ($n$=1,2,3,...). From (1.4) we have that $v=nh/2\pi mr$. By comparing this expression for $v$ with that in (1.1), we find that

$$r_n = \frac{\varepsilon_0 h^2}{\pi q^2 m} n^2 \equiv a_0 n^2 \quad (n = 1,2,3,\cdots) \tag{1.5}$$

In particular, the radius of the smallest allowable orbit is $r_1=a_0$ (*Bohr orbit*). Substituting (1.5) into (1.2), we find the allowable (quantized) values of the energy of the electron:

$$E_n = -\frac{mq^4}{8\varepsilon_0{}^2 h^2}\frac{1}{n^2} \equiv -\frac{\kappa}{n^2} \quad (n = 1,2,3,\cdots) \tag{1.6}$$

We notice that $E_n \to 0$ for $n\to\infty$, thus for $r\to\infty$. At this limit the electron dissociates itself from the atom and subsequently moves as a free particle. (Find the allowable values $F_n$ of the Coulomb force exerted on the electron and show that $F_n\to 0$ when $n\to\infty$.) The energy difference $E_\infty - E_1 = |E_1| = \kappa$ is called *ionization energy* of the hydrogen atom and represents the minimum energy necessary in order to detach the electron from the atom and set it free. Note that once freed the electron may assume *any* energy! That is, the quantization of energy does not concern freely moving electrons but only those constrained to move within a definite quantum system such as an atom, a molecule, a crystal, etc. It is therefore incorrect to assume that the energy is *always* quantized!

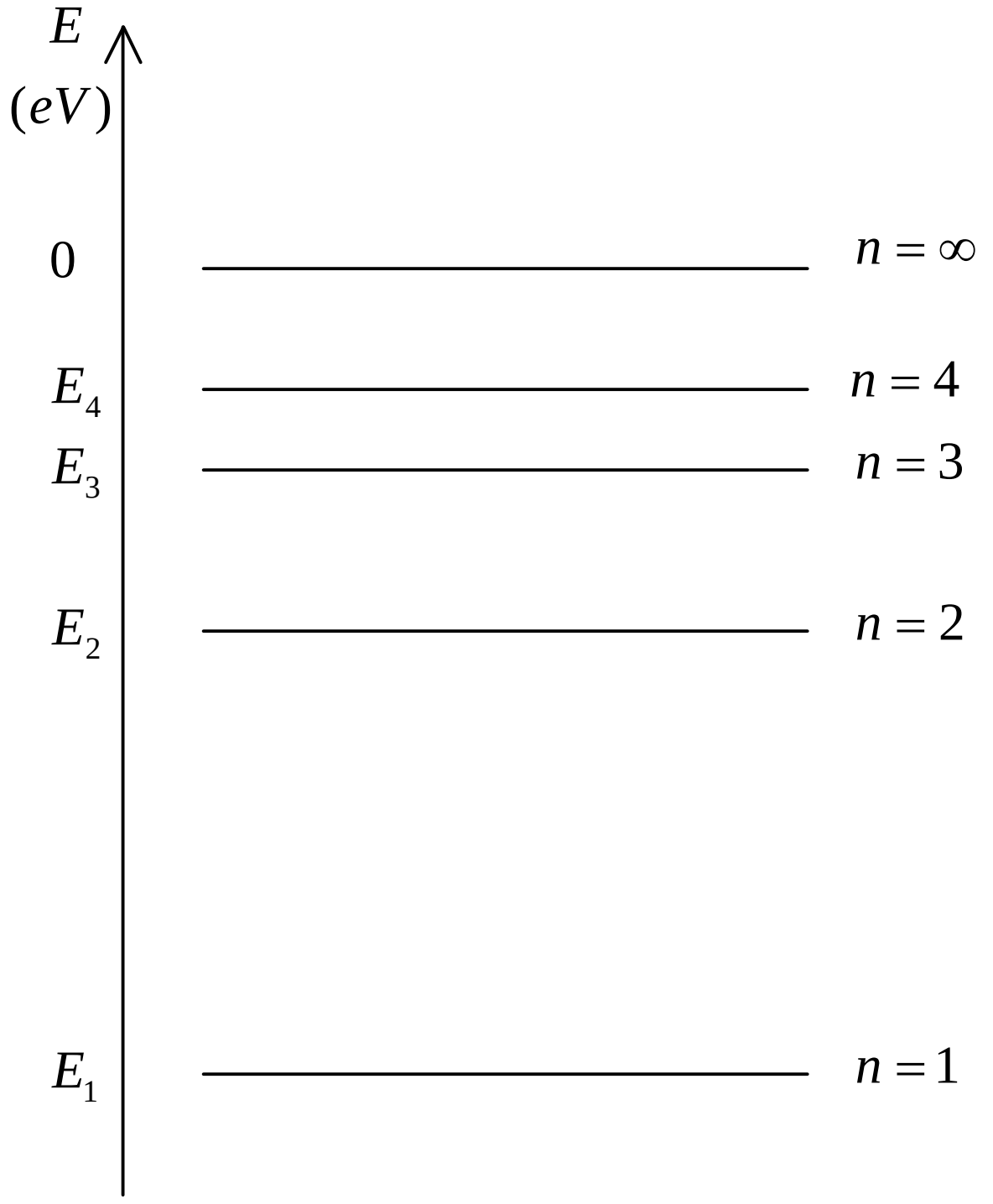

Fig. 1.2. Energy-level diagram for the hydrogen atom.

A very useful quantum concept is that of an *energy-level diagram* (Fig. 1.2). We draw a vertical axis, positively oriented upward, and we agree that its points will represent energy values expressed in *electronvolts*, *eV* ($1eV=1.6\times 10^{-19} J$). For any given value $E$ of the energy, we draw a horizontal line (*energy level*) that intersects the vertical axis at the point corresponding to that energy. In particular, the allowable energy levels for the electron in the hydrogen atom are drawn according to Eq. (1.6). With regard to its energy the electron may thus occupy *any one* of these energy levels; it will never be found, however, at an energy state that is *in between* neighboring allowable levels!

The lowest energy level $E_1$ corresponds to the *ground state* of the hydrogen atom, while the higher levels $E_2$, $E_3$, ..., represent *excited states* of the atom. We notice that the energy levels get denser as we move up the energy axis [we can explain this by evaluating the difference $\Delta E_n=E_{n+1}-E_n$ from (1.6) and by noticing that $\Delta E_n$ decreases as $n$ increases]. As $n\to\infty$, the energy $E$ tends to vary in a *continuous* manner up to the limit $E=0$ (free electron) and quantization of energy gradually disappears.

In addition to interpreting the stability of the hydrogen atom, the Bohr model is able to explain the *line spectrum* of emission or absorption of the atom, i.e., the fact that the atom selectively emits and absorbs specific frequencies of electromagnetic radiation. As we have seen, the electron is allowed to describe discrete circular orbits around the nucleus, of radii $r_1$, $r_2$, $r_3$, ..., and with corresponding energies $E_1$, $E_2$, $E_3$, ... Let us now assume that the electron makes a transition from an orbit $r_a$ *directly* to another orbit $r_b$, where $a\neq b$. Two possibilities exist: (*a*) If $a>b$, then, according to (1.6), $E_a>E_b$. The electron *falls* to a lower-energy orbit and, in the process, the atom emits *a single* photon. (*b*) If $a<b$, then $E_a<E_b$. The electron *absorbs* a photon and is *excited* to a higher-energy orbit. In any case, the frequency of the emitted or absorbed photon is

$$f=\frac{|E_a-E_b|}{h}=\frac{\kappa}{h}\left|\frac{1}{b^2}-\frac{1}{a^2}\right| \tag{1.7}$$

The wavelength of the emitted or absorbed radiation is given by the relation

$$\frac{1}{\lambda}=\frac{f}{c}=\frac{\kappa}{hc}\left|\frac{1}{b^2}-\frac{1}{a^2}\right| \tag{1.8}$$

where $c$ is the speed of light in empty (or, almost empty) space. The line spectrum of the hydrogen atom is related to the fact that both $f$ and $\lambda$ may assume discrete, rather than arbitrarily continuous, values. This, in turn, is a consequence of quantization of energy and angular momentum of the electron.

## 1.5 Multielectron Atoms

Although successful for hydrogen, the Bohr model cannot fully explain the structure of atoms having two or more electrons. The study of such atoms necessitates the use of Quantum Mechanics, a mathematically complex theory. In this theory, classical notions such as the position or the orbit of an electron are devoid of physical meaning given that, in view of the *uncertainty principle*, it is impossible to determine the cor-responding physical quantities with arbitrary precision during an experiment. Other physical quantities, such as the energy and the angular momentum of an electron, may only assume specific, discrete values; that is, they are *quantized*. These values are determined with the aid of parameters called *quantum numbers*. The set of quantum numbers that can be determined in an experimental observation of an electron represents a *quantum state* of this electron. In a sense, the state of the electron is the maximum information we can obtain about it within the limits of the uncertainty principle.

As mentioned above, a quantum state is characterized by specific (i.e., quantized) values of physical quantities such as energy and angular momentum. Mathematically, the state of an electron is expressed in the form of a *wave function* that satisfies a certain differential equation, called the *Schrödinger equation* [2,3]. This function depends on a set of parameters, which are precisely the quantum numbers we mentioned previously. The number and the possible values of the quantum numbers vary in accordance with the particular physical system to which the electron belongs.

The state of an electron in a multielectron atom is determined by a set of four quantum numbers $(n, l, m_l, m_s)$, which can take on the following values [2,3]:

$$n = 1, 2, 3, \ldots$$

$$l = 0, 1, 2, \ldots, (n-1) \qquad \text{(for a given } n\text{)}$$

$$m_l = 0, \pm 1, \pm 2, \ldots, \pm l \qquad \text{(for a given } l\text{)}$$

$$m_s = \pm \frac{1}{2}$$

The quantum number $l$ determines the magnitude of the angular momentum $\vec{L}$ of the electron, while the quantum number $m_l$ determines the direction of the angular momentum (or, to be more exact, the projection of $\vec{L}$ onto the $z$-axis). The quantum number $m_s$ determines the direction of the electron *spin* ("spin up" or "spin down"). Finally, the pair $(n, l)$ determines the energy of the electron. Analytically:

$$|\vec{L}| = \sqrt{l(l+1)}\,\hbar \,, \qquad L_z = m_l\,\hbar \qquad \left(\hbar = \frac{h}{2\pi}\right)$$

$$S_z = m_s\,\hbar = \pm\frac{\hbar}{2} \,, \qquad E = E(n,l)$$

According to the *Pauli exclusion principle,*

> *no two electrons in a quantum system (e.g., an atom, a molecule, a crystal, etc.) may be in the same quantum state, i.e., may share the same set of quantum numbers.*

Thus, for example, if two electrons in an atom have the same values of $n$, $l$ and $m_l$, then necessarily they will have different $m_s$ (+ ½ and – ½). Because of this principle, it is not possible for the entire set of electrons in an atom to occupy the lowest allowable energy level, given that this level does not have a sufficient number of quantum states to accommodate all electrons.[5]

---

[5] Exceptions to this rule are hydrogen (H, 1) and helium (He, 2).

The quantum states having the same value of $n$ (but different combinations of $l$, $m_l$ and $m_s$) constitute a *shell*. For $n$=1,2,3,4,..., the shells are denoted by $K$, $L$, $M$, $N$,..., respectively. The states of the $n$th shell having the same value of $l$ (but different combinations of $m_l$ and $m_s$) constitute the *subshell* $(n, l)$. For $l$=0,1,2,3,..., the subshells are named *s*, *p*, *d*, *f*,..., respectively. For a given value of $n$, there are $n$ different values of $l$; namely, $l$= 0,1,2,...,($n$–1). Therefore, the $n$th shell is divided into $n$ subshells.

The subshell $(n, l)$ is also indicated by the value of $n$ followed by the symbol corresponding to the value of $l$. Thus, for example, the subshell $(n, l)$=(2,1) is written 2$p$. The number of electrons in a subshell is indicated by a superscript next to the symbol of the subshell. For example, $2p^6$ indicates that there are 6 electrons in the 2$p$ subshell.

| *Shells* ($n$) | *Subshells* ($n, l$) | |
|---|---|---|
| $n$=1 | $l$=0 | $\Rightarrow$ 1$s$ |
| $n$=2 | $l$=0,1 | $\Rightarrow$ 2$s$, 2$p$ |
| $n$=3 | $l$=0,1,2 | $\Rightarrow$ 3$s$, 3$p$, 3$d$ |
| $n$=4 | $l$=0,1,2,3 | $\Rightarrow$ 4$s$, 4$p$, 4$d$, 4$f$ |

The *electronic capacity* of an atomic subshell is the maximum number of electrons this subshell can accommodate without violating the Pauli exclusion principle. The capacities of the various subshells are determined as follows: Since all electrons in a subshell occupy states with the same $l$ (and, of course, the same $n$), these states must have different combinations of $m_l$ and $m_s$. We want to find the maximum number of distinct pairs ($m_l$, $m_s$) corresponding to a given $l$. For any value of $l$, the quantum number $m_l$ can assume (2$l$+1) different values, namely, $0, \pm 1, \pm 2, ..., \pm l$. Each of these values may be combined with two values of $m_s$ (+ ½ and – ½). We thus have a total of 2(2$l$+1) different pairs ($m_l$, $m_s$) for any given $l$. In accordance with the Pauli exclusion principle, therefore, *the subshell* $(n, l)$ *may accommodate at most* 2(2$l$+1) *electrons.* Analytically:

| | | | | |
|---|---|---|---|---|
| $s$ | $\Rightarrow$ | $l$=0 | $\Rightarrow$ | 2 electrons |
| $p$ | $\Rightarrow$ | $l$=1 | $\Rightarrow$ | 6 electrons |
| $d$ | $\Rightarrow$ | $l$=2 | $\Rightarrow$ | 10 electrons |
| $f$ | $\Rightarrow$ | $l$=3 | $\Rightarrow$ | 14 electrons |

Now, the $K$ shell ($n$=1) has only the 1$s$ subshell, thus may accommodate 2 electrons, while the $L$ shell ($n$=2) has the 2$s$ and 2$p$ subshells, thus may accommodate 2+6=8 electrons. The $M$ shell ($n$=3) has the 3$s$, 3$p$ and 3$d$ subshells, thus may accommodate 2+6+10=18 electrons, etc. As a rule, however, *the last (or outermost) shell of an atom may not be occupied by more than 8 electrons*.

Given an atom with atomic number $Z$, its $Z$ electrons are distributed to the various subshells in an ordered manned, starting with 1$s$ and continuing with 2$s$, 2$p$, 3$s$, 3$p$, etc, until the entire stock of electrons has been exhausted. The *electronic configura-* *tion* of an atom (with certain small deviations, such as, e.g., in transition metals) is described by means of the following general scheme (the superscript next to the symbol of a subshell indicates the number of electrons in that subshell):

$$1s^2\ 2s^2\ 2p^6\ 3s^2\ 3p^6\ 3d^{10}\ 4s^2\ 4p^6\ ...$$

In most cases, the last (outermost) subshell is not filled up to its maximum capacity. Let us see some examples:

Sodium (Na, 11): $1s^2\ 2s^2\ 2p^6\ 3s^1$

Silicon (Si, 14): $1s^2\ 2s^2\ 2p^6\ 3s^2\ 3p^2$

Germanium (Ge, 32): $1s^2\ 2s^2\ 2p^6\ 3s^2\ 3p^6\ 3d^{10}\ 4s^2\ 4p^2$

Nickel (Ni, 28): $1s^2\ 2s^2\ 2p^6\ 3s^2\ 3p^6\ 3d^8\ 4s^2$

Notice that, in nickel, the 4*s* subshell, which belongs to the last shell, is completed before the 3*d* subshell has itself a chance to be completed. Which general rule would be violated if we transferred the two 4*s* electrons to the 3*d* subshell?

As we mentioned earlier, the energy of an electron in an atom depends on the pair of quantum numbers $(n, l)$. This means that electrons occupying states with the same values of $n$ and $l$ have the same energy. We conclude that *electrons belonging to the same subshell have the same energy.*[6] Each subshell is therefore characterized by an energy value common to all its electrons. We say that *each subshell corresponds to an electronic energy level* and that the electrons of the subshell *occupy* the corresponding energy level. An atomic *energy-level diagram* is represented as in Fig. 1.3.

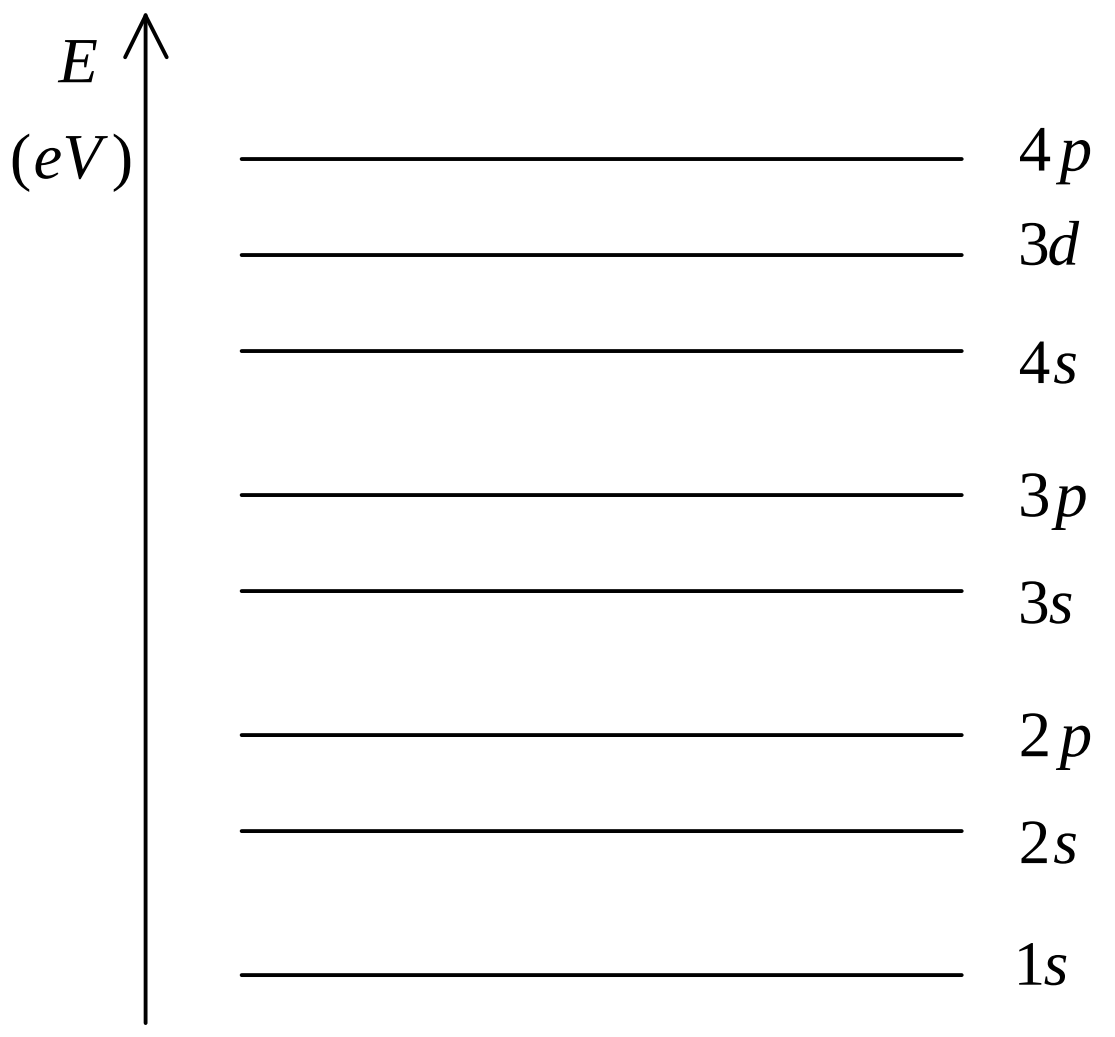

Fig. 1.3. Energy-level diagram for a multielectron atom.

[6] An exception is hydrogen, where the energy of its single electron depends only on $n$ ($E_n = -\kappa / n^2$), thus is a property related to *shells* rather than to subshells.

We note that this ordering of energy levels is not absolute and may present small deviations for certain atoms (e.g., the 4*s* level may be above the 3*d* level). Note also that the distances between successive energy levels are a characteristic property of each atom; these distances, therefore, vary from one atom to another.

The above analysis is, of course, oversimplified! Because of the complexity of the electromagnetic interactions within multielectron atoms, it is not absolutely correct to speak of energies of particular electrons but it would be more appropriate to refer to the (quantized) energy *of the atom as a whole*. Thus, in place of electronic energy levels we should draw *atomic* energy levels. The approximate model we have described assumes that every electron in an atom can move more or less independently of the remaining electrons within the electromagnetic field these electrons and the nucleus create [2,3]. We may thus define an average energy for each electron and express the energy of the atom as the sum of all individual electronic energies. The terms “atomic levels” and “molecular levels” will henceforth be understood to refer to *electronic* energy levels of atoms and molecules, respectively.

## 1.6 Molecules

Atoms often have the tendency to join together to form *molecules*. This is not done in an arbitrary manner but obeys certain logic: the ensuing quantum system must be *more stable* compared to the system of individual atoms that existed previously. This means that the merging is energetically favorable, in the sense that the potential energy of the new system (molecule) is *smaller* than that of the initial system of isolated atoms. Thus, an amount of energy must be spent in order to separate again the atoms of a molecule.

How exactly should we define a molecule? We might think of it as a group of two or more atoms held together by electromagnetic forces. But, from the moment two atoms interact, they cease to constitute autonomous, individual entities since each atom is influenced by the presence of the other; in particular, the motions and energies of the electrons in these atoms are altered. After the formation of the molecule, does it still make sense to say that an electron belongs to either atom $A$ or atom $B$? An entirely opposite view is to simply regard the molecule as a group of two or more nuclei surrounded by electrons in a manner that makes the whole structure stable.

Although one might perhaps find the latter view satisfactory, a compromise between the two views is possible by observing the following: When two atoms join to form a molecule, the electrons of their inner shells (which shells are completely filled in accordance with the exclusion principle) are not affected significantly, since the interactions they are subject to originate mainly from the individual ions to which these inner electrons belong. Those that *are* affected are the electrons of the *outermost* unfilled shells, i.e., the *valence electrons*, which are subject to the influence of both ions, so that it is no longer possible to absolutely determine to which atom each of these electrons belongs. To those outermost electrons the molecule owes the chemical bond that holds the two atoms together.

Let us now assume that $N$ similar atoms merge to form an *N-atomic molecule* (e.g., the diatomic molecules $H_2$ and $O_2$ for $N$=2, the triatomic molecule $O_3$ for $N$=3, etc.).

This molecule is a novel quantum system having a structure that is different from that of the initial atoms. In particular, *the molecule possesses at least N times as many electronic energy levels as those possessed by the individual atoms*. This is logical in view of the fact that the molecular energy levels must accommodate $N$ times as many electrons in comparison to the atomic levels. Now, as we saw in the previous section, the electronic energy levels of an atom correspond to the subshells 1*s*, 2*s*, 2*p*, 3*s*, 3*p*, etc. When $N$ similar atoms coalesce to form an $N$-atomic molecule, *each atomic energy level gives rise to N or more molecular levels*. We say that, in the process of formation of the molecule, each atomic energy level *splits* into $N$ or more molecular levels (in an energy-level diagram these levels are closely spaced). The qualitative diagram of Fig. 1.4 concerns the *diatomic* molecule ($N$=2).

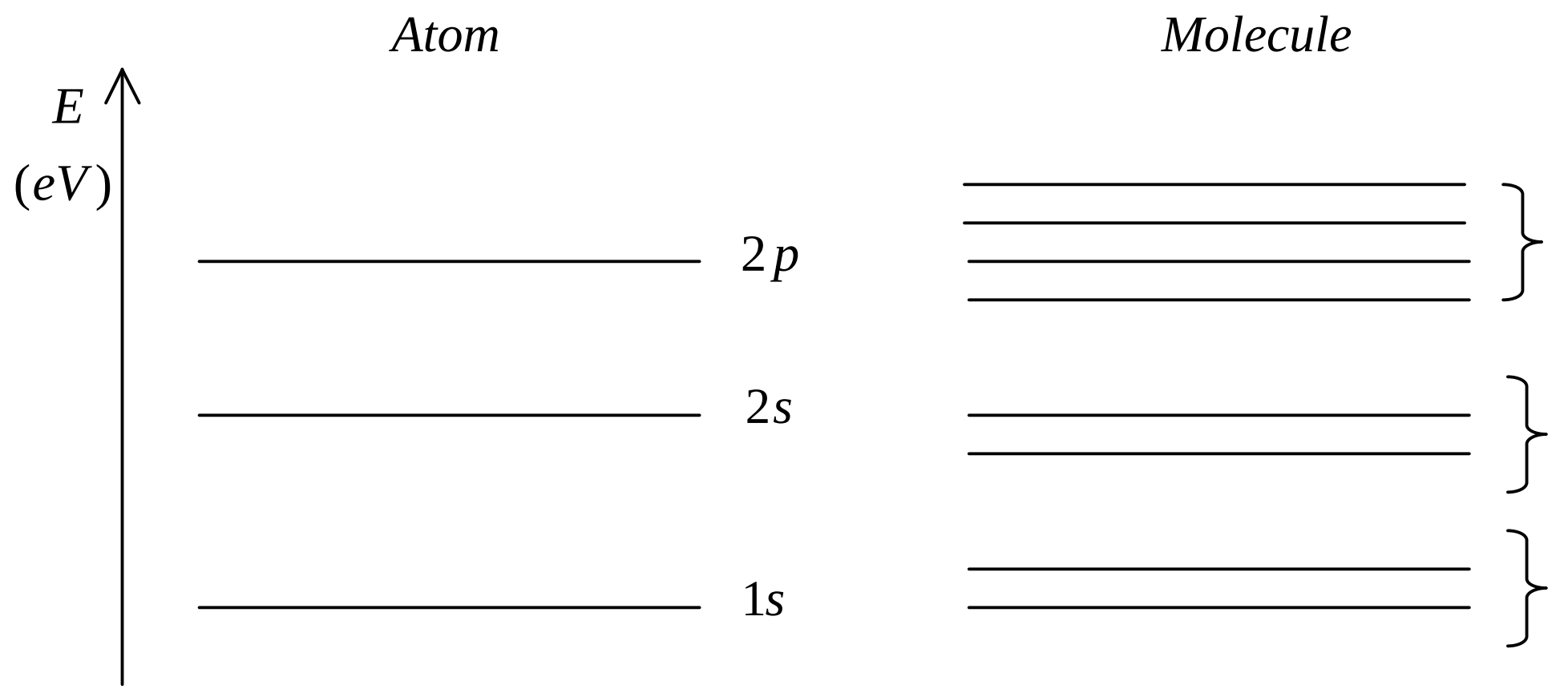

Fig. 1.4. Energy-level diagrams for an atom (left) and the corresponding diatomic molecule (right).

## 1.7 Energy Bands of Crystalline Solids

As mentioned earlier, the stability of several solid substances is due to their crystalline structure. A *crystal* is a regular array of atoms (or molecules, or ions) built by canonical repetition of a fundamental structural unit in three dimensions. The reason for the formation of crystals is basically the same as that for molecules: the result is energetically favorable. That is, the potential energy of the crystal is lower than that of the system of the constituent atoms, had they been isolated from each other.

In crystals, like in molecules, the electrons in the atoms can be distinguished into inner-shell electrons and outermost, or *valence*, electrons. The inner electrons are closer to the corresponding atomic nuclei, interacting with them relatively strongly and not being influenced appreciably by the neighboring atoms (or ions) in the crystal. Thus, to a large extent, these electrons retain the properties they would have in isolated atoms (e.g., their energies are almost the same and their quantum states are described by the same quantum numbers). On the other hand, the outermost electrons are influenced much by the neighboring atoms and thus their properties (in particular, their energies) differ considerably from those in the corresponding isolated atoms. One could say that, in a sense, these latter electrons belong to the whole crystal, not to individual atoms. The valence electrons are responsible for the bonding between the atoms, as well as for most physical properties of the crystal (such as its electrical and its thermal conductivity).

Let us consider a crystal composed of *N* similar atoms (or molecules, or ions). You can imagine it as a gigantic "molecule" containing a huge number of atoms arranged regularly in space. According to our discussion in Sec. 1.6, as the crystal is formed, each atomic energy level splits into *N* (or more) closely spaced levels. The positions of these levels and the spacing between them in an energy diagram depend on the distance between the atoms in the crystal, as shown in Fig. 1.5.

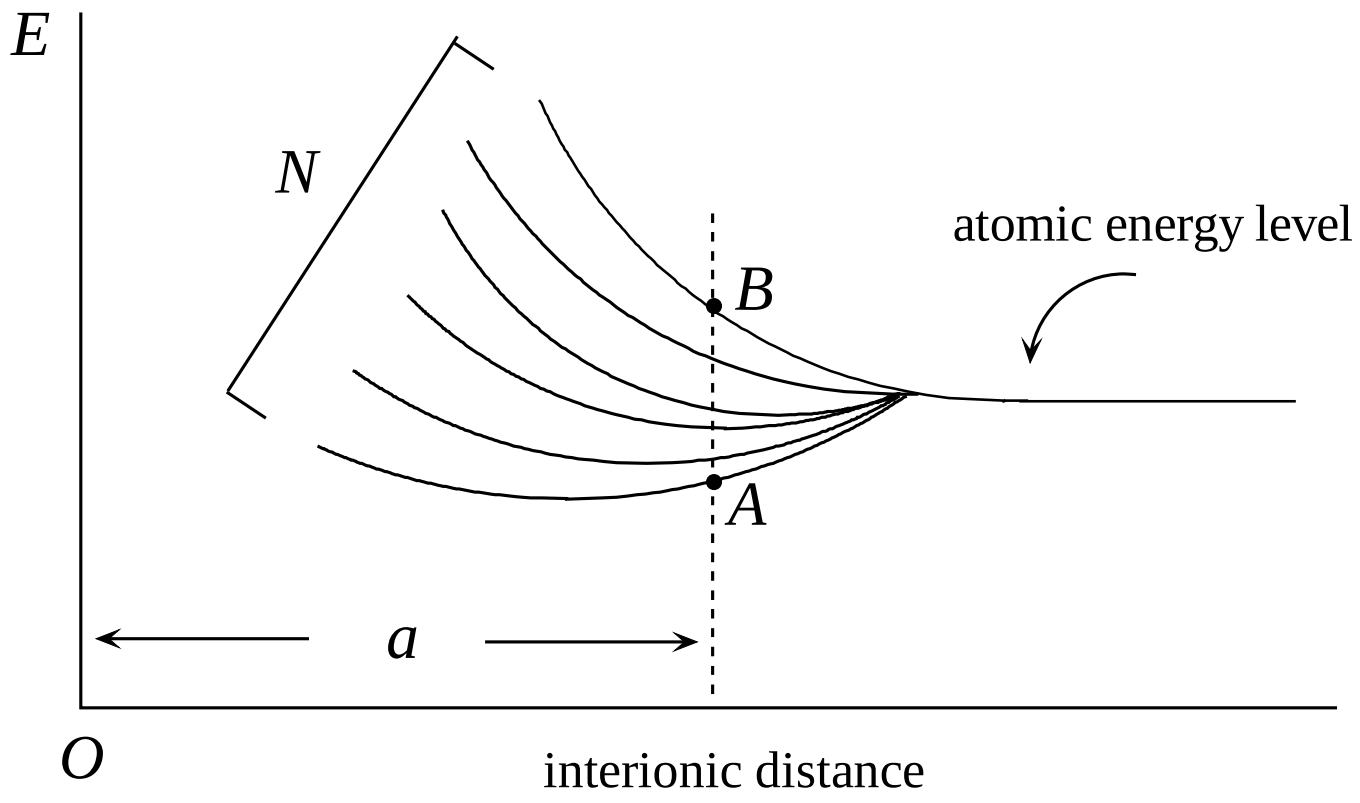

Fig. 1.5. Splitting of an atomic energy level into *N* crystalline levels.

The horizontal axis in the diagram of Fig. 1.5 represents the distance between two neighboring atoms in the crystal, while the vertical axis represents the energy of any of the *N* crystalline levels. The diagram can be interpreted as follows: Imagine that, initially, the *N* atoms are largely separated, so that they do not interact with each other (isolated atoms). We consider a specific electronic energy level, common to all atoms. As the atoms get closer in order to form a crystal structure, they begin to interact. From that moment on we are not dealing with a set of isolated atoms but rather with a novel quantum system, the crystal. In particular, the valence electrons of the atoms are no longer associated with specific atoms but constitute a unified system of electrons belonging to the whole crystal. At the same time, the considered atomic energy level splits into at least *N* crystalline levels, which determine the possible values of the energy of the electrons. Each of these levels corresponds to a value of the energy that is not fixed but varies with the distance between the ions. We can thus draw *N* curves, one for each level. Notice that, for a given interionic distance *a*, the energies of the allowable crystalline levels range from *A* to *B* in the diagram of Fig. 1.5. The energy distance *AB* varies with the distance between the atoms.

Now, the number *N* of atoms in crystals is very large, close to $10^{23}$ atoms $/cm^3$. Thus the *N* energy levels are so close to each other that it is impossible to distinguish one from another. We say that these levels form a continuous *energy band*. The *width* of a band (the energy distance *AB* in the diagram of Fig. 1.5) varies with the distance between the atoms. Figure 1.6 shows the energy band corresponding to a given atomic energy level, for a given distance $a$ between the atoms (or ions) in the crystal.

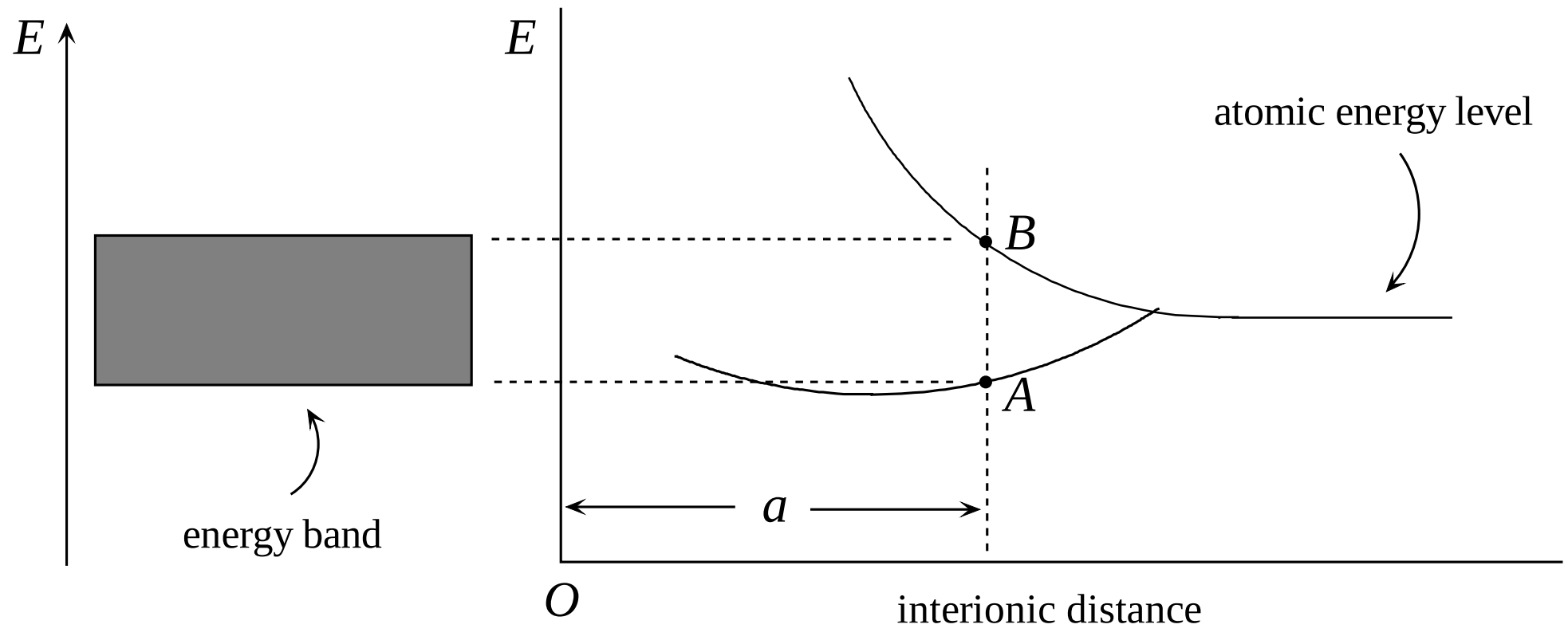

Fig. 1.6. Formation of an energy band in a crystal.

Generally speaking, one may say that each atomic energy level gives rise to a corresponding crystalline energy band, as Fig. 1.7 shows. This simple picture, however, is not absolutely correct with regard to the upper bands, occupied by the outermost (valence) electrons of the atoms.

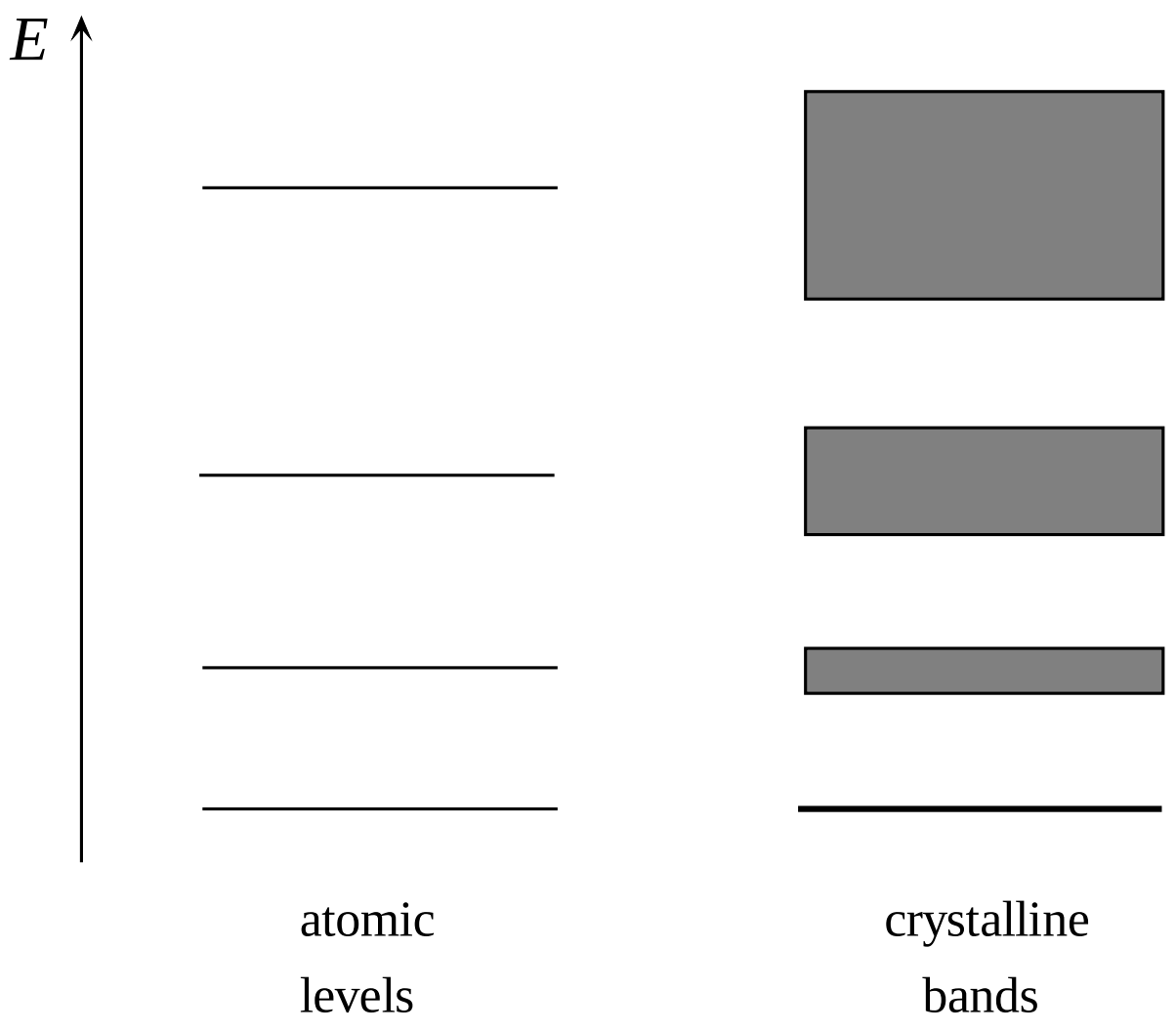

Fig. 1.7. Atomic energy levels (left) and the corresponding crystalline bands (right).

We observe that *the width of the bands increases as we move from the lowest to the highest band* (the lowest band does not differ much from an ordinary energy level). This can be explained as follows: Higher bands correspond to higher atomic energy levels which, in turn, correspond to outer atomic subshells. On the other hand, lower bands correspond to lower atomic energy levels, hence to inner atomic subshells. This means that the higher bands are occupied by the outer-shell electrons of the atoms, while the lower bands are occupied by the inner-shell electrons. As we have already mentioned, the inner electrons are mostly affected by the corresponding nuclei and are hardly aware of the presence of the neighboring atoms (or ions) in the crystal. Thus their energies do not differ significantly from those in the isolated atoms, with the effect that the energy bands these electrons occupy do not differ much in width from the corresponding atomic energy levels. On the other hand, the outer electrons (in particu-lar, the valence electrons) interact more strongly with the neighboring atoms and their energies are modified considerably, with the effect that the atomic energy levels to which these electrons belong are widened into energy bands during the formation of the crystal, the widening increasing as we move up the energy axis.

It should be noted that *the width of each band is independent of the size of the crystal, that is, of the total number of atoms in the crystal lattice*. Indeed, this width depends only on the *concentration* (density) of the atoms, i.e., the number of atoms *per unit volume*. This suggests that an atom in the crystal is influenced only by its closest neighbors, while it is practically unaffected by more distant atoms in the lattice.

We recall that each atomic energy level (which corresponds to an atomic subshell) can accommodate a specific number of electrons limited by the Pauli exclusion principle. Similarly, each crystalline energy band can accommodate a definite number of electrons, depending on the number of available quantum states in the band. As we know, the exclusion principle does not allow two or more electrons of the system to occupy the same state. This means that different electrons in the crystal cannot possess exactly the same set of quantum numbers. Therefore, the number of electrons in an energy band cannot exceed the number of quantum states in the band, each state corresponding to a given set of allowable quantum numbers.

The filling of energy bands with electrons follows the same logic as the filling of atomic energy levels (or, correspondingly, of atomic subshells). Thus, the lowest band is filled first, then the band immediately above it, etc., until the full stock of electrons is exhausted. In a non-excited state of the crystal, there is an uppermost band that contains electrons. This energy band is occupied by the valence electrons of the atoms and every allowable band above it is empty. Two possibilities exist:

(*a*) If the uppermost-occupied band is *not completely filled*, it is called the *conduction band* (Fig. 1.8).

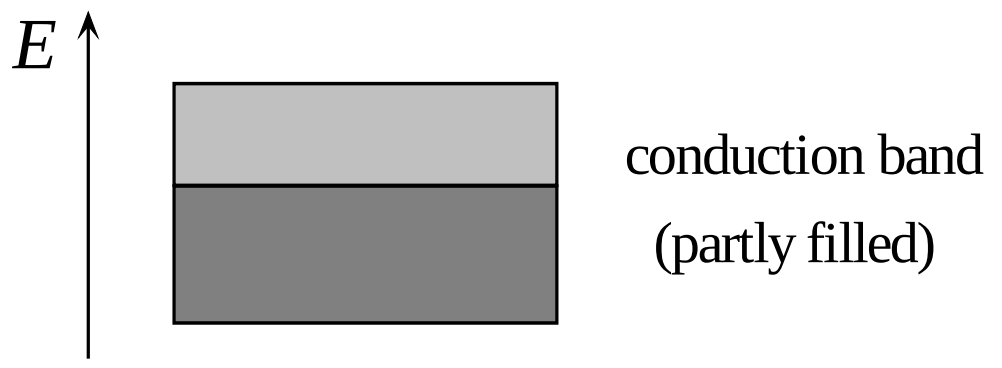

Fig. 1.8. Conduction band as an uppermost-occupied band.

(*b*) If the uppermost-occupied band is *full*, it is called the *valence band*, and the *empty* band just above it is then called the *conduction band* (Fig. 1.9). The energy region between the valence and the conduction band contains no allowable quantum states and constitutes the *forbidden band*. The energy width $E_G$ of the forbidden band is called the *energy gap*.

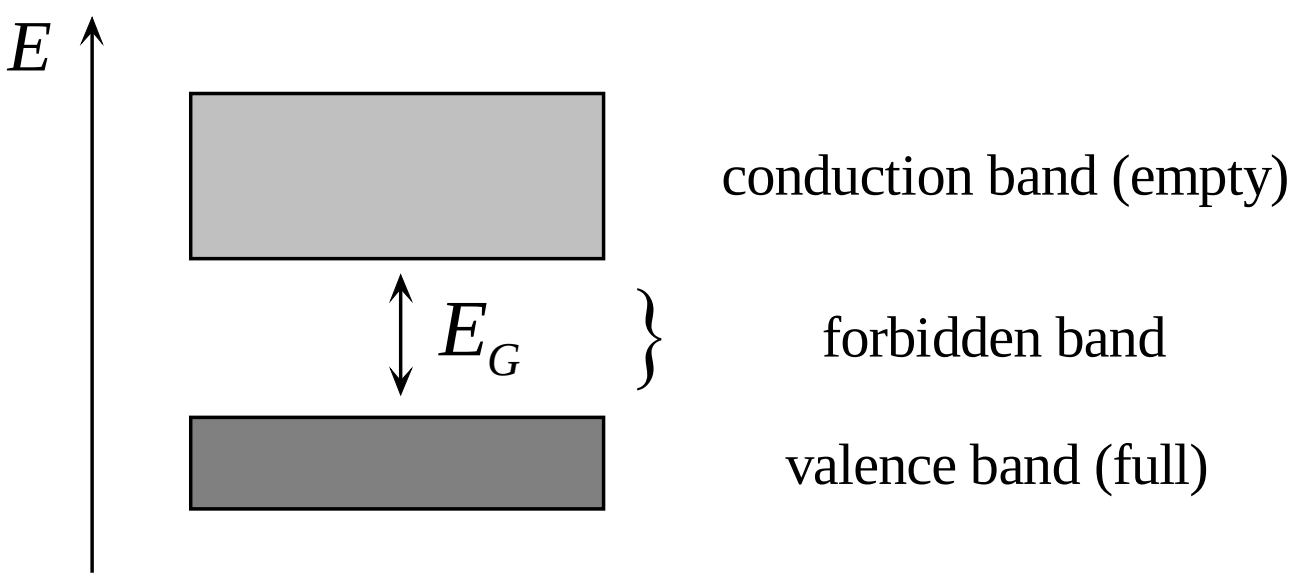

Fig. 1.9. Valence, conduction, and forbidden bands; the energy gap is $E_G$.

Note that in case (*b*) the conduction band is an *allowable* energy region for the electrons, despite the fact that this band is empty in the fundamental (non-excited) state of the crystal. Indeed, if given sufficient energy, some electrons of the valence band may be excited into the conduction band. On the contrary, *no electron may be excited to the forbidden band*, given that this energy region does not contain allowable electronic energy levels and quantum states. The energy gap $E_G$ represents the least energy required to excite an electron from the valence band to the conduction band. We will explain the physical significance of this process in the next chapter.

## 1.8 Band Formation in Tetravalent Crystals

To understand the properties of insulators and semiconductors, it is useful to examine the band formation in *tetravalent crystals*, i.e., crystals composed of similar tetravalent atoms. Being covalent structures (cf. Sec. 1.2) these crystals are not expected, under normal conditions, to exhibit substantial electrical and thermal conductivities since they do not possess a significant number of free electrons.

The last (outermost) shell of an atom of a tetravalent element is of the form

$$n s^2 \, n p^2 \quad (n = 2, 3, 4, \ldots)$$

*filled subshell* (pointing to $ns^2$)    *unfilled subshell* (pointing to $np^2$)

Thus, for $n = 2, 3, 4,$ respectively, we have the following configurations:

Carbon (C, 6): $1s^2\, 2s^2\, 2p^2$

Silicon (Si, 14): $1s^2\, 2s^2\, 2p^6\, 3s^2\, 3p^2$

Germanium (Ge, 32): $1s^2\, 2s^2\, 2p^6\, 3s^2\, 3p^6\, 3d^{10}\, 4s^2\, 4p^2$

The uppermost-occupied atomic energy level ($n p$) is not full since the corresponding subshell has only 2 electrons out of its total capacity of 6 electrons. One would expect that, by analogy, the uppermost-occupied crystalline energy band be also unfilled, since that band results from the widening of a partly occupied energy level. Things are not quite that simple, however! The fact that these crystals are not good electrical conductors indicates that their uppermost-occupied bands are *completely filled*; that is, each of these bands is a *valence band* (we will explain this in more detail in the next chapter). How does this happen?

Let us imagine that $N$ atoms of a tetravalent element (where $N \sim 10^{23}$), which atoms are initially infinitely separated from each other, begin to come closer in order to form a crystal [1,4]. Once the atoms are close enough to interact with each other, the atomic energy levels ($n\,s$) and ($n\,p$) begin to widen into energy bands. The band ($n\,s$) is full while the band ($n\,p$) is partly filled. As the atoms get closer, these bands are widened so much that they overlap, forming a single band. Finally, when the atoms reach the appropriate interionic distance for the formation of the crystal, the aforementioned single band splits again, this time into two *new* bands with different capacities from the initial ($n\,s$) and ($n\,p$) bands. All valence electrons of the atoms in the crystal (4 electrons per atom) are accommodated in the lower band, which is now a *completely filled* valence band, while the upper, *empty* band is a conduction band.

## References for Chapter 1

## QUESTIONS

**1.** Show that the energy distance between successive energy levels of the hydrogen atom decreases as we move up the energy-level diagram.

**2.** Evaluate the frequency and the wavelength of the radiation emitted by a hydrogen atom during a transition of its electron from the orbit $n$ to the orbit $(n–1)$, where $n>1$.

**3.** (*a*) Show that the radius $25a_0$ (where $a_0$ is the Bohr radius) corresponds to an allowable orbit for the electron in a hydrogen atom, and evaluate the angular momentum and the energy of the electron in that orbit. (*b*) At what distance from the nucleus does the electron have an angular momentum equal to $3h/\pi$ ?

**4.** Find the possible frequencies of radiation emitted by an excited hydrogen atom when the electron, initially moving on the third Bohr orbit, finally – albeit not necessarily directly – returns to the fundamental first orbit.

**5.** A hydrogen atom is excited by absorbing a photon; it then returns in two steps to its fundamental state, emitting two photons of wavelengths $\lambda_1$ and $\lambda_2$ . Find the wavelength $\lambda$ of the photon that was absorbed by the atom.

**6.** A moving electron hits a hydrogen atom and excites it from the fundamental to the third energy level. The atom then returns in two steps to its initial state, emitting two photons. (*a*) Find the minimum speed of the electron that hit the atom. (*b*) Find the frequencies of the two photons.

**7.** Ionization of a hydrogen atom can be produced by collision with a moving electron or by absorption of a photon. (*a*) What must be the minimum speed of the electron? (*b*) What must be the maximum wavelength of the photon?

**8.** A photon hits a hydrogen atom and causes ionization to the latter. The liberated electron then falls onto another hydrogen atom and excites it from the fundamental to the immediately higher energy level. Find the minimum frequency of the photon that hit the first hydrogen atom.

**9.** As we know, the lowest energy level of a many-electron atom is the 1*s* level (i.e., the energy level corresponding to the 1*s* subshell). Why then doesn't the totality of electrons in the atom occupy this particular level?

**10.** Draw a comparative energy-level diagram (not necessarily an exact one!) for the electrons of the oxygen (O) atom and those of the ozone ($O_3$) molecule.

**11.** Why do crystals have energy bands instead of energy levels like atoms and molecules? Why are higher-energy bands wider than lower-energy bands?

**12.** A band in a crystal has its full quota of electrons, while another band of the same crystal is partly filled. Which band is wider? Explain.

**13.** Consider two diamond crystals, a small one and a big one. Compare the widths of corresponding energy bands of the two crystals.

**14.** In what ways do the free electrons affect the stability and the physical properties of metals?

**15.** Give examples that demonstrate that the correspondence between the energy levels in each of a set of identical atoms, and the energy bands of a crystal composed from these atoms, is not perfect.

# CHAPTER 2

# ELECTRICAL CONDUCTIVITY OF SOLIDS

## 2.1 Introduction

Electrical conductivity is one of the most important properties of solids. On the basis of it we distinguish among three types of solids: *conductors* (or *metals*), *insulators*, and *semiconductors*. The last ones have intermediate conductivity compared to insulators and conductors. In contrast to metals, whose conductivity *decreases* with temperature, the conductivity of semiconductors *increases* with temperature.

As we will see below, the electrical conductivity of a crystalline solid is intimately related to the formation of the energy bands of the solid, or, more specifically, those bands occupied by the valence electrons of the atoms in the crystal. As a rule,

> *a band completely filled with electrons (not having, that is, unoccupied quantum states) does not contribute to conductivity, in contrast to a band that is partly filled.*

Thus, metals are characterized by the presence of partly occupied bands, while in insulators all occupied bands are completely filled. At relatively low temperatures, the electrical behavior of pure semiconductors is similar to that of insulators, since at these temperatures all bands of a semiconductor are full. The electrical conductivity of a semiconductor can be increased significantly, however, either by raising the temperature or by doping the material with suitable impurity atoms.

## 2.2 Conductors and Insulators

*Metals* (or *conductors*) are characterized by a *partly occupied* uppermost energy band; i.e., their highest occupied band is a *conduction band* (Fig. 2.1).

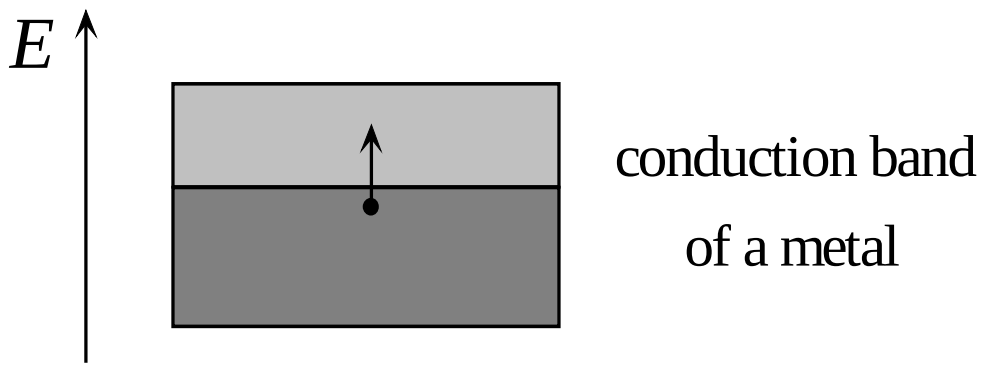

Fig. 2.1. Conduction band of a metal.

In essence, the conduction band is the totality of possible energy values assumed by the valence electrons of the atoms in the crystal; in other words, it is the allowable energy region for the valence electrons. These electrons have been freed from the atoms to which they belonged, those atoms thus left as *positive ions*. (The ionization process utilizes part of the energy liberated during the formation of the crystal; no additional external energy is therefore needed for this purpose.) The detached valence electrons move more or less freely in between the positive ions of the crystal lattice without be-ing affected appreciably by the ions; for this reason they are called *free electrons*. To these electrons the metal owes its electrical conductivity as well as a significant part of its thermal conductivity (another part is due to the vibrations of the ions that compose the lattice). Even more importantly, the presence of the free electrons guarantees the stability of the lattice (without them, the repulsive forces between the positive ions would make the crystal disintegrate!) [1].

The electrical conductivity of a metal can be described as follows: In the presence of an electric field in the interior of the metal, the free electrons gain energy and begin to accelerate. Part of this energy is lost, of course, due to collisions of the electrons with the ions. Finally, however, the electrons acquire a constant average velocity in a direction opposite to the electric field. This oriented motion of the electrons constitutes an *electric current*.

From a somewhat different viewpoint, the electrical conductivity of a metal is due to the fact that, under the influence of an electric field, the electrons of the uppermost-occupied energy levels in the conduction band are able to jump to higher *vacant* levels in this band without violating the Pauli exclusion principle (Sec. 1.5). The energy increase of the electrons is related to their acceleration by the electric field and is macroscopically perceived as an electric current. Notice that, if the energy band of the valence electrons were fully occupied, an excitation of these electrons to higher energy levels within the band would be impossible since all such levels would already be occupied and any extra electrons onto them would violate the Pauli principle.

The above analysis also explains why metals are *opaque* to visible light. This happens because the free electrons may absorb photons in the visible region of the spectrum of electromagnetic radiation and be excited to one of the many vacant higher-energy levels within the conduction band.

As examples of conductors, let us examine the cases of sodium and magnesium:

Sodium (Na, 11): $1s^2\ 2s^2\ 2p^6\ 3s^1$

Magnesium (Mg, 12): $1s^2\ 2s^2\ 2p^6\ 3s^2\ (3p^0)$

Sodium is a conductor since, in the process of the formation of the crystal, the half-filled atomic energy level 3*s* is broadened into a correspondingly partially filled conduction band. In magnesium, on the other hand, there is an *overlapping* of the bands that result from the broadening of the 3*s* atomic level (which is completely filled) and the 3*p* level (which is empty), the result being a partly occupied conduction band.[1]

At the other extreme, *insulators* are substances devoid of appreciable conductivity. A typical example is diamond, the crystalline structure of which is built with carbon atoms (C) tied to each other by covalent bonds of great strength. Since carbon is a tetravalent element, the diamond crystal possesses a *fully occupied valence band* and an *empty conduction band* (see Sec. 1.8), as seen in Fig. 2.2.

---

[1] In fact, a band overlapping of this sort occurs in all metals, even if the uppermost-occupied atomic energy level is not full.

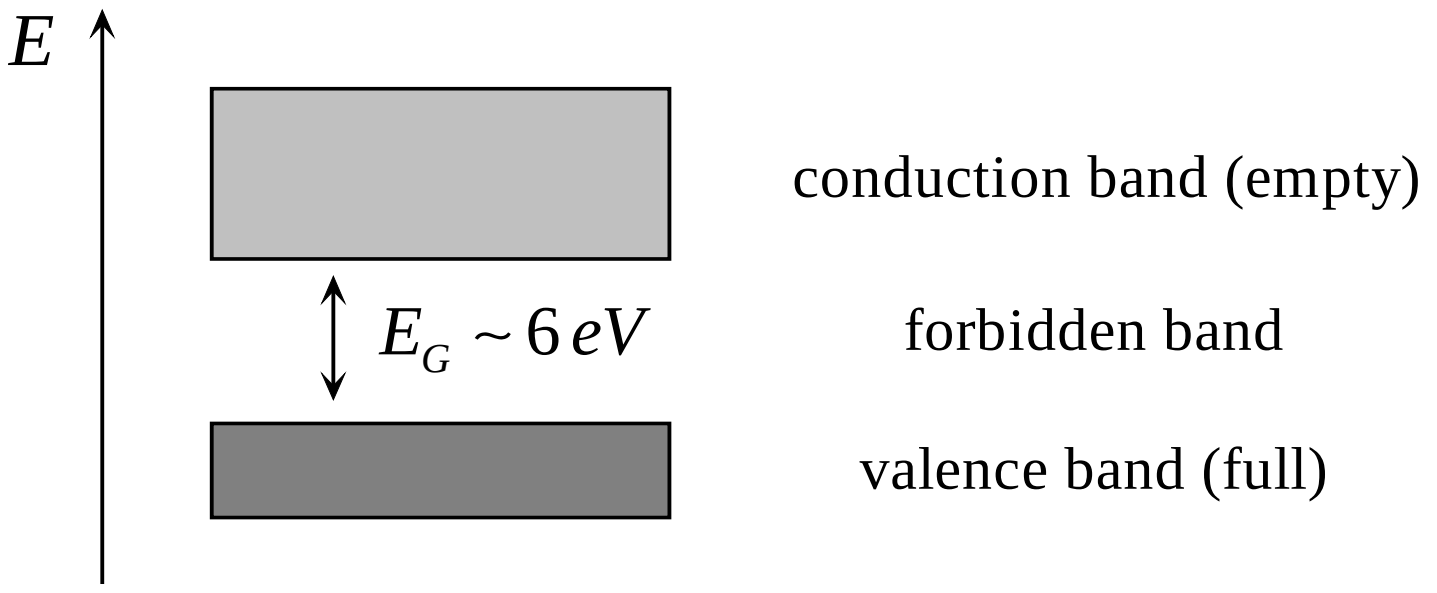

Fig. 2.2. Band structure of an insulator.

The valence band exactly accommodates the totality of valence electrons of the atoms in the crystal (4 electrons per atom). The *energy gap* $E_G$ represents the minimum energy needed in order for an electron of the valence band to be excited to the conduction band.[2] Physically, $E_G$ is the least energy required for "breaking" a covalent bond in the crystal and releasing a valence electron. The liberated electron will now belong to the energy region of the conduction band. We stress that under no circumstances may an electron be excited to the forbidden band, since that energy region contains no allowable energy levels and corresponding quantum states.

Given that all states in the valence band are occupied, it would be impossible to accelerate a valence electron by means of an electric field so as to generate an electric current. Indeed, an acceleration of the electron would result in an increase of its energy, thus an excitation of the electron to a higher energy level within the valence band. This, however, would violate the Pauli exclusion principle since all levels in this band are already occupied. The only allowable excitation of the valence electron is to the empty conduction band, a process that physically corresponds to the breaking of a covalent bond with a simultaneous liberation of an electron. This demands an energy $E \geq E_G$ , where $E_G$ is of the order of 6 $eV$ for diamond. Energy of this magnitude cannot be supplied by an electric field of typical strength. For this reason, diamond has minimal (practically zero) electrical conductivity; i.e., it is an insulator.

Because of the large value of its energy gap $E_G$ , diamond cannot absorb photons in the visible region of the spectrum, which photons have energies $1.5 - 3\ eV < E_G$ (the absorption of such a photon would excite an electron of the valence band into the forbidden band!). This explains the *transparency* of the diamond crystal. (The deep blue color of some diamonds is due to the presence of boron atoms within the crystal structure.)

## 2.3 Semiconductors

Compared to metals and insulators, *semiconductors* have intermediate conductivity. Typical examples are *silicon* (Si) and *germanium* (Ge). Like diamond, they are covalent solids composed of tetravalent atoms. Thus, at their fundamental (i.e., non-excited) state they possess a fully occupied valence band (Fig. 2.3). But, in contrast to diamond, the energy gap $E_G$ in semiconductors is relatively small, of the order of 1 $eV$.

---

[2] Note carefully that we are talking here about a move of the electron on the energy diagram, not in space!

Silicon (Si, 14): $1s^2\,2s^2\,2p^6\,3s^2\,3p^2$ ($E_G = 1.21\,eV$)

Germanium (Ge, 32): $1s^2\,2s^2\,2p^6\,3s^2\,3p^6\,3d^{10}\,4s^2\,4p^2$ ($E_G = 0.78\,eV$)

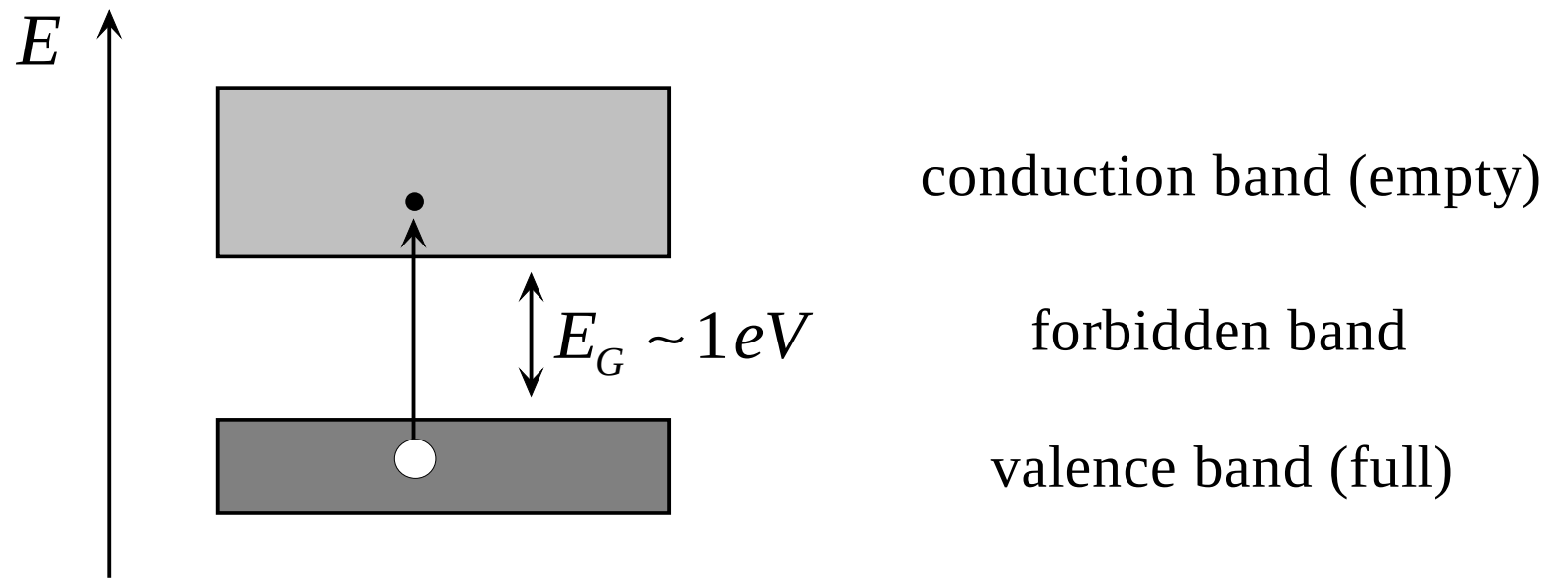

Fig. 2.3. Band structure of a semiconductor.

The valence band contains the valence electrons (4 per atom) of the atoms in the crystal. At very low temperatures (close to absolute zero, $T \sim 0\,K$) this band is fully occupied, which physically means that all valence electrons participate in covalent bonds and there are no free electrons in the crystal (the conduction band is empty). Thus, at those temperatures semiconductors behave like insulators. At higher temperatures, however (e.g., at room temperature, $T \sim 300\,K$) the electrons in the uppermost energy levels in the valence band receive sufficient thermal energy to jump over the relatively small energy gap $E_G$ and be excited to the conduction band.

Physically, the energy gap $E_G$ represents the least energy needed for the breaking of a covalent bond and the liberation of a valence electron. Under the influence of an electric field the freed electrons then move in an oriented way, like the free electrons in metals, having energies in the region of the conduction band.

With regard to their optical properties, semiconductors are *opaque* due to the fact that their energy gap is smaller than the energies of the photons in the visible region of the spectrum ($E_G < 1.5 - 3\,eV$). Electrons in the highest levels of the valence band can thus absorb photons in that spectral region and be excited to the conduction band; in other words, they are able to “escape” from the covalent bonds to which they belong and become free electrons. This results in an increase of conductivity of the semiconductor, an effect called *photoconductivity*.

Every liberated valence electron leaves behind an *incomplete* (“broken”) covalent bond, which constitutes a *hole* in the crystal of the semiconductor. Equivalently, every electron of the valence band that is excited to the conduction band leaves a vacant quantum state in the valence band, also called a *hole* (Fig. 2.3). The hole behaves like a *positively* charged particle since it is created by the *absence* of an electron from a previously electrically neutral atom. Moreover, as we will see below, a transport of the hole is possible within the crystal, the hole thus contributing to the conductivity of the semiconductor. In Fig. 2.4 we see a simplified, two-dimensional representation of the crystal lattice of a semiconductor (only the 4 valence electrons of each atom are shown). Each atom forms 4 covalent bonds with its 4 closest neighbors.

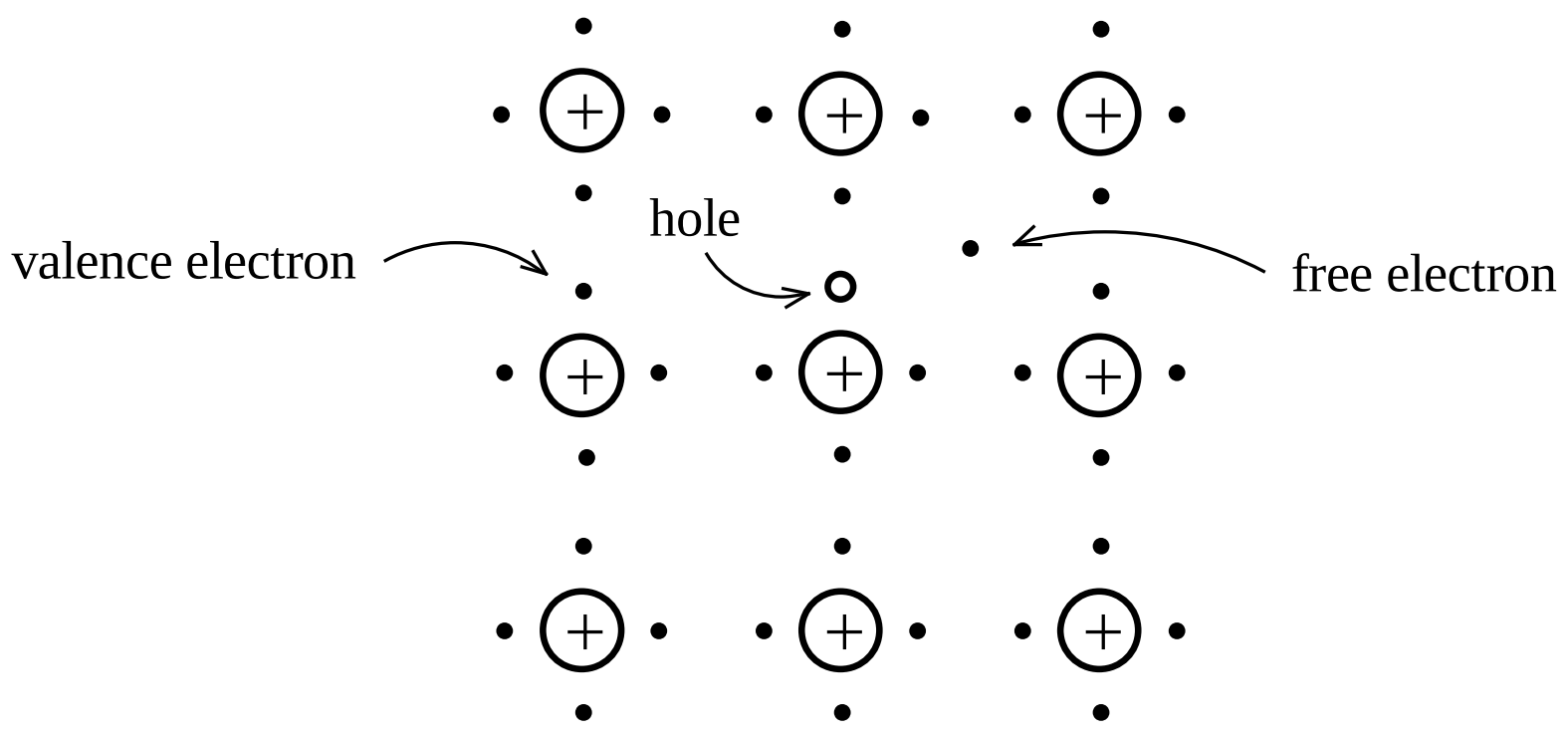

Fig. 2.4. Simplified, two-dimensional picture of the crystal lattice of a semiconductor.

Let us review some important physical concepts regarding semiconductors:

*Valence band:* It is the energy region formed by the totality of energy levels occupied by the *bound* valence electrons of the atoms in the crystal (the electrons that participate in covalent bonds).

*Conduction band:* It is the energy region allowable for the *free* electrons (those that have broken the covalent bonds to which they belonged).

*Energy gap* ($E_G$): It is the least energy required for the breaking of a covalent bond and the liberation of a valence electron.

*Hole:* It is an incomplete (broken) covalent bond in the crystal, corresponding to a vacant quantum state in the valence band.

Let us now examine the way in which a hole contributes to the conductivity of the semiconductor. When an incomplete bond exists at some location of the crystal lattice, it is relatively easy for a valence electron of a neighboring atom, under the action of an electric field, to abandon its own bond and cover that hole. (Given that the electron remains in the valence band, no energy amount $E_G$ is now needed.) This electron leaves behind a new hole (an incomplete bond). It is thus as if the initial hole were transferred to a new location, moving *opposite* to the direction of motion of the valence electron which left its bond to cover the initial hole. The new hole may, in turn, be covered by a valence electron of another neighboring atom, this process resulting in a further transport of the initial hole in a direction opposite to that of the valence electrons, and so forth. Given that the hole is in essence the *absence* of a (negatively charged) electron, we can regard it as equivalent to a *positively* charged particle of charge equal in magnitude to the charge of the electron. In conclusion:

- Holes can be regarded as "real", positively charged particles whose direction of motion is *opposite* to that of the valence electrons when the latter, under the action of an electric field, leave the covalent bonds to which they belong in order to cover neighboring incomplete bonds.

- The conductivity of a semiconductor is due to the motion of both the *free* electrons and the holes.

- The free electrons have energies in the region of the conduction band, while the holes belong to the energy region of the valence band.

It often happens that a *free* electron covers an incomplete bond. This process of *recombination of an electron-hole pair* corresponds to the transition of an electron from the conduction band to the valence band, resulting in the occupation of a vacant state (a hole) in the valence band by that electron. The total number of both (free) electrons and holes thus reduces by one unit.

## 2.4 Ohm's Law for Metals[3]

As we have already mentioned, the valence electrons of the atoms of a metal are easily detached from the atoms to which they belong (by using part of the energy liberated during the formation of the crystal), becoming *free electrons* with energies in the region of the conduction band. Their characterization as "free" indicates that these electrons are not subject to forces of any appreciable strength as they move within the crystal lattice, provided of course that they do not come too close to the ions of the lattice. The motion of the electrons is only disturbed by their occasional collisions with the ions, which results in deceleration or change of direction of motion of the electrons.

When there is no applied electric field in the interior of the metal[4] ($\vec{E} = 0$), the motion of the free electrons is random and in all directions so that, macroscopically, no electric current exists within the metal. The situation changes, however, if there is an electric field $\vec{E} \neq 0$ inside the metal. The field then exerts forces on the free electrons, compelling them to accelerate. The speed of the electrons would increase indefinitely (which fact would certainly infuriate Einstein!) if collisions with ions did not occur. Due to these collisions, the electrons lose part of their kinetic energy (which is absorbed by the crystal lattice producing Joule heating of the metal) until the electrons finally attain a constant average velocity $\vec{\upsilon}$ (*drift velocity*).

Since electrons are negatively charged, their direction of motion is *opposite* to $\vec{E}$. As we will see in Chapter 6, however, the motion of a negative charge in some direction is equivalent to the motion of a *positive* charge of equal magnitude, in the *opposite* direction. (For example, the two charges produce the same magnetic field and are subject to the same force by an external magnetic field.) We may thus *conventionally* assume that the mobile charges are *positive*, equal to $+q$, and their direction of motion is the *opposite* of the actual direction of motion of the electrons. Hence, conventionally, the drift velocity $\vec{\upsilon}$ of the charges will be assumed to have the *same* direction as the electric field $\vec{E}$. According to theoretical calculations [1-5] and consistently with experimental observations, for relatively small values of $\vec{E}$ the drift velocity is proportional to the field strength:

$$\vec{\upsilon} = \mu \vec{E} \tag{2.1}$$

The coefficient $\mu$ is called the *mobility* of the electron in the considered metal. (As we will see in Sec. 2.6, the coefficient $\mu$ is temperature-dependent.) This oriented motion of the electrons constitutes an *electric current*.

---

[3] See also Appendix C.

[4] When there is no risk of confusion, we will use the symbol $E$ for either the energy or the electric field strength.

We now consider an elementary section of a conducting material, in the shape of a thin wire of infinitesimal length *dl* and cross-sectional area *S* (Fig. 2.5). The volume of the wire is $dv = S dl$. The wire is carrying a current *I*.

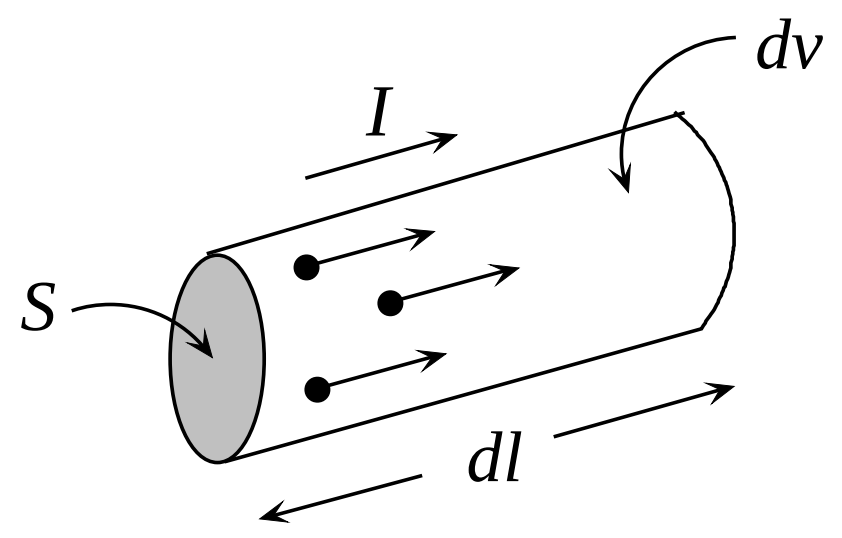

Fig. 2.5. Metal wire of infinitesimal length *dl*, carrying a current *I*.

We call *dN* the number of free electrons passing through the cross-section *S* in time *dt* and finally occupying the volume *dv* of the wire. The (conventionally positive) charge that passes through the cross-section *S* in time *dt* and occupies the volume *dv* is thus equal to $dQ = q dN$, where *q* is the absolute value of the charge of the electron. We observe that an electron travels a distance *dl* along the wire within time *dt*. Thus the drift speed of the electrons is $\upsilon = dl/dt$. Finally, the current on the wire is $I = dQ/dt$.

The *current density* at the cross-section *S* is

$$J = \frac{I}{S} = \frac{1}{S}\frac{dQ}{dt} = \frac{qdN}{Sdt} = \frac{qdNdl}{Sdldt} = q\frac{dN}{dv}\frac{dl}{dt} \tag{2.2}$$

The quantity

$$n = \frac{dN}{dv} \tag{2.3}$$

is the *electronic density* (or *free-electron concentration*) of the considered metal; it represents the concentration of free electrons (number of electrons per unit volume) in the material.[5] Also, by (2.1), $dl/dt = \upsilon = \mu E$, where $\upsilon$ and $E$ are the magnitudes of the corresponding vectors. Thus (2.2) is written as

$$J = q n \upsilon = q n \mu E \tag{2.4}$$

The product

$$\boxed{\sigma = q n \mu} \tag{2.5}$$

is called the *conductivity* of the metal. Equation (2.4) is finally written:

$$J = \sigma E \tag{2.6}$$

or, in vector form,

[5] In reality, only part of the free electrons contributes to electrical conduction; see Appendix C.

$$\boxed{\vec{J} = \sigma \vec{E}} \qquad (2.7)$$

where $\vec{J}$ is a vector of magnitude $J$, oriented in accordance with the *conventional* direction of the current (this will be further explained in Chapter 6). Equation (2.7) expresses the *general form of Ohm's law*. It is an empirical relation for metals, valid when the electric field $E$ is not too strong.

The general relation (2.7) was proven for an infinitesimal section of a metal and, in this sense, is independent of the shape or the dimensions of the metal. We now consider the special case of a metal wire of finite length $l$ and constant cross-section $S$, carrying a constant current $I$ (Fig. 2.6). We call $V$ the voltage (potential difference) between the two ends of the wire.

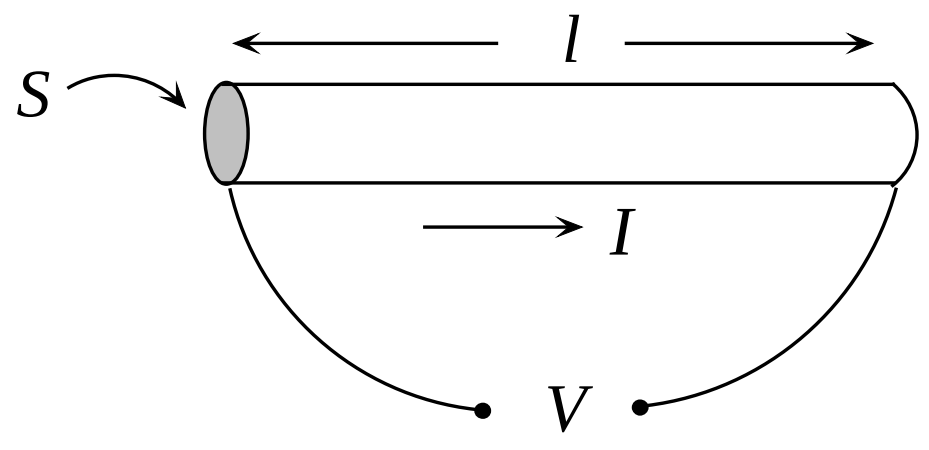

Fig. 2.6. Metal wire of finite length $l$, carrying a current $I$ and subject to a voltage $V$.

The current density $J=I/S$ is constant along the wire since both $I$ and $S$ are constant. From Ohm's law (2.6) it then follows that the electric field $E$ is also constant along the wire; that is, the electric field inside the wire is *uniform*. As will be shown in Chapter 6, the magnitude of the field is given in this case by $E=V/l$. Thus, by using (2.6) we have:

$$I = JS = \sigma ES = \sigma \frac{V}{l} S = \frac{V}{l/\sigma S}$$

We define the *resistivity* $\rho$ of the metal and the *resistance* $R$ of the wire by the relations

$$\rho = \frac{1}{\sigma} \;, \quad R = \frac{l}{\sigma S} = \frac{\rho l}{S} \qquad (2.8)$$

We thus obtain the *special form of Ohm's law*,

$$\boxed{I = \frac{V}{R}} \qquad (2.9)$$

We remark that the resistivity $\rho$ is a property of the conducting material, regardless of its shape or its dimensions, whereas the resistance $R$ is a property of the specific wire and depends on its geometrical characteristics. Note also that the special form (2.9) of Ohm's law is valid for a wire of *constant* cross-sectional area (explain this).

## 2.5 Ohm's Law for Semiconductors

As we know, the mobile charges in a semiconductor are the free electrons of the conduction band and the holes in the valence band. We call $n$ and $p$ the concentrations of electrons[6] and holes, respectively, where by *concentration* we generally mean the number of similar things (electrons, holes, atoms, etc.) per unit volume. In a *pure* or *intrinsic* semiconductor (that is, a semiconductor crystal without any impurity atoms) the number of electrons must be equal to the number of holes, given that each hole appears after the liberation of a valence electron from the covalent bond to which it belongs. Thus, for an intrinsic semiconductor we have that

$$n = p \equiv n_i \tag{2.10}$$

The common value $n_i$ of the two concentrations in a pure semiconductor is called *intrinsic concentration*. As we will see later on, Eq. (2.10) is generally not valid for a semiconductor having impurities.

When an electric field $\vec{E}$ exists inside a semiconductor, the motion of both electrons and holes is oriented according to this field and two parallel current densities $\vec{J}_n$ and $\vec{J}_p$ appear in the crystal. Both density vectors are in the direction of $\vec{E}$. To understand this, let us consider the motion of an electron and a hole inside an electric field $\vec{E}$, such as that existing in the interior of a parallel-plate capacitor (Fig. 2.7).

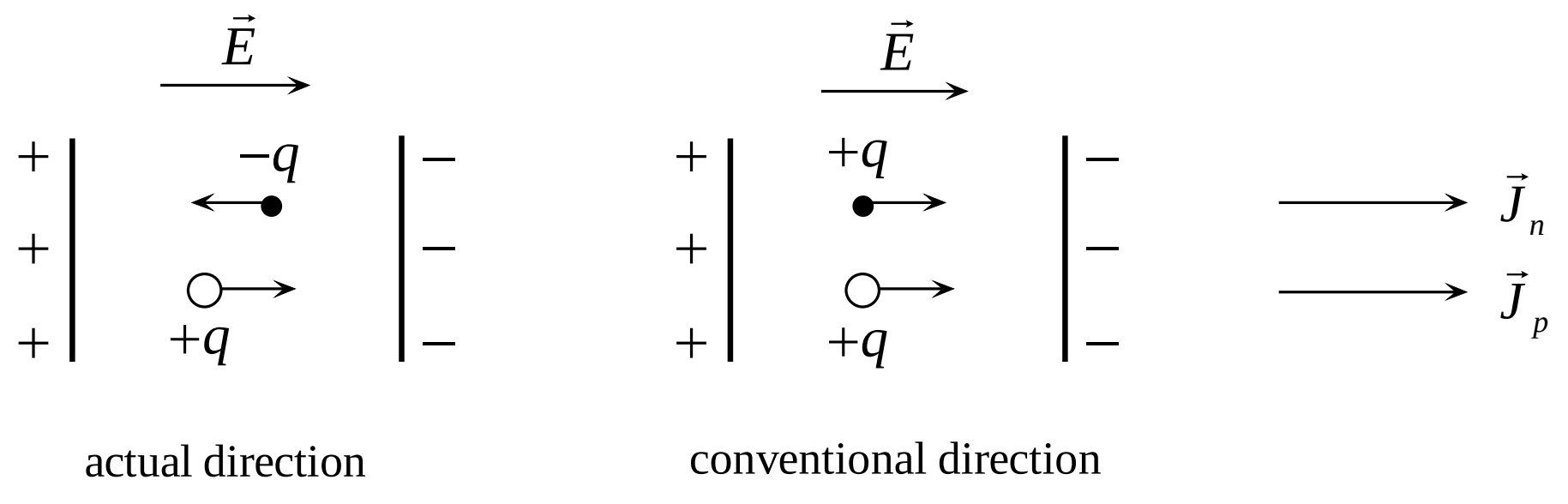

Fig. 2.7. Current densities of electrons and holes: actual and conventional directions of motion of charges.

In reality, the positively charged hole moves in the direction of $\vec{E}$ while the electron, being negatively charged, moves in the direction opposite to the field. Conventionally, however, the motion of the electron may be interpreted as the motion of a *positive* charge in the *opposite* direction, i.e., in the direction of $\vec{E}$. We conclude that the currents generated by the motions of electrons and holes are both in the direction of $\vec{E}$.

The currents $\vec{J}_n$ and $\vec{J}_p$ separately obey Ohm's law:

$$\vec{J}_n = \sigma_n \vec{E}\,, \qquad \vec{J}_p = \sigma_p \vec{E} \tag{2.11}$$

---

[6] By "electrons" we will henceforth mean the *free* electrons of the conduction band.

where the corresponding conductivities are

$$\sigma_n = q n \mu_n \ , \qquad \sigma_p = q p \mu_p \tag{2.12}$$

The $n$ and $p$ represent the concentrations of electrons and holes, respectively, while $\mu_n$ and $\mu_p$ are the *mobilities* of electrons and holes in the considered semiconductor; by $q$ we denote the absolute value of the charge of the electron. As found experimentally, $\mu_p$ is somewhat smaller than $\mu_n$ (can you explain this physically?). The total current is

$$\vec{J} = \vec{J}_n + \vec{J}_p = (\sigma_n + \sigma_p)\vec{E}$$

or

$$\boxed{\vec{J} = \sigma \vec{E}} \tag{2.13}$$

where

$$\boxed{\sigma = \sigma_n + \sigma_p = q n \mu_n + q p \mu_p} \tag{2.14}$$

is the *total conductivity* of the material. For a *pure* semiconductor,

$$\sigma_i = q n_i (\mu_n + \mu_p) \tag{2.15}$$

where we have taken (2.10) into account. Equations (2.13) and (2.14) express Ohm's law for a semiconductor.

We note that, in general, the conductivity $\sigma$ is a measure of the conducting ability of a material that obeys Ohm's law. Indeed, the larger is $\sigma$, the greater is the current for a given value of the electric field. It is now clear why, under normal conditions, a metal is much more conductive than a pure semiconductor: by comparing the conductivities (2.5) and (2.15) we see that $\sigma_{metal} >> \sigma_i$ . This is due to the fact that the electronic density $n$ of the metal is much larger than the intrinsic concentration $n_i$ of the semiconductor ($n >> n_i$) since the metal has many more mobile charges (free electrons) at its disposal, compared to the number of electrons and holes in the semiconductor. The disadvantage of the semiconductor is, of course, the presence of the forbidden band; in other words, the need for covalent bonds to be broken before any mobile charges may appear in the crystal.

## 2.6 Temperature Dependence of Conductivity

A fundamental difference between metals and semiconductors has to do with the way the conductivity of each substance is affected by the change of temperature. It has been observed that a raise of temperature produces an increase of resistance (thus a *decrease* of conductivity) in metals. On the other hand, an increase of temperature causes an *increase* of conductivity in semiconductors. To understand these effects we must examine the way in which the various factors appearing in the expressions for the conductivity are affected by temperature in each case.

*A. Metals*

The conductivity of a metal is given by Eq. (2.5): $\sigma = qn\mu$. The charge $q$ of the electron is, of course, independent of temperature. In metals, the electronic density $n$ (number of free electrons per unit volume) is fixed, independent of temperature, given that the number of free electrons in the crystal (equal to the total number of valence electrons of the atoms in the lattice, all of which atoms have been ionized) is determined from the outset and is therefore not affected by temperature changes. However, the mobility $\mu$ of the electrons *decreases* with temperature, for the following reason: An increase of temperature causes an increase of the amplitude of vibration of the ions composing the lattice, hence results in an increased probability of collisions of free electrons with ions. This makes it more difficult for the electrons to move in between the ions, with the result that the average velocity of the electrons is decreased for a given value of an applied electric field. Thus, according to (2.1), the electron mobility $\mu$ decreases with temperature. We conclude that

- *by increasing the temperature the conductivity of a metal is decreased.*

*B. Intrinsic semiconductors*

The conductivity of a pure semiconductor is given by Eq. (2.15): $\sigma_i = qn_i(\mu_n + \mu_p)$. An increase in temperature causes an increase of the number of electron-hole pairs in the crystal, since more and more covalent bonds are broken and more and more electrons of the valence band are excited to the conduction band, leaving holes behind. This results in an increase of the intrinsic concentration $n_i$. The mobilities $\mu_n$ and $\mu_p$ are reduced somewhat with the increase of temperature but not enough to match the increase of $n_i$. We conclude that

- *by increasing the temperature the conductivity of a semiconductor is increased.*

At sufficiently high temperatures the conductivity of a semiconductor becomes comparable to that of a metal. Of course, at ordinary temperatures metals are incomparably more conductive than semiconductors.[7] The latter substances, however, have the advantage of possessing *two* kinds of mobile charges, namely, electrons and holes. This makes semiconductors extremely useful in electronics technology.

*A note on superconductors*

The phenomenon of *superconductivity* is interesting from both the theoretical and the practical point of view. The difference between an ordinary metal and a superconducting one becomes apparent by comparing the curves representing the change of resistivity with absolute temperature (Fig. 2.8).

---

[7] For a good conductor, $n \approx 10^{22}$ electrons/$cm^3$. For an intrinsic semiconductor at room temperature (300 $K$), $n_i \approx 10^{10} - 10^{13}$ electrons/$cm^3$.

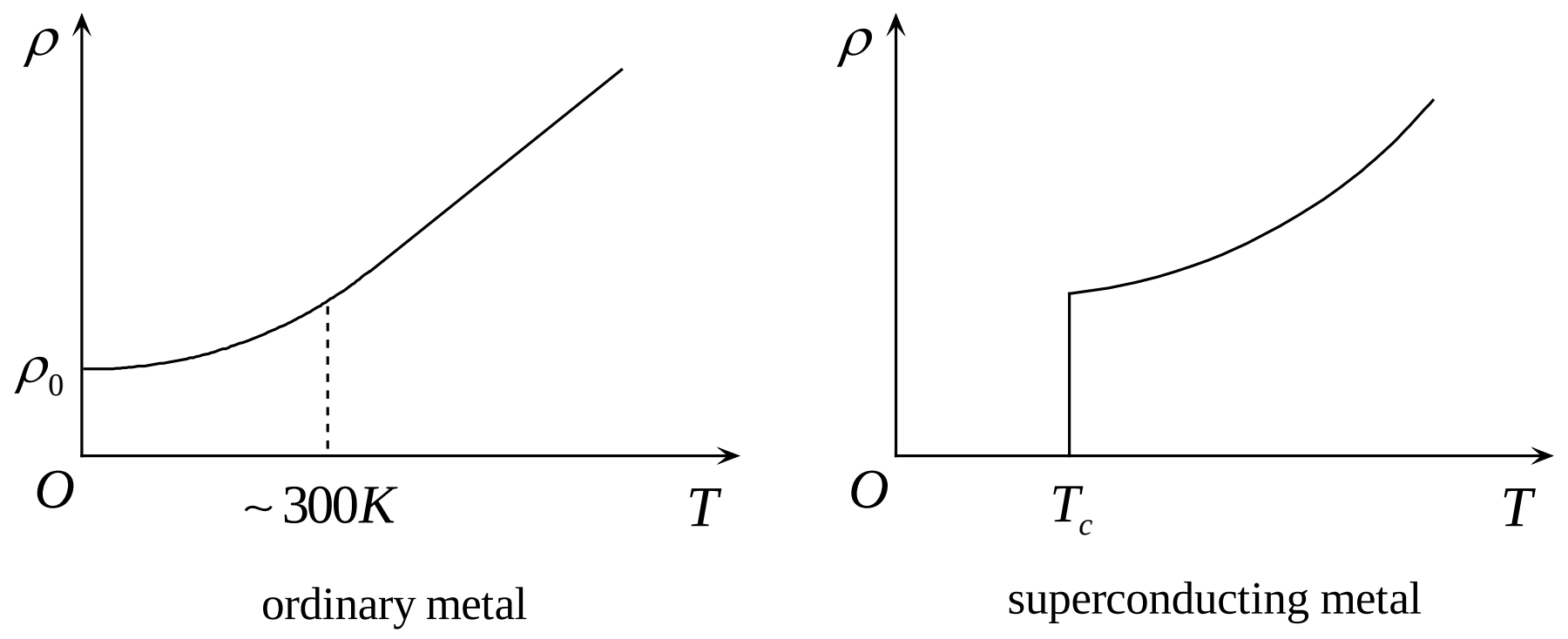

Fig. 2.8. Change of resistivity with temperature in metals and superconductors.

As mentioned above, the electrical resistance of a metal is due to vibrations of the ions composing the crystal lattice. As the temperature $T$ decreases, the amplitude of vibration becomes smaller and so does the resistivity $\rho$ of the material. As found experimentally, at ordinary temperatures ($T \sim 300\,K$) the resistivity is proportional to the absolute temperature. If this were to be the case for all $T$, the resistance of the metal should vanish for $T \to 0$. In reality, however, this does not occur. The reason is that, in addition to the vibrations of the ions there are other factors contributing to the resistance of the metal, such as, e.g., imperfections or impurities in the crystal lattice. At very low temperatures these factors are predominant over ionic vibrations, the latter tending to die out as $T$ approaches absolute zero. The resistivity thus tends to a finite value $\rho_0$ as $T \to 0$.

Things are different with *superconductors*, the resistivity of which *vanishes abruptly* when the temperature drops below a *critical temperature* $T_c$ characteristic for the given material (at temperatures above $T_c$ a superconductor behaves as an ordinary metal). For most natural superconductors (e.g., mercury or lead) the critical temperature is a few degrees above absolute zero, which hardly makes these substances useful in applications. Compounds have been discovered, however, exhibiting superconducting properties at much higher temperatures (exceeding $130\,K$). These discoveries have opened new possibilities in superconductor technology. Applications include (but are not limited to) the construction of superconducting magnets for creating very strong magnetic fields, the manufacturing of magnetometers for measuring extremely weak magnetic fields (these are useful devices in medical research), the storage of electrical energy without losses by using superconducting rings, etc.

## 2.7 Semiconductors Doped with Impurities

The conductivity of a semiconductor is increased significantly by the addition of suitable *impurities*. A doping with impurities spoils the balance of concentrations ($n=p$) between electrons and holes that exists in the intrinsic (pure) semiconductor. We thus distinguish two types of semiconductors with impurities; namely, *n-type* semiconductors if $n>p$, and *p-type* semiconductors if $p>n$. In $n$-type doping the electrons are the *majority carriers* and the holes are the *minority carriers*, while the converse is true for $p$-type doping.

*A. Semiconductors with n-type doping*

Imagine that, in a crystal of pure germanium (Ge) or silicon (Si) we replace a few tetravalent atoms with atoms of a *pentavalent* element such as phosphorus (P, 15) or arsenic (As, 33). The pentavalent element is called *donor* since its atoms offer an extra valence electron compared to the atoms of the intrinsic semiconductor, thus contributing to the conductivity of the crystal. The 4 of the 5 valence electrons of the donor atom form 4 covalent bonds with 4 neighboring atoms Ge or Si, while the 5$^{th}$ electron is unpaired and is easily freed from the donor atom, the related ionization energy being of the order of $0.01\,eV$. Thus, with the addition of a donor we succeed in increasing the number of free electrons in the crystal.

The doped semiconductor is a novel quantum system whose energy-band diagram is expected to differ in some respects from that of the intrinsic semiconductor. Where in the diagram will the 5$^{th}$ valence electron of the donor atom be accommodated *before* it is detached from the atom to which it belongs? Certainly not in the valence band, since this band is already filled at low temperatures. Nor can the electron be in the conduction band, given that it has not yet been freed from the donor atom. The only remaining possibility is that this 5$^{th}$ valence electron of the donor is on a *new* energy level $E_D$ that appears inside the forbidden band, just below the conduction band. A very small amount of energy (about $0.01\,eV$) is needed in order for the electron to be excited to the conduction band, as seen in Fig. 2.9 ($E_V$ represents the top level of the valence band, while $E_C$ is the bottom level of the conduction band).

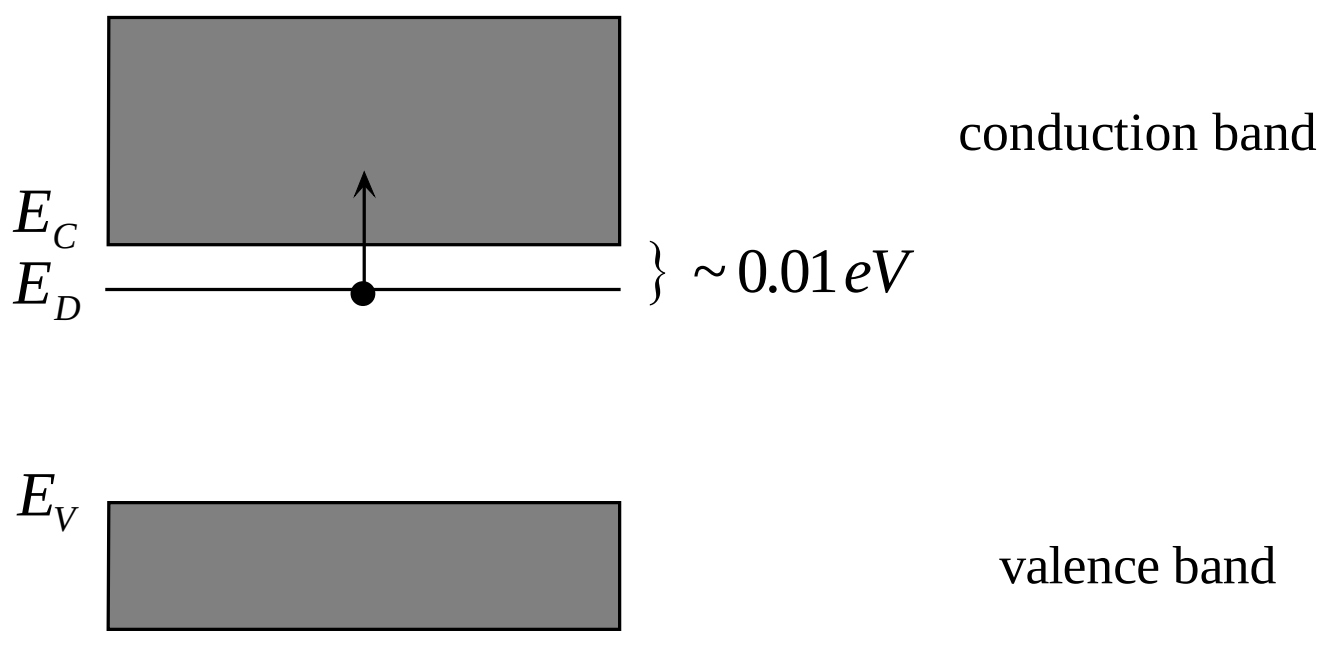

Fig. 2.9. Band structure of an *n*-type semiconductor.

The addition of donor impurities in an intrinsic semiconductor results not only in the increase of the number of free electrons in the conduction band but also in the *decrease of the number of holes* in the valence band. This happens because the free-electron surplus leads to an increased rate of recombination of electrons with holes.

*B. Semiconductors with p-type doping*

In a crystal of pure Ge or Si we replace a few tetravalent atoms with atoms of a *trivalent* element such as boron (B, 5), gallium (Ga, 31) or indium (In, 49). The trivalent element is called *acceptor* since its atoms, having one less valence electrons compared to the atoms of the intrinsic semiconductor, may accept the offer of one electron from the atoms of the semiconductor. The acceptor atom forms 3 covalent bonds with 4 neighboring atoms Ge or Si, the 4$^{th}$ bond being incomplete. This bond can be completed by a valence electron of some nearby atom Ge or Si, the electron leaving behind a new hole in the crystal (the required energy for this process is of the hind a new hole in the crystal (the required energy for this process is of the order of 0.01 *eV*). Thus, by adding acceptor impurities we manage to increase the number of holes in the crystal.

The acceptor introduces a new, *vacant* energy level $E_A$ inside the forbidden band, just above the valence band. By receiving a small amount of energy (about 0.01 *eV*) an electron of the valence band can easily be excited to this vacant level, leaving a hole behind (Fig. 2.10).

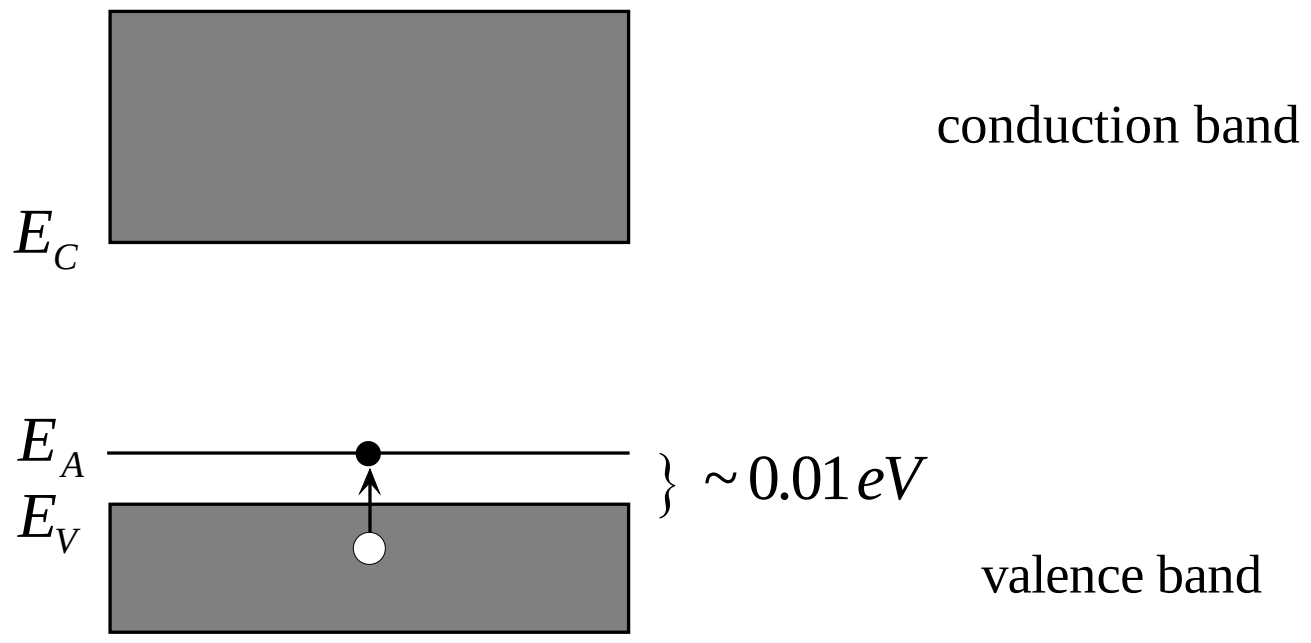

Fig. 2.10. Band structure of a *p*-type semiconductor.

The addition of acceptor impurities in an intrinsic semiconductor not only increases the number of holes in the valence band but also *decreases the number of electrons* in the conduction band due to an increased rate of recombination of electron-hole pairs. It is impressive that even a small amount of doping (say, one donor or acceptor atom for every $10^6$ atoms of intrinsic semiconductor) can increase the conductivity significantly (by about 10 times) at normal temperatures.

## 2.8 Mass-Action Law

In the previous section we mentioned that the addition of impurities to an intrinsic semiconductor causes an increase of concentration of one type of charge carriers (electrons or holes) with a simultaneous decrease in the concentration of the other type of carriers. (In an intrinsic semiconductor the two concentrations assume a common value, equal to the intrinsic concentration $n_i$.) This process does not take place in an arbitrary manner but obeys a certain physical law, called the *mass-action law* [1,2].

We call $n$ and $p$ the concentrations of electrons and holes, respectively, in a semiconductor of any type. When the semiconductor is in pure form (that is, before its doping with impurities) the two concentrations are equal: $n=p=n_i$. After the doping, however, we have that $n \neq p$ (except in a special case to be seen in Sec. 2.9). It should be stressed that the values of the concentrations $n$, $p$ and $n_i$ are measured in conditions of *thermal equilibrium*, that is, at a given, constant temperature. The mass-action law can be stated as follows:

> *Under thermal equilibrium, the product np of concentrations of electrons and holes in a given semiconductor is constant, independent of the kind or the amount of impurity doping in the semiconductor.*

This can be expressed mathematically as follows:

$$(np)_{doped\ semiconductor} = (np)_{intrinsic\ semiconductor} \Rightarrow np = n_i n_i \Rightarrow$$

$$\boxed{np = n_i^2} \qquad (2.16)$$

We note that the intrinsic concentration $n_i$ depends on the temperature, increasing with it (see Sec. 2.6). Therefore, Eq. (2.16) is valid *for a given, constant temperature*.

By means of the mass-action law we can easily explain why the doping of a pure semiconductor increases the conductivity of the material. The conductivity of a semiconductor (whether doped or not) is given by the general expression (2.14):

$$\sigma = qn\mu_n + qp\mu_p$$

Since the difference between $\mu_n$ and $\mu_p$ is relatively small, we can make the approximation

$$\mu_n \approx \mu_p \equiv \mu$$

Then,

$$\sigma = q\mu\,(n + p) \qquad (2.17)$$

We notice that the conductivity depends on the sum of concentrations of electrons and holes. By using (2.16), we have:

$$(n+p)^2 = (n-p)^2 + 4np = (n-p)^2 + 4n_i^2 \Rightarrow$$
$$(n+p)^2 = (n-p)^2 + \text{constant}$$

Thus (2.17) yields

$$\sigma = q\mu\sqrt{(n-p)^2 + \text{const.}} \qquad (2.18)$$

where the constant quantity inside the square root is equal to $4n_i^2$ and depends on the temperature. From (2.18) we deduce that the conductivity $\sigma$ assumes a *minimum* value when $n-p = 0 \Leftrightarrow n = p$, which, in turn, occurs when the semiconductor is *intrinsic*. Thus, by the slightest doping we will have $n-p \neq 0 \Leftrightarrow n \neq p$, therefore the value of $\sigma$ will increase (provided that the temperature is constant). An increase of the amount of doping increases the difference $|n-p|$, hence also the value of $\sigma$ according to (2.18).

In practice, the amount of doping in a semiconductor is minimal, of the order of 0.0001% . This sounds strange, given that an increased doping would result in a greater conductivity. Let us not forget, however, that the usefulness of semiconductors doesn't lie so much on the degree of their conductivity but rather on the *manner* these substances conduct electricity, by means of *two* kinds of charge carriers. A much heavier doping would increase the majority carriers significantly but, at the same time, it would make the minority carriers almost disappear!

*Application:* We will evaluate the concentrations of *minority* carriers in *n*-type and *p*-type semiconductors, given the intrinsic concentration $n_i$ and the concentrations $N_D$ and $N_A$ of donor and acceptor atoms, respectively. (The temperature is assumed given and constant.)

(*a*) *n-type semiconductor:* We assume that almost all donor atoms are ionized, that is, they have contributed their 5th valence electron, which is now a free electron with energies in the region of the conduction band. On the other hand, almost all electrons in that band are due to the donor, given that the majority of free electrons that preexisted in the pure semiconductor have already been recombined with holes in the valence band. It is thus reasonable to make the approximation $n \approx N_D$. By using the mass-action law (2.16) we can now find the concentration of holes, which constitute the minority carriers:

$$p = \frac{n_i^2}{N_D} \tag{2.19}$$

(*b*) *p-type semiconductor:* Almost all holes in the valence band are due to the acceptor; hence, $p \approx N_A$. By using Eq. (2.16) we find the concentration of electrons, which are the minority carriers in this case:

$$n = \frac{n_i^2}{N_A} \tag{2.20}$$

## 2.9 Semiconductors with Mixed Impurities

In a doped semiconductor it is possible for donor impurities to coexist with acceptor impurities. This composite semiconductor will be of type *n* or type *p*, depending on whether the majority carriers are electrons or holes, respectively. We call $N_D$ the concentration of (pentavalent) donor atoms and $N_A$ the concentration of (trivalent) acceptor atoms. We distinguish the following cases:

(*a*) If $N_D=N_A$, then $n=p=n_i$, where $n_i$ is the intrinsic concentration in the pure semiconductor. Thus, the extra electrons from the donor exactly annihilate the extra holes from the acceptor and the semiconductor behaves like an *intrinsic* one.

(*b*) If $N_D>N_A$, then $n>p$ and the semiconductor is of *type n*. The majority carriers are the electrons. If there were no acceptor atoms in the crystal, the electron concentration would be almost equal to the concentration of donor atoms (cf. Application at the end of Sec. 2.8). The acceptor atoms, however, eliminate part of the electrons. Assuming that $N_D>>N_A$, we can make the approximation $n \approx N_D - N_A$. By the mass-action law (2.16) we find the concentration of holes, which are the minority carriers:

$$p = \frac{n_i^2}{N_D - N_A} \tag{2.21}$$

(*c*) If $N_A>N_D$, then $p>n$ and the semiconductor is of *type p*. The concentration of holes (majority carriers) is $p \approx N_A - N_D$ (by assuming that $N_A>>N_D$), while the electron concentration is found with the aid of (2.16):

$$n = \frac{n_i^2}{N_A - N_D} \tag{2.22}$$

## 2.10 Diffusion Currents in Semiconductors

The origin of *diffusion currents* is different from that of currents obeying Ohm's law. Diffusion currents constitute a *statistical* phenomenon and are not related to the existence of an electric field. They are due to a *nonuniform distribution* of charge carriers (electrons or holes) in the crystal so that the concentrations $n$ and $p$ vary from one point to another:

$$n = n(x, y, z) , \quad p = p(x, y, z)$$

This inhomogeneity results in a transport of electrons or holes from regions of greater concentration to regions of lesser concentration in order for the distribution of carriers to finally become uniform. This oriented motion of charges constitutes a diffusion current.

The diffusion-current densities $\vec{J}_p$ and $\vec{J}_n$ for holes and electrons, respectively, are given by *Fick's law*:

$$\vec{J}_p = -qD_p \vec{\nabla} p \ , \qquad \vec{J}_n = +qD_n \vec{\nabla} n \tag{2.23}$$

where $q$ is the absolute value of the charge of the electron, and where $D_p$ and $D_n$ are the *diffusion constants* for holes and electrons, respectively. We note that the direction of $\vec{J}_n$ is *opposite* to the direction of motion of the electrons (for the positively charged holes, $\vec{J}_p$ is in their direction of motion). In the case of a uniform distribution of carriers, the concentrations $p$ and $n$ are constant so that $\vec{J}_p = 0$ and $\vec{J}_n = 0$.

If the $n$ and $p$ vary in only one direction (say, in the $x$-direction), then $n = n(x)$, $p = p(x)$, and the diffusion currents (2.23) take on the algebraic form

$$J_p = -qD_p \frac{dp}{dx} \ , \qquad J_n = +qD_n \frac{dn}{dx} \tag{2.24}$$

The choice of signs in Eqs. (2.23) and (2.24) must be consistent with the following physical requirement:

> *The direction of motion of the carriers is from regions of higher concentration to regions of lower concentration.*

Let us thus check the correctness of our signs. We assume, for simplicity, that $n = n(x)$ and $p = p(x)$.

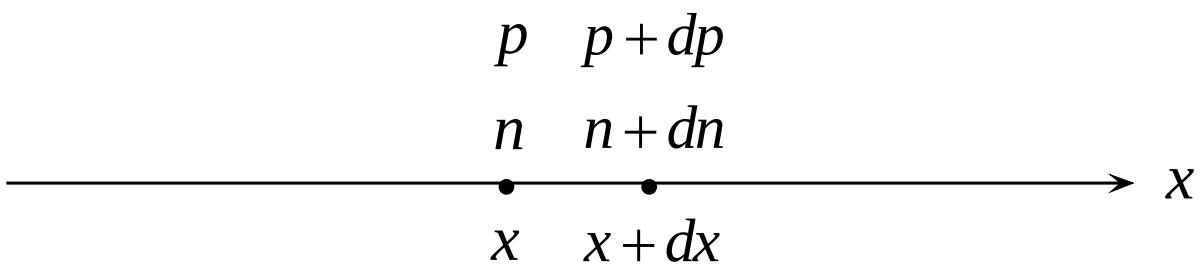

Fig. 2.11. Electron and hole concentrations at two points an infinitesimal distance *dx* apart.

We consider a distribution of electrons and holes along the *x*-axis (Fig. 2.11). Let *n* and *p* be the corresponding concentrations at the location *x*, at some given moment. At that same moment the concentrations at a nearby location *x*+*dx* are *n*+*dn* and *p*+*dp*, where the *dn* and *dp* represent the changes of concentration as one moves from *x* to *x*+*dx*. Without loss of generality, we assume that *dn*>0 and *dp*>0, so that *n*+*dn*>*n* and *p*+*dp*>*p*. This means that both concentrations increase in the positive direction of the *x*-axis. Therefore both the electrons and the holes must move in the *negative* direction of that axis, i.e., from the location of higher concentration to the location of lower concentration. We must now demonstrate that the signs in Eq. (2.24) conform to this requirement.

In the case of the holes we have that

$$\frac{dp}{dx} > 0 \quad \overset{(2.24)}{\Rightarrow} \quad J_p < 0$$

Thus, according to (2.24), the current $\vec{J}_p$ is in the negative direction of the *x*-axis. Since holes are positively charged, their direction of motion coincides with that of $\vec{J}_p$; that is, the holes move in the negative direction. This is in agreement with the prediction made above.

For the electrons we have that

$$\frac{dn}{dx} > 0 \quad \overset{(2.24)}{\Rightarrow} \quad J_n > 0$$

Thus, according to (2.24), the current $\vec{J}_n$ is in the positive direction of the *x*-axis. But, as will be explained in Chapter 6, the direction of motion of the (negatively charged) electrons is opposite to that of the current, which means that the electrons move in the negative direction. Again, this agrees with the prediction made previously.

## References for Chapter 2

## QUESTIONS

**1.** Give a description of the energy bands of metals, insulators and semiconductors. On the basis of the energy-band diagram, explain the electrical conductivity of each of these types of solid.

**2.** Explain why diamond is transparent while sodium and germanium are opaque. (Energies of photons in the visible spectrum: 1.5 - 3 *eV*.)

**3.** Consider three different crystals. Crystal *A* absorbs all electromagnetic radiation up to and including optical frequencies; crystal *B* absorbs radiation whose photons have energies of at least 5.9 *eV*; crystal *C* absorbs radiation with energies of at least 0.8 *eV*. Make a qualitative diagram of the upper energy bands for each crystal and describe the electrical and the optical properties of these crystals. (Energies of photons in the visible spectrum: 1.5 - 3 *eV*.)

**4.** Consider two crystals. Crystal *A* absorbs all electromagnetic radiation up to and including optical frequencies, while crystal *B* absorbs radiation whose photons have energies of at least 0.9 *eV*. The crystals are brought from a low-temperature region to a high-temperature region. What effect will this transfer have on the conductivity of each crystal?

**5.** Describe the *physical* significance of each of the following concepts:
*a.* Conduction band of a metal.
*b.* Valence and conduction bands of a semiconductor.
*c.* Energy gap in a semiconductor.
*d.* Hole in a semiconductor.

**6.** By using the general form of Ohm’s law, derive the familiar form, $I=V/R$, of this law for a metal wire of constant cross-section.

**7.** On the basis of Ohm’s law, explain why a metal is much more conductive than an intrinsic semiconductor at normal temperatures.

**8.** Describe the effect of temperature changes on the conductivity of metals and semiconductors. How do superconductors differ from ordinary metals in this respect?

**9.** Describe the physical mechanism by which an *n*-type or a *p*-type doping contributes to the conductivity of a semiconductor. What are the minority carriers in each case?

**10.** On the basis of the mass-action law, explain why by doping a pure semiconductor with impurities the conductivity of the substance is increased.

**11.** Consider a sample of unit volume of a crystal of pure germanium (Ge). The number of mobile electrons in the sample is equal to $\alpha$. While keeping the temperature constant, we replace $N$ atoms of Ge with phosphorus atoms (P, 15) and another $10\,N$ atoms of Ge with boron atoms (B, 5). What will be the number of electrons after the doping process is completed?

**12.** As argued in Sec. 2.2, diamond is transparent since its energy gap $E_G$ exceeds the energy of photons of visible light, making it impossible for such photons to be absorbed by the crystal. Instead of being absolutely transparent, however, some diamonds have a deep blue color due to the presence of boron atoms $(\mathrm{B}, 5)$ in the crystal. In what way do the boron impurities alter the energy-band diagram of pure diamond? Make a qualitative diagram for the blue diamond, taking into account that $E_G \simeq 6\, eV$ and that the photons of the "red" region of the optical spectrum have energies of the order of $1.7\, eV$. (Assume approximately that light consists of a "red" and a "blue" component.)

**13.** In what respect are diffusion currents different from currents obeying Ohm's law? How are the signs in the expressions (2.24) for diffusion currents justified?

## CHAPTER 3

# DISTRIBUTION OF ENERGY

### 3.1 Some Basic Concepts from Statistical Physics

By *system* we mean a set of identical particles such as electrons, atoms, molecules, or even holes in a semiconductor. The manner in which the various systems exchange energy with their surroundings constitutes the main subject of *Thermodynamics*. In Thermodynamics one is not particularly interested in the microscopic properties of the particles that compose the system. Macroscopic physical concepts such as heat or entropy are defined as experimentally measurable quantities without immediate connection with the internal structure of the system.

The macroscopic behavior of a system, however, *does* depend on the microscopic properties of the constituent particles. For example, the system of free electrons in a metal behaves differently from the system of molecules of an ideal gas. It is thus necessary to connect the macroscopic behavior of a system to its microscopic structure. This is essentially the subject of *Statistical Physics*.

Many properties of a system (such as, e.g., its temperature) are only defined if the system is in a state of *thermal equilibrium*. This means that there is no exchange of heat between the system and its environment, as well as no heat exchanges among the various parts of the system. (We recall that heat is a form of energy *exchange* that, in contrast to work, cannot be expressed macroscopically as force $\times$ displacement.) In a state of thermal equilibrium the system is characterized by a definite, constant absolute temperature $T$. In general, a system that does not interact (hence exchanges no energy) with its surroundings is said to be *isolated*.

We now consider an isolated system consisting of a large number of identical particles. We assume that the energy of each particle is quantized and may take on certain values $E_1$, $E_2$, $E_3$, ... , characteristic for this system. We say that each particle may *occupy* one of the available energy levels $E_1, E_2, E_3, \ldots$ , of the system. We also assume that the system occupies *unit volume*. Hence, all physical quantities concerning this system will be specified *per unit volume*. At some instant the particles are distributed to the various energy levels so that $n_i$ particles (per unit volume) occupy the level $E_i$ (which means that each of these $n_i$ particles has energy $E_i$), as seen in Fig. 3.1.

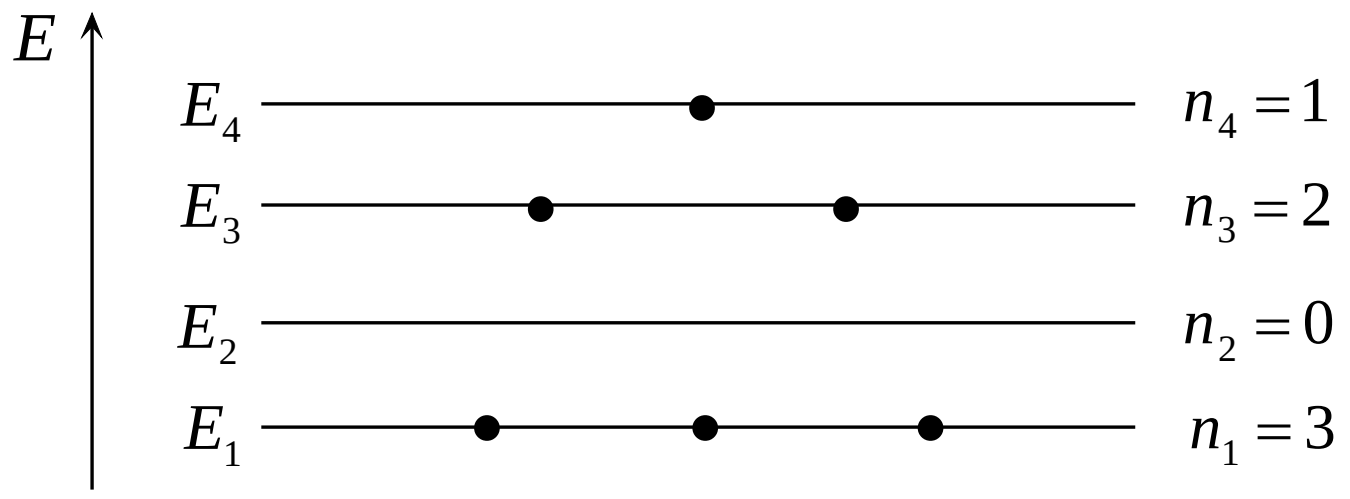

Fig. 3.1. Identical particles occupying available energy levels of an isolated system.

The total number of particles in the system is equal to

$$n = \sum_i n_i \tag{3.1}$$

while the total energy of the system is[1]

$$U = \sum_i n_i E_i \tag{3.2}$$

The ordered set $(n_1,\ n_2,\ n_3,\ ...) \equiv (n_i)$ constitutes a *partition* and defines a *microstate* of the system, compatible with the macroscopic state determined by the number $n$ of particles, the total energy $U$, etc. By the expression (3.2) we implicitly assume that the particles do not interact (or, at least, do not interact too strongly) with one another, so that we may define an average energy separately for each particle. This is approximately true for the molecules of ideal gases, as well as for the free electrons in metals.

Since the system is isolated, the $n$ and $U$ are constant. However, Eqs. (3.1) and (3.2) do not determine the partition $(n_i)$ uniquely, given that different partitions $(n_i)$, $(n_i')$, $(n_i'')$, etc, may correspond to the same values of $n$ and $U$. Now, for given $n$ and $U$ there is a *most probable* partition (microstate). When the system is in that state of maximum probability, we say that it is in *statistical equilibrium* (in Thermodynamics the term *thermal equilibrium* is used). When an isolated system reaches a state of statistical equilibrium, it tends to remain in that state – unless, of course, it is disturbed by some external action. Furthermore, as mentioned previously, in a state of equilibrium the system has a well-defined, constant temperature $T$. As a rule, we will always assume that the systems we consider are in statistical equilibrium.

Assume now that the particles in the system have energies that vary *continuously* from $E_1$ to $E_2$ ($E_1 \leq E \leq E_2$) instead of taking on discrete values $E_1, E_2, E_3, ...$ This is the case for the free electrons in a metal – their energies varying continuously within the limits of the conduction band – as well as for the molecules of an ideal gas that occupies a large volume. In this case there is an infinite number of energy levels varying between the limit values $E_1$ and $E_2$ . The distribution of the particles of the system among these levels is now described with the aid of a function $n(E)$, to be called the *occupation density*, defined as follows:

> *The product $n(E)dE$ represents the number of particles, per unit volume, whose energies have values between $E$ and $E+dE$ .*

One may say that the occupation density $n(E)$ expresses the *distribution of energy* in the system. More accurately, for a given value $E$ of the energy, the corresponding value $n(E)$ describes the "tendency" of the particles in the system to occupy energy levels in the vicinity of $E$: a larger $n(E)$ means a larger number of particles in the energy region between $E$ and $E+dE$ .

It is not hard to see that the total number $n$ of particles in the system, per unit volume, is equal to

[1] We assume that the values $E_i$ of the energy are characteristic of the specific *kind* of system and do *not* depend on the volume of the system.

$$n = \int_{E_1}^{E_2} n(E)\,dE \qquad (3.3)$$

In the case of metals, *n* represents the number of free electrons per unit volume; that is, the *electronic density* (2.3) of the metal.

In analogy to the situation with atomic electrons, the quantum state of a particle in the system is described with the aid of a set of quantum numbers, characteristic of the particular kind of system. In general, to every value *E* of the energy (that is, to every energy level) there correspond many different quantum states. Some of them will be occupied by particles while others will be vacant. In a manner similar to the definition of the occupation density *n*(*E*), we now define the *density of states N*(*E*) as follows:

> *The product N*(*E*)*dE represents the number of states, per unit volume of the system, whose energies have values between E and E+dE.*

Like the occupation density, the density of states is only defined if the energies of the particles vary in a continuous manner. It is also obvious that we cannot expect to find any particles in an energy region where there are no allowable quantum states. Therefore, *n*(*E*)=0 when *N*(*E*)=0. The converse is *not* true, given that there may exist allowable energy regions where all states are vacant (this is, e.g., the case with the upper part of the conduction band of a metal).

## 3.2 Maxwell-Boltzmann Distribution Law for an Ideal Gas

An important problem in Statistical Physics is the distribution of energy in an *ideal monatomic gas*. Since the gas molecules consist of a single atom, their energy is purely *translational kinetic* (there is no intermolecular potential energy, nor is there the rotational or the vibrational kinetic energy typical of a composite molecule). The molecular energy levels are thus given by the relation $E_i = ½\, m v_i^2$, where *m* is the mass of a molecule and where $v_i$ are the possible values of the velocity of the molecules. For given physical conditions, each level $E_i$ is occupied by all molecules having a common speed $v_i$.

The gas is a quantum system confined within the limited space of its container. According to Quantum Mechanics, the energy of the molecules is quantized and therefore the $v_i$ and $E_i$ take on discrete values, as suggested by the use of the index *i*. But, when the volume *V* occupied by the gas is large, we can approximately assume that the molecular kinetic energy $E = ½\, mv^2$ is not quantized but varies in a *continuous* fashion. The energy distribution in the system, therefore, involves the concepts of occupation density and density of states, defined in the previous section. As can be shown [1,2], the density of states is given by the expression[2]

$$N(E) = \frac{2\pi}{h^3}(2m)^{3/2} E^{1/2} \qquad (3.4)$$

[2] Since the energy *E* is purely kinetic, we have that $E \geq 0$; thus the presence of *E* inside a square root is acceptable.

Regarding the occupation density $n(E)$, we recall that it is defined by demanding that the product $n(E)dE$ represents the number of molecules, per unit volume, having energies between $E$ and $E+dE$. As is found [1,2], when the gas is in statistical equilibrium,

$$n(E) = \frac{2\pi n}{(\pi kT)^{3/2}} E^{1/2} e^{-E/kT} \tag{3.5}$$

where $n$ is the concentration of the molecules (number of molecules per unit volume) and $T$ is the absolute temperature. Note that *there is no limit to the number of molecules that can occupy a given quantum state*. In other words, the molecules of the ideal gas do not obey the Pauli exclusion principle.

The *average* (*kinetic*) *energy* of the molecules at temperature $T$ is given by [1,2]

$$\langle U \rangle = \frac{3}{2} kT \tag{3.6}$$

The constant $k$ appearing in Eqs. (3.5) and (3.6) is called the *Boltzmann constant* and is equal to

$$k = 8.62 \times 10^{-5} \, eV/K = 1.38 \times 10^{-23} \, J/K \tag{3.7}$$

If $N$ is the total number of molecules in the gas, the total energy of the system is equal to $N\langle U \rangle$. Thus, if $V$ is the volume occupied by the gas, the *total energy per unit volume* of the system is

$$U = \frac{N}{V}\langle U \rangle = n\langle U \rangle = \frac{3}{2} nkT \tag{3.8}$$

Notice that, according to (3.6),

> *the absolute temperature T of an ideal gas is a measure of the average kinetic energy of the molecules in a state of statistical equilibrium.*

In particular, *the kinetic energy of the molecules vanishes at absolute zero* ($T$=0). As we will see later on, an analogous statement is *not* valid for the free electrons in a metal, despite the superficial similarities of the latter system with the molecules of an ideal gas.

### 3.3 Quantum Statistics

We would now like to examine quantum systems that are more microscopic, such as the free electrons in a metal. The Maxwell-Boltzmann theory, which made successful predictions in the case of ideal gases, proves to be less suitable for the study of electronic systems. Let us explain why.

The Maxwell-Boltzmann theory is essentially a *classical* theory. Although we regarded the gas molecules as quantum particles (for example, we assumed that they occupy quantum states), in essence we treated them as classical particles since we ignored one of the most important principles of quantum theory; namely, the *uncertainty principle*. [Don't be deceived by the presence of the quantum constant $h$ in Eq. (3.4); the basic result (3.5) for the occupation density may be derived by entirely classical methods, without recourse to Quantum Mechanics.] Such an omission of quantum principles is not allowable in the case of electrons, given their exceedingly microscopic nature in comparison to gas molecules. The treatment of such profoundly quantum problems is the subject of *Quantum Statistics*.

In Quantum Statistics, *identical* particles that *interact* with one another are considered *indistinguishable*. By "identical particles" we mean particles that may replace one another without any observable effects in the macroscopic state of the system. (For example, the free electrons in a metal are identical particles since it doesn't matter *which* individual electrons occupy an energy level; it only matters *how many* electrons occupy that level.) In Classical Mechanics, where the notion of the trajectory of a particle is physically meaningful, it is possible to distinguish identical particles that interact by simply following the path of each particle in the course of an experiment. We say that classical particles are *distinguishable*. This is the view adopted by the Maxwell-Boltzmann theory for the molecules of ideal gases.

Things are not that simple, however, for systems of extremely microscopic particles such as, e.g., the electrons in a metal, given that the uncertainty principle does not allow a precise knowledge of the trajectories of such purely quantum particles (in quantum theory the notion of the trajectory is meaningless). Therefore, when *identical* quantum particles interact with one another, it is impossible to distinguish one from another during an experiment. We say that interacting identical particles are *indistinguishable*. (Identical particles that do *not* interact are considered distinguishable.)

Thus, Quantum Statistics is the enhancement of the corresponding classical theory by taking into account the implications of the uncertainty principle. According to the quantum theory, there are two kinds of fundamental particles in Nature, which follow separate statistical laws of distribution of energy when they are grouped to form systems of identical and indistinguishable particles:

- The particles that obey the Pauli exclusion principle are called *fermions* and they follow the *Fermi-Dirac distribution law*.
- The particles that do *not* obey the Pauli exclusion principle are called *bosons* and they follow the *Bose-Einstein distribution law*.

As has been observed,

> *particles having half-integer spins* (e.g., electrons) *are fermions*, while *particles with integral spins* (e.g., photons) *are bosons*.

Accordingly,

*two or more identical fermions may not occupy the same quantum state in a system*, whereas *an arbitrary number of identical bosons may occupy the same quantum state.*

Given that even the molecules of ideal gases are quantum systems consisting of various kinds of fermions (electrons, protons, neutrons, not to mention quarks!), we may wonder whether the Maxwell-Boltzmann distribution law has any use after all. Well, what keeps the classical theory in the game is the fact that, *for systems in which the uncertainty principle can be ignored*, both Fermi-Dirac and Bose-Einstein statistics reduce to Maxwell-Boltzmann statistics. Such a semi-classical system is an ideal gas of low density (i.e., having a small concentration of molecules) at a high temperature. In this case, quantum effects are not significant and the use of classical statistical methods leads to correct physical predictions.

## 3.4 Fermi-Dirac Distribution Law

Fermi-Dirac statistics applies to systems of identical and indistinguishable particles that obey the Pauli exclusion principle; that is, to systems of *fermions*. The free electrons in a metal are an important example of such a system. Although the energies of the electrons are quantized, we may approximately regard these energies as varying continuously within the limits of the conduction band. This approximation is valid when the volume of space within which the motion of the electrons takes place is relatively large (a similar condition is valid for the molecules of an ideal gas).

As we know, the mobile electrons in a metal are characterized as *free* because of their ability to move in between the positive ions without being subject to forces of appreciable strength (except, of course, when the electrons accidentally collide with the ions). In general, a free particle has constant potential energy that may arbitrarily be assigned zero value. The energy $E$ of a free electron is thus *purely kinetic*, which means that $E \geq 0$. We will therefore assume that the energy of a free electron in the metal may take on all values from 0 up to $+\infty$. (The upper limit is, of course, purely theoretical since the energy of an electron in the interior of a metal may not exceed the *work function* of that metal, equal to the minimum energy required for the “escape” of the electron from the crystal.)

Let $N(E)$ be the *density of states* in the conduction band of the metal. We recall that this function is defined so that the product $N(E)dE$ is equal to the number of quantum states (per unit volume) with energies between $E$ and $E+dE$ (equivalently, equal to the number of states belonging to all energy levels between $E$ and $E+dE$ in the conduction band). As can be shown [1-5] the function $N(E)$ is given by the expression

$$N(E) = \frac{4\pi}{h^3}(2m)^{3/2} E^{1/2} \equiv \gamma E^{1/2} \tag{3.9}$$

where $m$ is the mass of the electron. By comparing (3.9) with (3.4) we observe that the density of states for the electrons in a metal is twice that for the molecules of an ideal gas. This is due to the two possible orientations of the electron spin, that is, the two possible values of the quantum number $m_s$ $(=\pm 1/2)$. This consideration does not appear in the Maxwell-Boltzmann distribution since the classical theory does not take into account purely quantum concepts such as that of the spin of a particle.

To find the distribution of energy for the free electrons in a metal, we must determine the *occupation density* $n(E)$. As we know, this function is defined so that the product $n(E)dE$ represents the number of free electrons (per unit volume of the metal) with energies between $E$ and $E+dE$ (equivalently, the number of electrons occupying the energy levels between $E$ and $E+dE$ in the conduction band). It is not hard to see that, because of the Pauli exclusion principle, the number of electrons in this elementary energy interval cannot exceed the number of available quantum states in that interval:

$$n(E)\,dE \le N(E)\,dE \quad \Rightarrow \quad 0 \le \frac{n(E)}{N(E)} \le 1$$

We observe that the quotient $n(E)/N(E)$ satisfies the necessary conditions in order to represent probability. We thus define the *probability function* $f(E)$ by

$$f(E) = \frac{n(E)}{N(E)} \quad \Leftrightarrow \quad n(E) = f(E)\,N(E) \tag{3.10}$$

The function $f(E)$ represents the *fraction of states of energy E that are occupied by electrons*, or, equivalently, the *occupation probability* for any state of energy $E$.

The analytical expression for $f(E)$ is given by the *Fermi-Dirac distribution function*

$$\boxed{f(E) = \frac{1}{1+e^{(E-E_F)/kT}}} \tag{3.11}$$

where $T$ is the absolute temperature, $k$ is the Boltzmann constant (3.7), and $E_F$ is a parameter called the *Fermi energy* (or *Fermi level*, on an energy-level diagram) for the considered metal. We note that, although the present discussion concerns free electrons in metals, the expression (3.11) is generally valid *for all systems of fermions*.

By combining (3.10), (3.11) and (3.9) we can now write an expression for the occupation density $n(E)$, which quantity determines the distribution of energy for the free electrons in the metal:

$$n(E) = f(E)N(E) = \frac{\gamma E^{1/2}}{1+e^{(E-E_F)/kT}} \tag{3.12}$$

The physical significance of the Fermi energy $E_F$ can be deduced from (3.11) after making the following mathematical observations:

- For $T \to 0$, $\lim_{T\to 0^+}\left[e^{(E-E_F)/kT}\right] = \begin{cases} \infty, & E > E_F \\ 0, & E < E_F \end{cases}$

- For $T > 0$, $e^{(E-E_F)/kT} = 1$ when $E = E_F$

Therefore,

$$\text{for } T = 0 \quad \Rightarrow \quad f(E) = \begin{cases} 0, & E > E_F \\ 1, & E < E_F \end{cases} \tag{3.13}$$

while

$$\text{for } T > 0 \quad \Rightarrow \quad f(E_F) = \frac{1}{2} \tag{3.14}$$

These are physically interpreted as follows:

1. For $T$=0, *all* states with energies $E<E_F$ (i.e., all states up to the Fermi level) are *occupied* by electrons, while *all* states with $E>E_F$ are *empty*.
2. For $T$ >0, *half* the states with energy $E=E_F$ are occupied. That is, the occupation probability of any state on the Fermi level is equal to 50%.

We notice that the function $f(E)$ is discontinuous for $E=E_F$ when $T$=0. Hence, the occupation probability on the Fermi level is indeterminate for $T$=0. Figure 3.2 shows the graph of $f(E)$ for $T$= 0 and $T$ > 0. A diagram of this form applies, in general, to any system of fermions (not just to free electrons in metals).

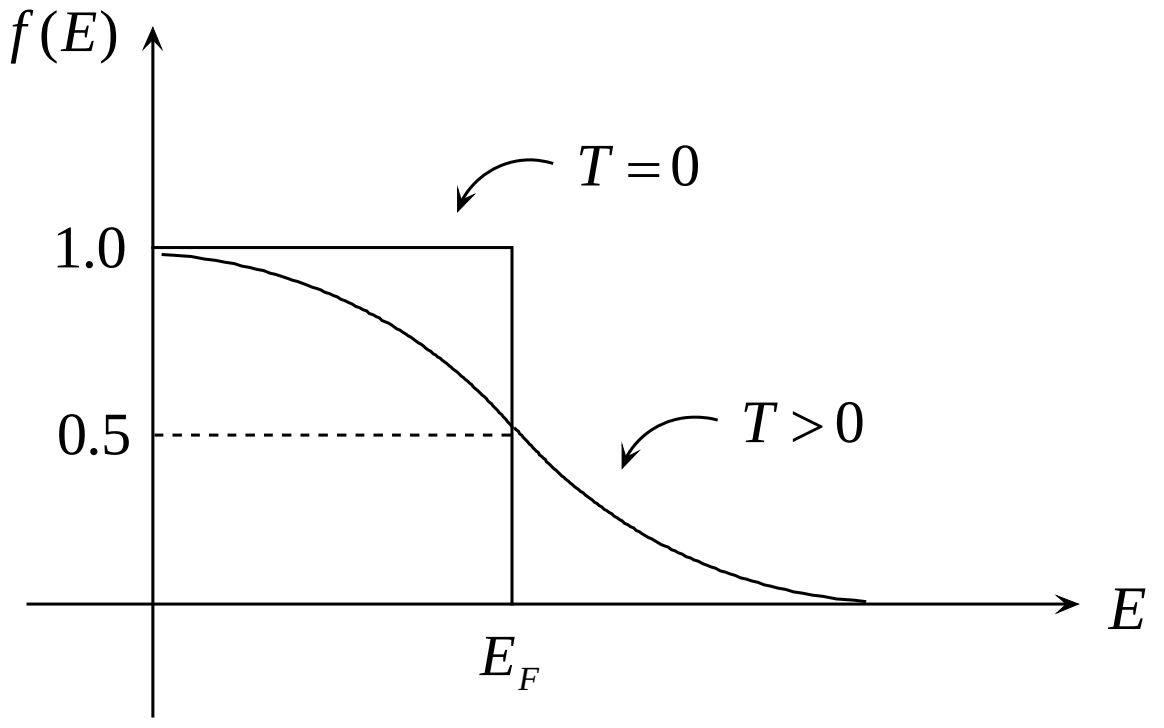

Fig. 3.2. Diagrammatic representation of the Fermi-Dirac distribution function for $T$=0 and $T$>0.

### 3.5 Fermi Energy of a Metal

As we saw in the previous section, the Fermi energy $E_F$ places an upper limit to the energies of the free electrons in a metal at $T$=0. Since the energy of a free electron is purely kinetic, we can write:

$$E_F = (E_{kinetic})_{max} \quad \text{for} \quad T=0 \tag{3.15}$$

That is,

> *the Fermi energy of a metal represents the maximum kinetic energy of the free electrons at absolute zero* $(T = 0)$.

Therefore, at $T$=0, all quantum states in the conduction band ranging from the lowest energy level $E$=0 up to the Fermi level $E=E_F$ are occupied by the free electrons, while all states above $E_F$ are empty. The diagram in Fig. 3.3 shows the conduction band of the metal for $T$=0.

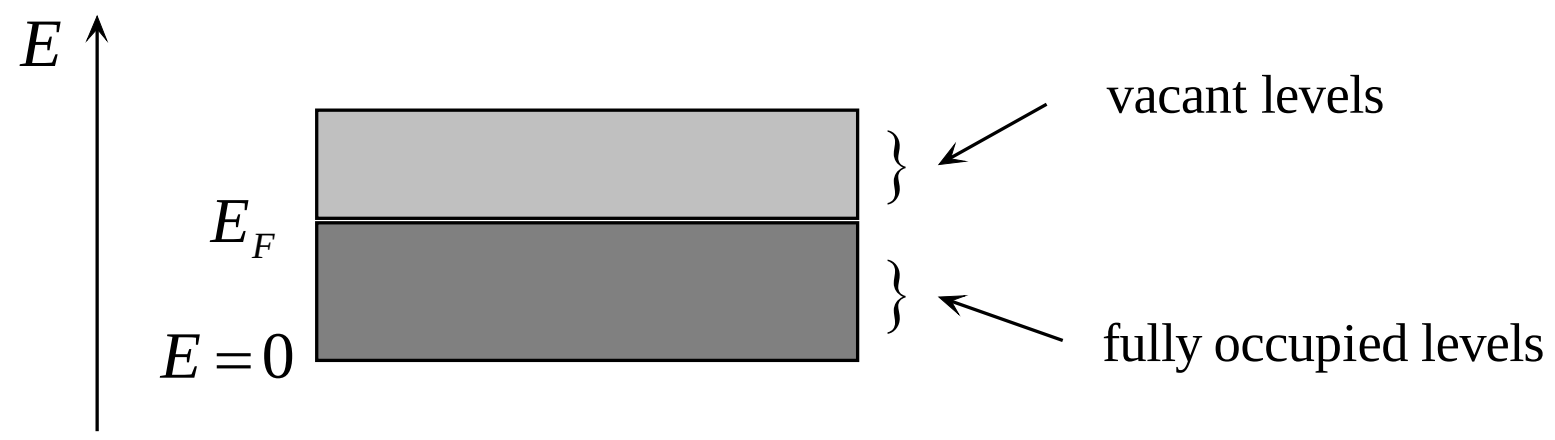

Fig. 3.3. Conduction band of a metal for $T$=0.

We notice a fundamental difference of the Fermi-Dirac theory for electrons from the classical theory for ideal gases. According to the latter theory, all gas molecules must have zero (kinetic) energy at absolute zero. On the other hand, at $T$=0 the free electrons in a metal have (kinetic) energies ranging from zero up to the Fermi energy. This occurs because the electrons, being fermions, obey the Pauli exclusion principle which does not allow all of them to occupy the lowest energy level $E$=0, given that this level does not possess a sufficient number of quantum states to accommodate all electrons. At temperatures $T>0$, however, by receiving thermal energy, some free electrons acquire (kinetic) energies greater than $E_F$. These electrons then occupy energy levels above the Fermi level within the conduction band. As we saw in Sec. 3.4, on the Fermi level itself, *half* the available quantum states are occupied for $T>0$.

We now describe a method for evaluating the Fermi energy $E_F$ of the system of mobile electrons in a metal. Let $n$ be the electronic density of the metal (number of free electrons per unit volume) and let $n(E)$ be the occupation density of the Fermi-Dirac distribution. These two quantities are related by Eq. (3.3):

$$n=\int_{E_1}^{E_2} n(E)\,dE=\int_0^{\infty} n(E)\,dE \qquad (3.16)$$

where here we have put $E_1$=0 and $E_2=+\infty$ (cf. Sec. 3.4). Using the expression (3.12) for $n(E)$, we have:

$$n=\int_0^{\infty}\frac{\gamma E^{1/2}}{1+e^{(E-E_F)/kT}}\,dE \qquad (3.17)$$

If we could calculate the integral in (3.17) analytically, the only thing to do would be to solve the result for $E_F$ and thus express the Fermi energy as a function of $n$ and $T$. Since, however, handling the above integral is not an easy task, we will restrict ourselves to something much easier; namely, we will evaluate $E_F$ for the special case where $T$=0. From (3.10), (3.9) and (3.13) we have that, at this temperature,

$$n(E)=f(E)N(E)=\begin{cases} 0\,, & E>E_F \\ \gamma E^{1/2}\,, & 0\le E<E_F \end{cases} \qquad (3.18)$$

Substituting (3.18) into (3.16), we find:

$$n = \int_0^{E_F} n(E)\,dE + \int_{E_F}^{\infty} n(E)\,dE = \int_0^{E_F} \gamma E^{1/2} dE + 0 \;\Rightarrow$$

$$n = \frac{2}{3}\gamma E_F^{\;3/2} \qquad (3.19)$$

so that

$$\boxed{E_F = \left(\frac{3n}{2\gamma}\right)^{2/3}} \qquad (3.20)$$

We observe that the Fermi energy of the metal at $T$=0 depends only on the concentration $n$ of free electrons and is independent of the dimensions of the crystal (i.e., of the total number of ions). As can be proven (see [4], Sec. 9-3) the value of $E_F$ that we have found does not change much at higher temperatures. Thus, although derived for $T$=0, relation (3.20) will be assumed valid *for all T*. Typical values of $E_F$ for metals range from about $3\,eV$ to $12\,eV$.

## 3.6 Fermi-Dirac Distribution for an Intrinsic Semiconductor

In an intrinsic semiconductor the electronic system of interest consists of the valence electrons of the atoms; specifically, the electrons that participate in covalent bonds as well as those that are free. In terms of energy, the aforementioned two groups of electrons belong to the valence band and the conduction band, respectively. The distribution of energy to the electrons is determined by the occupation density $n(E)$, which is related to the density of states $N(E)$ and the probability function $f(E)$ by

$$n(E) = f(E)\,N(E) \qquad (3.21)$$

As we know, the product $n(E)dE$ represents the number of electrons, per unit volume of the material, with energies between $E$ and $E+dE$.

The form of the function $N(E)$, analogous to the expression (3.9) for metals, depends on the energy region within which this function is defined (see Fig. 3.4) [3-5].

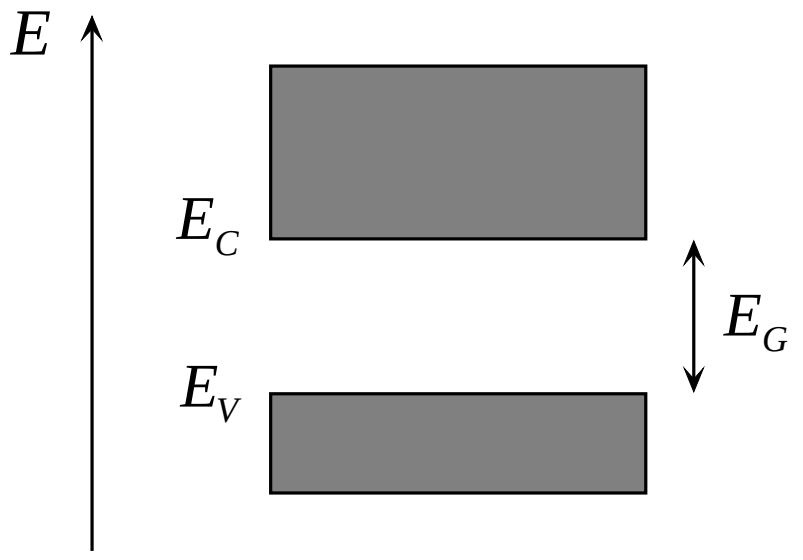

Fig. 3.4. Energy bands of an intrinsic semiconductor.

*a*. In the *conduction band*,

$$N(E) = \gamma (E - E_C)^{1/2}\ , \qquad E \geq E_C \tag{3.22}$$

*b*. In the *valence band*,

$$N(E) = \gamma (E_V - E)^{1/2}\ , \qquad E \leq E_V \tag{3.23}$$

*c*. In the *forbidden band* of a pure semiconductor there are no allowable quantum states; therefore,

$$N(E) = 0\ , \qquad E_V < E < E_C \tag{3.24}$$

The probability function for the electrons is given by the Fermi-Dirac formula:

$$f(E) = \frac{1}{1 + e^{(E - E_F)/kT}} \tag{3.25}$$

where $E$ admits values in the above-mentioned three energy regions. We would now like to find the probability function $f_p(E)$ for the *holes* in the valence band of a semiconductor. We think as follows: A quantum state at an energy level $E$ in the valence band is either occupied by an electron or “occupied” by a hole. If $f(E)$ and $f_p(E)$ are the corresponding occupation probabilities, then

$$f(E) + f_p(E) = 1 \quad \Leftrightarrow \quad f_p(E) = 1 - f(E) \tag{3.26}$$

Substituting the expression (3.25) for $f(E)$, we find that

$$f_p(E) = \frac{e^{(E - E_F)/kT}}{1 + e^{(E - E_F)/kT}} \tag{3.27}$$

Physically, the function $f_p(E)$ represents the fraction of states of energy $E$ that are *not* occupied by electrons, or, equivalently, the probability of non-occupation of a state of energy $E$.

## 3.7 Fermi Energy in Semiconductors

The Fermi energy of an *intrinsic* semiconductor is given by Eq. (3.28), below, where the meaning of the symbols is shown in Fig. 3.5 [3-5].

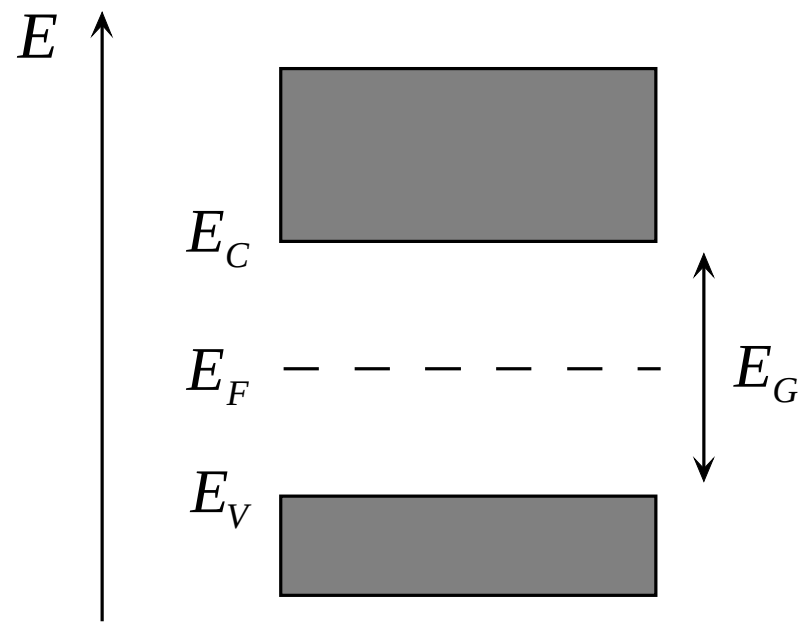

Fig. 3.5. Position of the Fermi level in an intrinsic semiconductor.

$$E_F = \frac{E_V + E_C}{2} \tag{3.28}$$

We write:

$$E_F = \frac{E_V + (E_V + E_G)}{2} = E_V + \frac{E_G}{2} \tag{3.29}$$

This means that

> *the Fermi level of an intrinsic semiconductor is located at the center of the forbidden band.*

Furthermore, $E_F$ is *independent of temperature*, as well as independent of the dimensions of the crystal (that is, of the number of atoms in the lattice).

How should we interpret the presence of $E_F$ inside the forbidden band of a pure semiconductor? Must we conclude that there *is*, after all, some allowable energy level inside an energy region we used to consider inaccessible to the electrons? No! Generally speaking, the Fermi energy $E_F$ is only a parameter of the Fermi-Dirac distribution law and *does not necessarily represent an allowable energy level for the electrons*. That is, the Fermi level may or may not contain allowable quantum states. In metals, $E_F$ is an allowable energy level since it is located inside the conduction band. This is not the case for intrinsic semiconductors, where the Fermi level is located inside the forbidden band.

We note that the presence of the Fermi level $E_F$ inside the forbidden band is absolutely consistent with the general physical interpretation of the Fermi energy given in Sec. 3.4. Let us explain why:

(*a*) For $T>0$ we know that $f(E_F)=1/2$. That is, half the states of the Fermi level are occupied by electrons. In our case, however, the level $E_F$ is located inside the forbidden band; hence it may not possess allowable quantum states. Thus on the Fermi level we have the following situation:

$$\tfrac{1}{2} \times 0 \text{ states} \;\Rightarrow\; 0 \text{ electrons}$$

which is reasonable, given that no energy level inside the forbidden band of an *intrinsic* semiconductor may contain electrons.

(*b*) For $T=0$, all *allowable* energy levels below $E_F$ are completely filled while all *allowable* levels above $E_F$ are empty. But, allowable levels immediately below and above $E_F$ exist in the valence and the conduction band, respectively. Hence, all levels in the valence band are fully occupied by the atomic valence electrons, while no energy level within the conduction band contains electrons. Physically this means that, for $T=0$, all covalent bonds are intact and there are no free electrons in the crystal.

The fact that the level $E_F$ is at the center of the forbidden band reflects an obvious symmetry between electrons and holes in an intrinsic semiconductor. This symmetry is expressed by the relation

$$n = p = n_i \quad \text{(pure semiconductor)} \qquad (3.30)$$

In a sense, the Fermi level "keeps equal distances" from the energy bands occupied by free electrons and holes, the two charge carriers being equally important in an intrinsic semiconductor.

You may guess now how the Fermi level of a pure semiconductor will be affected if we dope the crystal with impurities. The doping will spoil the electron-hole balance expressed by Eq. (3.30). In an *n*-type semiconductor the majority carriers are the electrons in the conduction band, while in a *p*-type semiconductor the majority carriers are the holes in the valence band. The Fermi level will then shift *toward the band occupied by the majority carriers* in each case. Thus, in an *n*-type semiconductor the Fermi level moves closer to the conduction band, while in a *p*-type semiconductor it moves closer to the valence band, as shown in Fig. 3.6.

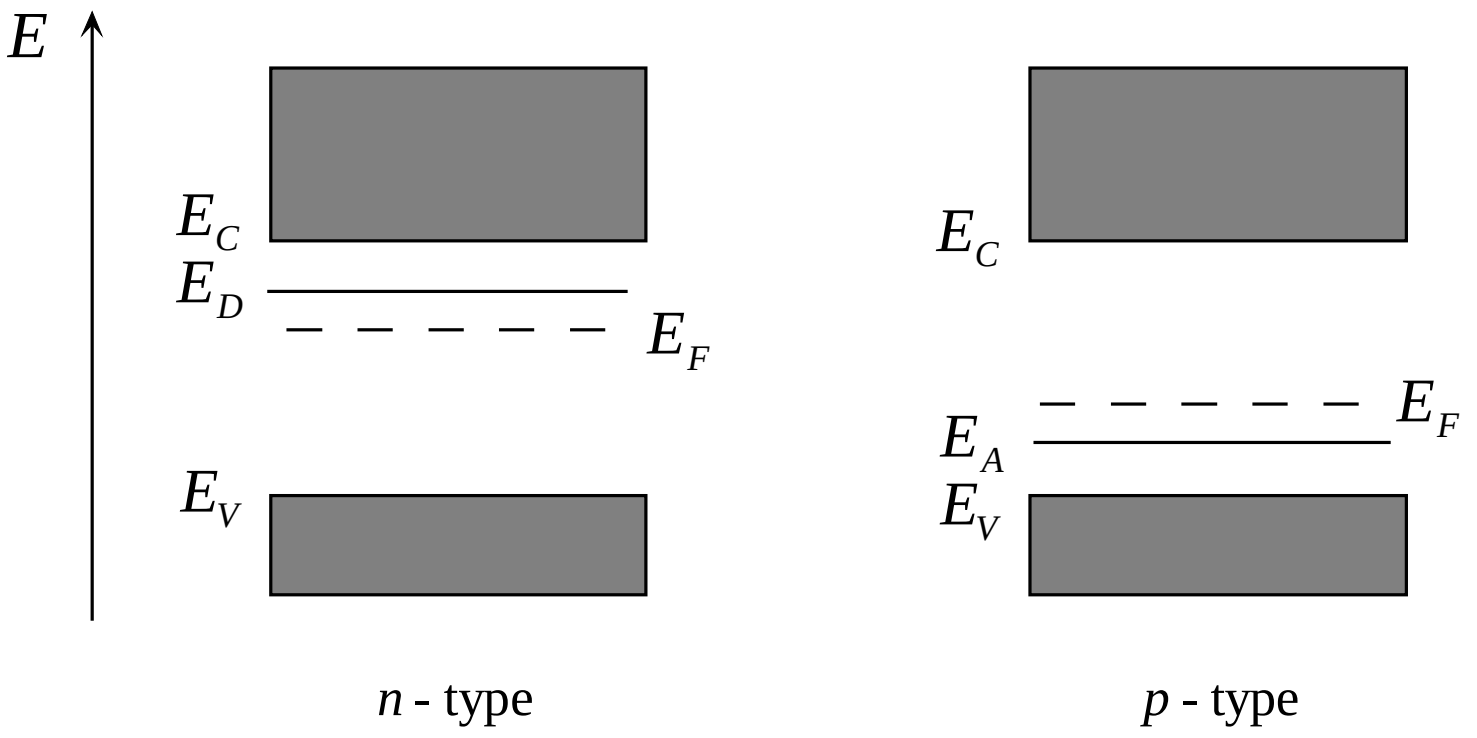

Fig. 3.6. Position of the Fermi level in semiconductors of types *n* and *p*.

In contrast to an intrinsic semiconductor, where $E_F$ is independent of temperature (the Fermi level always lies at the center of the forbidden band), in doped semiconductors $E_F$ changes with temperature. Specifically, as $T$ increases, $E_F$ moves *toward the center of the forbidden band*. This happens because, by the increase of temperature more and more covalent bonds are "broken" in the crystal, which results in an increase of concentration of intrinsic carriers (both electrons and holes) relative to the carriers contributed by the impurity atoms. The concentrations of electrons and holes thus progressively become equal and the semiconductor tends to return to its intrinsic state, with a simultaneous shift of the Fermi level toward the middle of the energy gap. Conversely, as $T \to 0$, the Fermi level $E_F$ passes *above* the donor level $E_D$ for *n*-type doping, or *below* the acceptor level $E_A$ for *p*-type doping.

The value of $E_F$ also depends on the concentration of impurity atoms. Adding more donor (acceptor) atoms in an *n*-type (*p*-type) semiconductor results in a further shift of the Fermi level toward the conduction (valence) band. In cases of extremely high doping, i.e., for $N_D > 10^{19}$ donor atoms $/cm^3$ or $N_A > 10^{19}$ acceptor atoms $/cm^3$, the Fermi level may even move into the conduction band or the valence band, respectively!

*Note: Fermi energy of a p-n junction*

The *p-n junction* [3,5] is in essence a semiconductor crystal one side of which is doped with acceptor atoms while the other side is doped with donor atoms, this making the structure look like a *p*-type semiconductor in contact with an *n*-type semiconductor.

If the two sides of the semiconductor are considered as separate crystals, their Fermi levels will be different (in the *p*-type crystal the level $E_F$ will be closer to the valence band, while in the *n*-type crystal $E_F$ will be closer to the conduction band). If, now, the two crystals are brought to contact in order to form a single, unified structure, this new quantum system will possess a single value of the Fermi energy. Thus a single Fermi level $E_F$ , common to the *p* and *n* sides of the crystal, will appear in the energy-band diagram, as shown in Fig. 3.7.

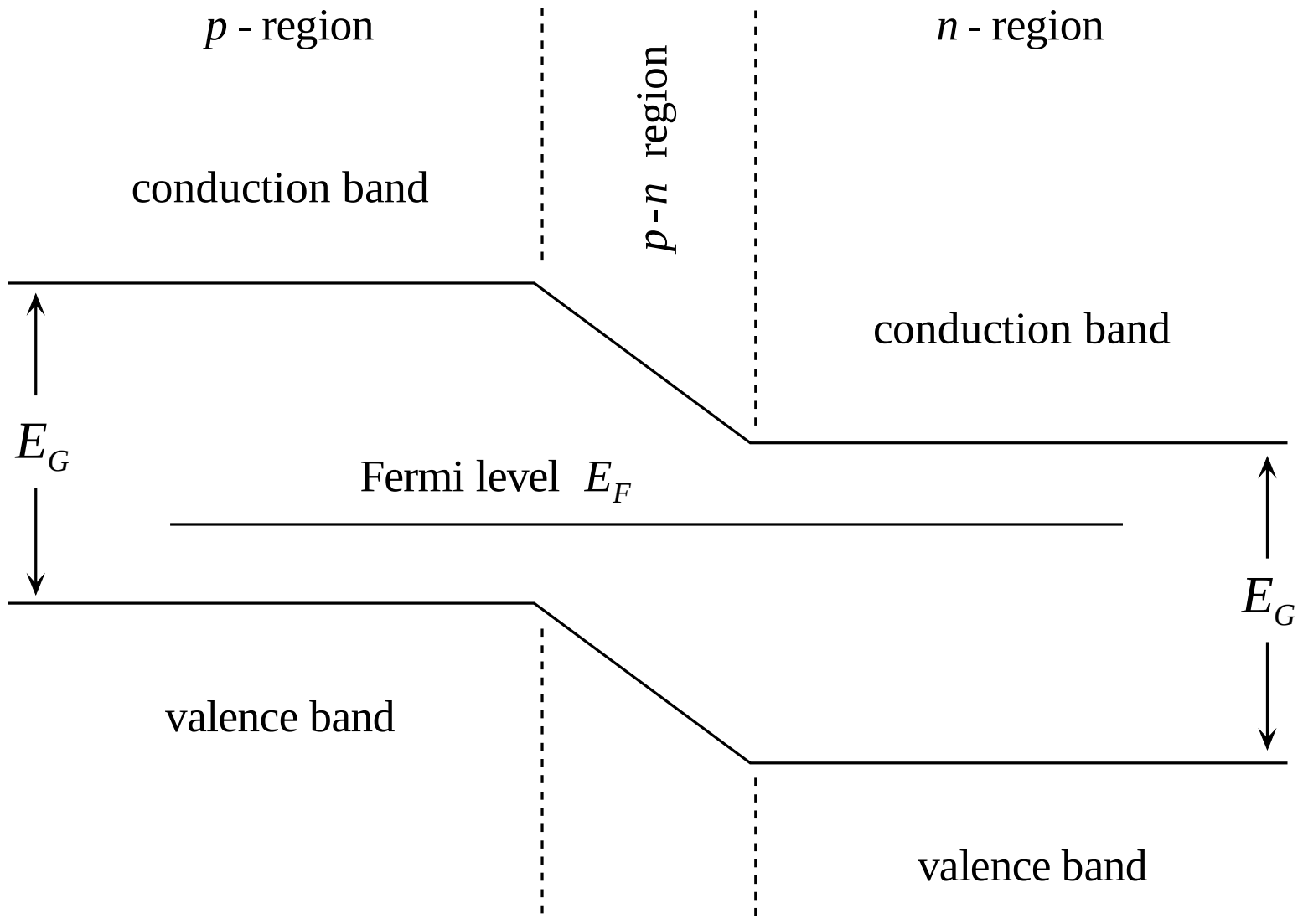

Fig. 3.7. Fermi level in a *p-n* junction.

The *p-n* junction is of central importance in Electronics [3] as it constitutes a fundamental structural element of electronic devices such as the semiconductor diode and the transistor.

## References for Chapter 3

# QUESTIONS

**1.** By using the expression (3.5) for the occupation density, verify Eq. (3.3) for the case of an ideal gas. *Hint:* Set $E_1 = 0$, $E_2 = \infty$, and use the integral formula

$$\int_0^\infty E^{1/2} e^{-E/kT} dE = \frac{1}{2}\left(\pi (kT)^3\right)^{1/2}$$

**2.** What is the fundamental difference between the classical Maxwell-Boltzmann theory and Quantum Statistics? In your opinion, which theory is the most general of the two?

**3.** Imagine a bizarre world in which the electrons would be bosons while the photons would be fermions. (*a*) What would be your grade in a Chemistry class? (*b*) What would be the cost of a laser pointer? [*Hint:* (*a*) Bosons do not obey the Pauli exclusion principle. All atomic electrons would therefore tend to occupy the lowest energy level, that is, the very first subshell. What would then be the structure of an atom? Would there be any chemical reactions? (*b*) A laser beam is a huge system of *identical* photons, i.e., photons in (almost) the same quantum state. Would such a beam exist if the photons obeyed the Pauli exclusion principle?]

**4.** Can the occupation density exceed the density of states in a system of fermions? How about a system of bosons?

**5.** What is the physical significance of the Fermi-Dirac distribution function? What is the physical significance of the Fermi energy? Suggest a method for deriving the probability function for holes in a semiconductor. What is the physical significance of that function?

**6.** Derive an expression for the Fermi energy $E_F$ of a metal. What is the physical significance of $E_F$ in this case? How does the situation differ in comparison to the classical theory of ideal gases?

**7.** Justify physically the presence of the Fermi level inside the forbidden band of a pure semiconductor. (Examine the cases $T = 0$ and $T > 0$ separately.)

**8.** Consider an *n*-doped semiconductor crystal. Describe the modification of the Fermi level of the system if (*a*) we add more donor atoms; (*b*) we add acceptor atoms; (*c*) we increase the temperature.

**9.** Consider an *n*-doped semiconductor crystal. We recall that the donor introduces a new energy level $E_D$ in the forbidden band, very close to the conduction band. At absolute temperature $T \to 0$, the level $E_D$ is occupied by the $5^{th}$ valence electron of the donor atom (at very low temperatures the donor atoms are not ionized). Show that, in the limit $T \to 0$, the Fermi level $E_F$ of the system passes *above* $E_D$. [*Hint:* Remember the physical significance of $E_F$ for $T = 0$.]

**10.** The Fermi energy of a metal is known, equal to $E_F$. The mobility of the electrons in this metal is $\mu$. Show that the resistivity of the metal is equal to

$$\rho = \frac{3}{2q\gamma\mu} E_F^{-3/2}$$

where $q$ is the absolute value of the charge of the electron and $\gamma$ is the constant defined in Eq. (3.9). [*Hint:* Find the conductivity $\sigma = 1/\rho$ of the metal (cf. Sec. 2.4).]

**11.** Consider two metals $M_1$ and $M_2$ . For the electron mobilities and the Fermi energies of these metals we are given that $\mu_1 = 4\mu_2$ and $E_{F,2} = 4\, E_{F,1}$ . The resistivity of $M_2$ is $\rho_2 = 1.5 \times 10^{-8}\, \Omega.m$ . Find the resistivity $\rho_1$ of $M_1$ . [*Answer:* $\rho_1 = 3 \times 10^{-8}\, \Omega.m$ ]

**12.** The electronic density of a metal is known, equal to $n$ . The external conditions are such that, according to the classical theory, the average kinetic energy of the air molecules is very close to zero. Determine the *maximum* kinetic energy of the free electrons in the metal according to the quantum theory.

**13.** Consider a crystal of an intrinsic semiconductor. In the energy-band diagram the Fermi level lies $0.4\, eV$ above the valence band. Determine the maximum wavelength of radiation absorbed by the crystal. Given: $h = 6.63 \times 10^{-34}\, J.s$ ; $c = 3 \times 10^8\, m/s$ ; $1\, eV = 1.6 \times 10^{-19}\, J$ . [*Answer:* $\lambda_{max} = 15.54 \times 10^{-7}\, m$]

CHAPTER 4

# ELEMENTS OF FIELD THEORY

## 4.1 Vector Fields and Vector Operators

We consider the standard Euclidean space $R^3$ with Cartesian coordinates $(x, y, z)$. We call $(\hat{u}_x, \hat{u}_y, \hat{u}_z)$ the unit vectors[1] in the directions of the corresponding axes (Fig. 4.1).

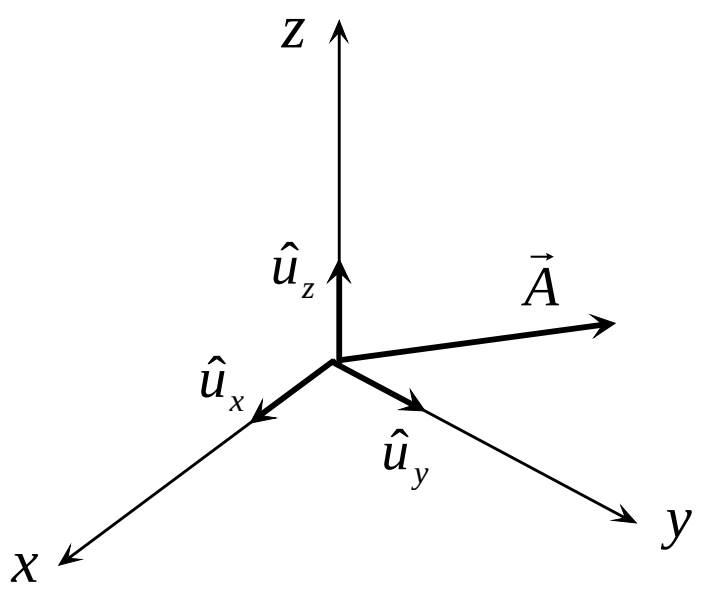

Fig. 4.1. A vector in 3-dimensional space.

A vector $\vec{A}$ in this space is written as

$$\vec{A} = A_x \hat{u}_x + A_y \hat{u}_y + A_z \hat{u}_z \equiv (A_x, A_y, A_z) \qquad (4.1)$$

where $A_x$, $A_y$, $A_z$ are the *rectangular components* of $\vec{A}$. The *magnitude* of $\vec{A}$ is defined as the non-negative quantity

$$|\vec{A}| = (A_x{}^2 + A_y{}^2 + A_z{}^2)^{1/2} \qquad (4.2)$$

Let $\vec{B} \equiv (B_x, B_y, B_z)$ be a second vector, and let $\theta$ be the angle between $\vec{A}$ and $\vec{B}$ where, by convention, $0 \le \theta \le \pi$. The *scalar ("dot") product* and the *vector ("cross") product* of $\vec{A}$ and $\vec{B}$ are defined, respectively, by the equations

$$\vec{A} \cdot \vec{B} = A_x B_x + A_y B_y + A_z B_z = |\vec{A}||\vec{B}|\cos\theta \qquad (4.3)$$

and

[1] The usual notation $(\vec{i}, \vec{j}, \vec{k})$ should better be avoided in Electromagnetism since it may cause confusion (the symbol *i* appears in complex quantities, while $\vec{k}$ denotes a wave vector).

$$\vec{A}\times\vec{B} = (A_y B_z - A_z B_y)\hat{u}_x + (A_z B_x - A_x B_z)\hat{u}_y + (A_x B_y - A_y B_x)\hat{u}_z$$
$$= \begin{vmatrix} \hat{u}_x & \hat{u}_y & \hat{u}_z \\ A_x & A_y & A_z \\ B_x & B_y & B_z \end{vmatrix} \qquad (4.4a)$$

with

$$|\vec{A}\times\vec{B}| = |\vec{A}||\vec{B}|\sin\theta \qquad (4.4b)$$

Notice that $\vec{A}\cdot\vec{B} = \vec{B}\cdot\vec{A}$, while $\vec{A}\times\vec{B} = -\vec{B}\times\vec{A}$.

Given a vector $\vec{A}$, the *unit vector* $\hat{u}$ in the direction of $\vec{A}$ can be expressed as follows:

$$\hat{u} = \frac{\vec{A}}{|\vec{A}|} \equiv \left( \frac{A_x}{|\vec{A}|}\ ,\ \frac{A_y}{|\vec{A}|}\ ,\ \frac{A_z}{|\vec{A}|} \right) \qquad (4.5)$$

By using (4.2), it can be shown that $|\hat{u}| = 1$.

*Exercise:* (*a*) Show that $|\vec{A}| = \sqrt{\vec{A}\cdot\vec{A}}$ and $\vec{A}\times\vec{A} = 0$. (*b*) Show that

$$\vec{A}\cdot(\vec{B}\times\vec{C}) = \vec{B}\cdot(\vec{C}\times\vec{A}) = \vec{C}\cdot(\vec{A}\times\vec{B}) = \begin{vmatrix} A_x & A_y & A_z \\ B_x & B_y & B_z \\ C_x & C_y & C_z \end{vmatrix} \qquad (4.6)$$

(*c*) Show that two nonzero vectors $\vec{A}$ and $\vec{B}$ are mutually *perpendicular* if and only if $\vec{A}\cdot\vec{B} = 0$; they are *parallel* to each other if and only if $\vec{A}\times\vec{B} = 0$. (*d*) Show that $\vec{A}\times\vec{B}$ is perpendicular to both $\vec{A}$ and $\vec{B}$. (*e*) Show that $\hat{u}_x \times \hat{u}_y = \hat{u}_z$.

A *scalar field* in $R^3$ is a mapping $\Phi$: $R^3 \to R$. Scalar fields are represented by functions $\Phi(\vec{r}) = \Phi(x, y, z)$, where

$$\vec{r} = x\hat{u}_x + y\hat{u}_y + z\hat{u}_z \equiv (x, y, z) \qquad (4.7)$$

is the *position vector* of a point in $R^3$, relative to the origin of coordinates of our space.

A *vector field* in $R^3$ is a mapping $\vec{A}$: $R^3 \to R^3$. Vector fields are represented by functions of the form

$$\vec{A}(\vec{r}) = \vec{A}(x, y, z) = A_x(x, y, z)\hat{u}_x + A_y(x, y, z)\hat{u}_y + A_z(x, y, z)\hat{u}_z$$
$$\equiv (A_x, A_y, A_z) \qquad (4.8)$$

Now, let $\Phi(x, y, z)$ be a scalar field. When the $x, y, z$ change by $\Delta x, \Delta y, \Delta z$, respectively, the value of the function $\Phi$ changes by

$$\Delta\Phi = \Phi(x+\Delta x, y+\Delta y, z+\Delta z) - \Phi(x, y, z) = \Phi(\vec{r} + \Delta\vec{r}\,) - \Phi(\vec{r}\,) \qquad (4.9)$$

where $\vec{r} \equiv (x, y, z)$ and $\Delta\vec{r} \equiv (\Delta x, \Delta y, \Delta z)$. On the other hand, the *differential* of $\Phi$ is

$$d\Phi = \frac{\partial\Phi}{\partial x}dx + \frac{\partial\Phi}{\partial y}dy + \frac{\partial\Phi}{\partial z}dz \qquad (4.10)$$

where $dx = \Delta x$, $dy = \Delta y$, $dz = \Delta z$. In general, $d\Phi \neq \Delta\Phi$ (unless $\Phi$ is a linear function). For very small changes $dx, dy, dz$, however, we can make the approximation $d\Phi \simeq \Delta\Phi$.

Consider the *vector operator*

$$\vec{\nabla} = \hat{u}_x \frac{\partial}{\partial x} + \hat{u}_y \frac{\partial}{\partial y} + \hat{u}_z \frac{\partial}{\partial z} \equiv \left( \frac{\partial}{\partial x}, \frac{\partial}{\partial y}, \frac{\partial}{\partial z} \right) \qquad (4.11)$$

Given a scalar function $\Phi(x, y, z)$, we define the vector field

$$grad\,\Phi = \vec{\nabla}\Phi = \frac{\partial\Phi}{\partial x}\hat{u}_x + \frac{\partial\Phi}{\partial y}\hat{u}_y + \frac{\partial\Phi}{\partial z}\hat{u}_z \equiv \left( \frac{\partial\Phi}{\partial x}, \frac{\partial\Phi}{\partial y}, \frac{\partial\Phi}{\partial z} \right) \qquad (4.12)$$

Now, we notice that (4.10) may be written in scalar-product form:

$$d\Phi = \left( \frac{\partial\Phi}{\partial x}, \frac{\partial\Phi}{\partial y}, \frac{\partial\Phi}{\partial z} \right) \cdot (dx, dy, dz)$$

Setting $d\vec{r} \equiv (dx, dy, dz)$ and taking (4.12) into account, we have:

$$\boxed{d\Phi = \left(\vec{\nabla}\Phi\right) \cdot d\vec{r}} \qquad (4.13)$$

This is a three-dimensional generalization of the familiar relation $df(x) = f'(x)\,dx$.

We call $\theta$ the angle between the vectors $\vec{\nabla}\Phi$ and $d\vec{r}$, and we set

$$d\vec{r} = |d\vec{r}\,|\,\hat{u} = (dl)\,\hat{u}$$

where $dl = |d\vec{r}\,|$ and where $\hat{u}$ is the unit vector in the direction of $d\vec{r}$. Equation (4.13) is then written:

$$d\Phi = (dl)\,\hat{u} \cdot \vec{\nabla}\Phi = \left|\vec{\nabla}\Phi\right| dl \cos\theta \qquad (4.14)$$

By (4.14) we can define the *rate of change of $\Phi$ in the direction of $\hat{u}$*:

$$\frac{d\Phi}{dl} = \hat{u} \cdot \vec{\nabla}\Phi = \left|\vec{\nabla}\Phi\right| \cos\theta \qquad (4.15)$$

We notice that the rate of change is maximum when $\theta=0$, i.e., when the displacement $d\vec{r}$ is in the direction of $\vec{\nabla}\Phi$. Therefore,

> *the vector grad Φ determines the direction in which the rate of change of the function Φ is maximum.*

On the other hand, the rate of change of $\Phi$ vanishes when $\theta=\pi/2$, i.e., when $d\vec{r}$ is *normal* to $\vec{\nabla}\Phi$. This leads us to the following geometrical statement:

> *The vector grad Φ is normal to the surface Φ(x,y,z) = C (where C is a constant), at each point of this surface.*

Indeed, let $d\vec{r}$ be an infinitesimal vector tangent to this surface at some point of the surface. In the direction of $d\vec{r}$, $\Phi(x,y,z)=const.$ $\Rightarrow$ $d\Phi = (\vec{\nabla}\Phi)\cdot d\vec{r} = 0$, so that $\vec{\nabla}\Phi \perp d\vec{r}$. This condition is valid for every $d\vec{r}$ tangent to the surface $\Phi(x,y,z)=const.$ Thus, the vector $\vec{\nabla}\Phi$ is normal to this surface at each point of the surface.

Given a vector field $\vec{A}(\vec{r}) \equiv (A_x, A_y, A_z)$, we define the *scalar* field $div\,\vec{A}$ and the *vector* field $rot\,\vec{A}$ by the relations [1,2]

$$div\,\vec{A} = \vec{\nabla}\cdot\vec{A} = \frac{\partial A_x}{\partial x} + \frac{\partial A_y}{\partial y} + \frac{\partial A_z}{\partial z} \qquad (4.16)$$

$$rot\,\vec{A} = \vec{\nabla}\times\vec{A} = \left(\frac{\partial A_z}{\partial y} - \frac{\partial A_y}{\partial z}\right)\hat{u}_x + \left(\frac{\partial A_x}{\partial z} - \frac{\partial A_z}{\partial x}\right)\hat{u}_y + \left(\frac{\partial A_y}{\partial x} - \frac{\partial A_x}{\partial y}\right)\hat{u}_z \qquad (4.17)$$

Equation (4.17) is written, symbolically, in the form[2]

$$rot\,\vec{A} = \vec{\nabla}\times\vec{A} = \begin{vmatrix} \hat{u}_x & \hat{u}_y & \hat{u}_z \\ \dfrac{\partial}{\partial x} & \dfrac{\partial}{\partial y} & \dfrac{\partial}{\partial z} \\ A_x & A_y & A_z \end{vmatrix} \qquad (4.18)$$

As can be proven, the following vector identities are valid:

$$rot\,(grad\,\Phi) = \vec{\nabla}\times\vec{\nabla}\Phi = 0 \qquad (4.19)$$

[2] One must be careful when developing the determinant since, e.g., $(\partial/\partial x)A_y \neq A_y(\partial/\partial x)$! As a rule, the differential operator is placed *on the left* of the function to be differentiated.

$$div(rot\,\vec{A}) = \vec{\nabla}\cdot(\vec{\nabla}\times\vec{A}) = 0 \qquad (4.20)$$

Also,

$$div(grad\,\Phi) = \vec{\nabla}\cdot\vec{\nabla}\Phi = \frac{\partial^2\Phi}{\partial x^2} + \frac{\partial^2\Phi}{\partial y^2} + \frac{\partial^2\Phi}{\partial z^2} = \nabla^2\Phi \qquad (4.21)$$

where we have introduced the *Laplace operator* :

$$\nabla^2 = \vec{\nabla}\cdot\vec{\nabla} = \frac{\partial^2}{\partial x^2} + \frac{\partial^2}{\partial y^2} + \frac{\partial^2}{\partial z^2} \qquad (4.22)$$

## 4.2 Integral Theorems

Some regions of space possess a *boundary*, whereas others do not. For example, a spherical region in $R^3$ is bounded by a spherical surface, while a circular disk on the plane is bounded by its circular border. In general, the boundary of an *n*-dimensional region ($n$=1,2,3) is an ($n$–1)-dimensional region. In the case of a one-dimensional region such as a segment of a curve ($n$=1), the 0-dimensional boundary consists of the two end points of the segment.

But, what is the boundary of a spherical surface or of a circle? The answer is that these boundaries simply do not exist! According to a theorem in Topology, *the boundary of a region is a region without a boundary.*

There is a fundamental theorem in Differential Geometry, called (general) *Stokes' theorem*, which in general terms states the following:

> *The integral of the "derivative" of a field, over a region Ω possessing a boundary ∂Ω, equals the integral of the field itself over the boundary ∂Ω of Ω.*

The term *"derivative"* may refer to an ordinary derivative like *d*/*dx*, to a *grad*, to a *div*, or to a *rot*. Symbolically,

$$\int_{\Omega} \text{"derivative" of the field} = \int_{\partial\Omega} \text{field} \qquad (4.23)$$

Let us see some examples:

1. The boundary of a line segment ($ab$) is the set of end points $\{a, b\}$. Let $f(x)$ be a function defined in ($ab$). Relation (4.23) is written, in this case, as

$$\int_a^b f'(x)\,dx = \int_a^b \frac{df}{dx}\,dx = \left[\,f(x)\right]_a^b = f(b) - f(a) \qquad (4.24)$$

which is the familiar *Newton-Leibniz formula* of integral calculus.

2. Let $C$ be a curve in $R^3$, connecting points $a$ and $b$ (Fig. 4.2). We consider an infinitesimal displacement $\vec{dl} \equiv (dx, dy, dz)$ on $C$, oriented in the assumed direction of traversing the curve (say, from $a$ to $b$).

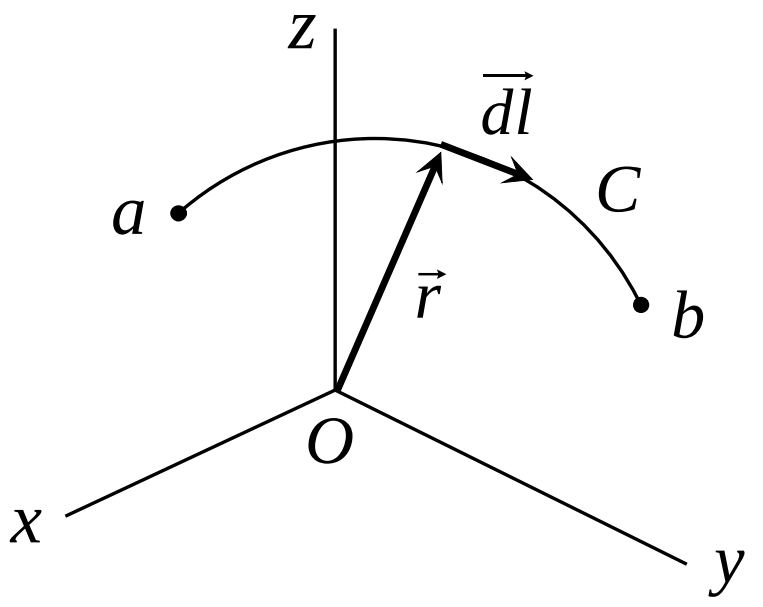

Fig. 4.2. Infinitesimal displacement on a curve in space.

If $\vec{A}(\vec{r}) = \vec{A}(x, y, z)$ is a vector field in $R^3$, the value of the *line integral* $\int_a^b \vec{A} \cdot \vec{dl}$ will depend, in general, on the choice of the path $C$ joining $a$ and $b$. Moreover, for a *closed* path $C$ (where the points $a$ and $b$ coincide) the closed line integral $\oint_C \vec{A} \cdot \vec{dl}$ will be generally different from zero. Consider now the special case where the field $\vec{A}$ is the *grad* of some scalar function $\Phi(x,y,z)$: $\vec{A} = \vec{\nabla} \Phi$. We then have:

$$\int_a^b (\vec{\nabla} \Phi) \cdot \vec{dl} = \int_a^b d\Phi = \Phi(b) - \Phi(a) \qquad (4.25)$$

$$\oint_C (\vec{\nabla} \Phi) \cdot \vec{dl} = 0 \qquad (4.26)$$

where we have taken (4.13) into account ($\vec{dl}$ here plays the role of $d\vec{r}$, since both represent an infinitesimal displacement in space). We notice that the value of the line integral in (4.25) depends only on the end points $a$ and $b$ of the path $C$ and is *independent* of the curve $C$ itself.

3. Consider a volume $V$ enclosed by a surface $S$ (Fig. 4.3). This *closed* surface constitutes the boundary of $V$. We let $dv$ be a volume element in the interior of $S$, and we let $da$ be an area element of $S$. At each point of $S$ we draw a vector $\vec{da}$ of magnitude $da$, normal to $S$ at the considered point. By convention, this vector, representing a *surface element*, is directed *outward*, i.e., toward the *exterior* of $S$.

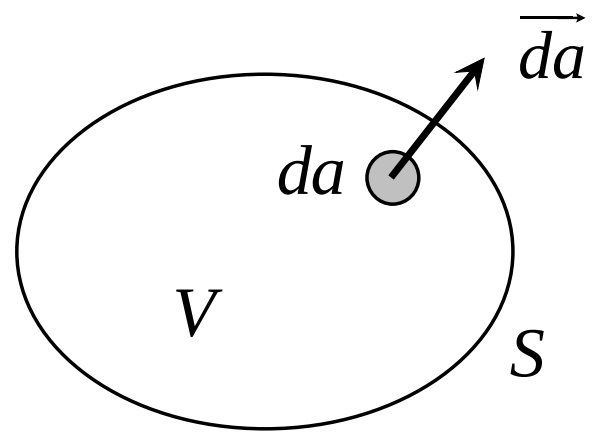

Fig. 4.3. A volume $V$ bounded by a closed surface $S$.

Let now $\vec{A}(\vec{r})$ be a vector field defined everywhere in *V* and on *S*. According to *Gauss' theorem* [1-3],

$$\boxed{\int_V (\vec{\nabla}\cdot\vec{A})\,dv = \oint_S \vec{A}\cdot\overrightarrow{da}} \qquad (4.27)$$

A surface integral of the form $\int_S \vec{A}\cdot\overrightarrow{da}$ (where, in general, the surface *S* may be open or closed) is called the *flux* of the vector field $\vec{A}$ through *S*.

4. Consider an *open* surface *S* bounded by a *closed* curve *C* (Fig. 4.4). We arbitrarily assign a positive direction of traversing *C* and we consider an element $\overrightarrow{dl}$ of *C* oriented in the positive direction. We also consider a surface element $\overrightarrow{da}$ of *S*, normal to *S*. We can choose, if we wish, the opposite direction for $\overrightarrow{da}$, provided that we simultaneously reverse the direction of $\overrightarrow{dl}$, i.e., the positive direction of traversing *C*. (Since *S* is an open surface, it is meaningless to say that $\overrightarrow{da}$ is pointing either "inward" or "outward".) The *relative* direction of $\overrightarrow{dl}$ and $\overrightarrow{da}$ is determined by the *right-hand rule*: if we rotate the fingers of our right hand in the positive direction of traversing the curve *C* (which direction is consistent with that of $\overrightarrow{dl}$), our extended thumb points in the direction of $\overrightarrow{da}$.

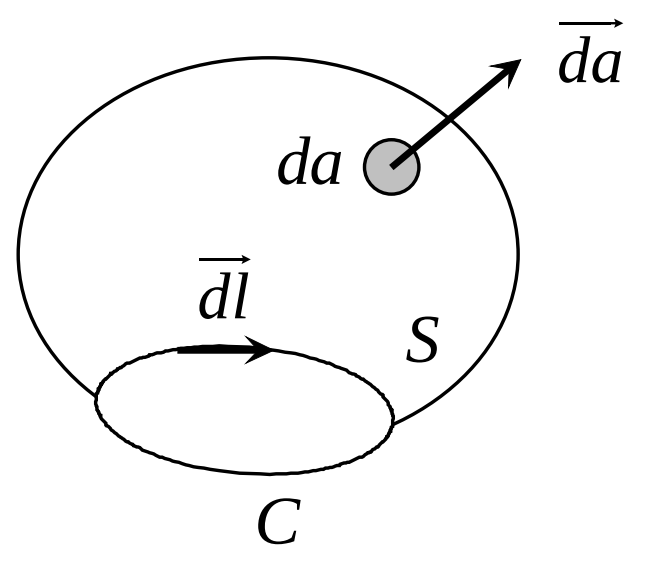

Fig. 4.4. An open surface *S* bounded by a closed curve *C*.

Now, if $\vec{A}(\vec{r})$ is a vector field defined everywhere on *S* and on *C*, then, according to the special form of *Stokes' theorem* [1-3],

$$\boxed{\int_S (\vec{\nabla}\times\vec{A})\cdot\overrightarrow{da} = \oint_C \vec{A}\cdot\overrightarrow{dl}} \qquad (4.28)$$

## 4.3 Irrotational and Solenoidal Vector Fields

A vector field $\vec{A}(\vec{r})$ is said to be *irrotational* if

$$\vec{\nabla}\times\vec{A} = 0 \qquad (4.29)$$

Then, under appropriate topological conditions[3], there exists a scalar function $\Phi(\vec{r})$ such that

$$\vec{A} = \vec{\nabla}\Phi \qquad (4.30)$$

[Notice that, then, $\vec{\nabla}\times\vec{A} = \vec{\nabla}\times\vec{\nabla}\Phi = 0$, in view of (4.19).] Furthermore, the value of a line integral of an irrotational field, along a curve connecting two points *a* and *b*, depends only on the limit points *a* and *b* of the curve (not on the curve itself), while any *closed* line integral of the field vanishes. Indeed:

$$\int_a^b \vec{A}\cdot\overrightarrow{dl} = \int_a^b (\vec{\nabla}\Phi)\cdot\overrightarrow{dl} = \int_a^b d\Phi = \Phi(b) - \Phi(a),$$

which is independent of the path $a\to b$. Moreover, by (4.29) and by Stokes' theorem (4.28), we have:

$$\oint \vec{A}\cdot\overrightarrow{dl} = 0$$

A vector field $\vec{B}(\vec{r})$ is said to be *solenoidal* if

$$\vec{\nabla}\cdot\vec{B} = 0 \qquad (4.31)$$

A vector function $\vec{A}(\vec{r})$ then exists such that

$$\vec{B} = \vec{\nabla}\times\vec{A} \qquad (4.32)$$

[Notice that, then, $\vec{\nabla}\cdot\vec{B} = \vec{\nabla}\cdot(\vec{\nabla}\times\vec{A}) = 0$, in view of (4.20).] Furthermore, the value of a surface integral (the *flux*) of a solenoidal field, on an open surface *S* bounded by a closed curve *C*, depends only on the border *C* of *S* (not on the surface *S* itself), while any *closed* surface integral of the field vanishes. Indeed, by using (4.32) and Stokes' theorem (4.28), we have:

$$\int_S \vec{B}\cdot\overrightarrow{da} = \int_S (\vec{\nabla}\times\vec{A})\cdot\overrightarrow{da} = \oint_C \vec{A}\cdot\overrightarrow{dl} \qquad (4.33)$$

Hence,

$$\int_{S_1} \vec{B}\cdot\overrightarrow{da} = \int_{S_2} \vec{B}\cdot\overrightarrow{da} \quad \textit{for any } S_1 \textit{ and } S_2 \textit{ having a common border } C \qquad (4.34)$$

Moreover, if *S* is a *closed* surface enclosing a volume *V*, relation (4.31) and Gauss' theorem (4.27) yield

$$\oint_S \vec{B}\cdot\overrightarrow{da} = \int_V (\vec{\nabla}\cdot\vec{B})\,dv = 0 \qquad (4.35)$$

---

[3] The spatial domain in which the components of $\vec{A}$ are differentiable functions must be *simply connected* [1,2,4].

Geometrical meaning

- *An irrotational vector field cannot have closed field lines: its field lines must be open.*

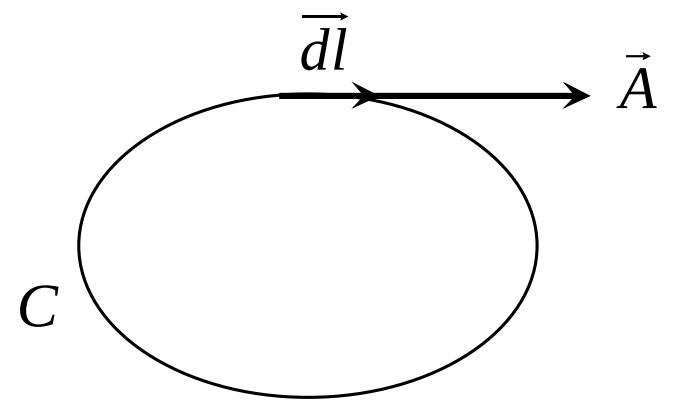

Fig. 4.5. A hypothetical closed field line *C*.

Indeed, let us assume that the irrotational field $\vec{A}(\vec{r})$ possesses a closed field line $C$ (Fig. 4.5). At each point of the field line, the field $\vec{A}$ is *tangent* to the line. If $\overrightarrow{dl}$ is an infinitesimal segment of the field line, we can consider that $\overrightarrow{dl}$ is tangent to the line, thus collinear with $\vec{A}$. Therefore we will have:

$$\oint_C \vec{A}\cdot\overrightarrow{dl} = \oint_C |\vec{A}||\overrightarrow{dl}| > 0$$

which is impossible, given that, for any irrotational field, $\oint \vec{A}\cdot\overrightarrow{dl} = 0$.

- *The field lines of a solenoidal field cannot have a beginning or an end in any finite region of space; either they are closed or they begin at infinity and end at infinity.*

Indeed, assume that a number of field lines of a solenoidal field $\vec{B}$ begin at some point of space. Consider a closed surface $S$ surrounding this point. Then, the *flux* of the field through $S$, proportional, by convention, to the number of field lines of $\vec{B}$ crossing $S$ (see, e.g., [3,5]) is

$$\oint_S \vec{B}\cdot\overrightarrow{da} \neq 0$$

which is impossible since it contradicts (4.35).

Physical meaning

- *A time-independent irrotational force field is conservative*

(this will be explained in the next section).

- *A solenoidal field cannot have isolated sources (poles).*

Indeed, the integral $\oint_S \vec{B}\cdot\overrightarrow{da}$ is a measure of the total strength of sources of a field $\vec{B}$ in the interior of a closed surface *S* (the field lines of $\vec{B}$ begin or end at these sources); see [3,5]. For a solenoidal field, however, the above integral vanishes on any closed surface *S*. Hence no field sources may exist inside *S* and, indeed, anywhere in space.

## 4.4 Conservative Force Fields

A *static* (time-independent) force field $\vec{F}(\vec{r})$ is *conservative* if the produced work $W_{AB}$ in moving a test particle from a point *A* to another point *B* in the field is independent of the path joining these points. Equivalently, the work done on the particle along any *closed* path *C* is zero:

$$W_{AB}=\int_A^B \vec{F}\cdot\overrightarrow{dl}\ \ \textit{independent of path}\ \Leftrightarrow\ \oint_C \vec{F}\cdot\overrightarrow{dl}=0 \tag{4.36}$$

Let *S* be an open surface bounded by the closed curve *C*. By Stokes' theorem and by (4.36) we have:

$$\oint_C \vec{F}\cdot\overrightarrow{dl}=\int_S(\vec{\nabla}\times\vec{F})\cdot\overrightarrow{da}=0$$

This relation must be valid for *every* open surface bounded by *C*. Thus we must have:

$$\boxed{\vec{\nabla}\times\vec{F}=0} \tag{4.37}$$

We conclude that

*a conservative force field is necessarily irrotational.*

From (4.36) it also follows that [2] there exists some scalar function such that $\vec{F}(\vec{r})$ is the *grad* of that function. We write:

$$\boxed{\vec{F}=-\vec{\nabla}U} \tag{4.38}$$

The function $U(\vec{r})=U(x,y,z)$ represents the *potential energy* of the test particle at the point $\vec{r}\equiv(x,y,z)$ of the field. [The negative sign in (4.38) is purely a matter of convention and has no special physical significance.]

The work $W_{AB}$ is written:

$$W_{AB}=\int_A^B \vec{F}\cdot\overrightarrow{dl}=-\int_A^B(\vec{\nabla}U)\cdot\overrightarrow{dl}=-\int_A^B dU\ \ \Rightarrow$$

$$W_{AB}=U(\vec{r}_A)-U(\vec{r}_B)\equiv U_A-U_B \tag{4.39}$$

Now, according to the *work-energy theorem*,

$$W_{AB} = T_B - T_A \tag{4.40}$$

where $T$ is the kinetic energy of the test particle. By combining (4.39) and (4.40), it is not hard to show that [4]

$$T_A + U_A = T_B + U_B \tag{4.41}$$

The sum $(T+U)$ represents the *total mechanical energy* of the test particle. Equation (4.41) then expresses the *principle of conservation of mechanical energy*:

> *In a conservative force field the total mechanical energy of a test particle is constant during the motion of the particle in the field.*

## References for Chapter 4

[4] Notice that, if we didn't put a negative sign in (4.38), this sign would inevitably appear in (4.41), compelling us to define the total mechanical energy as a *difference* rather than as a sum.

## QUESTIONS

**1.** Show that the *grad* is a linear operator: $\vec{\nabla}(f+g)=\vec{\nabla}f+\vec{\nabla}g$ , for any functions $f(x,y,z)$ and $g(x,y,z)$. Also show that $\vec{\nabla}(kf)=k\vec{\nabla}f$ , where $k$ is a constant.

**2.** Show that the *grad* satisfies the *Leibniz rule*: $\vec{\nabla}(fg)=g\vec{\nabla}f+f\vec{\nabla}g$ , for any functions $f(x,y,z)$ and $g(x,y,z)$. We say that the *grad* operator is a *derivation* on the set of all differentiable functions in $R^3$.

**3.** Prove the vector identities (4.19) and (4.20).

**4.** Give the physical and the geometrical significance of the concepts of an irrotational and a solenoidal vector field.

**5.** (*a*) Show that a conservative force field is necessarily irrotational. (*b*) Can a time-dependent force field $\vec{F}(\vec{r},t)$ be conservative, even if it happens to be irrotational? (*Hint:* Is work along a given curve a uniquely defined quantity in this case? See [4].)

## CHAPTER 5

# STATIC ELECTRIC FIELDS

### 5.1 Coulomb's Law and Electric Field

Consider two electric charges $q_1$, $q_2$ a distance $r$ apart (Fig. 5.1). Let $\hat{r}$ be the unit vector in the direction from $q_1$ to $q_2$. We call $\vec{F}_{12}$ the electric force exerted *by* $q_1$ *on* $q_2$. According to *Coulomb's law*, this force is given in S.I. units by the expression[1]

$$\vec{F}_{12} = \frac{1}{4\pi\varepsilon_0} \frac{q_1 q_2}{r^2} \hat{r} \qquad (5.1)$$

where $\varepsilon_0$=8.85×10$^{-12}$ $C^2/N.m^2$. The force is attractive if the charges have opposite signs ($q_1 q_2$<0), while it is repulsive for charges of the same sign ($q_1 q_2$>0). We note that, in particular, the charge of an electron is $-q_e$, where $q_e$=1.6×10$^{-19}$ $C$.

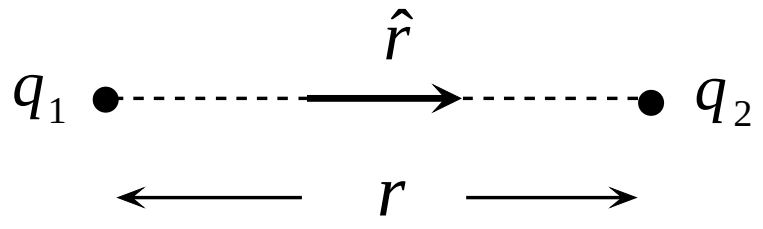

Fig. 5.1. Two charges a distance *r* apart.

We say that an electric field exists in a region of space if any *stationary* test charge $q_0$ in this region is subject to a force that, in general, varies from point to point in the region. Let $\vec{F}$ be the force on $q_0$ at some given point. We define the *electric field* at that point as the force per unit charge,

$$\vec{E} = \frac{\vec{F}}{q_0} \qquad (5.2)$$

In S.I. units the magnitude of the electric field is expressed in $N/C$, as follows from the above definition.

The field $\vec{E}$ that exerts the force $\vec{F}$ on the test charge $q_0$ is produced by some system of charges that does not contain $q_0$. According to Coulomb's law, the force on $q_0$ is proportional to $q_0$. Thus the quotient $\vec{F}/q_0$ is eventually independent of $q_0$. That is, the vector $\vec{E}$ defined in (5.2) expresses a property *of the electric field itself* and is independent of the test charge used to determine the field. It is also obvious that the direction of the electric field is that of the force on a *positive* charge.

---

[1] See Appendix A.

*Example:* Electric field produced by a point charge $q$

We consider a *positive* test charge $q_0$ at a point $P$ a distance $r$ from $q$ (Fig. 5.2). We call $\hat{r}$ the unit vector in the direction from $q$ to $q_0$. We draw the cases $q>0$ and $q<0$ separately.

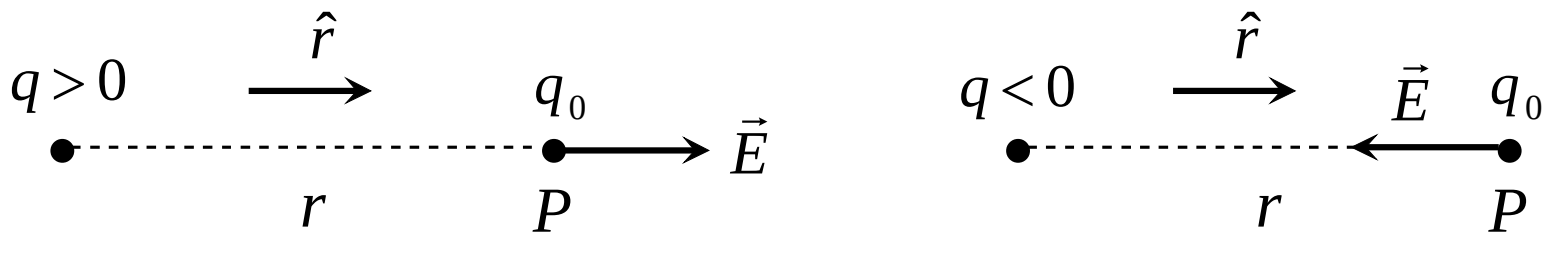

Fig. 5.2. Electric fields produced by a positive and by a negative charge $q$; the test charge $q_0$ is positive.

According to (5.1), the Coulomb force exerted on $q_0$ by $q$ is

$$\vec{F} = \frac{1}{4\pi\varepsilon_0} \frac{q_0 q}{r^2} \hat{r}$$

Therefore the electric field at point $P$ is $\vec{E} = \vec{F} / q_0 \Rightarrow$

$$\boxed{\vec{E} = \frac{1}{4\pi\varepsilon_0} \frac{q}{r^2} \hat{r}} \tag{5.3}$$

An electric field of the form (5.3) is called a *Coulomb field*.

More generally, we consider a set of point charges $q_1, q_2, \ldots$, as well as a test charge $q_0$ at point $P$ (Fig. 5.3). We call $r_1, r_2, \ldots$, the distances of $q_0$ from the corresponding charges.

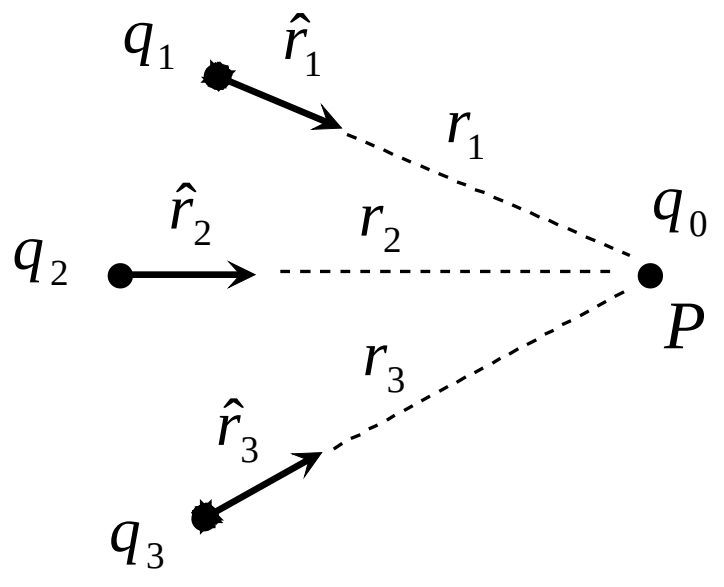

Fig. 5.3. A set of point charges exerting electric forces on a test charge $q_0$.

The total force on $q_0$ is

$$\vec{F} = \sum_i \vec{F}_i = \sum_i \frac{1}{4\pi\varepsilon_0} \frac{q_0 q_i}{r_i^2} \hat{r}_i = q_0 \frac{1}{4\pi\varepsilon_0} \sum_i \frac{q_i}{r_i^2} \hat{r}_i$$

Hence the electric field at point $P$ is $\vec{E} = \vec{F} / q_0 \Rightarrow$

$$\vec{E} = \frac{1}{4\pi\varepsilon_0} \sum_i \frac{q_i}{r_i^2} \hat{r}_i = \sum_i \vec{E}_i \qquad (5.4)$$

where $\vec{E}_i$ is the Coulomb field (5.3) due to $q_i$. We note that the *principle of superposition* is valid, according to which

> *the electric field due to a set of charges equals the vector sum of the fields due to each charge separately.*

An electric field is said to be *static* if the field vector $\vec{E}$ is *time-independent* at all points (this vector may change, however, from one point to another). Thus, for a static field, $\partial\vec{E}/\partial t = 0 \Leftrightarrow \vec{E} = \vec{E}(\vec{r})$. In particular, the static electric field produced by a set of point charges *at rest* with respect to an inertial observer is called *electrostatic*.

On the other hand, an electric field is said to be *uniform* if, at any time, the field is *spatially constant*, i.e., has the same vector value everywhere.

## 5.2 Gauss' Law

An immediate consequence of Coulomb's law is Gauss' law. Mathematically, it corresponds to the first of the Maxwell equations (Chap. 9).

Consider a volume *V* bounded by a closed surface *S* (Fig. 5.4). An element of *S* is represented by a vector $\overrightarrow{da}$ normal to *S*, directed *outward* and having magnitude $|\overrightarrow{da}|=da$, where $da$ denotes an elementary area (see Sec. 4.2).

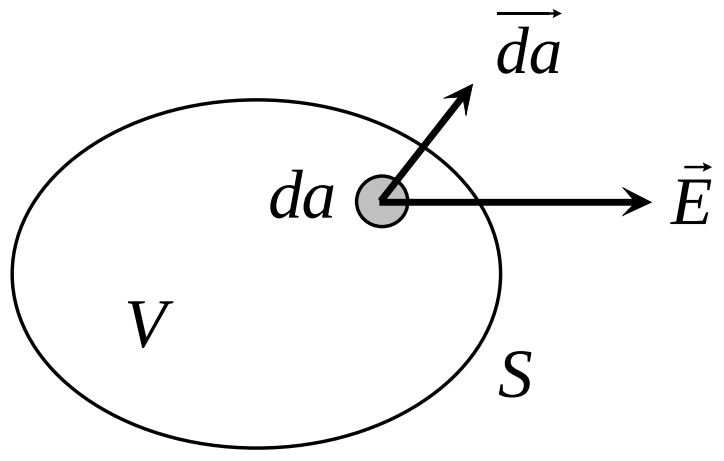

Fig. 5.4. A closed surface *S* lying inside an electric field.

The surface *S* lies inside an electric field $\vec{E}(\vec{r})$. This field is produced by a system of charges located both in the interior and in the exterior of *S*. We call $Q_{in}$ the *total charge enclosed by S*. We also define the *flux of* $\vec{E}$ *through S* as the surface integral $\oint_S \vec{E} \cdot \overrightarrow{da}$. According to *Gauss' law*, this flux depends only on the *internal* charge $Q_{in}$. Specifically,

$$\boxed{\oint_S \vec{E} \cdot \overrightarrow{da} = \frac{Q_{in}}{\varepsilon_0}} \qquad (5.5)$$

It should be noted carefully that, although in (5.5) only the total charge in the interior of *S* appears, the vector $\vec{E}$ represents the electric field due to *all* charges, both inside *and* outside of *S*! The external charges, however, do not contribute to the flux through the closed surface *S*. This can be explained as follows:

First, we define an *electric field line* as a curve at every point of which the electric field vector is a *tangent* vector. The flux of the electric field through *S* is, by convention, proportional to the "number" of field lines passing through *S*. Furthermore, the field lines of the Coulomb field of a point charge either begin or end at the charge, depending on whether the charge is positive or negative, respectively.[2] Now, for a charge in the *exterior* of *S*, every field line of its own Coulomb field crosses the surface *twice*, once entering *S* and once going out. Hence, whatever flux goes in comes out and, as a result, the total flux crossing the closed surface is zero. That is, an external charge does not contribute to the total flux through *S*. On the contrary, every field line of a charge in the *interior* of *S* crosses the surface only *once*, and thus it contributes a non-zero flux through *S*.

We mentioned earlier that Gauss' law is an immediate consequence of Coulomb's law (see, e.g., [1,3]). We now demonstrate the converse, i.e., that Gauss' law yields Coulomb's law. To this end, we consider a point charge *q* placed at the center of an imaginary spherical surface *S* of radius *r* (Fig. 5.5). Without loss of generality, we assume that the charge *q* is positive.

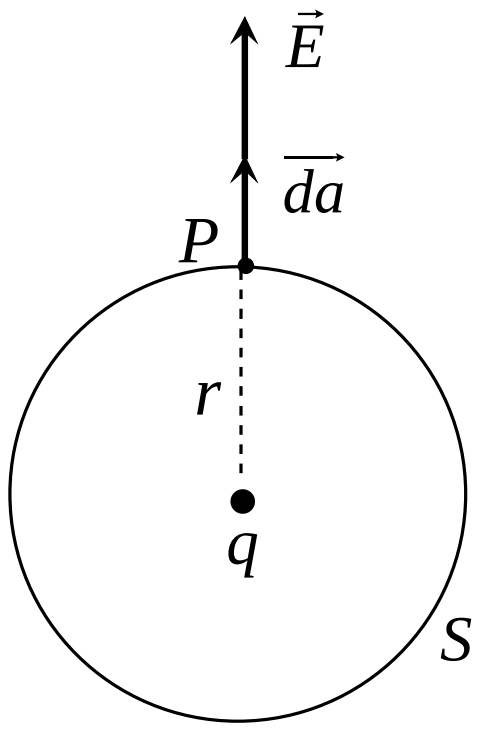

Fig. 5.5. A positive charge *q* at the center of a spherical surface *S*.

Let *P* be any point of the spherical surface. By symmetry, the electric field $\vec{E}$ at *P*, produced by *q*, will be normal to *S*, thus parallel to the surface element $\overrightarrow{da}$. Furthermore, the magnitude $E=|\vec{E}\,|$ of the electric field will be constant over the surface *S*. By applying (5.5) with $Q_{in}= q$, we have:

$$\frac{q}{\varepsilon_0} = \oint_S \vec{E}\cdot\overrightarrow{da} = \oint_S |\vec{E}\,||\overrightarrow{da}\,| = |\vec{E}\,| \oint_S da = E\,(4\pi r^2) \;\Rightarrow\; E = \frac{q}{4\pi\varepsilon_0 r^2}$$

If $\hat{r}$ is a unit vector normal to *S* at *P*, directed toward the exterior of *S* (in other words, having the direction of $\overrightarrow{da}$), we write:

[2] Of course, an *infinite* number of field lines begin or end at a given charge *q*! By *"number"* of lines we actually refer to the *density* of these lines, which is conventionally proportional to *q*. (See, e.g., [1,2].)

$$\vec{E} = E\,\hat{r} = \frac{1}{4\pi\varepsilon_0}\frac{q}{r^2}\,\hat{r}$$

which is precisely relation (5.3) for the Coulomb field. Finally, if $q_0$ is a test charge placed at *P*, the force exerted on $q_0$ by the field $\vec{E}$ is

$$\vec{F} = q_0\vec{E} = \frac{1}{4\pi\varepsilon_0}\frac{q_0 q}{r^2}\,\hat{r}$$

This is, of course, Coulomb's law.

It is evident from the above simple example that Gauss' law in the integral form (5.5) can be very useful in problems with a *high degree of symmetry*. In such cases we may choose the closed surface *S* so that the electric field has a constant value *E* on *S*, which allows us to take the quantity *E* out of the integral in (5.5). Unfortunately, however, problems with symmetry are an exception rather than a rule in Electricity! For this reason we will try to convert Eq. (5.5) into an equivalent equation where the electric field to be determined will appear in terms of its *derivative*, rather than its integral. We seek, that is, a *differential* equation equivalent to Gauss' law.

An electric field is produced by a system of charges which at a given time are located at certain positions relative to an inertial observer. Charges are usually visualized as isolated pointlike quantities, as is evident from the form of Eqs. (5.1), (5.3) and (5.4). In physical reality, however, we more often encounter *continuous distributions of charge* rather than distributions of discrete point charges. This leads us to the concept of density of charge, defined as follows:

Let *dv* be an elementary volume at some point *P* of space with coordinates (*x*,*y*,*z*) relative to our frame of reference. We call *dq* the elementary charge contained within *dv*. The *charge density* at point *P* is[3]

$$\rho(\vec{r}) = \rho(x,y,z) = \frac{dq}{dv} \tag{5.6}$$

where $\vec{r}$ is the position vector of *P*. For variable *P*, the density $\rho(\vec{r})$ is a scalar function. Note that, in the case of a *point charge* at *P*, the value of this function is *infinite* at that point since $dv=0$ while $dq\neq 0$ there. We will therefore *not* use the concept of charge density for point charges.[4] The total charge within a finite volume *V* is

$$Q_{in} = \int dq = \int_V \rho(\vec{r})\,dv \tag{5.7}$$

We now return to Gauss' law (5.5). By Gauss' integral theorem (4.27), the left-hand side of Eq. (5.5) is written

[3] This is the *volume* charge density. For charge distributed over a surface, a *surface* charge density may also be defined [1,3].
[4] This would require the use of an "exotic" function, the Dirac "delta function" [1,4-6].

$$\oint_S \vec{E} \cdot \vec{da} = \int_V (\vec{\nabla} \cdot \vec{E})\, dv$$

By using the integral representation (5.7) for $Q_{in}$, Eq. (5.5) then takes the form

$$\int_V (\vec{\nabla} \cdot \vec{E})\, dv = \int_V \frac{1}{\varepsilon_0} \rho\, dv$$

For this to be true for an *arbitrary* volume *V*, the integrands on the two sides of the above equation must be equal:

$$\boxed{\vec{\nabla} \cdot \vec{E} = \frac{\rho}{\varepsilon_0}} \qquad (5.8)$$

Relation (5.8) expresses *Gauss' law in differential form*. It relates the *div* of the electric field at some point of space with the charge density at that point.

Gauss' law (5.8) is of general validity, both for static and for time-dependent electric fields $\vec{E}(\vec{r},t)$. Thus the charge density is generally assumed to be of the form $\rho(\vec{r},t)$. We now show that

> *in a region of space where a static electric field exists, the distribution of charge is static (time-independent).*

Indeed, from (5.8) we have that

$$\frac{\partial \rho}{\partial t} = \varepsilon_0 \frac{\partial}{\partial t} (\vec{\nabla} \cdot \vec{E}) = \varepsilon_0 \vec{\nabla} \cdot \frac{\partial \vec{E}}{\partial t}$$

where we have used the fact that the partial derivatives of a continuous function commute with one another. Now, in the case of a static $\vec{E}$-field we have $\partial \vec{E} / \partial t = 0$, so that $\partial \rho / \partial t = 0$. That is, the charge distribution, represented by $\rho$, must be static.

The converse is *not* true, however. Indeed, in a region of space with a static distribution of charge (or even with no charge at all) the electric field can still change with time if there is a time-dependent distribution of charge *outside* the region.

## 5.3 Electrostatic Potential

As we have seen, Gauss' law (5.8) is a consequence of Coulomb's law (5.1); it relates the *div* of the electric field with the distribution of charge in some region of space. From Coulomb's law also follows a relation for the *rot* of the *electrostatic* field.

*Proposition:* In an electrostatic field $\vec{E}(\vec{r})$ the line integral $\int_a^b \vec{E} \cdot \vec{dl}$ is independent of the path connecting the points *a* and *b*. Equivalently, for every closed path *C* within the field,

$$\oint_C \vec{E} \cdot \overrightarrow{dl} = 0 \tag{5.9}$$

It follows that *the electrostatic field is irrotational:*

$$\boxed{\vec{\nabla} \times \vec{E} = 0} \tag{5.10}$$

*Proof:* We recall that an electrostatic field is a static (time-independent) electric field produced by a system of charges that are *at rest* with respect to an inertial observer.[5] Let us examine first the simple case of the electrostatic field produced by a single point charge $q$. Consider a curve within this field, extending from $a$ to $b$, as well as a point of this curve having position vector $\vec{r}$ relative to $q$ (Fig. 5.6).

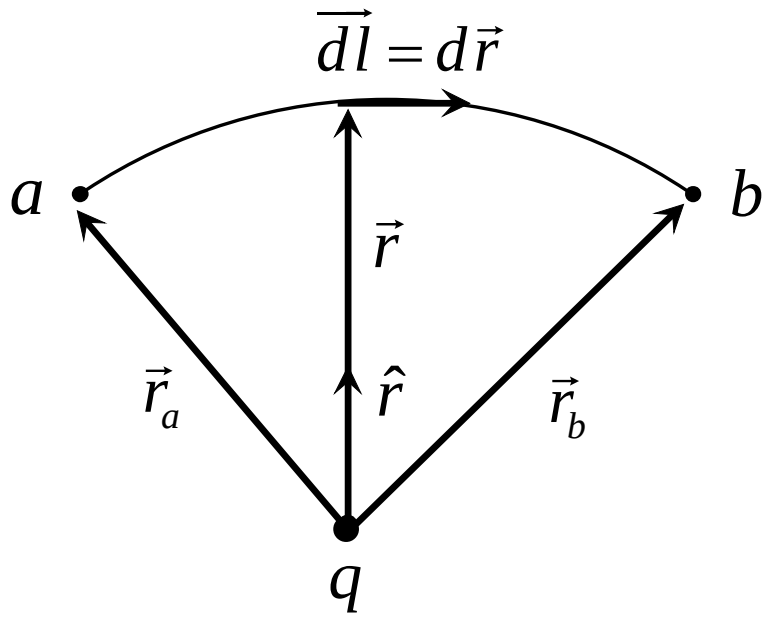

Fig. 5.6. A curve *ab* inside the electric field produced by a point charge *q*.

The Coulomb field of $q$ at the considered point is

$$\vec{E} = \frac{1}{4\pi\varepsilon_0} \frac{q}{r^2} \hat{r} \equiv E(r)\, \hat{r} \tag{5.11}$$

where $r = |\vec{r}\,|$. But, $\hat{r} = \frac{\vec{r}}{|\vec{r}|} = \frac{\vec{r}}{r}$ , so that

$$\vec{E} = \frac{E(r)}{r} \vec{r} \tag{5.12}$$

Let now $\overrightarrow{dl}$ be an infinitesimal, positively oriented (i.e., from $a$ to $b$) section of the curve. The element $\overrightarrow{dl}$ represents an infinitesimal change of the position vector $\vec{r}$ along the curve: $\overrightarrow{dl} = d\vec{r}$ . Hence,

$$\vec{E} \cdot \overrightarrow{dl} = \vec{E} \cdot d\vec{r} = \frac{E(r)}{r} \vec{r} \cdot d\vec{r}$$

But,

$$\vec{r} \cdot d\vec{r} = \frac{1}{2} d(\vec{r} \cdot \vec{r}) = \frac{1}{2} d(|\vec{r}\,|^2) = \frac{1}{2} d(r^2) = \frac{1}{2}(2r dr) \;\Rightarrow$$

[5] In Chapter 9 we will learn that an electric field (even a static one) can be "generated" by other means; specifically, by a time-dependent magnetic field.

$$\vec{r}\cdot d\vec{r} = r\,dr \tag{5.13}$$

Thus,

$$\vec{E}\cdot\overrightarrow{dl} = \frac{E(r)}{r}\,r\,dr = E(r)\,dr$$

and

$$\int_a^b \vec{E}\cdot\overrightarrow{dl} = \int_a^b E(r)\,dr = \frac{q}{4\pi\varepsilon_0}\int_a^b \frac{dr}{r^2} = \frac{q}{4\pi\varepsilon_0}\left(\frac{1}{r_a}-\frac{1}{r_b}\right) \tag{5.14}$$

regardless of the curve joining *a* and *b*. In the case of a *closed* curve, the points *a* and *b* coincide and the right-hand side of (5.14) vanishes. Thus (5.9) is indeed valid for the Coulomb field of a point charge. Now, if we have several charges $q_1$, $q_2$,..., at rest with respect to the observer, the total electric field they produce is the vector sum of the fields due to each charge separately, as required by the principle of superposition: $\vec{E} = \vec{E}_1 + \vec{E}_2 + \cdots$. Therefore,

$$\oint \vec{E}\cdot\overrightarrow{dl} = \oint \vec{E}_1\cdot\overrightarrow{dl} + \oint \vec{E}_2\cdot\overrightarrow{dl} + \cdots = 0 + 0 + \cdots = 0$$

We conclude that relation (5.9) is valid for any electric field produced by a static distribution of charge; that is, for any *electrostatic* field.

We now consider an open surface *S* inside an electrostatic field, bounded by a closed curve *C*. By combining (5.9) with Stokes' theorem (4.28), we have:

$$\int_S (\vec{\nabla}\times\vec{E})\cdot\overrightarrow{da} = \oint_C \vec{E}\cdot\overrightarrow{dl} = 0$$

If the surface integral on the left is to vanish for *any* open surface *S* bounded by *C*, the integrand $\vec{\nabla}\times\vec{E}$ itself must vanish; i.e., Eq. (5.10) must be satisfied. This means that the electrostatic field is irrotational.

*Comment:* As we have said, Gauss' law (5.8) is a mathematical expression of Coulomb's law. On the other hand, relation (5.10) has to do not as much with Coulomb's law itself as with the fact that the Coulomb force (5.1) is a *central* force, i.e., is of the form $\vec{F} = F(r)\,\hat{r}$. This reflects in the form (5.11) of the Coulomb field.

Since the electrostatic field $\vec{E}(\vec{r})$ is irrotational, it can be expressed as the *grad* of a scalar field (cf. Sec. 4.3). We write:

$$\boxed{\vec{E} = -\vec{\nabla}V} \tag{5.15}$$

(the negative sign is a matter of convention and has no special physical significance). The function $V(\vec{r}) = V(x,y,z)$ is called the *electric potential*. We note that, for a given field $\vec{E}(\vec{r})$, the potential *V* is not uniquely defined since the functions $V(\vec{r})$ and $V'(\vec{r}) = V(\vec{r}) + C$ (where *C* is a constant quantity) yield the same vector function $\vec{E}(\vec{r})$ when substituted into (5.15).

For an elementary displacement $\overrightarrow{dl} = d\vec{r}$ , we have:

$$\vec{E}\cdot\overrightarrow{dl} = -(\vec{\nabla}V)\cdot\overrightarrow{dl} = -\,dV \quad \Leftrightarrow \quad dV = -\vec{E}\cdot\overrightarrow{dl} \tag{5.16}$$

Thus, along a curve from $a$ to $b$, $\int_a^b \vec{E}\cdot\overrightarrow{dl} = -\int_a^b dV \;\Rightarrow$

$$\boxed{\int_a^b \vec{E}\cdot\overrightarrow{dl} = V_a - V_b} \tag{5.17}$$

Notice that the line integral (5.17) is independent of the path connecting *a* and *b*, as required. The quantity ($V_a - V_b$) represents the *potential difference* (also called *voltage* in electric-circuit theory) between the points $a$ and $b$. In S.I. units, electric potential is measured in *Volts* (*V*).

*Example:* Potential of the Coulomb field

The Coulomb field produced by a point charge *q* is given by (5.11), while its line integral from *a* to *b* is given by (5.14). Combining the latter relation with (5.17), we have:

$$\int_a^b \vec{E}\cdot\overrightarrow{dl} = \frac{1}{4\pi\varepsilon_0}\left(\frac{q}{r_a} - \frac{q}{r_b}\right) = V_a - V_b$$

Therefore the electric potential at a point in the field is of the form

$$V(r) = \frac{1}{4\pi\varepsilon_0}\frac{q}{r} + C$$

where *r* is the distance of that point from *q* and where *C* is an arbitrary constant quantity. By arbitrarily assuming that *V*=0 at infinity (i.e., $V \to 0$ for $r \to \infty$) we have that $C=0$ and that, therefore,

$$\boxed{V(r) = \frac{1}{4\pi\varepsilon_0}\frac{q}{r}} \tag{5.18}$$

We notice that the Coulomb potential assumes a constant value over a spherical surface of radius *r*, centered at *q*.

If now we have a system of stationary charges $q_1$, $q_2$,..., the potential of their electrostatic field will be the sum of the potentials due to each charge separately [this follows from the superposition principle for the electric field, in combination with Eq. (5.15)]. Thus, if $r_i$ is the distance of a point *P* from the charge $q_i$, the electric potential at *P* will be

$$V = \frac{1}{4\pi\varepsilon_0}\sum_i \frac{q_i}{r_i} \tag{5.19}$$

As we already know, the *electric field lines* are curves whose orientation coincides with the direction of the electric field $\vec{E}$ at every point of space. Thus, at every point $P$ the vector $\vec{E}$ is tangent to the *unique* field line passing through $P$. Since the electrostatic field is irrotational, its field lines *cannot* be closed (Sec. 4.3). These lines begin at positive charges and/or terminate at negative charges.

An *equipotential surface* is a surface over which the electric potential $V$ assumes a constant value. In other words, it is the locus of points with coordinates $(x,y,z)$ at which $V(x,y,z)=C$, for a given constant $C$. We now show that

*an equipotential surface intersects normally the electric field lines.*

Indeed: Given that $\vec{E} = -\vec{\nabla} V$ and that the field $\vec{E}$ is tangent to the electric field lines everywhere, these lines will be directed parallel to $\vec{\nabla} V$ at all points in the field. On the other hand, the *grad* of $V$ is normal to the surface $V(x,y,z)=C$, at every point of this surface (Sec. 4.1). It follows that the field lines themselves are normal to the equipotential surface.

In the special case of the Coulomb field of a point charge $q$, the equipotential surfaces are spherical surfaces centered at $q$, according to (5.18). The electric field lines extend radially from $q$ and thus are normal to the equipotential surfaces.

## 5.4 Poisson and Laplace Equations

As we saw in previous sections, the electrostatic field $\vec{E}(\vec{r})$ obeys two fundamental differential equations:

$$\vec{\nabla} \cdot \vec{E} = \frac{\rho}{\varepsilon_0} \qquad (5.20)$$

$$\vec{\nabla} \times \vec{E} = 0 \qquad (5.21)$$

Gauss' law (5.20) is an immediate consequence of Coulomb's law, while (5.21) states that the electrostatic field is irrotational (among other things, this means that this field cannot have closed field lines).

None of the above equations can by itself determine the field $\vec{E}$. Indeed, according to a theorem by *Helmholtz* [1,3] a complete determination of a static vector field requires simultaneous knowledge of *both* the *div* and the *rot* of the field, together with suitable *boundary conditions* (for example, the value of $\vec{E}$ must approach zero as we get further away from the charges that create the field).

Equation (5.21) yields a partial solution to the problem:

$$\exists\, V(\vec{r}):\ \vec{E} = -\vec{\nabla} V \qquad (5.22)$$

where $V$ is the electric potential. To determine $V$ we substitute (5.22) into (5.20):

$$\vec{\nabla}\cdot(-\vec{\nabla}V) = -\nabla^2 V = \frac{\rho}{\varepsilon_0} \;\Rightarrow$$

$$\boxed{\nabla^2 V = -\frac{\rho}{\varepsilon_0}} \tag{5.23}$$

where $\nabla^2$ is the Laplace operator [see Eq. (4.22)]. Equation (5.23) is called the *Poisson equation*. In a region with no electric charges (although *outside* this region there *are* charges, those that create the electric field in the first place!) we have $\rho(\vec{r}) = 0$ at all points, so that (5.23) reduces to the *Laplace equation*:

$$\boxed{\nabla^2 V \equiv \frac{\partial^2 V}{\partial x^2} + \frac{\partial^2 V}{\partial y^2} + \frac{\partial^2 V}{\partial z^2} = 0} \tag{5.24}$$

Generally speaking, for a given charge distribution $\rho(\vec{r})$ and for given boundary conditions for *V*, the differential equation (5.23) uniquely determines the potential $V(\vec{r})$. The electric field is then the $-grad$ of the potential, according to (5.22).

## 5.5 Electrostatic Potential Energy

In Sec. 4.4 we defined the concept of a *conservative force field* $\vec{F}(\vec{r})$ and we showed that a field of this kind is necessarily *irrotational*: $\vec{\nabla}\times\vec{F} = 0$. A scalar function $U(\vec{r})$ then exists such that the force on a test particle at the point $\vec{r} \equiv (x, y, z)$ of the field is expressed as the $-grad$ of *U* at that point: $\vec{F} = -\vec{\nabla}U$. The function *U* is called the *potential energy* of the particle.

We now show that

*the electrostatic field is conservative.*

Indeed, let *q* be a test charge inside an electrostatic field $\vec{E}(\vec{r})$. The force exerted on *q* by the field is

$$\vec{F}(\vec{r}) = q\,\vec{E}(\vec{r}) = q\,(-\vec{\nabla}V) = -\vec{\nabla}(qV) \equiv -\vec{\nabla}U(\vec{r})$$

where $V(\vec{r})$ is the electrostatic potential and where (5.15) has been used. We observe that a *potential energy* of *q* can be defined by

$$\boxed{U(\vec{r}) = q\,V(\vec{r})} \tag{5.25}$$

This is precisely the condition in order that the field be conservative. It follows that the *total mechanical energy* (*T+U*) of the test charge *q* is constant in time. Moreover, the field produces no work on *q* when the charge describes a *closed* path. Equiva-lently, the work done on $q$ when it moves from point $A$ to point $B$ in the field is independent of the specific curve joining these points, as follows from (4.39) and (5.25):

$$W_{AB} = U_A - U_B = q\,V_A - q\,V_B = q\,V_{AB} \qquad (5.26)$$

where $V_{AB} = V_A - V_B$ is the potential difference between the two points.

*Comment:* Note that the potential $V$ is a property of the electrostatic field itself, while the potential energy $U$ is a property of the charge $q$ within this field.

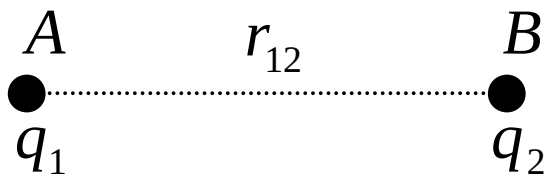

Fig. 5.7. Two mutually interacting charges.

Consider now two point charges $q_1$ and $q_2$ , a distance $r_{12}$ apart (Fig. 5.7). The charges are located at points $A$ and $B$, respectively. The potential energy of $q_1$ in the Coulomb field produced by $q_2$ is

$$U_{12} = q_1 V_2(A) = q_1 \left( \frac{1}{4\pi\varepsilon_0} \frac{q_2}{r_{12}} \right)$$

while the potential energy of $q_2$ in the Coulomb field of $q_1$ is

$$U_{21} = q_2 V_1(B) = q_2 \left( \frac{1}{4\pi\varepsilon_0} \frac{q_1}{r_{12}} \right)$$

We observe that $U_{12} = U_{21} = U$, where

$$U = \frac{1}{4\pi\varepsilon_0} \frac{q_1 q_2}{r_{12}} \qquad (5.27)$$

The quantity $U$ represents the *potential energy of the system of charges* $q_1, q_2$ . More generally, the potential energy of a system of charges $q_1, q_2, \ldots$, is given by the expression

$$U = \frac{1}{4\pi\varepsilon_0} \sum_{i<j} \frac{q_i q_j}{r_{ij}} \qquad (5.28)$$

Physically, this quantity represents the work necessary in order to compose a system of mutually interacting charges in some limited region of space, by bringing each (initially isolated) charge from infinity. (We arbitrarily assume that $U = 0$ when $r_{ij} = \infty$.)

By using (5.25) we can define the S.I. unit of potential as follows:

$$1\,Volt = 1\,Joule\,/\,Coulomb \quad \text{(or, } 1\,V = 1\,J\,/\,C\text{, for short)}$$

We notice that

$$1V = 1\frac{N \cdot m}{C} \quad \Rightarrow \quad 1\frac{N}{C} = 1\frac{V}{m} \qquad (\textit{unit of electric field strength})$$

## 5.6 Metallic Conductor in Electrostatic Equilibrium

As we learned in Chapter 2, the electrical conductivity of a metal owes itself to the *free electrons* of the metal. These are valence electrons of the atoms of the metal that have been detached from the atoms to which they belong and move freely inside the crystal lattice. The metallic conductor is in *electrostatic equilibrium* if there is no *macroscopic* motion of charge in its interior and on its surface (with the exception of irregular thermal motions of the electrons).

A conductor in electrostatic equilibrium exhibits the following characteristics (see Fig. 5.8):

1. The electric field in the interior of the conductor is zero, while just outside the conductor the field is normal to the surface of the conductor.
2. The charge density in the interior of the conductor is zero. Thus, nonzero net charge may only exist on the surface of the conductor.
3. The surface of the conductor is an equipotential surface. Furthermore, the potential in the interior of the conductor is constant (has the same value everywhere) and equal to the potential on the surface.

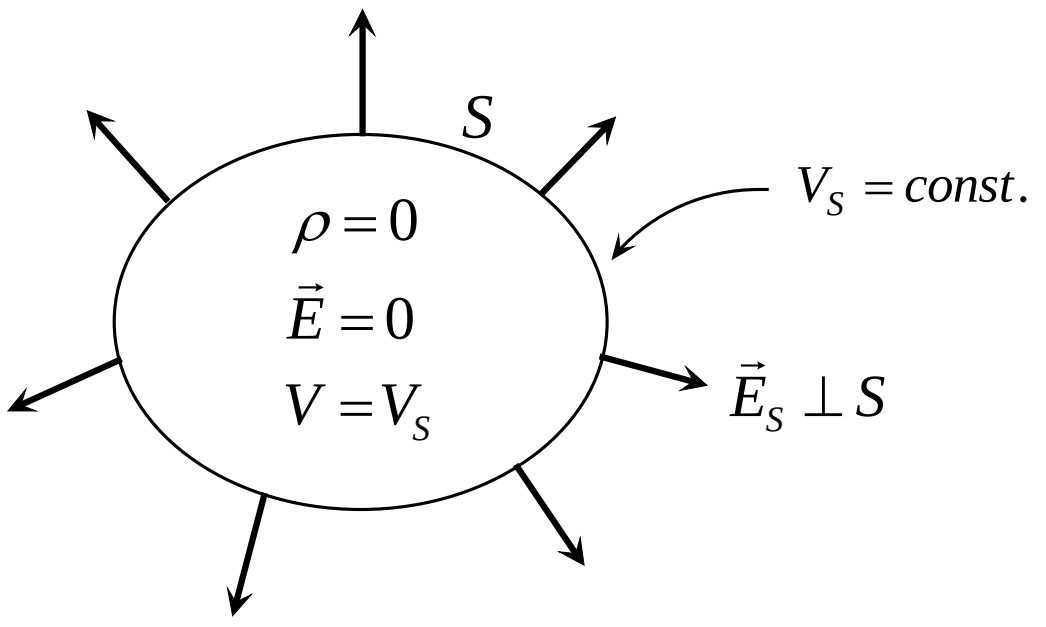

Fig. 5.8. Electrostatic properties on the surface *S* and in the interior of a conductor.

*Proof:* (1) If there were a nonzero electric field in the interior of the conductor it would set the free charges in motion; thus no state of electrostatic equilibrium would exist inside the conductor. On the other hand, if the electric field just outside the conductor were not normal to the surface of the conductor, the field would have a component tangent to the surface, which would set in motion the free electrons on the surface. (2) In the interior of the metal, $\vec{E} = 0$. Then, according to Gauss' law (5.8), it must be $\rho = 0$ everywhere inside the conductor. (3) Since the electric field just outside the metal is normal to the surface *S* of the metal, this surface intersects normally the electric field lines, thus is an equipotential surface (see Sec. 5.3) having a constant potential $V_S$. We consider a point $a$ on the surface and a point $b$ in the interior of the conductor, as well as an arbitrary path joining these points and totally located inside the conductor. The potential difference between $a$ and $b$ is found by using (5.17) and by taking into account that $\vec{E}=0$ in the interior of the metal:

$$V_a - V_b = \int_a^b \vec{E}\cdot\overrightarrow{dl} = 0 \quad \Rightarrow \quad V_b = V_a = V_S$$

since $V_a=V_S$ at any point $a$ on the surface. Hence $V_b=V_S$, at any point $b$ in the interior of the conductor.

We now consider a conductor carrying a surface charge $Q$. Let $V$ be the (constant) electric potential throughout the space occupied by the conductor, *where we agree that* $V=0$ *when the conductor carries no net charge* ($Q=0$). As can be shown [1,3] the ratio $Q/V$ is a constant quantity, independent of the charge $Q$ on the conductor. This ratio is called the *capacitance* of the conductor:

$$C = \frac{Q}{V} \tag{5.29}$$

The S.I. unit of capacitance is the *Farad* ($F$). Obviously, $1F=1C\cdot V^{-1}$.

A more complex system is that which consists of two conductors carrying opposite charges $+Q$ and $-Q$. If $V_1$ and $V_2$ are the corresponding electric potentials of these conductors, we can show (see Prob. 4) that $V_1>V_2$. Thus the potential difference $\Delta V=V_1-V_2$ is positive. This system is called a *capacitor* and its capacitance is defined by

$$C = \frac{Q}{V_1 - V_2} = \frac{Q}{\Delta V} \tag{5.30}$$

*Application:* As can be shown [1,3] the electric field in the interior of a parallel-plate capacitor is uniform, is normal to the plates and is directed from the positive to the negative plate. Show that the magnitude of this field is

$$\boxed{E = \frac{\Delta V}{l}} \tag{5.31}$$

where $l$ is the perpendicular distance between the plates and where $\Delta V$ is the potential difference between them. The above relation is generally valid for *any uniform electrostatic field.* [*Hint:* Apply Eq. (5.17) along an electric field line extending from the positive to the negative plate. Make the most convenient choice of a field line by taking into account that the plates are equipotential surfaces.]

## References for Chapter 5

# QUESTIONS

**1.** Suggest a physical process by which one may create (*a*) a static but non-uniform electric field; (*b*) a uniform but non-static electric field.

**2.** In Gauss' law in integral form (5.5) the electric field on a closed surface *S* is associated with the total charge in the interior of *S*. Will the field on *S* be affected if we remove all charges in the *exterior* of this surface?

**3.** (*a*) Show that the electric field in a region of space where a non-static distribution of charge $\rho(\vec{r},t)$ exists *cannot* be static. (*b*) In a region of space the distribution of charge is static. Does the electric field in that region have to be static?

**4.** Justify the principle of superposition for the electric potential: "At any point of space, the electric potential due to a system of charges equals the algebraic sum of potentials due to each charge separately".

**5.** The electric field is a physical quantity having a uniquely defined value that is measurable at every point of space. Is the same true with regard to the electric potential? Thus, can the potential be regarded as an absolute physical quantity? How about potential *difference*?

**6.** Starting with Coulomb's law, derive an expression for the potential of the Coulomb field produced by a point charge *q*.

**7.** What is an equipotential surface? Show that every such surface intersects normally the electric field lines. Determine the equipotential surfaces of the Coulomb field produced by a point charge *q*.

**8.** (*a*) Show that the electrostatic field is conservative and derive an expression for the potential energy of a charge *q* inside this field. (*b*) Find the potential energy of a hydrogen atom when the electron is a distance *r* from the nucleus (proton).

**9.** Consider a closed surface *S*. The electric field in the interor of *S* is zero, while on *S* itself the field acquires non-zero values and is directed normal to *S* at all points of this surface. Show that (*a*) *S* is an equipotential surface; (*b*) the interior of *S* is a space of constant potential, equal to that on *S*; (*c*) the total charge in the interor of *S* is zero.

**10.** By means of a simple example show that Coulomb's law follows directly from Gauss' law.

## PROBLEMS

**1.** Consider a closed surface $S$ inside a uniform electrostatic field $\vec{E}$. (*a*) Show that the total electric flux through $S$ is zero. (*b*) Show that the total electric charge in the interior of $S$ is zero.

***Solution:*** Let $V$ be the volume enclosed by $S$. By Gauss' law, the total flux through $S$ is proportional to the total charge $Q_{in}$ enclosed by $S$. Using Gauss' integral theorem and taking into account that $\vec{E}$ is a constant vector, we have:

$$\frac{Q_{in}}{\varepsilon_0} = \oint_S \vec{E} \cdot \overrightarrow{da} = \int_V (\vec{\nabla} \cdot \vec{E})\, dv = 0 \quad (\text{since } div\vec{E} = 0)$$

**2.** Is it possible for an electrostatic field of the form $\vec{E} = F(x,y,z)\,\hat{u}_x$ to exist? What do you conclude regarding the potential $V$ of such a field?

***Solution:*** An electrostatic field must be irrotational:

$$\vec{\nabla} \times \vec{E} = 0 \;\Rightarrow\; \begin{vmatrix} \hat{u}_x & \hat{u}_y & \hat{u}_z \\ \dfrac{\partial}{\partial x} & \dfrac{\partial}{\partial y} & \dfrac{\partial}{\partial z} \\ F & 0 & 0 \end{vmatrix} = 0 \;\Rightarrow\; \frac{\partial F}{\partial z}\,\hat{u}_y - \frac{\partial F}{\partial y}\,\hat{u}_z = 0$$

Now, for a vector to vanish, each component of it must be zero:

$$\frac{\partial F}{\partial y} = \frac{\partial F}{\partial z} = 0 \;\Rightarrow\; F = F(x), \quad \text{so that } \vec{E} = F(x)\,\hat{u}_x$$

Then, given that $\vec{E} = -\vec{\nabla} V$,

$$\frac{\partial V}{\partial x}\,\hat{u}_x + \frac{\partial V}{\partial y}\,\hat{u}_y + \frac{\partial V}{\partial z}\,\hat{u}_z = -F(x)\,\hat{u}_x$$

Equating corresponding coefficients on the two sides, we have:

$$\frac{\partial V}{\partial x} = -F(x), \quad \frac{\partial V}{\partial y} = \frac{\partial V}{\partial z} = 0 \;\Rightarrow\; V = V(x), \quad \frac{dV}{dx} = -F(x)$$

**3.** Prove the following statements with regard to an electrostatic field: (*a*) the electric field lines are oriented in the direction of *maximum decrease* of the electric potential; (*b*) a *positive* charge that is initially at rest tends to move in the direction of *decreasing* potential, while a negative charge moves in the opposite way; (*c*) any charge (positive or negative) tends to move in the direction in which its potential energy is *decreasing*.

***Solution:*** (*a*) The orientation of the field lines is determined everywhere by the direction of the electric field $\vec{E}$, which is tangent to these lines. Now, for an elementary displacement $d\vec{r}$ within the electric field, the corresponding change of the potential is

$$dV = (\vec{\nabla} V) \cdot d\vec{r} = -\vec{E} \cdot d\vec{r}$$

In particular, for a displacement $d\vec{r}$ along a field line, in the direction of orientation of the line, the element $d\vec{r}$ has the direction of $\vec{E}$ and the scalar product $\vec{E}\cdot d\vec{r}$ attains its maximum value. The change $dV$ of the potential thus admits an *absolutely* maximum *negative* value. That is, the *decrease* of the potential is greatest along a field line, in the direction of orientation of the latter.

($b$) The force on the charge is $\vec{F}=q\vec{E}$. If $q$>0, the force is in the direction of $\vec{E}$, thus in the direction of maximum decrease of $V$. This will therefore be the direction of motion of a positive charge that is initially at rest (a negative charge will move in the opposite direction, i.e., that of increasing $V$).

($c$) For an elementary displacement $d\vec{r}$ within the electric field, the change of potential energy of a charge $q$ is

$$dU=(\vec{\nabla}U)\cdot d\vec{r}=-\vec{F}\cdot d\vec{r}$$

If $q$ is initially at rest, the displacement $d\vec{r}$ will be in the direction of the force $\vec{F}$ and the scalar product $\vec{F}\cdot d\vec{r}$ will attain a maximum value. Thus the change $dU$ of the potential energy of $q$ will admit an *absolutely* maximum *negative* value. The charge $q$ will therefore move in the direction of maximum *decrease* of its potential energy, *regardless of the sign of q*!

**4.** Two charged conductors $A$ and $B$ carry net charges $+Q$ and $-Q$, respectively. Show that the electric potential of $A$ is greater than that of $B$.

***Solution:*** We recall that the electric potential assumes a constant value at all points occupied by a conductor (whether on its surface or in its interior). Let $V_A$ and $V_B$ be the potentials of the two conductors. Along an arbitrary path connecting $A$ with $B$,

$$V_A-V_B=\int_A^B \vec{E}\cdot\overrightarrow{dl} \qquad (1)$$

where $\vec{E}$ is the electric field along this path. Without loss of generality (given that the value of the integral is independent of the choice of path) we may assume that we move from $A$ to $B$ along an electric field line. Such a line is always oriented from the positively charged conductor $A$ to the negatively charged conductor $B$, in accordance with the orientation of the field $\vec{E}$ (can you justify this?). Along the chosen path the vectors $\vec{E}$ and $\overrightarrow{dl}$ are in the same direction, so that $\vec{E}\cdot\overrightarrow{dl}>0$. The integral in (1) thus assumes a positive value and, therefore, $V_A-V_B>0 \;\Rightarrow\; V_A>V_B$.

**5.** In the interior of a conductor there is a cavity that contains no charges (Fig. 5.9). The conductor is assumed to be in electrostatic equilibrium. ($a$) Show that the electric field within the cavity is zero. ($b$) Show that the total charge on the surface of the cavity is zero. ($c$) We now place a charge $Q$ inside the cavity. Find the total charge induced on the wall of the cavity, as well as the total charge on the surface of the conductor.

***Solution:*** ($a$) Since the cavity contains no charges, there can be no electric field lines beginning or ending *inside* the cavity. Also, since the conductor is in electrostatic equilibrium, the charge density in its interior is zero everywhere, which means that there are no nonzero charges there as well. Thus there are no field lines beginning or ending *inside* the conductor either. Any field line must therefore begin and end on the wall of the cavity, directed from a positive to a negative charge.

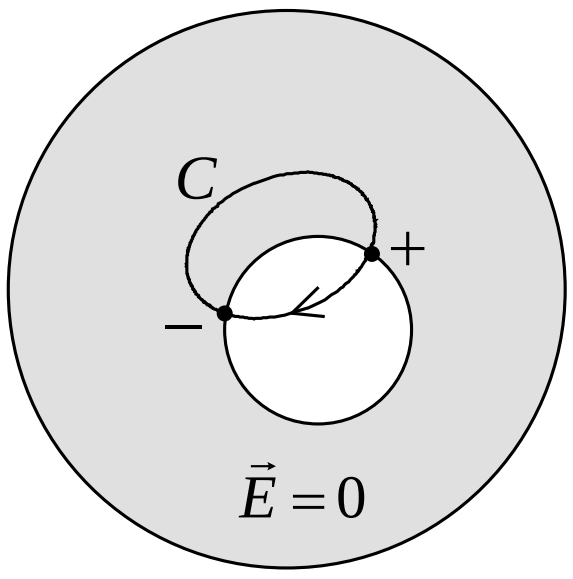

Fig. 5.9. A conductor with a cavity in its interior.

We consider a closed path $C$ such that a section of it lies inside the cavity and coincides with an electric field line. The electric field $\vec{E}$ inside the cavity is thus tangential at every point of that section of $C$. Taking into account that $\vec{E}=0$ in the interior of the conductor, we have:

$$\oint_C \vec{E}\cdot\overrightarrow{dl} = \int_{cavity} \vec{E}\cdot\overrightarrow{dl} + \int_{conductor} \vec{E}\cdot\overrightarrow{dl} = \int_{cavity} |\vec{E}||\overrightarrow{dl}| \;\; > 0$$

which cannot be correct, given that $\oint \vec{E}\cdot\overrightarrow{dl}=0$ for any electrostatic field. The assumption we made, that there is a nonzero electric field inside the cavity, was therefore wrong. That is, we must have $\vec{E}=0$ inside the cavity.

(*b*) We consider a closed surface $S$ inside the conductor, surrounding the cavity. At every point of $S$ we have that $\vec{E}=0$. Let $Q$ be the total charge on the surface of the cavity. The surface $S$ encloses no other charges, given that the charge density both in the interior of the conductor and in the interior of the cavity is zero. By Gauss' law and by taking into account that $Q_{in}=Q$ and that $\vec{E}=0$ on $S$, we have:

$$\oint_S \vec{E}\cdot\overrightarrow{da} = 0 = \frac{Q}{\varepsilon_0} \quad\Rightarrow\quad Q=0 \;\; \textit{on the surface of the cavity}$$

Note that the above results are valid even if there exists a nonzero electric field in the *exterior* of the conductor. That is, the cavity is *electrically isolated* from the outside world, being "protected", so to speak, by the conductor surrounding it.

(*c*) We consider again a closed surface $S$ inside the conductor, surrounding the cavity. At every point of $S$ we have that $\vec{E}=0$, as before. Let $Q_{wall}$ be the total induced charge on the wall of the cavity. The total charge enclosed by $S$ is $Q_{in}=Q+Q_{wall}$. By Gauss' law and by taking into account that $\vec{E}=0$ on $S$, we have:

$$\oint_S \vec{E}\cdot\overrightarrow{da} = 0 = \frac{Q_{in}}{\varepsilon_0} \quad\Rightarrow\quad Q_{in}=Q+Q_{wall}=0 \quad\Rightarrow\quad Q_{wall}=-Q$$

Now, the conductor is electrically neutral and there is no net electric charge in its interior. So, since the wall of the cavity carries a charge $-Q$, there must necessarily be a charge $+Q$ on the *surface* of the conductor. This surface charge makes it known to us that there is a charge $Q$ inside the cavity!

**6.** A solid metal sphere of radius $R$ carries a total positive charge $Q$ uniformly distributed over its surface (Fig. 5.10). Determine the electric potential both inside and outside the sphere. (Assume that $V=0$ at an infinite distance from the sphere.)

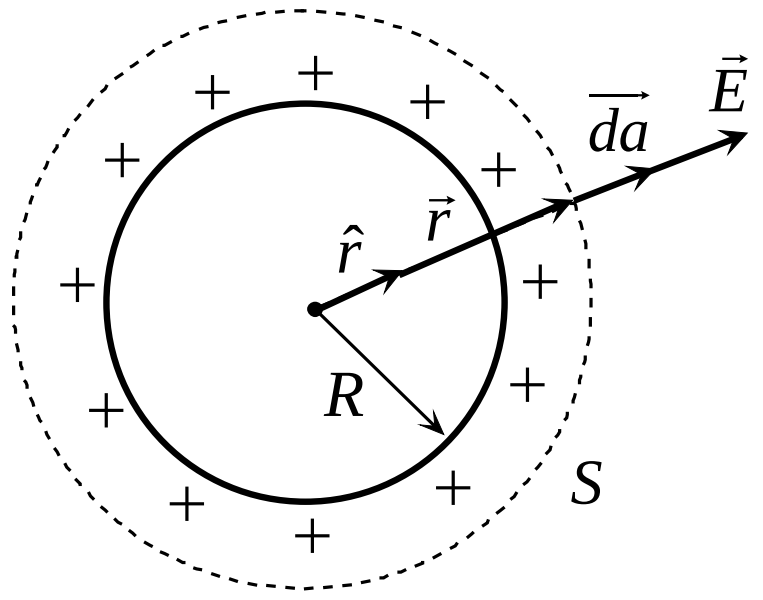

Fig. 5.10. A metal sphere of radius $R$, carrying a total positive charge $Q$ on its surface.

***Solution:*** By the spherical symmetry of the problem the electric field $\vec{E}$ outside the sphere is radial and directed outward (since $Q>0$), while its magnitude $|\vec{E}|=E$ is constant over any spherical surface concentric with the sphere. Hence the external field is of the form

$$\vec{E}=E(r)\,\hat{r}\ ,\quad r>R \tag{1}$$

where $r$ is the distance from the center of the sphere. Since the solid sphere is conducting, it will be $\vec{E}=0$ in its interior, i.e., for $r<R$. To find the electric field in the exterior of the sphere ($r>R$) we apply Gauss' law on a spherical surface $S$ of radius $r>R$, concentric with the sphere. The total charge enclosed by $S$ is $Q_{in}=Q$:

$$\oint_S \vec{E}\cdot\overrightarrow{da}=\frac{Q_{in}}{\varepsilon_0}=\frac{Q}{\varepsilon_0}\quad \text{where}\quad \vec{E}\cdot\overrightarrow{da}=[E(r)\,\hat{r}]\cdot[(da)\,\hat{r}]=E(r)\,da \ \Rightarrow$$

$$\frac{Q}{\varepsilon_0}=\oint_S E(r)\,da=E(r)\oint_S da=E(r)(4\pi r^2)\ \Rightarrow$$

$$E(r)=\frac{1}{4\pi\varepsilon_0}\frac{Q}{r^2}\ ,\quad r>R \tag{2}$$

From (1) and (2) we see that the electric field in the exterior of the sphere is the same as the Coulomb field of a hypothetical point charge $Q$ placed at the center of the sphere! As can be shown, relation (2) is valid for $r=R$ also, thus yielding the electric field on the surface of the sphere (note that the field is *noncontinuous* at $r=R$).

We now seek the electric potential $V(r)$. In general, for a displacement from position $a$ to position $b$ within the electric field, we have:

$$\int_a^b \vec{E}\cdot d\vec{r}=V_a-V_b\ ,$$

which is independent of the path joining the end points $a$ and $b$. Now,

$$\vec{E}\cdot d\vec{r}=E(r)\,\hat{r}\cdot d\vec{r}=\frac{E(r)}{r}\,\vec{r}\cdot d\vec{r}=\frac{E(r)}{r}\,r\,dr=E(r)\,dr$$

where use has been made of Eq. (5.13). Hence,

$$V_a - V_b = \int_a^b E(r)\,dr \qquad (3)$$

where $E(r)$ is given by relation (2) for $r \geq R$, while $E(r)=0$ for $r<R$. For the potential in the exterior of the sphere, we choose point $a$ to be on a spherical surface of radius $r \geq R$ where the potential is $V(r)$ (equipotential surface), while point $b$ is assumed to be at infinity ($r=\infty$) where $V_\infty=0$:

$$V(r) - V_\infty = \int_r^\infty E(r)\,dr = \frac{Q}{4\pi\varepsilon_0}\int_r^\infty \frac{dr}{r^2} = -\frac{Q}{4\pi\varepsilon_0}\left(0-\frac{1}{r}\right) \;\Rightarrow$$

$$V(r) = \frac{1}{4\pi\varepsilon_0}\frac{Q}{r} \;, \quad r \geq R$$

Notice again that the potential in the exterior of the sphere is the same as the Coulomb potential due to a point charge $Q$ placed at the center of the sphere. For the potential in the interior of the sphere, we choose point $a$ to be on a spherical surface of radius $r<R$ where the potential is $V(r)$; we take point $b$ on the surface of the sphere where $V(R)=Q/4\pi\varepsilon_0 R$; and, we use (3) by taking into account that $E(r)=0$ for $r<R$:

$$V(r) - V(R) = \int_r^R E(r)\,dr = 0 \;\Rightarrow$$

$$V(r) = V(R) = \frac{Q}{4\pi\varepsilon_0 R} \;, \quad r \leq R$$

We notice that the space occupied by the sphere is a space of constant potential.

*Exercise:* Show that, if in place of the solid sphere we had a uniformly charged *spherical shell*, our results would be exactly the same! (To evaluate the electric field in the empty space bounded by the shell, use again Gauss' law, observing that there are no electric charges in this region.)

**7.** A *spherical capacitor* consists of an inner conducting sphere of radius $a$ and charge $+Q$, surrounded by a concentric conducting spherical shell of radius $b$ and charge $-Q$ (Fig. 5.11). (*a*) Evaluate the capacitance of the system. (*b*) Show that the electric field in the exterior of the capacitor ($r>b$) is zero and determine the electric potential both inside ($a \leq r \leq b$) and outside ($r \geq b$) the capacitor (by $r$ we denote the distance from the center of the spheres).

***Solution:*** (*a*) Due to the spherical symmetry of the problem, the electric field in the interior of the capacitor ($a<r<b$) is of the form

$$\vec{E} = E(r)\,\hat{r} \qquad (1)$$

That is, the field is radial and directed from the inner to the outer sphere, while its magnitude $|\vec{E}|=E$ has a constant value over any spherical surface concentric with the two conducting spheres.

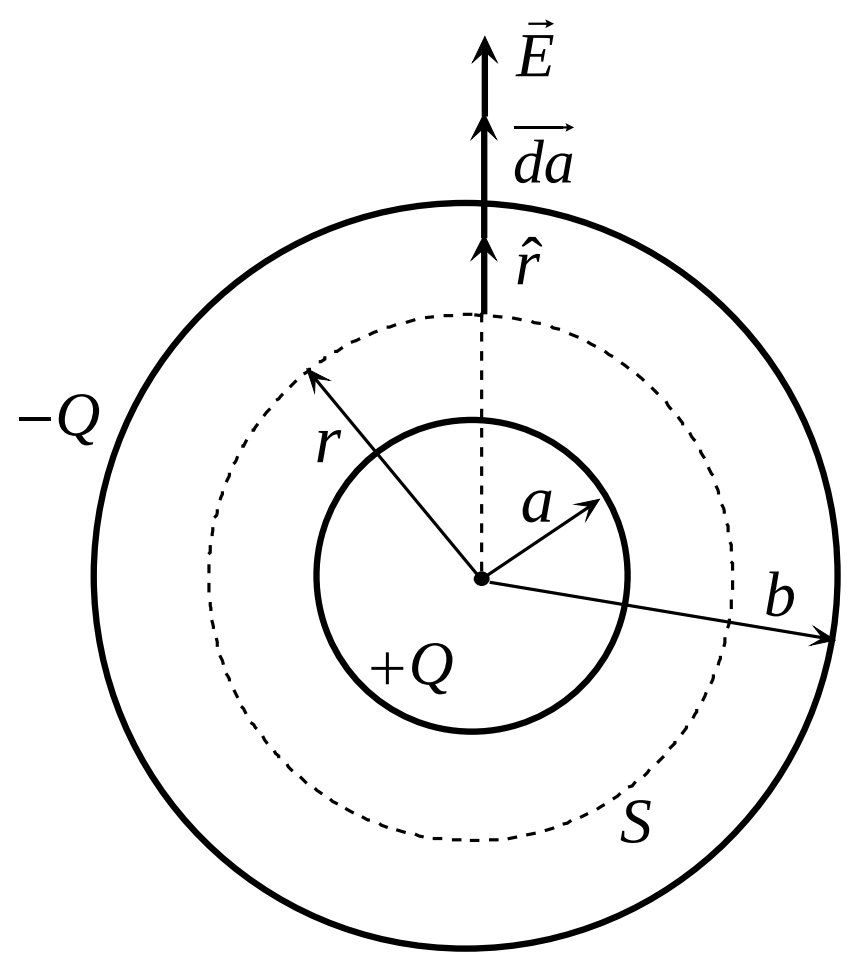

Fig. 5.11. A spherical capacitor carrying a charge $\pm Q$.

We apply Gauss' law for a spherical surface $S$ of radius $r$, noting that the total charge enclosed by $S$ is $Q_{in}=+Q$:

$$\oint_S \vec{E}\cdot\overrightarrow{da}=\frac{Q_{in}}{\varepsilon_0}\Rightarrow\ \frac{Q}{\varepsilon_0}=\oint_S[E(r)\hat{r}]\cdot[(da)\hat{r}]=\oint_S E(r)\,da=E(r)\oint_S da=E(r)(4\pi r^2)\Rightarrow$$

$$E(r)=\frac{1}{4\pi\varepsilon_0}\frac{Q}{r^2}\ ,\quad a<r<b \qquad (2)$$

The surfaces of the two spheres are equipotentials. To find the potential difference $V=V_a-V_b$ between the spheres, we consider an arbitrary path from the inner to the outer sphere:

$$V=V_a-V_b=\int_a^b \vec{E}\cdot d\vec{r}=\int_a^b E(r)\,dr \qquad (3)$$

where we have used the relation $\vec{E}\cdot d\vec{r}=E(r)\,dr$ (see Prob. 6). Substituting (2) into (3), we have:

$$V=\frac{Q}{4\pi\varepsilon_0}\int_a^b\frac{dr}{r^2}=\frac{Q}{4\pi\varepsilon_0}\left(\frac{1}{a}-\frac{1}{b}\right)=\frac{Q}{4\pi\varepsilon_0}\frac{(b-a)}{ab}$$

The capacitance is then $C=\dfrac{Q}{V}=4\pi\varepsilon_0\dfrac{ab}{(b-a)}$ .

(*b*) In the exterior of the capacitor ($r>b$) the electric field will again be of the form (1). We apply Gauss' law for a spherical surface $S$ of radius $r>b$, noting that the total charge enclosed by $S$ is now $Q_{in}=Q+(-Q)=0$:

$$\oint_S \vec{E}\cdot\overrightarrow{da}=\frac{Q_{in}}{\varepsilon_0}=0\ \Rightarrow\ E(r)(4\pi r^2)=0\ \Rightarrow\ E(r)=0$$

Thus $\vec{E}=0$ for $r>b$. Let now $V(r)$ be the potential on a spherical equipotential surface of radius $r>b$. For an arbitrary path connecting this surface to the outer conducting sphere, we have:

$$V(r) - V_b = \int_r^b \vec{E} \cdot d\vec{r} = 0 \;\;\Rightarrow\;\; V(r) = V_b = const.$$

By arbitrarily assuming that $V_b$=0, we have that $V(r)$=0 for $r \geq b$. In the interior of the capacitor ($a<r<b$) the electric field is given by (1) and (2). In this case we have (remembering that $V_b$=0):

$$V(r) - V_b = \int_r^b \vec{E} \cdot d\vec{r} = \int_r^b E(r)\, dr = \frac{Q}{4\pi\varepsilon_0} \int_r^b \frac{dr}{r^2} \;\;\Rightarrow$$

$$V(r) = \frac{Q}{4\pi\varepsilon_0} \left( \frac{1}{r} - \frac{1}{b} \right) \;, \quad a \leq r \leq b$$

## CHAPTER 6

# ELECTRIC CURRENT

### 6.1 Current Density

The term *electric current* refers to the *oriented* motion of electric charge. The orientation of the charge can be accomplished by applying an electric field or it may be the result of an uneven distribution of charge in a region of space (*diffusion current*; see Sec. 2.10). We note that the irregular thermal motion of, say, the free electrons in a metal, which takes place even without the presence of an electric field, does *not* constitute electric current since this motion is random and non-oriented.

Consider the totality of natural phenomena associated with *motion* (not just *presence*) of electric charge. In each case there are two factors to be considered; namely, the *sign* and the *direction of motion* of the charge. As a simple example, let us consider a positive charge $q$ moving to the right (as we view it) with velocity $\vec{\upsilon}$ :

$$q \bullet\!\!\longrightarrow \vec{\upsilon}$$

Let $A$ and $B$ be two neighboring points through which the charge passes at times $t$ and $t+dt$, respectively. Since $q>0$, within the time interval $dt$ the total charge at point $A$ decreases by $q$ while the total charge at $B$ increases by $q$. The end result would be the same, however, if a *negative* charge $-q$ were to move from $B$ to $A$ with velocity $-\vec{\upsilon}$ , thus to the *left* as we view it:

$$-\vec{\upsilon} \longleftarrow\!\!\bullet\; -q$$

(Since $B$ "loses" negative charge, it is as if it gains positive charge; at the same time, $A$ "gains" negative charge, which is as if it loses positive charge; exactly as before.) We thus conclude that

> *the motion of a negative charge in some direction is equivalent to the motion of a positive charge in the opposite direction.*

As an example, when a charge $q$ moves inside a magnetic field $\vec{B}$ with velocity $\vec{\upsilon}$ , it is subject to a magnetic force $\vec{F} = q(\vec{\upsilon} \times \vec{B})$ (see Chap. 7). The same force is exerted on a charge $-q$ moving in the opposite direction with velocity $-\vec{\upsilon}$ :

$$\vec{F} = (-q)[(-\vec{\upsilon}) \times \vec{B}] = q(\vec{\upsilon} \times \vec{B})$$

Similarly, the magnetic field produced by a charge $q$ moving with velocity $\vec{\upsilon}$ is the same as that produced by a charge $-q$ moving with velocity $-\vec{\upsilon}$ .

In natural phenomena associated with the *motion* of electric charges it is often convenient to regard the moving charges as positive. Thus, a negative charge $-q$ moving in some direction may be treated as if it were a positive charge $+q$ moving in the opposite direction. The direction of an electric current is taken to be the direction of motion of (actual or hypothetical) *positive* charges; it is called the *conventional direction* of the current. The conventional direction coincides with the actual direction of motion of positive charges while it is opposite to the direction of motion of negative charges.

We consider a region of space where electric charges are in motion. Let $da_\perp$ be an elementary surface (we will use the same symbol for its area) at some point $\vec{r} \equiv (x, y, z)$ of the region. We assume that an elementary quantity of charge $dq$ passes through $da_\perp$ within a time interval $dt$, moving with velocity $\vec{\upsilon}$ in a direction *normal* to $da_\perp$. We call

$$\hat{u} = \frac{\vec{\upsilon}}{|\vec{\upsilon}|} = \frac{\vec{\upsilon}}{\upsilon} \qquad (\text{where } \upsilon = |\vec{\upsilon}|)$$

the unit vector in the direction of motion (normal to $da_\perp$). The *current density* at the considered point is defined as[1]

$$\vec{J} = \frac{dq}{da_\perp dt}\,\hat{u} = \frac{dq}{\upsilon\, da_\perp dt}\,\vec{\upsilon} \tag{6.1}$$

We notice that the direction of $\vec{J}$ coincides with the direction of motion of *positive* charge ($dq$>0) while it is *opposite* to the direction of motion of *negative* charge. We also note that the current density $\vec{J}$ is unchanged if we simultaneously reverse the sign *and* the direction of motion of the charge. Thus, negative charge moving in a certain direction yields the same current density as positive charge of the same magnitude, moving in the opposite direction with equal speed.

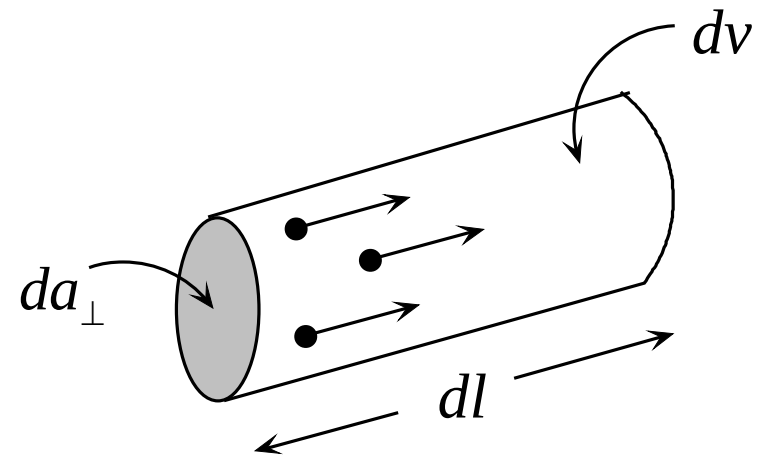

Fig. 6.1. An elementary volume *dv* within a current flow.

Let us now imagine that a quantity of charge $dq$ crosses normally the elementary surface $da_\perp$ and, within time $dt$, advances a distance $dl$ and finally occupies an elementary volume $dv$ equal to $dv = da_\perp\, dl = da_\perp(\upsilon dt) = \upsilon\, da_\perp\, dt$ (Fig. 6.1). Relation (6.1) is now written:

$$\vec{J} = \frac{dq}{dv}\,\vec{\upsilon}$$

The quantity $dq/dv$ represents the *density of moving charge*, $\rho_\kappa$ , at the considered location. Thus, finally,

[1] This is actually the *volume* current density. For a charge flow over a surface, a *surface* current density may also be defined [1,2].

$$\boxed{\vec{J} = \rho_\kappa \vec{\upsilon}} \tag{6.2}$$

Note carefully that the quantity $\rho_\kappa$ is the density of the *moving* charge, to which the current $\vec{J}$ owes its existence. It must *not* be confused with the *total* charge density $\rho$ due to *all* charges, whether moving or stationary (e.g., mobile electrons and stationary positive ions in a metal).

We now consider a surface *S* (which may be open or closed) through which electric charge passes (Fig. 6.2). Let $\overrightarrow{da}$ be an infinitesimal surface element, i.e., a vector normal to the surface at some point of it, the magnitude *da* of which vector is equal to an elementary area on *S*. We call *dq* the elementary charge passing through the area element *da* within time *dt*, and we call *dQ* the total charge crossing the entire surface *S* in this time interval. We write, symbolically,

$$dQ = \int_S dq$$

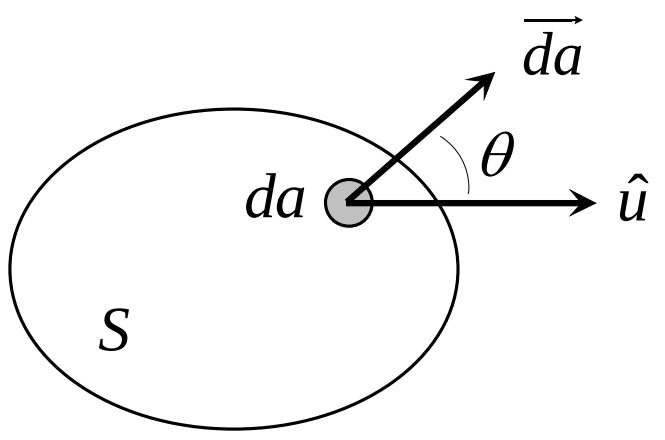

Fig. 6.2. A surface *S* through which electric charge passes.

Let $\vec{\upsilon}$ be the velocity of motion of the charge *dq* through *da*. As before, we call

$$\hat{u} = \frac{\vec{\upsilon}}{|\vec{\upsilon}|} = \frac{\vec{\upsilon}}{\upsilon}$$

the unit vector in the direction of motion. By convention, we will assume that $0 \le \theta \le \pi/2$ (where $\theta$ is the angle between $\overrightarrow{da}$ and $\hat{u}$), i.e., that the charge crosses the surface moving from one side of it to the other in the direction indicated by $\overrightarrow{da}$. This assumption is based on remarks we made earlier. That is, if the *actual* motion of the charge is the opposite of that indicated by $\overrightarrow{da}$, we may equivalently consider that an *opposite* charge *dq* crosses the surface in a direction *consistent* with that of $\overrightarrow{da}$. (For example, if positive charge enters a closed surface *S*, thus moving inconsistently with the *outward* surface element $\overrightarrow{da}$, we may consider that negative charge of equal magnitude flows out of *S*, now moving in a direction consistent with $\overrightarrow{da}$.) The scalar product

$$da_\perp = \overrightarrow{da} \cdot \hat{u} = da \cos\theta$$

represents the projection of $\overrightarrow{da}$ in the direction of $\hat{u}$ and gives the area of an elementary surface *normal* to $\hat{u}$. Because of the convention we made regarding $\theta$, we will have $da_\perp > 0$.

We define the *elementary current* through the area element $da$ as

$$dI = \frac{dq}{dt}$$

The *total current* through the entire surface *S* is then

$$I = \int_S dI = \frac{dQ}{dt} \quad (6.3)$$

Note that *I* is an *algebraic* value and may be positive or negative, depending on the sign of the electric charge passing through *S*.

The current density at the location of the element $da$ is

$$\vec{J} = \frac{dq}{da_\perp dt}\,\hat{u} = \frac{dI}{da_\perp}\,\hat{u} \quad (6.4)$$

The relation between the total current *I* and the current density is found by noticing that

$$\vec{J} \cdot \overrightarrow{da} = \frac{dI}{da_\perp}\,\hat{u} \cdot \overrightarrow{da} = \frac{dI}{da_\perp} da_\perp = dI$$

Equation (6.3) is then written:

$$I = \frac{dQ}{dt} = \int_S \vec{J} \cdot \overrightarrow{da} \quad (6.5)$$

Relation (6.5) may be viewed as the definition of the current density $\vec{J}$ on a surface *S*: it is a vector function defined at all points of *S* and such that its surface integral over *S* equals the total current passing through *S* at a given time.

Let us consider the special case where *S* is a *plane* surface and $\vec{J}$ is *constant* over *S* and *normal* to it at each point of the surface. The vectors $\vec{J}$ and $\overrightarrow{da}$ are then parallel and, if we assume that they are also in the same direction, $\vec{J} \cdot \overrightarrow{da} = |\vec{J}||\overrightarrow{da}| = J\,da$, where the value of $J$ is constant over *S*. Hence,

$$I = \int_S J\,da = J\int_S da = J\,S \quad \Rightarrow$$

$$J = \frac{I}{S} \quad (6.6)$$

where by *S* we here denote the total area of the plane surface. Relation (6.6) describes, for example, the current density in the interior of a metal wire of cross-sectional area *S*, carrying a current *I* (see Sec. 2.4).

## 6.2 Equation of Continuity and Conservation of Charge

We consider a closed surface $S$ enclosing a volume $V$ (Fig. 6.3). The surface element $\vec{da}$ is a vector normal to $S$ and directed *outward*. Thus, when we refer to electric charge passing through an elementary area $da$ in a direction consistent with $\vec{da}$ we mean charge *coming out* of $S$. (As we have mentioned, the actual physical situation may be the opposite; i.e., charge of the opposite sign may in fact be *going into S*.) As we remarked following Eq. (6.1), the current density $\vec{J}$ is a well-defined quantity, independent of whether we prefer to view the charge as "coming out" of the closed surface or "going into" it.

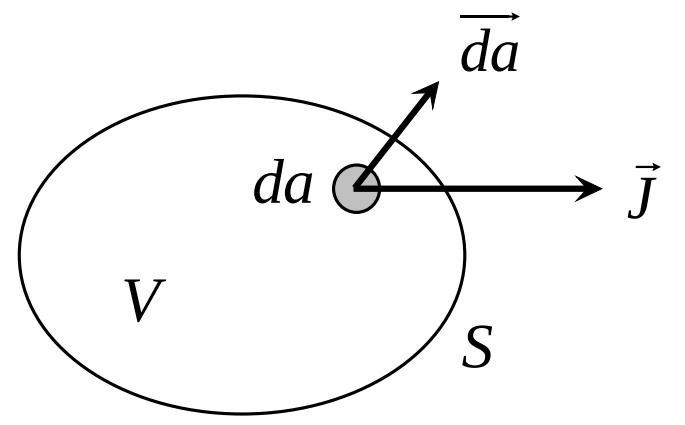

Fig. 6.3. Current density on a closed surface $S$ enclosing a volume $V$.

The total charge coming out of $S$ per unit time is given by Eq. (6.5):

$$I=\frac{dQ}{dt}=\oint_S \vec{J}\cdot\vec{da}=\int_V(\vec{\nabla}\cdot\vec{J})\,dv$$

where use has been made of Gauss' integral theorem. Since *charge is conserved* (it can neither vanish nor be created from zero), the charge appearing outside the volume $V$ within time $dt$ must be equal to the charge *lost* inside this volume within the same time interval. Thus the rate of change of the total charge contained in $V$ is

$$\frac{dQ_{in}}{dt}=-\frac{dQ}{dt}=-\int_V(\vec{\nabla}\cdot\vec{J})\,dv \qquad (6.7)$$

On the other hand, if $\rho(\vec{r},t)$ is the charge density inside $V$,

$$Q_{in}=\int_V \rho\, dv$$

The integral is eventually a function of $t$ only, as becomes evident by writing it more analytically as $\int_V \rho(x,y,z,t)\,dxdydz$. We thus have:

$$\frac{dQ_{in}}{dt}=\frac{d}{dt}\int_V \rho\, dv=\frac{\partial}{\partial t}\int_V \rho\, dv=\int_V \frac{\partial\rho}{\partial t}\,dv \qquad (6.8)$$

By comparing (6.7) and (6.8), we have:

$$-\int_V (\vec{\nabla}\cdot\vec{J})\,dv = \int_V \frac{\partial\rho}{\partial t}\,dv \;\Rightarrow\; \int_V (\vec{\nabla}\cdot\vec{J} + \frac{\partial\rho}{\partial t})\,dv = 0$$

In order for the integral on the right to vanish for an *arbitrary* choice of the volume *V*, the integrand itself must vanish:

$$\boxed{\vec{\nabla}\cdot\vec{J} + \frac{\partial\rho}{\partial t} = 0} \qquad (6.9)$$

Relation (6.9) is called the *equation of continuity*. It is evident from the physical reasoning leading to it that this equation expresses *conservation of charge*. In Chap. 9 we will derive Eq. (6.9) again by using the Maxwell equations.

## 6.3 Ohm's Law

As we know (Sec. 5.6), the electric field $\vec{E}$ in the interior of a metallic conductor in *electrostatic equilibrium* is zero. (Indeed, a field $\vec{E} \neq 0$ would put the free electrons of the conductor in motion; hence the condition of electrostatic equilibrium would be violated.) On the other hand, it is obvious that in the interior of a *current-carrying* conductor a nonzero electric field $\vec{E}$ must exist. This field exerts forces on the free electrons, putting them into accelerated motion. The speed of the electrons would increase unlimitedly with time if the electrons didn't lose part of their kinetic energy as a result of collisions with the ions of the metal. The free electrons thus finally acquire a constant average velocity $\vec{\upsilon}$. *By convention*, we regard the mobile electrons as *positively* charged particles, so that $\vec{\upsilon}$ is in the direction of $\vec{E}$ (in reality, of course, the opposite happens). As mentioned in Sec. 2.4, for relatively small values of the electric field $\vec{E}$ the average velocity of the electrons is proportional to the field strength:

$$\vec{\upsilon} = \mu\vec{E} \qquad (6.10)$$

The coefficient $\mu$ is called the *mobility* of the electron in the considered metal. As explained in Sec. 2.6, the mobility decreases with temperature, resulting in an increase of the electrical resistance of the metal.

Combining Eqs. (6.2) and (6.10), we have: $\vec{J} = \rho_\kappa \mu\vec{E}$, where $\rho_\kappa$ is the density of the *mobile* charge (this charge contains *only* the free electrons and *not* the stationary ions of the metal). We can write:

$$\rho_\kappa = q_e n \qquad (6.11)$$

where $q_e$ is the charge of the electron (conventionally assumed here to be positive) and where *n* is the *electronic density* of the metal (number of free electrons per unit volume). Hence, $\vec{J} = q_e n \mu \vec{E}$. The quantity

$$\sigma = q_e n \mu \qquad (6.12)$$

is the *conductivity* of the metal (see Sec. 2.4). We thus recover our familiar general form of *Ohm's law*:

$$\boxed{\vec{J} = \sigma \vec{E}} \tag{6.13}$$

Let us consider now the case of a metal wire of length $l$ and constant cross-sectional area $S$, carrying a constant current $I$ (Fig. 6.4). We call $\Delta V = V_1 - V_2$ the potential difference (voltage) between the ends of the wire.

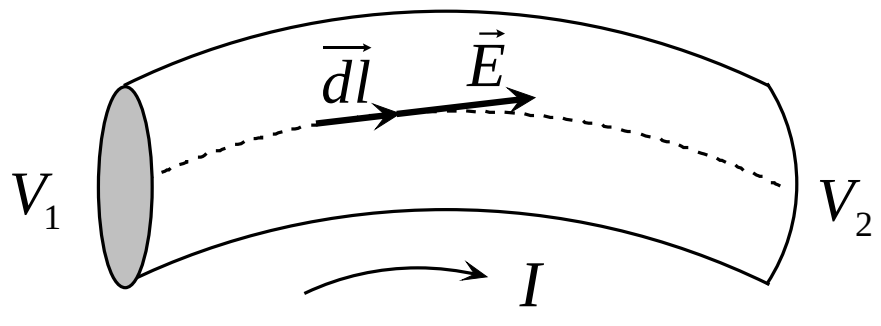

Fig. 6.4. Metal wire of length $l$, carrying a current $I$ and subject to a voltage $\Delta V = V_1 - V_2$.

The current density $\vec{J}$ within the wire is tangential to the axis of the wire and oriented in the direction of motion of the (conventionally positive) mobile charges. By definition, *this is the direction of the current I* in the wire. According to Ohm's law (6.13), this will also be the direction of the electric field $\vec{E}$ in the interior of the wire. Furthermore, the magnitude $E$ of the field is constant along the wire. Indeed, taking the magnitude of (6.13) and combining this with (6.6), we have:

$$E = \frac{J}{\sigma} = \frac{I}{\sigma S} = \text{constant} \tag{6.14}$$

given that the $I$, $\sigma$, $S$ are constant along the wire.

Let now $\vec{dl}$ be an elementary displacement on the axis of the wire, in the direction of $\vec{J}$ (thus of $\vec{E}$ also). To this displacement there corresponds an infinitesimal change of the electric potential,

$$dV = -\vec{E} \cdot \vec{dl} = -|\vec{E}|\,|\vec{dl}| = -E\,dl$$

Taking into account that $E$ is constant along the wire, we have:

$$\int_{V_1}^{V_2} dV = -\int_0^l E\,dl = -E \int_0^l dl \;\Rightarrow\; V_1 - V_2 \equiv \Delta V = E\,l \;\Rightarrow$$

$$\boxed{E = \frac{\Delta V}{l} = \frac{V_1 - V_2}{l}} \tag{6.15}$$

Note that $V_1 > V_2$ since, by definition, $E>0$ (see also Chap. 5, Prob. 3). If the wire happens to be *rectilinear*, the electric field $\vec{E}$ in its interior will be constant in magnitude and direction; in other words, it will be *uniform*. In this case, relation (6.15) represents the familiar expression for the strength of a uniform electrostatic field.

By comparing the expressions for $E$ in (6.14) and (6.15), it is not hard to show that

$$\boxed{I = \frac{\Delta V}{R}} \tag{6.16}$$

where $R$ is the *resistance* of the wire (measured in *Ohms*, $\Omega$):

$$R = \frac{l}{\sigma S} \tag{6.17}$$

Equation (6.16) is the *special form of Ohm's law* for metal wires of constant cross-section. Note that the wire is not required to be rectilinear (an assumption that was made for simplicity in Sec. 2.4).

*Comment:* The classical expression (6.12) for the conductivity $\sigma$ is only approximately correct since it implies that the totality of free electrons in the metal (the concentration of which electrons equals $n$) contributes to electrical conduction. This assumption, however, is not true in reality, as discussed in Appendix C.

## References for Chapter 6

## QUESTIONS

**1.** Consider a current of electrons and a current of holes in an *intrinsic* (pure) semiconductor. Both the electrons and the holes are moving under the influence of an electric field. Assuming that, approximately, the mobilities of electrons and holes are equal, compare the corresponding current densities for the two charge carriers (cf. Sec. 2.5).

**2.** Suppose you are given the equation of continuity (6.9) without any information regarding its physical content. What would you do to demonstrate that (6.9) expresses conservation of charge?

**3.** An electrically charged conductor is in a state of electrostatic equilibrium. By using Ohm's law and Gauss' law show that the charge must necessarily lie on the surface of the conductor.

**4.** We connect the two ends of a metal wire with the terminals (positive and negative) of a battery. What will be the direction of the electric field and the electric current in the wire? What is the *actual* direction of motion of the charge carriers? (The positive terminal of the battery is at a higher potential relative to the negative terminal. By convention, the direction of the current is that of the current-density vector.)

**5.** Show that, in order for the special form (6.16) of Ohm's law to be valid, the cross-sectional area of the metal wire must be *uniform*, i.e., constant along the wire. Under what conditions is the electric field inside the wire uniform?

## PROBLEMS

**1.** A metal wire carries a constant current $I$. Show that in the interior of the wire the electric field is static and the total charge density vanishes at all points. (See also Prob. 2 in Chap. 9.)

***Solution:*** We call $\rho$ the *total* charge density in the interior of the wire. This density is due to *all* charges, both the mobile electrons *and* the stationary positive ions (it must not be confused with the density $\rho_\kappa$ of the mobile electrons alone). We will make use of three fundamental laws:

Ohm's law: $\vec{J} = \sigma \vec{E}$ (1)

Gauss' law: $\vec{\nabla} \cdot \vec{E} = \dfrac{\rho}{\varepsilon_0}$ (2)

Equation of continuity: $\vec{\nabla} \cdot \vec{J} + \dfrac{\partial \rho}{\partial t} = 0$ (3)

Since the current $I$ is constant in time,

$$\frac{\partial \vec{J}}{\partial t} = 0 \overset{(1)}{\Rightarrow} \frac{\partial \vec{E}}{\partial t} = 0 \;\; (static\, \vec{E}) \overset{(2)}{\Rightarrow} \frac{\partial \rho}{\partial t} = 0 \overset{(3)}{\Rightarrow} \vec{\nabla} \cdot \vec{J} = 0$$

$$\overset{(1)}{\Rightarrow} \vec{\nabla} \cdot \vec{E} = 0 \overset{(2)}{\Rightarrow} \rho = 0$$

*Physical interpretation:* When the current in the wire is constant, the charge of the mobile electrons exactly counterbalances the opposite charge of the (stationary) ions of the metal so that the wire is electrically neutral everywhere in its interior.

*Comment:* When calculating the total charge density $\rho$, the electron must *always* be treated as *negatively* charged! This is because the quantity $\rho$ is related only to the *presence*, not the *motion*, of charges. On the contrary, in the relation $\vec{J} = \rho_\kappa \vec{\upsilon}$ for the *moving* electrons we are free to change the sign of $\rho_\kappa$ provided that we simultaneously invert the direction of $\vec{\upsilon}$, so that the current density $\vec{J}$ is left unchanged.

**2.** Prove *Kirchhoff's first rule*: In a region $R$ of space where the electric field is static, the total electric current through any closed surface is zero.

***Solution:*** Static electric field $\Rightarrow \partial \vec{E} / \partial t = 0, \; \forall \vec{r} \in R$. By Gauss' law,

$$\rho = \varepsilon_0 (\vec{\nabla} \cdot \vec{E}) \;\Rightarrow\; \frac{\partial \rho}{\partial t} = \varepsilon_0 (\vec{\nabla} \cdot \frac{\partial \vec{E}}{\partial t}) = 0, \; \forall \vec{r} \in R$$

Then, by the equation of continuity,

$$\vec{\nabla} \cdot \vec{J} + \frac{\partial \rho}{\partial t} = 0 \;\Rightarrow\; \vec{\nabla} \cdot \vec{J} = 0, \; \forall \vec{r} \in R$$

The volume integral of the equation on the right, in a volume $V$ bounded by a closed surface $S$, is

$$\int_V (\vec{\nabla} \cdot \vec{J})\, dv = 0$$

By using Gauss' integral theorem, the volume integral is transformed into an integral over the closed surface *S*. We thus have:

$$\oint_S \vec{J} \cdot \overrightarrow{da} = 0$$

The surface integral represents the total current passing through the closed surface *S* (since $\overrightarrow{da}$ is directed outward relative to *S*, we are actually speaking of the total current *exiting S*). Physically, the vanishing of this integral suggests that, as far as their absolute values are concerned, the total current going into *S* equals the total current coming out of *S* (a negative "outgoing" current is equivalent to a positive "ingoing" one). In the limit $V \to 0$, the closed surface *S* degenerates to a point. In the case of an electric network, such a point corresponds to a *junction* and Kirchhoff's first rule expresses the conservation of charge at each junction of the system. (Another rule, called "Kirchhoff's second rule", expresses the conservation of energy along any closed path in the network.)

# CHAPTER 7

# STATIC MAGNETIC FIELDS

## 7.1 The Magnetic Field and the Biot-Savart Law

As we know, an electric field is produced by a distribution of electric charges, regardless of their state of motion, and acts on *all* charges, even stationary ones. On the other hand, magnetic fields are produced by *moving* charges (i.e., *electric currents*) and act only on *moving* charges. Of course, the state of motion of a charge is dependent upon the observer: a charge that appears to be moving relative to an observer $A$ will seem to be at rest with respect to an observer $B$ traveling with the charge. Observer $A$ will record both an electric and a magnetic field, while observer $B$ will only detect an electric field.

The *magnetic force* on a charge $q$ moving with velocity $\vec{\upsilon}$ within a magnetic field $\vec{B}$ is given by

$$\vec{F}_m = q(\vec{\upsilon} \times \vec{B}) \qquad (7.1)$$

If an electric field $\vec{E}$ is also present, the total force (called the *Lorentz force*) on $q$ will be

$$\vec{F} = q[\vec{E} + (\vec{\upsilon} \times \vec{B})] \qquad (7.2)$$

The S.I. unit for the magnetic field strength is the *Tesla* ($T$). As follows from Eq. (7.1), $1\,T = 1\,N/(A \cdot m)$, where $1\,A \equiv 1\,Ampere = 1C/s$.

An important property exhibited by magnetic fields is that

> *magnetic forces produce no work on moving charges; thus a purely magnetic force cannot affect the speed of a charge.*

In other words, the magnetic force may only change the direction of motion of a charge without speeding it up or slowing it down. Indeed, the work done by a magnetic force on a charge $q$ as the latter traces a path from point $a$ to point $b$ within the magnetic field is given by

$$W_m = \int_a^b \vec{F}_m \cdot \overrightarrow{dl} = q \int_a^b (\vec{\upsilon} \times \vec{B}) \cdot \overrightarrow{dl} = 0$$

since

$$(\vec{\upsilon} \times \vec{B}) \cdot \overrightarrow{dl} = (\vec{\upsilon} \times \vec{B}) \cdot \frac{\overrightarrow{dl}}{dt} dt = [(\vec{\upsilon} \times \vec{B}) \cdot \vec{\upsilon}\,] dt = 0$$

Then, by the work-energy theorem, $\Delta T = W_m = 0 \Rightarrow T = constant \Rightarrow \upsilon = constant$, where $\upsilon$ is the speed and $T$ is the kinetic energy of $q$.

Assume now that in place of a single moving charge $q$ we have a current $I$ on a metal wire (Fig. 7.1). The wire is placed inside a static (time-independent) magnetic field $\vec{B}(\vec{r})$ which owes its existence to a system of stationary currents that do not contain the current $I$. Let $\vec{dl}$ be an element of the wire located at a point $\vec{r}$ relative to the origin $O$ of our coordinate system and oriented in the direction of motion of the *conventionally positive* mobile electrons of the metal (by definition, this is the direction of the current $I$ in the wire).

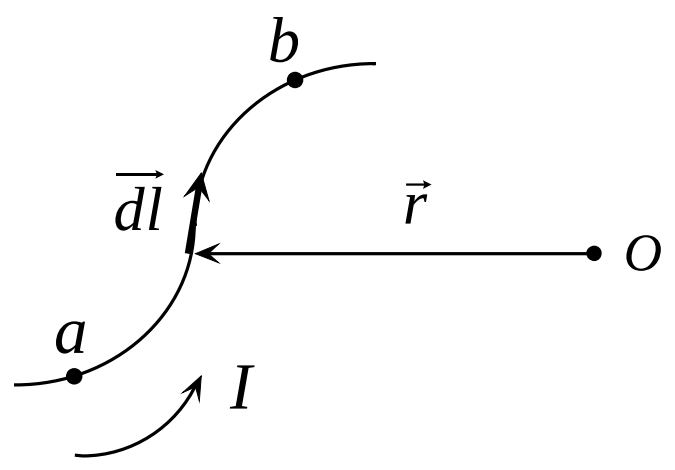

Fig. 7.1. A current-carrying wire inside a static magnetic field.

If an elementary charge $dq$ passes from point $\vec{r}$ of the wire within time $dt$, then $I=dq/dt$. In the time interval $dt$ the moving charge $dq$ covers the section $\vec{dl}$ of the wire. Therefore a charge $dq=Idt$, located instantaneously at the point $\vec{r}$ of the wire, is moving with velocity $\vec{\upsilon}=\vec{dl}/dt$ along the wire. Let now $\vec{B}$ be the magnetic field at the location $\vec{r}$ of $\vec{dl}$. The elementary magnetic force on the section $\vec{dl}$ of the wire is

$$d\vec{F}_m = dq\,(\vec{\upsilon}\times\vec{B}) = I\,dt\,(\frac{\vec{dl}}{dt}\times\vec{B}) = I\,(\vec{dl}\times\vec{B})$$

Thus the magnetic force on a finite section $ab$ of the wire is

$$\vec{F}_m = I\int_a^b \vec{dl}\times\vec{B}(\vec{r}) \qquad (7.3)$$

where the line integral on the right-hand side of (7.3) is to be evaluated along the wire. If $dx$, $dy$, $dz$ are the Cartesian components of $\vec{dl}$, the cross product in the integrand in (7.3) is given by the determinant expression

$$\vec{dl}\times\vec{B} = \begin{vmatrix} \hat{u}_x & \hat{u}_y & \hat{u}_z \\ dx & dy & dz \\ B_x & B_y & B_z \end{vmatrix}$$

We recall that a static distribution of charge, expressed by a time-independent charge density $\rho(\vec{r})$, produces an electrostatic field $\vec{E}(\vec{r})$. In an analogous way, a static current distribution represented by a time-independent current density $\vec{J}(\vec{r})$ is the source of a static magnetic field $\vec{B}(\vec{r})$ (the term *magnetostatic field* is also used). In particular, a static magnetic field can be produced by a system of currents $I_1$, $I_2$,..., which are constant in both magnitude and shape.

Let $I$ be such a current on a wire and let $\vec{dl}$ be an element of the wire, oriented in the direction of the current (Fig. 7.2). We consider a point $P$ of space at which we want to determine the value $\vec{B}(P)$ of the magnetic field produced by $I$. We let $\vec{R}$ be the position vector of $P$ relative to $\vec{dl}$ and we call $\hat{R} = \vec{R}/R$ (where $R=|\vec{R}|$) the unit vector in the direction of $\vec{R}$.

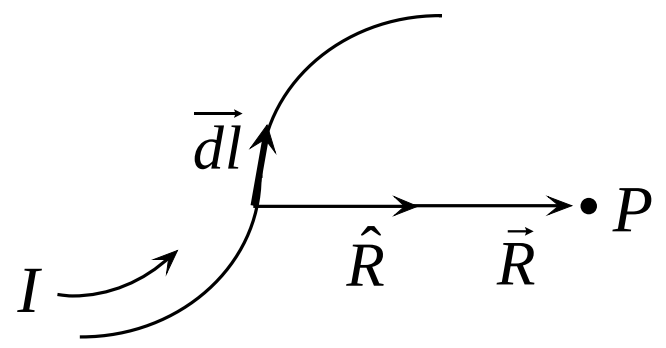

Fig. 7.2. A current *I* producing a magnetic field at point *P*.

As found by experiment, the small segment $\vec{dl}$ of the current contributes an elementary amount to the magnetic field at $P$, equal to[1]

$$d\vec{B} = \frac{\mu_0}{4\pi} I \frac{\vec{dl} \times \hat{R}}{R^2}$$

where $\mu_0 = 4\pi \times 10^{-7} N/A^2$. The total magnetic field at $P$, due to the entire current $I$ on the wire, is

$$\vec{B}(P) = \frac{\mu_0}{4\pi} I \int \frac{\vec{dl} \times \hat{R}}{R^2} \tag{7.4}$$

where the line integral is to be evaluated along the wire. Relation (7.4) expresses the *Biot-Savart law*. Its importance in Magnetism is analogous to that of Coulomb's law in Electricity.

## 7.2 Gauss' Law for Magnetism

As we know, the sources of the electric field are electric charges, regardless of their state of motion. According to Gauss' law, the electric flux through a closed surface $S$ is a measure of the number of sources in the interior of $S$; that is, of the total electric charge enclosed by $S$.

What are the sources of the *magnetic* field? As we saw in the previous section, this field can be produced by *moving* electric charges (or electric currents). On the other hand, there have never been detected any free (isolated) "magnetic charges" (or *magnetic poles*) in Nature. Among other things, this suggests that

*the total magnetic flux through a closed surface S is zero:*

---

[1] See Appendix A.

$$\boxed{\oint_S \vec{B} \cdot \overrightarrow{da} = 0} \qquad (7.5)$$

This can be interpreted geometrically as follows: the number of field lines of $\vec{B}$ entering $S$ equals the number of field lines exiting from it. Indeed, the surface $S$ does not enclose isolated magnetic sources at which new field lines could begin or end. Hence, any field line entering $S$ at some point must necessarily exit from $S$ at some other point. We conclude that

> *the magnetic field lines do not have a beginning or an end in any finite region of space; either they are closed or they begin at infinity and end at infinity.*

This is in contrast to the electric field lines, which begin and/or end at electric charges.

By using Gauss' integral theorem we can transform the surface integral in (7.5) into a volume integral. We thus have:

$$\int_V (\vec{\nabla} \cdot \vec{B})\, dv = 0$$

where $V$ is the volume enclosed by $S$. For this to be valid for an arbitrary volume $V$, the following differential equation must be satisfied:

$$\boxed{\vec{\nabla} \cdot \vec{B} = 0} \qquad (7.6)$$

This equation is the *differential form of Gauss' law* for the magnetic field. Note that there is no "magnetic charge density" on the right-hand side of (7.6), analogous to the electric charge density $\rho$ appearing in Gauss' law (5.8) for the electric field.

According to Eq. (7.6), *the magnetic field is solenoidal*. Therefore,

> *the magnetic flux* $\int_S \vec{B} \cdot \overrightarrow{da}$ *through an open surface S depends only on the border of S, thus is the same for all open surfaces sharing a common border*

(cf. Sec. 4.3; see also Prob. 4). This is not the case with electric flux, $\int_S \vec{E} \cdot \overrightarrow{da}$, since the space between two open surfaces sharing a common border may contain electric charges at which additional electric field lines begin or end. These lines will differentiate the electric flux passing through the two surfaces.

## 7.3 Ampère's Law

Consider a closed path (loop) $C$ in some region of space (Fig. 7.3). Assume that a positive direction of traversing $C$ has arbitrarily been assigned. Assume further that the loop $C$ encircles a set of currents $I_1$, $I_2$, ... (in other words, the aforementioned currents pass through $C$).

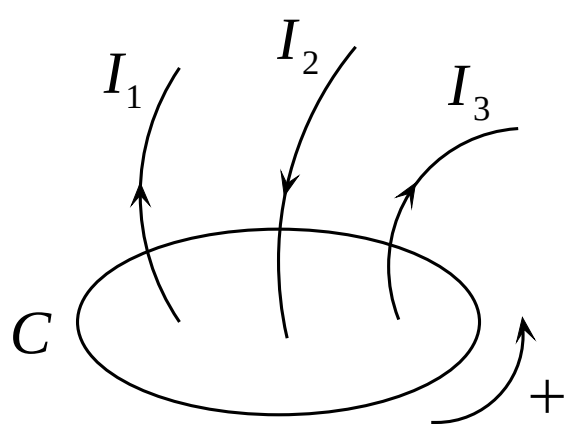

Fig. 7.3. An oriented loop $C$ encircling a set of currents.

The total current passing through the loop is the *algebraic* sum of $I_1$, $I_2$, … In this sum the signs of the currents are determined by the "right-hand rule", as follows: We rotate the fingers of our right hand in the positive direction of traversing the loop. A current is considered positive (negative) if its direction agrees with (is opposite to) the direction of our extended thumb. Thus the total current through the loop $C$ in Fig. 7.3 is $I_{in} = I_1 - I_2 + I_3$ (where the $I_1, I_2, I_3$ represent *magnitudes*, hence are *positive* quantities). We recall that the direction of a current $I$ is defined as the direction of the corresponding current density $\vec{J}$ or, equivalently, the direction of motion of (conventionally) *positive* charges (see Chap. 6).

Let us assume now that the loop $C$ is the border of an open surface $S$ (Fig. 7.4). Let $\overrightarrow{dl}$ be an element of $C$ oriented in the positive direction of traversing the loop, and let $\overrightarrow{da}$ be an element of $S$. The direction of $\overrightarrow{da}$ is defined consistently with that of $\overrightarrow{dl}$ according to the right-hand rule, as explained in Sec. 4.2.

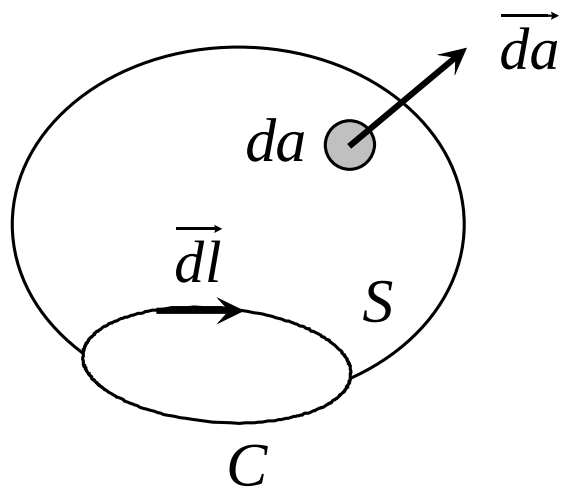

Fig. 7.4. An open surface $S$ bordered by an oriented loop $C$.

We call $I_{in}$ the total current through the loop $C$. There are possibly other currents in this region of space that do not go through $C$ and, therefore, are not contained in $I_{in}$. All currents are assumed to be constant in time; thus their combined effect is the creation of a static magnetic field $\vec{B}(\vec{r})$ in space. Note that this field owes its existence to *all* currents, both internal *and* external relative to $C$.

As follows by the Biot-Savart law [1-3] the value of the closed line integral of $\vec{B}$ along the loop $C$ *depends only on the total internal current* $I_{in}$ (even though $\vec{B}$ itself is also dependent on the external currents!). Specifically,

$$\boxed{\oint_C \vec{B} \cdot \overrightarrow{dl} = \mu_0 I_{in}} \qquad (7.7)$$

Equation (7.7) is the integral form of *Ampère's law*. Note the similarity in spirit between this law and Gauss' law in Electricity.

Now, by Stokes' theorem,

$$\oint_C \vec{B} \cdot \overrightarrow{dl} = \int_S (\vec{\nabla} \times \vec{B}) \cdot \overrightarrow{da}$$

Moreover, by "current through the loop $C$" we actually mean the current passing through *any* open surface *S* bounded by *C*. (To see why this current is independent of *S* for a given *C*, take into account that, in general, a current is not something that begins somewhere and ends somewhere else but it must always describe the closed path of a circuit. For a more rigorous proof, see Prob. 5.) Thus, if $\vec{J}(\vec{r})$ is the current density on *S*,

$$I_{in} = \int_S \vec{J} \cdot \overrightarrow{da}$$

As argued above, the value of $I_{in}$ is independent of *S* for a given *C*. The integral equation (7.7) is now written:

$$\int_S (\vec{\nabla} \times \vec{B}) \cdot \overrightarrow{da} = \int_S \mu_0 \vec{J} \cdot \overrightarrow{da}$$

For this to be true for any surface $S$ bounded by the closed curve $C$, we must have:

$$\boxed{\vec{\nabla} \times \vec{B} = \mu_0 \vec{J}} \qquad (7.8)$$

Equation (7.8) is the *differential form of Ampère's law*.

*Comments:*

1. Gauss' law (5.8) in Electricity and the corresponding law (7.6) in Magnetism are valid for both static and time-dependent fields $\vec{E}$ and $\vec{B}$. On the contrary, Ampère's law (7.8) is valid for a *static* field $\vec{B}$ only. Similarly, the relation $\vec{\nabla} \times \vec{E} = 0$ is only valid for an *electrostatic* field. As we will see in Chapter 9, the two equations for the *curl* of $\vec{E}$ and $\vec{B}$ must be revised in the case of time-dependent electromagnetic fields.
2. Both Gauss' law (7.6) and Ampère's law (7.8) originate from the Biot-Savart law [1-3]. Similarly, both Gauss' law (5.8) in Electricity and the relation $\vec{\nabla} \times \vec{E} = 0$ for *electrostatic* fields are direct consequences of Coulomb's law.

## References for Chapter 7

## QUESTIONS

**1.** It was mentioned in Sec. 7.1 that the magnetic field is produced by moving charges and acts only on moving charges. Can you justify this statement on the basis of the fundamental laws of Magnetism?

**2.** In the interior of a long straight tube there is a uniform magnetic field directed parallel to the axis of the tube. A charge enters the tube moving in a direction normal to the tube's cross-section. Will the charge ever exit from the tube?

**3.** A charged particle escaped from our laboratory and is now moving freely in the positive direction of the $x$-axis. We want to halt it by setting up a suitable field in the area. What kind of field should we use, an electric or a magnetic one? What must be the direction of that field if the particle is (*a*) a proton? (*b*) an electron?

**4.** An eccentric cook is wearing two caps, a smaller internal one and a larger external one, sewn together as to have a common opening edge. The cook enters our laboratory at a moment when we perform a measurement of a magnetic field existing in that area. Through which cap will the most magnetic flux pass?

**5.** We are at the central square of a small village. Just outside the village, extraterrestrials have set up a huge electric circuit carrying a constant current $I$. A scientist who takes her coffee at a cafe at the square perceives the unexpected existence of a magnetic field. To investigate its origin, she partitions the perimeter of the square into a very large number of elementary sections $\overrightarrow{dl_i}$ ($i$=1,2,...) and she then measures the value $\vec{B}_i$ of the magnetic field at each section. Her goal is to evaluate the sum $\sum_i \vec{B}_i \cdot \overrightarrow{dl_i}$. Knowing the reality of the situation, can you help her find the result without any effort? (*Hint:* In the limit $\overrightarrow{dl_i} \to 0$, the above sum may be replaced by a line integral.)

## PROBLEMS

**1.** The Coulomb field (5.3) is a special case of a radial field of the form $\vec{E}(\vec{r}) = E(r)\,\hat{r}$, where $r = |\vec{r}\,|$ and $\hat{r} = \vec{r}\,/\,r$ (the position vector $\vec{r}$ is taken relative to the location of the point charge that produces the field; by $\hat{r}$ we denote the unit vector in the direction of $\vec{r}$ ). Show that no radial magnetic field of the form $\vec{B}(\vec{r}) = B(r)\,\hat{r}$ may exist. What is the physical significance of this?

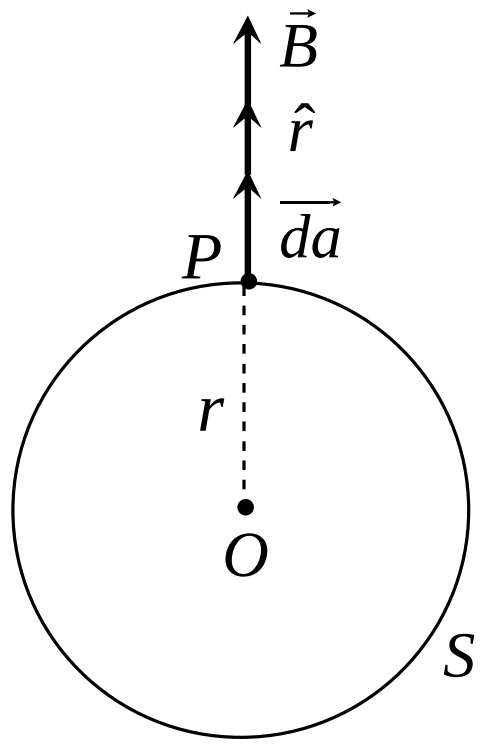

Fig. 7.5. Hypothetical radial magnetic field on a spherical surface *S* of radius *r*.

***Solution:*** Consider a spherical surface *S* of radius *r*, centered at some fixed reference point *O* (Fig. 7.5). Let $\vec{B}$ be the magnetic field vector at some point *P* of *S*. The position vector of *P* with respect to *O* is $\vec{r} = \overrightarrow{OP}$ . By Gauss' law for Magnetism, the total magnetic flux through *S* is zero:

$$\oint_S \vec{B} \cdot \overrightarrow{da} = 0 \;\Rightarrow\; \oint_S [B(r)\,\hat{r}] \cdot [(da)\,\hat{r}] = 0 \;\Rightarrow\; \oint_S B(r)\,da = 0$$

or, since $B(r)$ has a constant value on $S$,

$$B(r) \oint_S da = 0 \;\Rightarrow\; B(r)(4\pi r^2) = 0 \;\Rightarrow\; B(r) = 0 \;\Rightarrow\; \vec{B} = 0$$

Therefore the assumption $\vec{B}(\vec{r}) = B(r)\,\hat{r}$ cannot hold for $\vec{B} \neq 0$ . Indeed, if a magnetic field of this form existed, the magnetic flux through *S* would not vanish, which would indicate the presence of an isolated magnetic "charge" at the point *O* at which the radial magnetic field lines begin. But, as we know, no such magnetic poles exist!

**2.** Show that a nonzero total current passes through any (closed) field line of a static magnetic field.

***Solution:*** Let *C* be a field line of $\vec{B}$ . At each point of *C* the vector $\vec{B}$ is tangential, having the direction of the element $\overrightarrow{dl}$ of *C*. Hence,

$$\oint_C \vec{B} \cdot \overrightarrow{dl} = \oint_C |\vec{B}|\,|\overrightarrow{dl}| > 0 \qquad (1)$$

On the other hand, by Ampère's law,

$$\oint_C \vec{B} \cdot \overrightarrow{dl} = \mu_0\, I_{in} \qquad (2)$$

From (1) and (2) it follows that $I_{in} \neq 0$.

**3.** A long, thin, hollow metal cylinder of radius *R* carries a constant current *I* that runs parallel to the cylinder's axis. Determine the magnetic field both inside and outside the cylinder.

***Solution:*** We will first take a look at a simpler and more fundamental problem: the magnetic field produced by a long rectilinear current *I* (Fig. 7.6). As found by the Biot-Savart law, the magnetic field lines are circular, each circle having its center on the axis of the current and belonging to a plane normal to that axis. The magnetic field $\vec{B}$ is tangential at every point of a field line, its direction being determined by the right-hand rule; that is, if we rotate the fingers of our right hand in the direction of $\vec{B}$, our extended thumb will point in the direction of *I*.

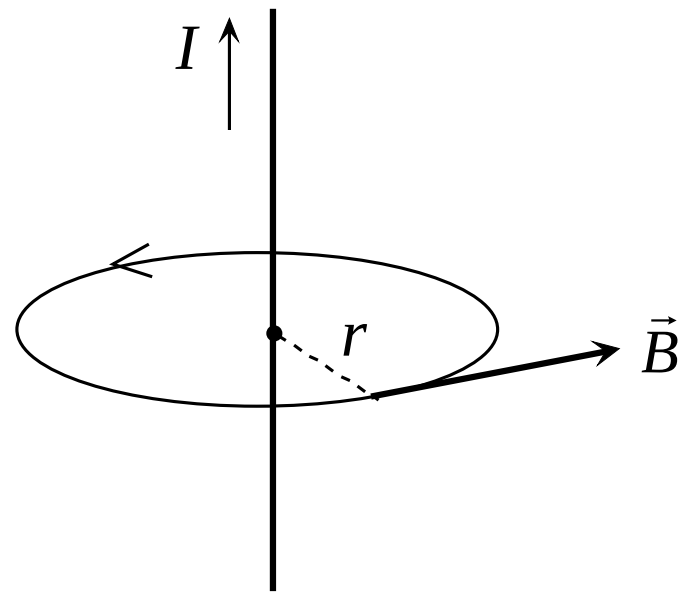

Fig. 7.6. Magnetic field line around a rectilinear current *I*.

By symmetry, the magnitude *B* of $\vec{B}$ is constant along a field line. If $r$ is the radius of this line, it can be shown that

$$\boxed{B(r) = \frac{\mu_0 I}{2\pi r}} \qquad (1)$$

By integrating $\vec{B}$ on a field line *C* we thus recover Ampère's law:

$$\oint_C \vec{B} \cdot \vec{dl} = \oint_C |\vec{B}||\vec{dl}| = \oint_C B\,dl = B \oint_C dl = B(2\pi r) = \mu_0 I = \mu_0 I_{in}$$

where we have used the fact that *B* is constant on *C*, equal to *B*(*r*).

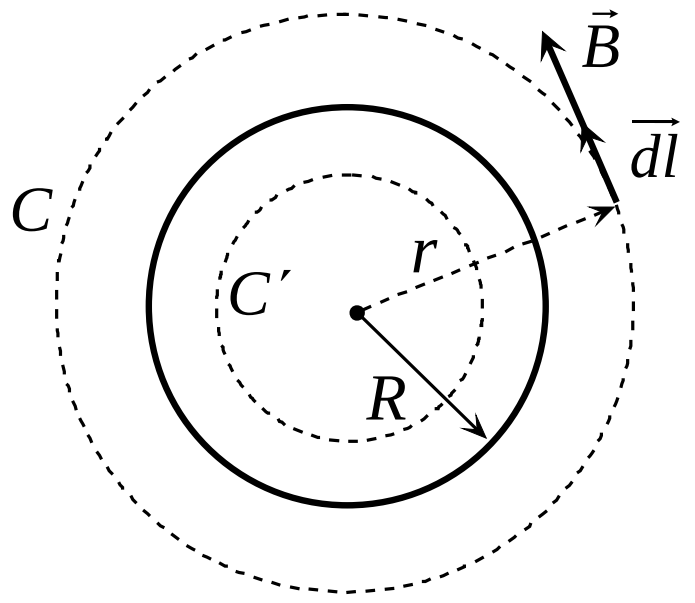

Fig. 7.7. Cross-section of a current-carrying metal cylinder of radius *R*.

Let us now return to our original problem. Consider a normal cross-section of the cylinder, as seen in Fig. 7.7. Imagine that the axis of the cylinder is normal to the page and that the current $I$ is directed *outward*, i.e., toward the reader. Since the problem has the same symmetry as the problem of the long rectilinear current (*cylindrical symmetry*), the field lines of the magnetic field $\vec{B}$ both inside and outside the cylinder will be circles normal to the cylinder and centered on its axis, while the magnitude $B$ of $\vec{B}$ will be constant along a field line. Let $C$ be a field line of radius $r>R$, i.e., a line external to the cylinder. By applying Ampère's law on the closed path $C$ and by noticing that $I_{in}=I$, we have:

$$\oint_C \vec{B}\cdot\overrightarrow{dl} = \mu_0 I_{in} \;\Rightarrow\; \mu_0 I = \oint_C |\vec{B}||\overrightarrow{dl}| = \oint_C B\,dl = B\oint_C dl = B(2\pi r) \;\Rightarrow$$

$$B(r) = \frac{\mu_0 I}{2\pi r} \;, \quad r > R \qquad (2)$$

By comparing (2) with (1) we notice that the magnetic field in the *exterior* of the cylinder is the same as that of a hypothetical rectilinear current $I$ flowing along the cylinder's axis! Now, to find the magnetic field in the *interior* of the cylinder, we apply Ampère's law on a loop $C'$ with $r<R$, noticing that $I_{in}=0$ in this case:

$$\oint_{C'} \vec{B}\cdot\overrightarrow{dl} = \mu_0 I_{in} \;\Rightarrow\; 0 = \oint_{C'} |\vec{B}||\overrightarrow{dl}| = \oint_{C'} B\,dl = B\oint_{C'} dl = B(2\pi r) \;\Rightarrow$$

$$B(r) = 0 \;, \quad r < R$$

That is, there is no magnetic field inside the cylinder.

**4.** By using Gauss' law for the magnetic field, show that an equal magnetic flux passes through any two open surfaces $S_1$, $S_2$ having a common border $C$.

***Solution:*** The magnetic flux through a surface $S$ is defined as the surface integral of the magnetic field $\vec{B}$ over $S$. According to Gauss' law, the field $\vec{B}$ is solenoidal. Now, as discussed in Sec. 4.3, the value of the surface integral of a solenoidal field, over an open surface $S$ bordered by a closed curve $C$, depends only on the border $C$ of $S$ and is the same for any two open surfaces sharing a common border $C$. This is exactly what we needed to prove.

It is instructive, however, to view an alternative proof of the above property. Let $S$ be a *closed* surface (Fig. 7.8). According to Gauss' law,

$$\oint_S \vec{B}\cdot\overrightarrow{da} = 0 \qquad (1)$$

The surface element $\overrightarrow{da}$ in the above integral is always directed *outward* relative to $S$, at all points of this surface. Imagine now that we partition $S$ into two *open* surfaces $S_1$ and $S_2$ by drawing a closed curve $C$ on $S$. Clearly, these two surfaces have a common border $C$.

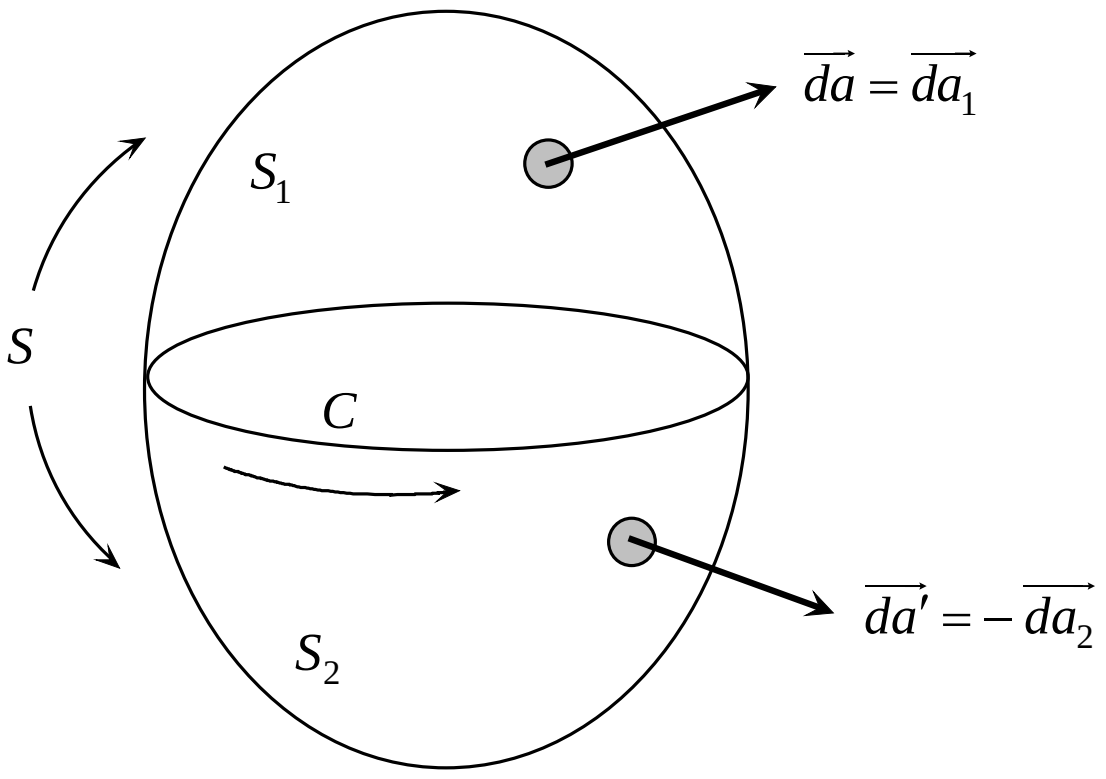

Fig. 7.8. A closed surface $S$ partitioned into two open surfaces $S_1$ and $S_2$ by the closed curve $C$.

Equation (1) is written

$$\oint_S \vec{B}\cdot\overrightarrow{da}=\int_{S_1}\vec{B}\cdot\overrightarrow{da}+\int_{S_2}\vec{B}\cdot\overrightarrow{da'}=0 \qquad (2)$$

where both $\overrightarrow{da}$ and $\overrightarrow{da'}$ point *outward* with respect to $S$. Now, if we assign a positive direction of traversing the closed curve $C$, only one of the above surface elements, say $\overrightarrow{da}$, will be oriented consistently with the direction of $C$ according to the right-hand rule, while the orientation of $\overrightarrow{da'}$ will be inconsistent with that of $C$. The element $-\overrightarrow{da'}$, however, will be properly oriented. By setting $\overrightarrow{da_1}\equiv\overrightarrow{da}$ and $\overrightarrow{da_2}\equiv-\overrightarrow{da'}$, relation (2) is finally rewritten as

$$\int_{S_1}\vec{B}\cdot\overrightarrow{da_1}-\int_{S_2}\vec{B}\cdot\overrightarrow{da_2}=0 \;\Leftrightarrow\; \int_{S_1}\vec{B}\cdot\overrightarrow{da_1}=\int_{S_2}\vec{B}\cdot\overrightarrow{da_2}\;;\ \text{Q.E.D.}$$

**5.** Consider a region of space in which the distribution of charge is static. Let $C$ be a closed curve in this region and let $S$ be any open surface bordered by $C$. We define the total current through $C$ as the surface integral of the current density over $S$:

$$I_{in}=\int_S \vec{J}\cdot\overrightarrow{da}$$

Show that, for a given $C$, the quantity $I_{in}$ has a well-defined value independent of the particular choice of $S$.

***Solution:*** Since the charge density $\rho$ inside the region is static ($\partial\rho/\partial t=0$), the equation of continuity, Eq. (6.9), reduces to $\vec{\nabla}\cdot\vec{J}=0$. This means that, within the considered region of space, the current density has the properties of a solenoidal field. In particular, the value of the surface integral of $\vec{J}$ over an open surface $S$ depends only on the border $C$ of $S$ and is the same for all surfaces sharing a common border $C$.

*Exercise:* Suggest an alternative proof, in the spirit of Prob. 4.

CHAPTER 8

# STATIC ELECTRIC AND MAGNETIC FIELDS IN MATTER

## 8.1 Electric and Magnetic Dipole Moments

Consider a system of charges $q_1$, $q_2$,…, the instantaneous positions of which are $\vec{r}_1, \vec{r}_2, \cdots$, respectively, relative to the origin $O$ of our coordinate system (Fig. 8.1). The *electric dipole moment* of the system is defined as the vector

$$\vec{p} = \sum_i q_i \vec{r}_i = q_1 \vec{r}_1 + q_2 \vec{r}_2 + \cdots \qquad (8.1)$$

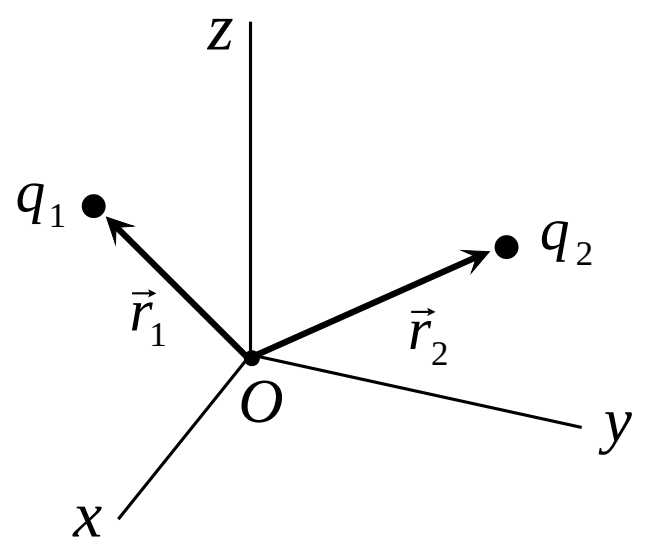

Fig. 8.1. A system of charges $q_1, q_2, ...$ in space.

In general, $\vec{p}$ depends on the choice of the reference point $O$ in space. In certain cases, however, $\vec{p}$ has an absolute significance independent of the choice of $O$. As an example, let us consider an *electric dipole* consisting of two opposite charges $q_1 = -q$ and $q_2 = +q$, a distance $s$ apart (Fig. 8.2). We call $\vec{s}$ the vector *from* $-q$ *to* $+q$. The electric dipole moment of the system is

$$\vec{p} = (-q)\vec{r}_1 + q\vec{r}_2 = q(\vec{r}_2 - \vec{r}_1) \;\Rightarrow$$

$$\boxed{\vec{p} = q\vec{s}} \qquad (8.2)$$

Obviously, $\vec{p}$ is independent of the location of the origin $O$ of our coordinate system.

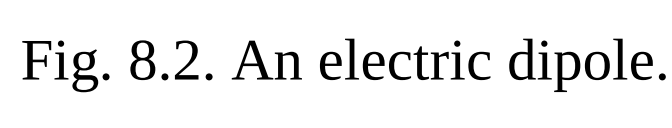

Fig. 8.2. An electric dipole.

Imagine now that we place the electric dipole inside a *uniform* electric field $\vec{E}$. The two charges will then be subject to forces $\vec{F}_1 = -q\vec{E}$, $\vec{F}_2 = +q\vec{E}$, which form a *couple* (Fig. 8.3). The dipole is thus subject to an external torque

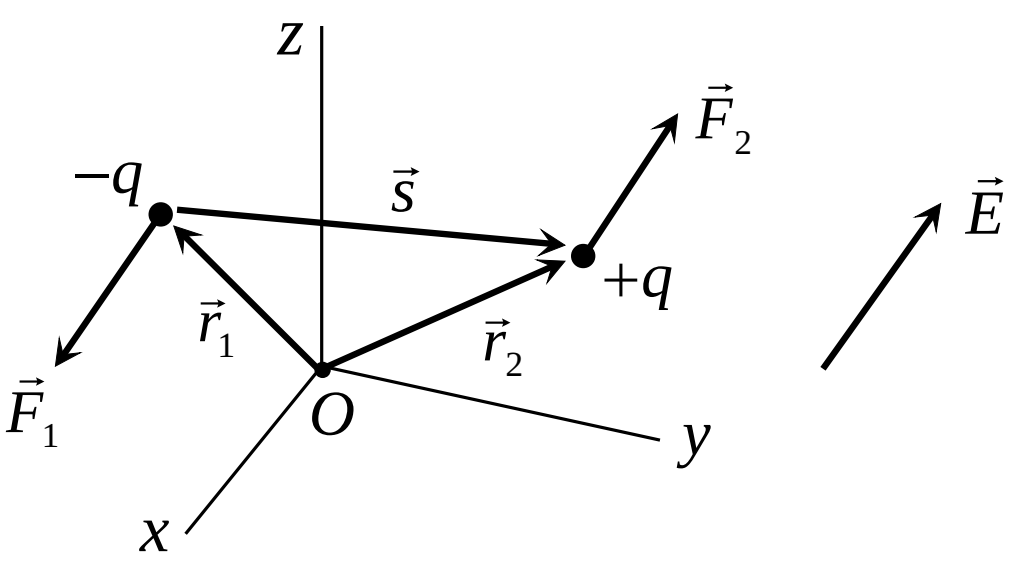

Fig. 8.3. Forces on an electric dipole placed inside a uniform electric field.

$$\vec{T} = (\vec{r}_1 \times \vec{F}_1) + (\vec{r}_2 \times \vec{F}_2) = q(\vec{r}_2 - \vec{r}_1) \times \vec{E} = q\vec{s} \times \vec{E} \quad \Rightarrow$$

$$\vec{T} = \vec{p} \times \vec{E} \tag{8.3}$$

(notice that this torque is independent of the reference point *O*). The torque $\vec{T}$ *tends to rotate the dipole so that its dipole moment* $\vec{p}$ *is oriented in the direction of the electric field* $\vec{E}$.

The term *magnetic dipole* is misleading, given that no individual magnetic poles exist! By this term we simply mean *a current-carrying loop*. The origin of this paradoxical name is the similarity in form between the mathematical expressions for the magnetic field produced by a current loop and the electric field produced by an electric dipole [1,2].

Let us thus consider a small plane loop of area $a$, carrying a current *I*. We define a vector $\vec{a}$ of magnitude $a$, normal to the plane of the loop and directed consistently with the direction of the current according to the right-hand rule (Fig. 8.4). The *magnetic dipole moment* of the current loop is the vector[1]

$$\boxed{\vec{m} = I\,\vec{a}} \tag{8.4}$$

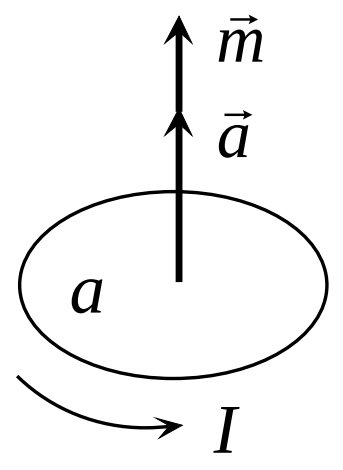

Fig. 8.4. A magnetic "dipole" and its dipole moment.

Notice that $\vec{m}$ is normal to the plane of the loop and directed in accordance with the direction of $I$ through the right-hand rule. As found by the Biot-Savart law [1,2] the magnetic field produced by a *circular* current loop has the direction of the magnetic dipole moment $\vec{m}$ at all points of the axis normal to the loop and passing through its center.

[1] We assume that *I* represents the magnitude of the current; hence $I > 0$.

Now, if a magnetic dipole is placed inside a magnetic field $\vec{B}$, it is subject to an external torque

$$\vec{T} = \vec{m} \times \vec{B} \tag{8.5}$$

which *tends to align the dipole moment* $\vec{m}$ *with the magnetic field* $\vec{B}$.

## 8.2 Electric Polarization

As we know, conductors owe their conductivity to free electrons, i.e., electrons that are not attached to specific atoms but are able to move freely inside the crystal lattice. On the other hand, in insulators (or *dielectrics*, as they are also called) all electrons – more importantly, the valence ones – are bound within the atoms or molecules to which they belong and are not easily detached from them. We will now study the behavior of a dielectric when an electric field $\vec{E}$ exists in its interior.

Let us first examine the case of a dielectric consisting of electrically neutral atoms. Ordinarily, an atom is not expected to exhibit electric dipole moment because of the symmetry of the charge distribution in it (the negative charge due to the electrons is symmetrically distributed around the positively-charged nucleus, hence no overall separation between positive and negative charge exists in the atom). If, however, the atom is placed into an electric field $\vec{E}$, a separation between positive and negative charge appears. Specifically, the nucleus undergoes a displacement in the direction of $\vec{E}$, while the electrons tend to shift in the opposite direction. The atom now looks like a small electric dipole whose dipole moment $\vec{p}$ is *in the direction of* $\vec{E}$. Moreover, it is observed that if the field $\vec{E}$ is not too strong, the dipole moment $\vec{p}$ is approximately proportional to $\vec{E}$. Since $\vec{p}$ owes its existence exclusively to $\vec{E}$ (it did not exist before), we say that this dipole moment has been *induced* by the electric field.

An analogous situation occurs if the dielectric consists of *nonpolar molecules*, i.e., molecules that do not possess a pre-existent electric dipole moment. An example is the $CO_2$ molecule (Fig. 8.5). Again, an electric field will cause an overall separation between positive and negative charge in the molecule.

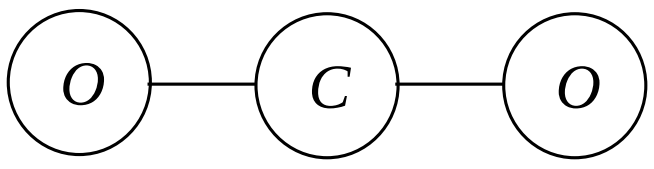

Fig. 8.5. A nonpolar molecule.

On the other hand, some dielectrics consist of *polar molecules* possessing *permanent* (pre-existent) electric dipole moment $\vec{p}$. Such is the case of the $H_2O$ molecule (Fig. 8.6). The electrons in this molecule spontaneously tend to accumulate closer to the oxygen atom, thus endowing this atom with negative charge while leaving the hydrogen atoms positively charged. If such a molecule is placed into an electric field $\vec{E}$, its dipole moment $\vec{p}$ will tend to be aligned with $\vec{E}$. The end result will thus be again a dipole moment oriented in the direction of $\vec{E}$, the only difference being that this time the moment pre-existed rather than was induced by $\vec{E}$.

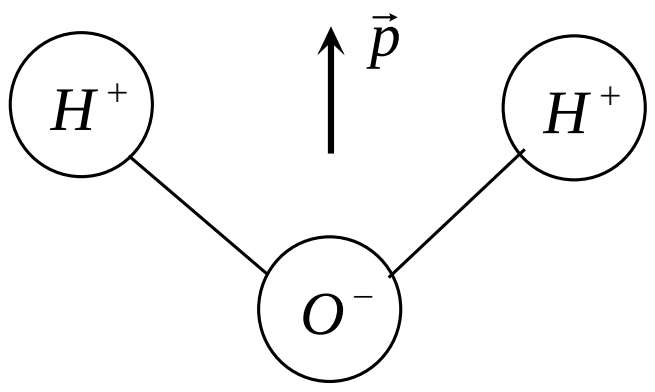

Fig. 8.6. A polar molecule.

In both polar and nonpolar dielectrics the effect of the presence of an electric field $\vec{E}$ in their interior is the appearance of a large number of small electric dipoles oriented in the direction of $\vec{E}$. We say that the electric field causes *electric polarization* to the dielectric. We define the *polarization vector* $\vec{P}$ as the electric dipole moment per unit volume of the material:

$$\vec{P} = \frac{d\,\vec{p}}{d\,v} \tag{8.6}$$

As a result of polarization, charges of a new kind appear in the interior as well as on the surface of the dielectric. They are called *polarization charges*. In contrast to free charges in conductors, polarization charges are *bound charges* associated with specific atoms or molecules and thus cannot move freely within the material. The reason for their existence is the disturbance of the initial internal electrostatic balance of the dielectric due to polarization caused by the applied electric field; these charges thus disappear as soon as the aforementioned field is removed. The *bound-charge density* in the interior of the dielectric is given by [1,2]

$$\rho_b = -\vec{\nabla} \cdot \vec{P} \tag{8.7}$$

where we have generally assumed that the polarization vector $\vec{P}$ may vary in space. In the case of a *uniform* polarization, the vector $\vec{P}$ is constant and $\rho_b=0$ in the interior of the dielectric. Thus, in this case the polarization charges are confined to the surface of the material.

If in the interior or on the surface of the dielectric there exist additional charges *that are not due to polarization*, these charges are called *free charges*[2] and their volume density is denoted $\rho_f$. Examples of free charges are: the mobile electrons on the metal plates of a capacitor (a dielectric substance is usually placed between these plates); external charges that have somehow been embedded into the dielectric; $Na^+$ and $Cl^-$ ions in saline water; etc.

In a region of space occupied by a polarized dielectric the *total charge density* is the sum

$$\rho = \rho_f + \rho_b \tag{8.8}$$

The electric field $\vec{E}$ inside the dielectric obeys Gauss' law:

[2] In spite of this name, it should not be assumed that free charges are necessarily mobile charges!

$$\vec{\nabla}\cdot\vec{E}=\frac{\rho}{\varepsilon_0} \;\Rightarrow\; \varepsilon_0(\vec{\nabla}\cdot\vec{E})=\rho_f+\rho_b=\rho_f-\vec{\nabla}\cdot\vec{P} \;\Rightarrow\; \vec{\nabla}\cdot(\varepsilon_0\vec{E}+\vec{P})=\rho_f$$

where we have used (8.7) for the polarization charge. We define the *electric displacement*

$$\vec{D}=\varepsilon_0\vec{E}+\vec{P} \qquad (8.9)$$

Gauss' law is now written:

$$\boxed{\vec{\nabla}\cdot\vec{D}=\rho_f} \qquad (8.10)$$

or, in integral form,

$$\boxed{\oint_S \vec{D}\cdot\overrightarrow{da}=Q_{f,in}} \qquad (8.11)$$

where $Q_{f,in}=\int_V \rho_f\, dv$ is the total *free* charge inside a volume *V* bounded by a closed surface *S*.

*Exercise:* By using Gauss' integral theorem (4.27), prove the equivalence between (8.10) and (8.11).

In a *linear dielectric* the polarization $\vec{P}$ is proportional to the electric field $\vec{E}$ at all points, provided that this field is not too strong. We write:

$$\vec{P}=\varepsilon_0\,\chi_e\,\vec{E} \qquad (8.12)$$

where the *dimensionless* factor (pure number) $\chi_e$ is called the *electric susceptibility* of the medium. The electric displacement $\vec{D}$ is then written:

$$\vec{D}=\varepsilon_0\vec{E}+\vec{P}=\varepsilon_0(1+\chi_e)\vec{E} \;\Rightarrow$$

$$\boxed{\vec{D}=\varepsilon\,\vec{E}} \qquad (8.13)$$

where $\varepsilon$ is called the *permittivity* of the medium, equal to

$$\varepsilon=\varepsilon_0(1+\chi_e) \qquad (8.14)$$

The dimensionless quantity

$$\kappa_e=\frac{\varepsilon}{\varepsilon_0}=1+\chi_e \qquad (8.15)$$

is the *relative permittivity*, or *dielectric constant*, of the medium. For all substances, $\chi_e\geq 0$, $\varepsilon\geq\varepsilon_0$ and $\kappa_e\geq 1$. For *free (empty) space*, $\chi_e=0$, $\varepsilon=\varepsilon_0$ and $\kappa_e=1$.

In the case of a linear and *homogeneous* dielectric, Gauss' law (8.11) is written (by taking into account that $\varepsilon = constant$ for a homogeneous medium):

$$\oint_S \varepsilon \vec{E} \cdot \overrightarrow{da} = \varepsilon \oint_S \vec{E} \cdot \overrightarrow{da} = Q_{f,in} \quad \Rightarrow$$

$$\oint_S \vec{E} \cdot \overrightarrow{da} = \frac{Q_{f,in}}{\varepsilon} \tag{8.16}$$

By combining (8.16) with the general form (5.5) of Gauss' law, we have:

$$\boxed{\oint_S \vec{E} \cdot \overrightarrow{da} = \frac{Q_{in}}{\varepsilon_0} = \frac{Q_{f,in}}{\varepsilon}} \tag{8.17}$$

where $Q_{f,in}$ is the total *free* charge enclosed by *S*, while $Q_{in}$ is the *total* charge (including *both* free and bound charges) in the interior of this surface.

*Comment:* It should be emphasized that $\vec{E}$ represents the *final* value of the electric field in the interior of the dielectric, *after* the polarization of the material and the appearance of the bound charge have taken place; it does *not* represent the *applied* external field $\vec{E}_0$ that caused the polarization in the first place. Let us view the situation in more detail: The external field $\vec{E}_0$ causes polarization inside the dielectric and induces bound charge. This charge produces an additional electric field in the dielectric, and the total field now induces additional bound charge, which in turn produces yet more electric field, etc., until the final values for $\vec{E}$ and the bound charge are established. Of course, this process takes place almost instantaneously!

## 8.3 Magnetization

When a material object is placed inside a magnetic field $\vec{B}$ it experiences magnetic polarization, or *magnetization*. This effect consists in the appearance of a large number of small magnetic dipoles in the interior of the object, which dipoles endow the material with a net magnetic dipole moment. We define the *magnetization vector* $\vec{M}$ as the magnetic dipole moment per unit volume of the material:

$$\vec{M} = \frac{d\vec{m}}{dv} \tag{8.18}$$

In contrast to the electric polarization $\vec{P}$, which is always in the direction of the electric field $\vec{E}$, the magnetization $\vec{M}$ may *either* have the direction of $\vec{B}$ *or* be opposite to it. In the former case the material is called *paramagnetic* (e.g., oxygen, sodium, aluminum, etc.) while in the latter case it is called *diamagnetic* (e.g., hydrogen, carbon dioxide, water, copper, etc.). For most substances the effect of the magnetization lasts as long as they are inside a magnetic field. Some materials, however, called *ferromagnetic* (e.g., iron, nickel, cobalt) retain a significant part of their magnetization long after the applied magnetic field has been removed.

Whether a medium is paramagnetic or diamagnetic depends on the result of a competition in its interior, between paramagnetic and diamagnetic effects that take place *simultaneously*. The paramagnetic effects are due to the alignment of *pre-existent* magnetic dipole moments in the direction of the applied magnetic field, which field they tend to *reinforce*. On the contrary, in diamagnetic effects (which are *universal*, occurring in all kinds of atom) the dipole moments are *induced* by the applied magnetic field and tend to *oppose* it. Atoms or crystals having no permanent magnetic dipole moment are diamagnetic (examples are covalent and ionic solids). Metals may be either paramagnetic or diamagnetic.

The induced magnetic dipole moments are due to the effect of $\vec{B}$ on the orbital motion of the atomic electrons of the material (you may imagine the motion of an electron about an atomic nucleus as being equivalent to a small current loop). This effect results in an additional magnetic dipole moment in a direction *opposite* to $\vec{B}$ [1].

The pre-existent (permanent) magnetic dipole moments may owe their existence to either of two factors: (*a*) the spins of the free electrons in a metal (the positive ions do not possess a permanent magnetic dipole moment, with the exception of transition metals such as Fe, Ni, Co, etc.); (*b*) atoms or ions having incomplete outer shells, thus possessing a permanent magnetic dipole moment (this is, e.g., the case with the positive ions of Fe, Ni and Co, in which the 3*d* subshell – which becomes an outer subshell as soon as the 4*s* electrons are dissociated from the atom to become free electrons – is incomplete; we note that, in general, complete shells do *not* contribute to the overall magnetic dipole moment of an atom [3]). The strongly paramagnetic character of the transition metals is due to the contribution of both the above-described factors under the action of an applied magnetic field.

As a result of magnetization, currents of a new kind appear in the interior as well as on the surface of the material. They are called *magnetization currents*. In contrast to currents of free charges flowing on conductors, magnetization currents are *bound currents*, in the sense that they do not constitute transfer of charge over long distances in the material but are cumulative (macroscopic) effects produced by a contribution of a large number of microscopic currents that are inseparably associated with specific atoms or molecules of the substance. With the exception of permanent magnets, magnetization currents cease to exist as soon as the magnetic field that caused the magnetization is removed. The *bound-current density* in the interior of the material is given by [1,2]

$$\vec{J}_b = \vec{\nabla} \times \vec{M} \tag{8.19}$$

where we have generally assumed that the magnetization vector $\vec{M}$ may vary in space. If the magnetization is *uniform*, the vector $\vec{M}$ is constant and $\vec{J}_b = 0$ in the interior of the material; magnetization currents are thus confined to the surface of the medium.

If in the interior or on the surface of the material there exist additional currents *that are not due to magnetization*, these are called *free currents* (since they generally represent an actual transfer of charge over long distances in the medium) and their cur-rent density is denoted $\vec{J}_f$. These currents may flow on metal wires, or may be due to the motion of ions inside some fluid, etc.

In a region of space occupied by a magnetized material the *total current density* is the sum

$$\vec{J} = \vec{J}_f + \vec{J}_b \tag{8.20}$$

The magnetic field $\vec{B}$ inside the medium obeys Ampère's law:

$$\vec{\nabla} \times \vec{B} = \mu_0 \vec{J} \Rightarrow \frac{1}{\mu_0}(\vec{\nabla} \times \vec{B}) = \vec{J}_f + \vec{J}_b = \vec{J}_f + (\vec{\nabla} \times \vec{M}) \Rightarrow \vec{\nabla} \times (\frac{1}{\mu_0}\vec{B} - \vec{M}) = \vec{J}_f$$

where we have used (8.19) for the magnetization current. We define the *auxiliary field* (with no special name!)

$$\vec{H} = \frac{1}{\mu_0}\vec{B} - \vec{M} \tag{8.21}$$

Ampère's law is now written:

$$\boxed{\vec{\nabla} \times \vec{H} = \vec{J}_f} \tag{8.22}$$

or, in integral form,

$$\boxed{\oint_C \vec{H} \cdot \vec{dl} = I_{f,in}} \tag{8.23}$$

where $I_{f,in} = \int_S \vec{J}_f \cdot \overrightarrow{da}$ is the total *free* current passing through an open surface *S* bounded by a closed curve *C*.

*Exercise:* By using Stokes' theorem (4.28), prove the equivalence between (8.22) and (8.23).

In a *linear* magnetized medium the magnetization $\vec{M}$ is proportional to the magnetic field $\vec{B}$ at all points, provided that this field is not too strong. Now, it would be logical to write a proportionality relation between $\vec{M}$ and $\vec{B}$ analogous to the relation (8.12) between $\vec{P}$ and $\vec{E}$ for dielectrics. In Magnetism, however, it is customary to relate $\vec{M}$ with the auxiliary field $\vec{H}$:

$$\vec{M} = \chi_m \vec{H} \tag{8.24}$$

where the *dimensionless* factor (pure number) $\chi_m$ is called the *magnetic susceptibility* of the medium. It is found that $\chi_m>0$ for paramagnetic media while $\chi_m<0$ for diamagnetic media. The auxiliary field $\vec{H}$ is now written:

$$\vec{H} = \frac{1}{\mu_0}\vec{B} - \vec{M} = \frac{1}{\mu_0}\vec{B} - \chi_m \vec{H} \;\Rightarrow\; \vec{B} = \mu_0 (1+\chi_m)\vec{H} \;\Rightarrow$$

$$\boxed{\vec{H} = \frac{1}{\mu}\vec{B}} \qquad (8.25)$$

where $\mu$ is called the *magnetic permeability* of the medium, equal to

$$\mu = \mu_0 (1+\chi_m) \qquad (8.26)$$

The dimensionless quantity

$$\kappa_m = \frac{\mu}{\mu_0} = 1+\chi_m \qquad (8.27)$$

is called the *relative permeability* of the medium and is always positive: $\kappa_m>0$ (for all substances, $|\chi_m|<1$, so that $1+\chi_m>0$). In particular, $\kappa_m>1$ for paramagnetic media and $\kappa_m<1$ for diamagnetic media. For *free (empty) space*, $\chi_m=0$, $\mu=\mu_0$ and $\kappa_m=1$.

In the case of a linear and *homogeneous* medium, Ampère's law (8.23) is written (by taking into account that $\mu=$ *constant* for a homogeneous medium):

$$\oint_C \frac{1}{\mu}\vec{B}\cdot\overrightarrow{dl} = \frac{1}{\mu}\oint_C \vec{B}\cdot\overrightarrow{dl} = I_{f,in} \quad\Rightarrow$$

$$\oint_C \vec{B}\cdot\overrightarrow{dl} = \mu I_{f,in} \qquad (8.28)$$

By combining (8.28) with the general form (7.7) of Ampère's law, we have:

$$\boxed{\oint_C \vec{B}\cdot\overrightarrow{dl} = \mu_0 I_{in} = \mu I_{f,in}} \qquad (8.29)$$

where $I_{f,in}$ is the total *free* current passing through the loop $C$, while $I_{in}$ is the *total* current (including *both* free and bound currents) passing through $C$.

## 8.4 Applications

Imagine that we charge a parallel-plate capacitor by connecting it with a certain source (e.g., a battery). Assume that, initially, there is only air between the plates. As soon as the charging process is completed, we disconnect the capacitor from the source. The charges $\pm Q_0$ on the metal plates are then fixed; they are *free* charges since they do not originate from the polarization of some dielectric medium. After an electrostatic equilibrium is established, there is a uniform electrostatic field $\vec{E}_0$ in the interior of the capacitor (Fig. 8.7). If $V_0$ is the potential difference between the plates, the capacitance of the system is $C_0= Q_0/V_0$.

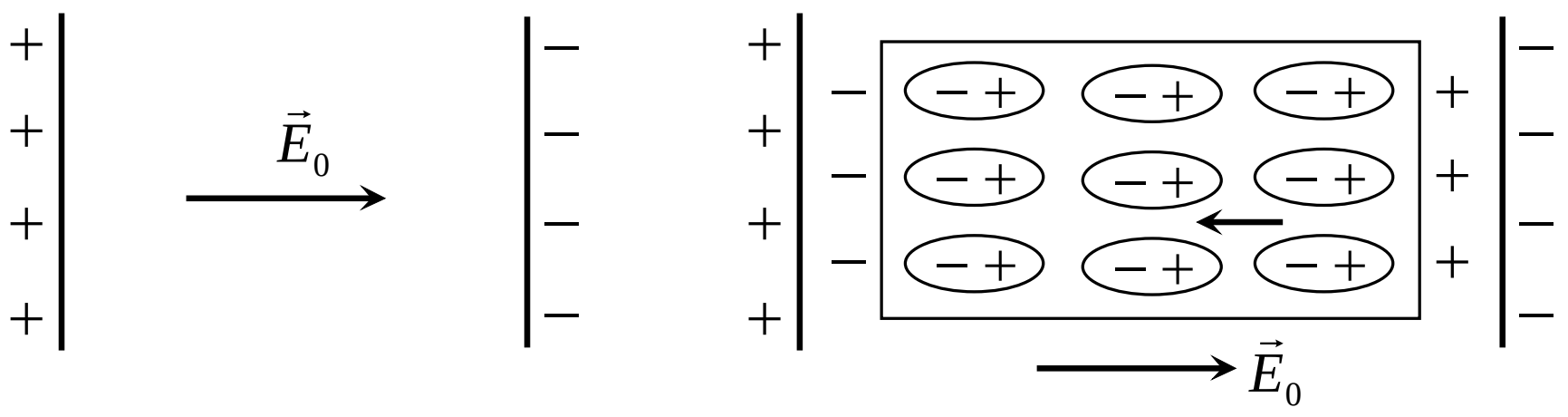

Capacitor without a dielectric

Capacitor with a dielectric

Fig. 8.7. Polarization of a dielectric inside a parallel-plate capacitor.

We now fill in the space between the plates with a dielectric material having a dielectric constant $\kappa$. The dielectric experiences polarization by the electric field of the capacitor and is transformed into an aggregate of little electric dipoles aligned with the field, as seen in Fig. 8.7. As a result, bound charges appear on the surface of the dielectric (in the interior of the material, bound charges cancel each other). The surface bound charges generate an additional electric field inside the dielectric, which field is *opposite* to the initial field $\vec{E}_0$ of the capacitor. Hence, by putting a dielectric in between the plates we have effectively *reduced the value of the electric field* inside the capacitor. As can be shown [1] the new value of the field is

$$E = \frac{E_0}{\kappa} \tag{8.30}$$

(remember that $\kappa > 1$). Given that $E_0 = V_0 / l$ and $E = V / l$ (where $l$ is the perpendicular distance between the plates) we conclude that the dielectric also has the effect of *reducing the potential difference* between the plates from its initial value $V_0$:

$$V = \frac{V_0}{\kappa} \tag{8.31}$$

What is the effect of the dielectric on the capacitance? By taking into account that the free charges $\pm Q_0$ on the plates of the capacitor are the same before and after the introduction of the dielectric, and by using the relations $C_0 = Q_0 / V_0$ and $C = Q_0 / V$ in combination with (8.31), we find that the new capacitance is

$$C = \kappa C_0 \tag{8.32}$$

In words, the dielectric causes an *increase of the capacitance*.

What would be the free charge on the capacitor after the introduction of the dielectric, had we not disconnected the source? In this case the potential difference between the plates has a fixed value (same before and after the introduction of the dielectric) equal to the voltage $V$ of the source. The relations $Q_0 = C_0 V$ and $Q = CV$, in combination with (8.32), yield

$$Q = \kappa Q_0 \qquad (8.33)$$

That is, by means of the dielectric we achieve an *increase of the (free) charge on the capacitor under constant voltage*.

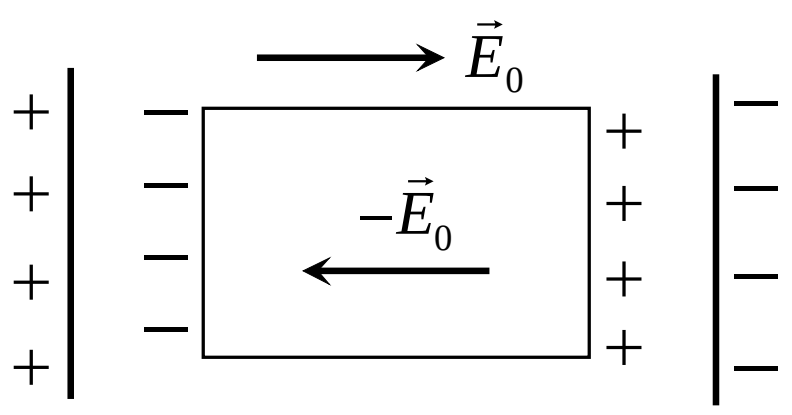

Fig. 8.8. A piece of metal between the plates of a capacitor.

For comparison, let us see what will happen if in place of the dielectric we put a piece of metal inside the capacitor (Fig. 8.8). The electric field of the capacitor causes a transfer of *free* charge in the metal. This charge is distributed on the surface of the metal and creates an additional electric field $-\vec{E}_0$ in the interior, in order for the net electric field inside the metal to be zero, as required by the electrostatic situation (cf. Sec. 5.6). We conclude that

> *a metal fully eliminates the electric field in its interior, while a dielectric only reduces the field without fully eliminating it.*

In contrast to electric polarization, the effect of which is to *reduce* the value of an applied electric field inside a medium, magnetization is a more complex effect that divides the set of all physical substances into two categories; namely,

1. *paramagnetic* materials ($\chi_m>0$), which favor an *increase* of the value of an applied magnetic field $\vec{B}$ in their interior (this is related to the fact that the magnetization $\vec{M}$ in these substances is *in the direction of* $\vec{B}$);
2. *diamagnetic* materials ($\chi_m<0$), which tend to *reduce* the value of an applied field $\vec{B}$ in their interior (in these substances the magnetization $\vec{M}$ is *opposite* to $\vec{B}$).

Thus, for example, one way to increase the magnetic field inside a solenoid is to put in its interior a core made of some paramagnetic material. It should be noted, however, that, for most substances, $|\chi_m|<<1$. In applications we therefore often set $\kappa_m= 1+\chi_m \approx 1$ and $\mu\approx\mu_0$ .

## References for Chapter 8

# QUESTIONS

**1.** What is a magnetic dipole? Should this term be interpreted literally? What justifies the standard use of the term?

**2.** Describe the behavior of an electric dipole inside an electric field, as well as the behavior of a magnetic dipole inside a magnetic field.

**3.** In what respect is the polarization mechanism in $CO_2$ different from that in $H_2O$?

**4.** What is the difference between free and bound charges, as well as between free and bound currents?

**5.** For what practical purposes do we place a dielectric in the interior of a capacitor?

**6.** A piece of metal and a piece of dielectric are placed inside an electrostatic field. In what respects will the responses of these substances to the field differ?

**7.** Show that the polarization (bound) charges of a uniformly polarized dielectric are confined to the surface of the dielectric. Similarly, show that the magnetization (bound) currents of a uniformly magnetized substance are confined to the surface of the substance.

**8.** (*a*) Consider a homogeneous linear dielectric in the interior of which there are no *free* charges. Show that the *total* electric charge inside the dielectric is zero. (*b*) Consider a homogeneous linear material in the interior of which there are no *free* currents. Show that the *total* current passing through any closed curve inside the material is zero.

**9.** Show that $\mu>\mu_0$ for paramagnetic substances while $\mu<\mu_0$ for diamagnetic substances.

**10.** The magnetic field in the interior of a linear medium is of the form $\vec{B}=B_0\,\hat{u}_z\ (B_0>0)$. What will be the directions of the field $\vec{H}$ and the magnetization $\vec{M}$ inside this medium if the medium is (*a*) paramagnetic, (*b*) diamagnetic ?

# PROBLEMS

**1.** A charge $+q$ is located at $\vec{r}_+ \equiv (1,-1,1)$, while a charge $-q$ is located at $\vec{r}_- \equiv (1,0,1)$. (*a*) Find the electric dipole moment of the system. (*b*) Find the torque exerted on the system by a uniform electric field $\vec{E} = E_0\,\hat{u}_x$.

***Solution:*** We have:

$$\vec{p} = q\,\vec{s} = q\,(\vec{r}_+ - \vec{r}_-) \equiv q\,(0,-1,0) \equiv (0,-q,0) \equiv -q\,\hat{u}_y \ . \quad \text{Also,}$$

$$\vec{E} = E_0\,\hat{u}_x \equiv (E_0,0,0) \quad \text{and} \quad \vec{T} = \vec{p}\times\vec{E} = \begin{vmatrix} \hat{u}_x & \hat{u}_y & \hat{u}_z \\ 0 & -q & 0 \\ E_0 & 0 & 0 \end{vmatrix} = qE_0\,\hat{u}_z \equiv (0,\,0,\,qE_0)$$

**2.** Consider a circular current loop of radius $R$, lying on the $xy$-plane and carrying a current $I$ (Fig. 8.9). The current is flowing clockwise as we look the $xy$-plane from the side of the positive $z$-semiaxis. Find the magnetic dipole moment of the loop, as well as the torque exerted on the loop by a uniform magnetic field $\vec{B} = B_0\,\hat{u}_x$.

***Solution:***

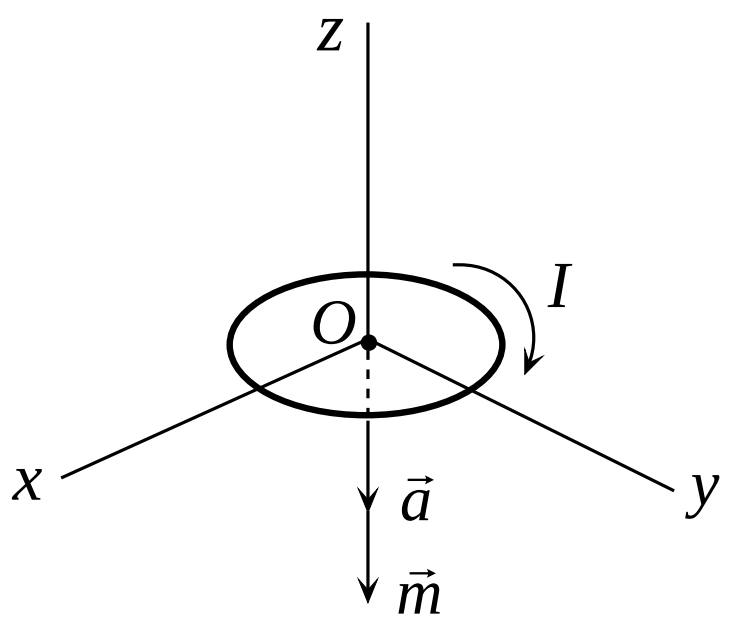

Fig. 8.9. A circular current loop on the $xy$-plane.

The area of the loop is $a = \pi R^2$ or, in vector form, $\vec{a} = -a\,\hat{u}_z = -\pi R^2\,\hat{u}_z$. Then,

$$\vec{m} = I\,\vec{a} = -I\pi R^2\,\hat{u}_z \equiv (0,0,-I\pi R^2)$$

Also,

$$\vec{B} = B_0\,\hat{u}_x \equiv (B_0,0,0)$$

Therefore,

$$\vec{T} = \vec{m}\times\vec{B} = \begin{vmatrix} \hat{u}_x & \hat{u}_y & \hat{u}_z \\ 0 & 0 & -I\pi R^2 \\ B_0 & 0 & 0 \end{vmatrix} = -I\pi R^2 B_0\,\hat{u}_y \equiv (0,\,-I\pi R^2 B_0,\,0)$$

Note that our results do not depend on the location of the center $O$ of the circular loop on the $xy$-plane.

**3.** Consider a system of charges $q_1$, $q_2$,… Let $\vec{p}$ be the electric dipole moment of the system with respect to some reference point $O$. Show that $\vec{p}$ is independent of the choice of $O$ if and only if the total charge of the system is zero.

***Solution:*** We consider two reference points $O$ and $O'$ (Fig. 8.10) and we call $\vec{p}$ and $\vec{p}\,'$ the corresponding electric dipole moments of the system relative to these points. We have:

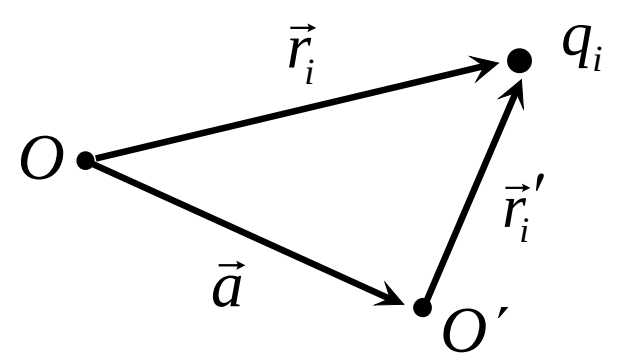

Fig. 8.10. A charge $q_i$ viewed from two reference points $O$ and $O'$.

$$\vec{p}=\sum_i q_i\,\vec{r}_i\ ,\quad \vec{p}\,'=\sum_i q_i\,\vec{r}_i\,'=\sum_i q_i\,(\vec{r}_i-\vec{a})\quad \text{where}\ \ \vec{a}=\overrightarrow{OO'}\,.\ \text{Then,}$$

$$\vec{p}=\vec{p}\,'\ \Leftrightarrow\ \sum_i q_i\,\vec{r}_i=\sum_i q_i\,(\vec{r}_i-\vec{a})\ \Leftrightarrow\ \sum_i q_i\,\vec{r}_i=\sum_i q_i\,\vec{r}_i-\left(\sum_i q_i\right)\vec{a}\ \Leftrightarrow\ \sum_i q_i=0$$

This condition is satisfied, in particular, in the case of an electric dipole.

**4.** Show that in the interior of a homogeneous linear dielectric the free-charge density $\rho_f$ and the polarization-charge density $\rho_b$ are related by

$$\rho_b=\left(\frac{\varepsilon_0}{\varepsilon}-1\right)\rho_f=-\frac{\chi_e}{1+\chi_e}\,\rho_f$$

What physical conclusion do you draw for the case where $\rho_f=0$ ?

***Solution:*** In general, $\vec{\nabla}\cdot\vec{D}=\rho_f$. Since the medium is homogeneous linear, $\vec{D}=\varepsilon\vec{E}$ with $\varepsilon=const$. Hence, $\vec{\nabla}\cdot(\varepsilon\vec{E})=\varepsilon\,(\vec{\nabla}\cdot\vec{E})=\rho_f\ \Rightarrow$

$$\boxed{\vec{\nabla}\cdot\vec{E}=\frac{\rho_f}{\varepsilon}}\qquad (1)$$

But, by Gauss' law, $\vec{\nabla}\cdot\vec{E}=\rho/\varepsilon_0$ where $\rho=\rho_f+\rho_b$ is the total charge density. By comparing with (1), we have:

$$\frac{\rho_f}{\varepsilon}=\frac{\rho}{\varepsilon_0}=\frac{\rho_f+\rho_b}{\varepsilon_0}\quad\Rightarrow\quad \rho_b=\left(\frac{\varepsilon_0}{\varepsilon}-1\right)\rho_f$$

Moreover, by taking into account that $\varepsilon=\varepsilon_0\,(1+\chi_e)$, we find: $\dfrac{\varepsilon_0}{\varepsilon}-1=-\dfrac{\chi_e}{1+\chi_e}$.

We notice that $\rho_b=0$ when $\rho_f=0$. That is, if there is no free charge in the interior of the dielectric, the bound charge distributes itself on the *surface* of the material. It also follows from Eq. (8.7) that the polarization of the dielectric is uniform in this case.

## CHAPTER 9

# TIME-DEPENDENT ELECTROMAGNETIC FIELDS

### 9.1 Introduction

Let us recall the basic laws for *static* (i.e., time-independent) electric and magnetic fields ($\partial\vec{E}/\partial t=0,\ \partial\vec{B}/\partial t=0$):

$$\begin{aligned}
&(a)\quad \vec{\nabla}\cdot\vec{E}=\rho/\varepsilon_0 &&\Leftrightarrow\quad \oint_S \vec{E}\cdot\overrightarrow{da}=Q_{in}/\varepsilon_0\\
&(b)\quad \vec{\nabla}\cdot\vec{B}=0 &&\Leftrightarrow\quad \oint_S \vec{B}\cdot\overrightarrow{da}=0\\
&(c)\quad \vec{\nabla}\times\vec{E}=0 &&\Leftrightarrow\quad \oint_C \vec{E}\cdot\overrightarrow{dl}=0\\
&(d)\quad \vec{\nabla}\times\vec{B}=\mu_0\vec{J} &&\Leftrightarrow\quad \oint_C \vec{B}\cdot\overrightarrow{dl}=\mu_0 I_{in}
\end{aligned}\tag{9.1}$$

We notice that (*a*) and (*c*) concern the $\vec{E}$-field alone, while (*b*) and (*d*) concern the $\vec{B}$-field alone. Hence the static laws might give one the impression that there are two independent fields in Nature, an electric and a magnetic. This means that, correspondingly, there should be two independent branches in Physics, namely, "Electricity" and "Magnetism". Maybe even two respective kinds of specialists: "electrologists" and "magnetologists"! One thing is certain, however: neither of these "specialists" would be able to explain to you how the energy produced in the Sun reaches the Earth as light and heat, or, by what physical mechanism you are able to listen to your favorite music on radio stations located several miles away.

The truth is that relations (9.1), in the form they are written, are *not* valid in the case of *time-dependent* fields. To be accurate, (*a*) and (*b*) continue to be valid but (*c*) and (*d*) take on new forms that *contain both fields* $\vec{E}$ *and* $\vec{B}$. It is thus revealed that

> *the electric and the magnetic field are not independent of each other but constitute two interrelated "components" of a single electromagnetic (e/m) field.*

(The term *"electromagnetic"* will henceforth be written *"e/m"*, for short.) The generalized version of Eq. (9.1) for time-dependent e/m fields constitutes the *Maxwell equations*. These equations led to one of the most important predictions of e/m theory (as well as of Theoretical Physics, in general): the existence of *electromagnetic (e/m) waves* that can travel even in empty space.

Of course, at a given time *t* the e/m field is represented by a pair of fields $(\vec{E},\vec{B})$, one called *electric* while the other called *magnetic*. Let us not forget, however, that the $\vec{E}$ and $\vec{B}$ do not evolve in time independently of each other but the change of one field influences the other. We also note that the separation of the e/m field into electric and magnetic components is dependent upon the state of motion of the observer: what is perceived as "electric" by one observer may be perceived as "magnetic" by another observer in relative motion with the former one (see, e.g., Problem 3). This idea played a key role in the creation of the Theory of Relativity.

## 9.2 Electromotive Force

The concept of the *electromotive force* (or *emf*, for short) is often a source of confusion to the student due to the diversity of situations where this concept applies, leading to a multitude of expressions for the emf (see, e.g., [1-6]). This section is based on ideas presented in [7,8].

Consider a region of space in which an e/m field exists. In the most general sense, any *closed* path $C$ (or *loop*) within this region will be called a *"circuit"* if a charge flow can be sustained on it. We arbitrarily assign a positive direction of traversing the loop $C$ and we consider an element $\vec{dl}$ of $C$ oriented in the positive direction.

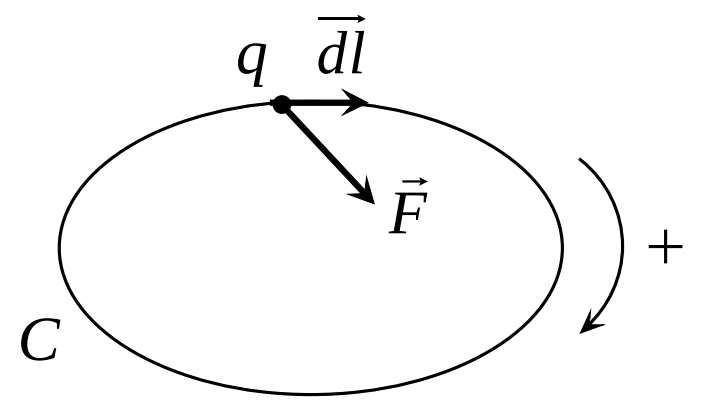

Fig. 9.1. Force on a test charge $q$ at a point of an oriented loop $C$.

Imagine now a test charge $q$ located at the position of the circuit element $\vec{dl}$, and let $\vec{F}$ be the force on $q$ at time $t$ (Fig. 9.1). This force is exerted by the e/m field itself as well as, possibly, by additional *energy sources* (such as batteries or some external mechanical action) that may contribute to the generation and preservation of a current flow around the loop $C$.[1] The *force per unit charge* at the location of $\vec{dl}$, at time $t$, is

$$\vec{f} = \frac{\vec{F}}{q} \qquad (9.2)$$

By measuring this force at all points of the loop, *at the same time t*, we can construct the closed line integral

$$\mathcal{E} = \oint_C \vec{f} \cdot \vec{dl} \qquad (9.3)$$

The quantity $\mathcal{E}=\mathcal{E}(t)$ is called the *electromotive force* (*emf*) *of the circuit C at time t*. Note that the sign of $\mathcal{E}$ is dependent upon our choice of the positive direction of traversing $C$; by changing this convention, the direction of $\vec{dl}$ is reversed and, accordingly, so does the sign of $\mathcal{E}$. On the contrary, the sign of $\mathcal{E}$ does *not* depend on the sign of the test charge $q$ since, by reversing the sign of $q$, the direction of $\vec{F}$ is also reversed leaving $\vec{f}$ unchanged, according to (9.2).

[1] Even in the case of external mechanical action, the direct action on $q$ is *electromagnetic* in nature!

Relation (9.3) constitutes the general definition of the emf. The expression for $\mathcal{E}$ may take on various forms, depending on the particular e/m situation. Here are some examples:

(*a*) Consider a circuit consisting of a closed wire $C$. The wire is moving inside a *static* magnetic field $\vec{B}(\vec{r})$. Let $\vec{\upsilon}$ be the velocity of the element $\overrightarrow{dl}$ of $C$ relative to our inertial frame of reference. A charge $q$ (say, a free electron) at the location of $\overrightarrow{dl}$ executes a composite motion, due to the motion of the loop $C$ itself relative to our frame, as well as the motion of $q$ *along* $C$. The total velocity of $q$ relative to us is $\vec{\upsilon}_{tot} = \vec{\upsilon} + \vec{\upsilon}'$, where $\vec{\upsilon}'$ is the velocity of $q$ in a direction parallel to $\overrightarrow{dl}$. The force exerted on $q$ by the magnetic field is

$$\vec{F} = q\,(\vec{\upsilon}_{tot} \times \vec{B}) = q\,(\vec{\upsilon} \times \vec{B}) + q\,(\vec{\upsilon}' \times \vec{B}) \;\Rightarrow$$
$$\vec{f} = \frac{\vec{F}}{q} = (\vec{\upsilon} \times \vec{B}) + (\vec{\upsilon}' \times \vec{B})$$

By (9.3), then, the emf of the circuit $C$ is

$$\mathcal{E} = \oint_C \vec{f} \cdot \overrightarrow{dl} = \oint_C (\vec{\upsilon} \times \vec{B}) \cdot \overrightarrow{dl} + \oint_C (\vec{\upsilon}' \times \vec{B}) \cdot \overrightarrow{dl}$$

But, since $\vec{\upsilon}'$ is parallel to $\overrightarrow{dl}$, we have that $(\vec{\upsilon}' \times \vec{B}) \cdot \overrightarrow{dl} = 0$. Thus, finally,

$$\mathcal{E} = \oint_C (\vec{\upsilon} \times \vec{B}) \cdot \overrightarrow{dl} \tag{9.4}$$

Note that the wire *need not maintain a fixed shape, size or orientation* during its motion. Note also that the velocity $\vec{\upsilon}$ may vary around the circuit.

By using (9.4) it can be proven that

$$\mathcal{E} = -\frac{d\Phi_m}{dt} \tag{9.5}$$

where $\Phi_m = \int \vec{B} \cdot \overrightarrow{da}$ is the *magnetic flux* through the wire $C$ at time $t$ (see Problem 9). Note carefully that relation (9.5) does not express any novel physical law; it is simply a direct consequence of the definition of the emf !

(*b*) Consider now a closed wire $C$ that is *at rest* inside a *time-varying* magnetic field $\vec{B}(\vec{r}, t)$. As experiments show, as soon as $\vec{B}$ starts changing, a current begins to flow in the wire. This is impressive given that the free charges in the (stationary) wire were initially at rest. And, as everybody knows, a magnetic field exerts forces on *moving* charges only! It is also observed experimentally that, if the magnetic field $\vec{B}$ stops varying with time, the current in the wire disappears. The only field that can put an initially stationary charge in motion and keep this charge moving is an *electric* field.

We are thus compelled to conclude that

- *a time-varying magnetic field is necessarily accompanied by an electric field.*

It is often said that a changing magnetic field *induces* an electric field. This is somewhat misleading since it gives the impression that the "source" of an electric field could be a magnetic field. Let us keep in mind, however, that the true sources of any e/m field are the electric charges and the electric currents.

So, let $\vec{E}(\vec{r},t)$ be the electric field accompanying the time-varying magnetic field $\vec{B}$. Consider again a charge $q$ at the position of the element $\overrightarrow{dl}$ of the wire. Given that the wire is now at rest (relative to our inertial frame), the velocity of $q$ will be due to the motion of the charge along the wire only, i.e., in a direction parallel to $\overrightarrow{dl}$: $\vec{\upsilon}_{tot}=\vec{\upsilon}'$ (since $\vec{\upsilon}=0$). The force on $q$ by the e/m field is

$$\vec{F}=q\,[\vec{E}+(\vec{\upsilon}_{tot}\times\vec{B})]=q\,[\vec{E}+(\vec{\upsilon}'\times\vec{B})] \;\Rightarrow$$
$$\vec{f}=\frac{\vec{F}}{q}=\vec{E}+(\vec{\upsilon}'\times\vec{B})$$

The emf of the circuit $C$ is now

$$\mathcal{E}=\oint_C \vec{f}\cdot\overrightarrow{dl}=\oint_C \vec{E}\cdot\overrightarrow{dl}+\oint_C (\vec{\upsilon}'\times\vec{B})\cdot\overrightarrow{dl}$$

But, as explained earlier, $(\vec{\upsilon}'\times\vec{B})\cdot\overrightarrow{dl}=0$. Thus, finally,

$$\mathcal{E}=\oint_C \vec{E}\cdot\overrightarrow{dl} \qquad (9.6)$$

(*c*) If the closed wire $C$ is located inside an *electrostatic* field $\vec{E}(\vec{r})$, then, by the fact that the electrostatic field is irrotational it follows that $\mathcal{E}=0$ (show this).

(*d*) As will be shown in Problem 7, in the case of a circuit $C$ consisting of an ideal battery (i.e., one with no internal resistance) connected to an external resistor, the emf in the direction of the current is equal to the voltage of the battery: $\mathcal{E}=V$. Moreover, the emf in this case represents the *work per unit charge* done by the source (battery).

*Exercise:* By the general definition (9.3), show that the emf has dimensions of electric potential.

## 9.3 The Faraday-Henry Law

In a region of space where a time-varying e/m field $(\vec{E},\vec{B})$ exists, we consider an arbitrary open surface *S* bounded by a closed curve *C*, as seen in Fig. 9.2. Physically, this curve may represent a conducting wire. The directions of the loop element $\overrightarrow{dl}$ and the surface element $\overrightarrow{da}$ (where the latter vector is normal to *S*) are related to each other through the familiar right-hand rule. The element $\overrightarrow{dl}$ is directed consistently with the chosen positive direction of traversing the loop *C*.

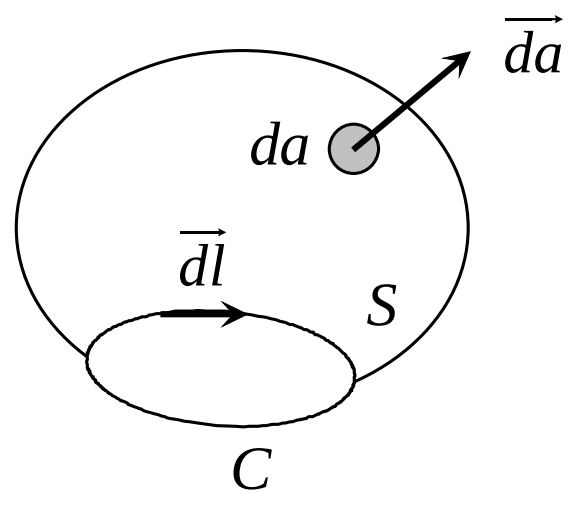

Fig. 9.2. An open surface *S* bordered by a closed curve *C*.

If the loop *C* is assumed stationary relative to the inertial observer, the emf along *C* at time *t* is given by (9.6):

$$\mathcal{E}(t) = \oint_C \vec{E}\cdot\overrightarrow{dl} \tag{9.7}$$

The magnetic flux through *S* at this instant is

$$\Phi_m(t) = \int_S \vec{B}\cdot\overrightarrow{da} \tag{9.8}$$

(Note that the signs of both $\mathcal{E}$ and $\Phi_m$ depend on the chosen positive direction of traversing the loop *C*.) Since the magnetic field $\vec{B}$ is *solenoidal*, the value of $\Phi_m$ for a given *C* is independent of the choice of the surface *S* bounded by *C*; that is, the same magnetic flux will pass through *any* open surface having the fixed closed curve *C* as its boundary. (Since the magnetic field lines are closed, the same number of field lines will pass through *any* open surface bounded by *C*.) Physically, this is related to the fact that no free magnetic poles exist.

According to the *Faraday-Henry law*,

$$\mathcal{E} = -\frac{d\Phi_m}{dt} \tag{9.9}$$

or, explicitly,

$$\boxed{\oint_C \vec{E}\cdot\overrightarrow{dl} = -\frac{d}{dt}\int_S \vec{B}\cdot\overrightarrow{da}} \tag{9.10}$$

That is,

> *at any moment t, the emf of the circuit C equals the negative of the rate of change of the magnetic flux passing through C.*

Equation (9.10) is the Faraday-Henry law in integral form. The negative sign on the right-hand sides of (9.9) and (9.10) expresses *Lenz's law*, which is a sort of "law of inertia" for Electromagnetism.

Note carefully that, despite the similarity of (9.9) with (9.5), the former equation now expresses a *true physical law* rather than being simply a mathematical consequence of the definition of the emf !

The integral law (9.10) can be re-expressed in differential form by using Stokes' theorem,

$$\oint_C \vec{E} \cdot \overrightarrow{dl} = \int_S (\vec{\nabla} \times \vec{E}) \cdot \overrightarrow{da}$$

and by noticing that, after the integration in (9.8) is performed, the magnetic flux $\Phi_m$ is a function of $t$ alone, so that

$$\frac{d\Phi_m}{dt} = \frac{\partial \Phi_m}{\partial t} \;\Rightarrow\; \frac{d}{dt}\int_S \vec{B} \cdot \overrightarrow{da} = \frac{\partial}{\partial t}\int_S \vec{B} \cdot \overrightarrow{da} = \int_S \frac{\partial \vec{B}}{\partial t} \cdot \overrightarrow{da}$$

Equation (9.10) is now written:

$$\int_S (\vec{\nabla} \times \vec{E}) \cdot \overrightarrow{da} = -\int_S \frac{\partial \vec{B}}{\partial t} \cdot \overrightarrow{da}$$

In order for this to be valid for any open surface *S* bounded by *C*, we must have:

$$\boxed{\vec{\nabla} \times \vec{E} = -\frac{\partial \vec{B}}{\partial t}} \tag{9.11}$$

Equation (9.11) is the Faraday-Henry law in differential form.

We notice that if $\partial \vec{B}/\partial t \neq 0$, then necessarily $\vec{E} \neq 0$. Hence, as mentioned in Sec. 9.2, *a time-varying magnetic field is always accompanied by an electric field* (or, as is often said, *induces* an electric field). If, however, $\vec{B}$ is *static* ($\partial \vec{B}/\partial t = 0$), then $\vec{E}$ is *irrotational*: $\vec{\nabla} \times \vec{E} = 0 \Leftrightarrow \oint \vec{E} \cdot \overrightarrow{dl} = 0$; this allows for the possibility that $\vec{E} = 0$.

We also note that the electric field induced by a time-dependent magnetic field *may have closed field lines*, since it is not irrotational.

If the closed curve $C$ is a conducting wire of total resistance $R$, and if the circuit does not contain external energy sources, then, during the time-change of the magnetic flux the wire will sustain a current equal to

$$I = \frac{\mathcal{E}}{R} = -\frac{1}{R}\frac{d\Phi_m}{dt} \tag{9.12}$$

(see [7]) where $\Phi_m$ is the magnetic flux passing through the circuit at time $t$. The sign of the algebraic value of $I$ determines the direction of the current with respect to the positive direction of traversing the circuit $C$.

Let us assume, for example, that the chosen positive direction of traversing $C$ is *clockwise* as $C$ is viewed from above (imagine that $C$ is a curve on this page). Then $\vec{da}$ is directed *into* the page, according to the right-hand rule. If more and more magnetic flux $\Phi_m$ per unit time is going *into* the page, then $d\Phi_m/dt > 0$ and $I < 0$, which means that the induced current flows *counterclockwise*. This results in the creation of an additional magnetic flux that *opposes* the increase of the external flux permeating the circuit. Indeed, this is the physical meaning of Lenz's law.

Sometimes the circuit itself is responsible for the time-change of the magnetic flux going through it. Let us assume that the circuit carries a variable current $I=I(t)$. This current creates a time-dependent magnetic field around the circuit. As follows from the Biot-Savart law [1] the magnetic flux passing through the circuit is proportional to the current $I$:

$$\Phi_m = L\,I \tag{9.13}$$

The coefficient $L$ is called the *self-inductance* (or simply the *inductance*) of the circuit and depends on the geometrical characteristics of the latter. To find the part of the emf that is *not* due to external energy sources, we substitute (9.13) into (9.9):

$$\mathcal{E}_m = -\frac{d\Phi_m}{dt} = -\frac{d}{dt}(L\,I) = -L\frac{dI}{dt} \tag{9.14}$$

where we have assumed that the shape of the circuit is fixed, so that $L$ is constant. As expected by Lenz's law, this additional emf *opposes* the time change of the current $I$; it is often called the *"back emf"*.

## 9.4 The Ampère-Maxwell Law

In a region of space where a time-dependent e/m field $(\vec{E}, \vec{B})$ exists, we consider an arbitrary open surface $S$ bounded by a given closed curve $C$ (see Fig. 9.2). The elec-tric flux through $S$ at time $t$ is $\int_S \vec{E}\cdot\overrightarrow{da}$. Also, if $\vec{J}(\vec{r},t)$ is the current density on $S$, the total current passing through $S$ at time $t$ is[2]

$$I_{in} = \int_S \vec{J}\cdot\overrightarrow{da} \tag{9.15}$$

The *Ampère-Maxwell law,* which generalizes Ampère's law for static fields (Sec. 7.3), is expressed in integral form as follows:

$$\boxed{\oint_C \vec{B}\cdot\overrightarrow{dl} = \mu_0 I_{in} + \varepsilon_0\mu_0 \frac{d}{dt}\int_S \vec{E}\cdot\overrightarrow{da}} \tag{9.16}$$

By Stokes' theorem,

$$\oint_C \vec{B}\cdot\overrightarrow{dl} = \int_S (\vec{\nabla}\times\vec{B})\cdot\overrightarrow{da}$$

Furthermore, the surface integral on the right-hand side of (9.16) is a function of $t$ alone, so that

$$\frac{d}{dt}\int_S \vec{E}\cdot\overrightarrow{da} = \frac{\partial}{\partial t}\int_S \vec{E}\cdot\overrightarrow{da} = \int_S \frac{\partial\vec{E}}{\partial t}\cdot\overrightarrow{da}$$

Hence, by also taking into account (9.15), equation (9.16) is written as

$$\int_S (\vec{\nabla}\times\vec{B})\cdot\overrightarrow{da} = \int_S \left(\mu_0 \vec{J} + \varepsilon_0\mu_0 \frac{\partial\vec{E}}{\partial t}\right)\cdot\overrightarrow{da}$$

In order for this to be true for any open surface $S$ bounded by the given closed curve $C$, we must have:

$$\boxed{\vec{\nabla}\times\vec{B} = \mu_0 \vec{J} + \varepsilon_0\mu_0 \frac{\partial\vec{E}}{\partial t}} \tag{9.17}$$

Relation (9.17) expresses the Ampère-Maxwell law in differential form.

In a region of space where no electric currents exist ($\vec{J}=0$), we have:

$$\vec{\nabla}\times\vec{B} = \varepsilon_0\mu_0 \frac{\partial\vec{E}}{\partial t} \tag{9.18}$$

---

[2] Explain why the arguments of Prob. 5 of Chap. 7 do not apply in this case; hence $I_{in}$ is generally dependent on the choice of $S$. (*Hint:* Is the charge distribution constant in time in the region of interest?) See also Appendix B.

We notice that, if $\partial \vec{E} / \partial t \neq 0$, then $\vec{B} \neq 0$. Hence, *a time-changing electric field is necessarily accompanied by a magnetic field* (or, as we say, *induces* a magnetic field). For historical reasons, the term

$$\vec{J}_d = \varepsilon_0 \frac{\partial \vec{E}}{\partial t} \qquad (9.19)$$

is called the *displacement current*. The name is deceptive, however, since $\vec{J}_d$ does *not* represent any actual motion of charges! More on the displacement current is said in Appendix B.

## 9.5 The Maxwell Equations

Let us summarize the laws associated with the e/m field:

$$\begin{array}{ll} (a) & \vec{\nabla} \cdot \vec{E} = \dfrac{\rho}{\varepsilon_0} \\ (b) & \vec{\nabla} \cdot \vec{B} = 0 \\ (c) & \vec{\nabla} \times \vec{E} = -\dfrac{\partial \vec{B}}{\partial t} \\ (d) & \vec{\nabla} \times \vec{B} = \mu_0 \vec{J} + \varepsilon_0 \mu_0 \dfrac{\partial \vec{E}}{\partial t} \end{array} \qquad (9.20)$$

Equations (*a*) and (*b*) express Gauss' law for the electric and the magnetic field, respectively. Specifically, relation (*a*) is a mathematical statement of Coulomb's law, while the physical meaning of (*b*) is the absence of free magnetic poles. Equation (*c*) expresses the Faraday-Henry law (or *law of e/m induction*, as is often called) while (*d*) is the Ampère-Maxwell law containing the so-called displacement current (Sec. 9.4).

The set of differential equations (9.20) represents the *Maxwell equations*. They are for Electromagnetism what Newton's laws are for Mechanics. We emphasize that the densities $\rho(\vec{r},t)$ and $\vec{J}(\vec{r},t)$ in (9.20) *contain all charges and all currents*, respectively, whether free or bound (see Chapter 8).

Relations (9.20) are the most general form of Maxwell's equations. There is, however, a more convenient form of these equations suitable for the study of e/m fields inside material substances that are subject to electrical polarization or magnetization. Consider a substance in the interior of which the e/m field is $(\vec{E},\vec{B})$. Assuming, in general, that the substance has both dielectric and magnetic properties, we call $\vec{P}$ and $\vec{M}$ the polarization and magnetization vectors, respectively. We also introduce the auxiliary fields (see Chap. 8)

$$\vec{D} = \varepsilon_0 \vec{E} + \vec{P} \ , \quad \vec{H} = \frac{1}{\mu_0}\vec{B} - \vec{M} \tag{9.21}$$

For a *linear* medium,

$$\vec{P} = \varepsilon_0 \chi_e \vec{E} \ , \quad \vec{M} = \chi_m \vec{H} \tag{9.22}$$

so that

$$\vec{D} = \varepsilon \vec{E} \ , \quad \vec{H} = \frac{1}{\mu}\vec{B} \tag{9.23}$$

where

$$\varepsilon = \varepsilon_0 (1 + \chi_e) \ , \quad \mu = \mu_0 (1 + \chi_m) \tag{9.24}$$

If now $\rho_f(\vec{r},t)$ and $\vec{J}_f(\vec{r},t)$ are the *free* charge and current densities, respectively, inside the material, the Maxwell equations take on the form

$$\begin{array}{ll} (a) & \vec{\nabla} \cdot \vec{D} = \rho_f \\ (b) & \vec{\nabla} \cdot \vec{B} = 0 \\ (c) & \vec{\nabla} \times \vec{E} = -\dfrac{\partial \vec{B}}{\partial t} \\ (d) & \vec{\nabla} \times \vec{H} = \vec{J}_f + \dfrac{\partial \vec{D}}{\partial t} \end{array} \tag{9.25}$$

Notice that the "displacement current" (which has nothing to do with an actual motion of charges!) is represented by the term $\partial \vec{D} / \partial t$ in the last equation. The presence of the electric displacement $\vec{D}$ justifies this term's misleading name.

From Maxwell's equations there follows a set of *boundary conditions* that the e/m field must satisfy at the interface of two different media. These conditions can be stated as follows:

1. The components of $\vec{E}$ *parallel* to the interface, as well as the component of $\vec{B}$ *normal* to this surface, are *always continuous*.
2. The component of $\vec{D}$ *normal* to the interface is *discontinuous* if there is *free* charge on this surface.
3. The components of $\vec{H}$ *parallel* to the interface are *discontinuous* if there is *free* current on this surface.

In the case of two *linear* media *1* and *2*, on the interface of which there are no *free* charges or currents, the boundary conditions relating the values of the e/m field at the two sides of the interface are written:

$$\begin{aligned} &(a) \quad \vec{E}_1^{\parallel} = \vec{E}_2^{\parallel} \\ &(b) \quad \vec{B}_1^{\perp} = \vec{B}_2^{\perp} \\ &(c) \quad \vec{D}_1^{\perp} = \vec{D}_2^{\perp} \;\Rightarrow\; \varepsilon_1 \vec{E}_1^{\perp} = \varepsilon_2 \vec{E}_2^{\perp} \\ &(d) \quad \vec{H}_1^{\parallel} = \vec{H}_2^{\parallel} \;\Rightarrow\; \frac{1}{\mu_1} \vec{B}_1^{\parallel} = \frac{1}{\mu_2} \vec{B}_2^{\parallel} \end{aligned} \tag{9.26}$$

The symbol || denotes a component parallel to the interface, while the symbol $\perp$ denotes a component normal to this surface. The indices 1 and 2 correspond to the values of a given component in the two media.

## 9.6 Conservation of Charge

As is well known from classical Mechanics, Newton's laws predict a number of *conservation laws* such as, e.g., conservation of mechanical energy of a particle subject to conservative forces, conservation of momentum of an isolated system of particles, etc. The Maxwell equations are also associated with certain conservation laws, such as the *conservation of charge* and the *conservation of energy* (Poynting's theorem). In particular, it was the need to comply with conservation of charge that prompted Maxwell to correct Ampère's law by adding the "displacement current" in the last of his four equations.

To demonstrate that Maxwell's equations satisfy conservation of charge, we consider the two non-homogeneous Maxwell equations (those that, in addition to the electric or/and the magnetic field, contain the "sources" $\rho, \vec{J}$ of the e/m field):

$$\begin{aligned} &(a) \quad \vec{\nabla} \cdot \vec{E} = \frac{\rho}{\varepsilon_0} \\ &(d) \quad \vec{\nabla} \times \vec{B} = \mu_0 \vec{J} + \varepsilon_0 \mu_0 \frac{\partial \vec{E}}{\partial t} \end{aligned} \tag{9.27}$$

We take the *div* of (*d*):

$$\vec{\nabla} \cdot (\vec{\nabla} \times \vec{B}) = \vec{\nabla} \cdot \left( \mu_0 \vec{J} + \varepsilon_0 \mu_0 \frac{\partial \vec{E}}{\partial t} \right) = \mu_0 (\vec{\nabla} \cdot \vec{J}) + \varepsilon_0 \mu_0 \left( \vec{\nabla} \cdot \frac{\partial \vec{E}}{\partial t} \right) \tag{9.28}$$

But, $\vec{\nabla} \cdot (\vec{\nabla} \times \vec{B}) = 0$, identically. Also, $\vec{\nabla} \cdot \frac{\partial \vec{E}}{\partial t} = \frac{\partial}{\partial t} (\vec{\nabla} \cdot \vec{E})$, given that partial derivatives commute with one another. Making substitutions into (9.28), eliminating $\mu_0$, and substituting $\vec{\nabla} \cdot \vec{E}$ from (9.27)(*a*), we are led to the *equation of continuity*:

$$\boxed{\vec{\nabla} \cdot \vec{J} + \frac{\partial \rho}{\partial t} = 0} \qquad (9.29)$$

which is the same as Eq. (6.9) derived in Chap. 6.

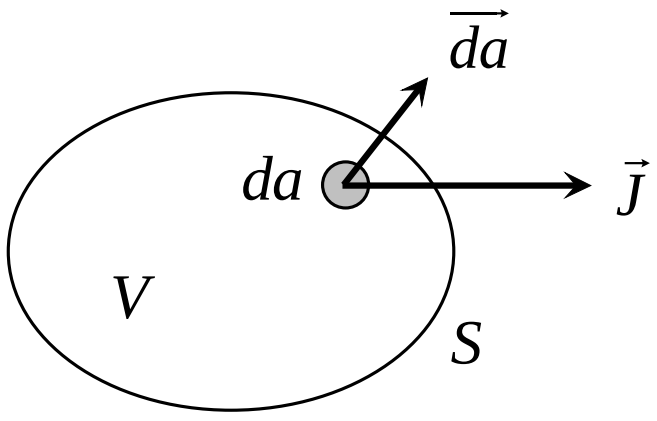

Fig. 9.3. Current density on a closed surface *S* enclosing a volume *V*.

To see the physical meaning of (9.29), we take the volume integral of this equation in a volume *V* bounded by a closed surface *S* (Fig. 9.3):

$$\int_V \frac{\partial \rho}{\partial t}\, dv = -\int_V (\vec{\nabla} \cdot \vec{J})\, dv \qquad (9.30)$$

Now,

$$\int_V \frac{\partial \rho}{\partial t}\, dv = \frac{\partial}{\partial t} \int_V \rho\, dv = \frac{d}{d\,t} \int_V \rho\, dv\,,$$

given that the integral $\int_V \rho\, dv$ is a function of $t$ alone. Also, by Gauss' theorem,

$$\int_V (\vec{\nabla} \cdot \vec{J})\, dv = \oint_S \vec{J} \cdot \overrightarrow{da}\,.$$

Thus (9.30) is rewritten as

$$\boxed{\frac{d}{d\,t} \int_V \rho\, dv = -\oint_S \vec{J} \cdot \overrightarrow{da}} \qquad (9.31)$$

We put

$$\int_V \rho\, dv = Q_{in}(t)\,, \quad \oint_S \vec{J} \cdot \overrightarrow{da} = I_{out}(t) \qquad (9.32)$$

where $Q_{in}(t)$ is the total charge inside the volume *V* at time *t*, while $I_{out}(t)$ is the total current *exiting* the closed surface *S* at that instant. Equation (9.31) is thus written more simply as

$$\frac{d}{d\,t} Q_{in}(t) = -I_{out}(t) \qquad (9.33)$$

We note the following:

1. If $Q_{in}$ *decreases* with time ($dQ_{in}/dt<0$) then $I_{out}$>0; therefore the net flow of charge through *S* is *outward*.
2. If $Q_{in}$ *increases* with time ($dQ_{in}/dt>0$) then $I_{out}$<0; therefore the net flow of charge through *S* is *inward*.
3. If there is no *net* charge flow through *S*, then $I_{out}$=0 and $dQ_{in}/dt=0$. That is, *the total charge inside the volume V is constant*.

The last remark expresses the *principle of conservation of charge*. As we see, this is a direct consequence of the Maxwell equations.

## 9.7 Electromagnetic Potentials

Consider now the two homogeneous Maxwell equations – those, that is, that do not contain "sources" (charges or currents):

$$\begin{aligned} &(b) \qquad \vec{\nabla}\cdot\vec{B}=0 \\ &(c) \qquad \vec{\nabla}\times\vec{E}+\frac{\partial\vec{B}}{\partial t}=0 \end{aligned} \tag{9.34}$$

Equation (*b*) is automatically satisfied by making the assumption that $\vec{B}(\vec{r},t)$ is the *rot* of a vector field $\vec{A}(\vec{r},t)$:

$$\vec{B}=\vec{\nabla}\times\vec{A} \tag{9.35}$$

[Indeed, then, $\vec{\nabla}\cdot\vec{B}=\vec{\nabla}\cdot(\vec{\nabla}\times\vec{A})=0$, identically.] Equation (*c*) is then written:

$$0=(\vec{\nabla}\times\vec{E})+\frac{\partial}{\partial t}(\vec{\nabla}\times\vec{A})=(\vec{\nabla}\times\vec{E})+\left(\vec{\nabla}\times\frac{\partial\vec{A}}{\partial t}\right) \;\Rightarrow\; \vec{\nabla}\times\left(\vec{E}+\frac{\partial\vec{A}}{\partial t}\right)=0$$

The last equality on the right is automatically true if we assume that the sum $\vec{E}+\dfrac{\partial\vec{A}}{\partial t}$ is the *grad* (or, for reasons of uniformity with Electrostatics, the *negative grad*) of a scalar field $\Phi(\vec{r},t)$:

$$\vec{E}+\frac{\partial\vec{A}}{\partial t}=-\vec{\nabla}\Phi \;\Rightarrow$$

$$\vec{E}=-\vec{\nabla}\Phi-\frac{\partial\vec{A}}{\partial t} \tag{9.36}$$

Equations (9.35) and (9.36) constitute a partial solution of Maxwell's equations since they satisfy two of them, namely, the homogeneous equations (9.34). The functions $\vec{A}(\vec{r},t)$ and $\Phi(\vec{r},t)$ are called the *electromagnetic potentials*.

These equations, however, do not determine the potentials $\vec{A}$ and $\Phi$ uniquely for given $\vec{E}$ and $\vec{B}$. Indeed, let $\vec{A}$ and $\Phi$ be two functions satisfying (9.35) and (9.36). We consider two new functions,

$$\vec{A}' = \vec{A} + \vec{\nabla}\chi \ , \quad \Phi' = \Phi - \frac{\partial \chi}{\partial t} \tag{9.37}$$

where $\chi(\vec{r},t)$ is an *arbitrary* scalar function. I leave this as an exercise to show that

$$\vec{\nabla} \times \vec{A}' = \vec{B} \ , \qquad -\vec{\nabla}\Phi' - \frac{\partial \vec{A}'}{\partial t} = \vec{E}$$

That is, if the potentials $\vec{A}$ and $\Phi$ satisfy (9.35) and (9.36) for given $\vec{E}$ and $\vec{B}$, then the new potentials $\vec{A}'$ and $\Phi'$ also satisfy these equations for the *same* $\vec{E}$ and $\vec{B}$.

*Exercise:* Write Eqs. (9.37) for the special choice $\chi(\vec{r},t) = \psi(\vec{r}) - Ct$ ($C$=const.). This choice is particularly convenient for static e/m fields.

## 9.8 The Energy of the E/M Field and the Poynting Vector

As we know, to create an e/m field we need a distribution of charge $\rho(\vec{r},t)$ or current $\vec{J}(\vec{r},t)$ in some region of space.[3] For a variety of reasons, building such a distribution requires the expenditure of energy. This energy is stored in the space where the e/m field exists and constitutes the *energy of the e/m field*.

For example, an electrostatic field $\vec{E}(\vec{r})$ is created by a static (i.e., time-independent) distribution of charge $\rho(\vec{r})$. To build such a distribution we need to do some work in order to overcome the repulsive Coulomb forces between the (positive or negative) charges that make up the distribution. On the other hand, a static magnetic field $\vec{B}(\vec{r})$ is due to a static current distribution $\vec{J}(\vec{r})$. The creation of such a distribution requires doing work, for the following reasons: (*a*) Current means charges in motion. Setting a particle in motion requires work equal to the kinetic energy acquired by the particle. (*b*) Energy is needed to build up a current against the "back emf" in a circuit having nonzero inductance (this emf opposes any increase of the current in the circuit; see Sec. 9.3).

Let $U=U(t)$ be the total energy of an e/m field at time $t$. We assume that this energy is distributed over the entire space where the e/m field exists. Let $dv$ be a volume element at the field point $\vec{r}$, and let $dU$ be the quantity of e/m energy contained in $dv$. We define the *energy density* of the e/m field to be the function

$$u(\vec{r},t) = \frac{dU}{dv} \tag{9.38}$$

[3] We will exclude from this discussion the case of a system of isolated point charges, each of which occupies zero volume and thus represents an infinite charge density $\rho$. See [1], Sec. 2.4.

so that

$$U(t)=\int u(\vec{r},t)\,dv \tag{9.39}$$

where the integration takes place over the entire space where the e/m field exists. We assume that a part of the total energy is due to the electric field and another part is due to the magnetic field, and we write $U=U_e+U_m$, where

$$U_e(t)=\int u_e(\vec{r},t)\,dv \;,\quad U_m(t)=\int u_m(\vec{r},t)\,dv \tag{9.40}$$

and where

$$u(\vec{r},t)=u_e(\vec{r},t)+u_m(\vec{r},t) \tag{9.41}$$

As can be proven [1,2,5,9] the general expressions for the energy densities are

$$\begin{aligned} u_e &= \frac{1}{2}\vec{E}\cdot\vec{D} \;,\quad u_m=\frac{1}{2}\vec{B}\cdot\vec{H} \\ u &= u_e+u_m=\frac{1}{2}(\vec{E}\cdot\vec{D}+\vec{B}\cdot\vec{H}) \end{aligned} \tag{9.42}$$

In a *linear medium*, $\vec{D}=\varepsilon\vec{E}$, $\vec{H}=\frac{1}{\mu}\vec{B}$, and so

$$\begin{aligned} u_e &= \frac{1}{2}\varepsilon E^2 ,\quad u_m=\frac{1}{2\mu}B^2 \\ u &= \frac{1}{2}\varepsilon E^2+\frac{1}{2\mu}B^2 \end{aligned} \tag{9.43}$$

where we have put $E=|\vec{E}|$, $B=|\vec{B}|$. For *empty space* we set $\varepsilon=\varepsilon_0$ and $\mu=\mu_0$.

We now define the *Poynting vector*

$$\boxed{\vec{N}=\vec{E}\times\vec{H}=\frac{1}{\mu}(\vec{E}\times\vec{B})} \tag{9.44}$$

where the last expression on the right is valid for a *linear* medium (for the *vacuum* we set $\mu=\mu_0$). To see the physical significance of this vector, we consider a volume $V$ bounded by a closed surface $S$ (Fig. 9.4). To simplify our analysis, we assume that there are no *free* charges inside $V$; thus there is no loss of e/m energy due to work done by the electric field on moving charges (we recall that the magnetic field produces no work on moving charges since the magnetic force on a charge is always normal to the charge's velocity).

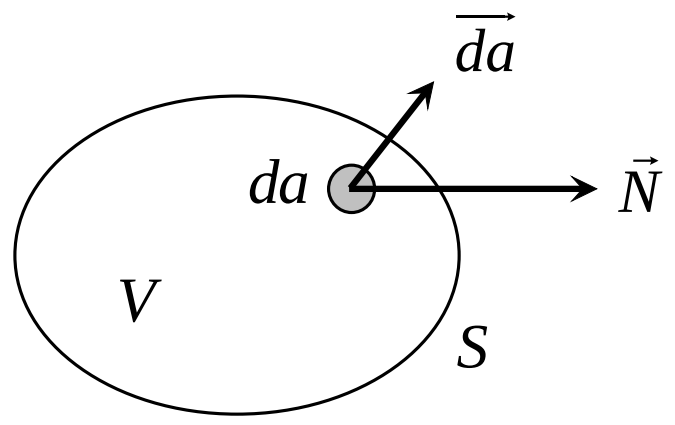

Fig. 9.4. Poynting vector on a closed surface *S* enclosing a volume *V*.

We now assume that, within an infinitesimal time interval *dt* there is a decrease *dU* of the total e/m energy *U* inside the volume *V*. Because energy cannot be lost, the amount *dU* of energy subtracted from *V* must appear *outside* the surface *S* within this time interval. This means that an amount of energy *dU* crosses *S* flowing outward in time *dt*. A portion of this energy passes through the elementary surface *da*. As can be proven [1,2,9],

> *the quantity* $\vec{N} \cdot \overrightarrow{da}$ *equals the energy per unit time exiting the volume V through the elementary surface da.*

We notice that $\vec{N}$ has dimensions of energy per unit time per unit area. This may remind you of the definition of the current density $\vec{J}$. Indeed, recall that (Sec. 6.1) the product $\vec{J} \cdot \overrightarrow{da}$ equals the charge per unit time crossing the surface element *da*. This analogy prompts us to seek some sort of "equation of continuity" for the energy of the e/m field, analogous to Eq. (6.9) for the electric charge.

We think as follows: The total energy per unit time *coming out* of *S* is given by the integral $\oint_S \vec{N} \cdot \overrightarrow{da}$. (If this integral has a *positive* value, the net transfer of energy through *S* is *outward*, while for a *negative* value of the integral the net flow of energy is actually *inward*.) Now, the energy coming out of *S* within time *dt* must be *subtracted* from the total amount of energy inside *V*. At time *t*, the total energy content *U*(*t*) of *V* is

$$U(t) = \int_V u(\vec{r}, t)\, dv$$

Conservation of energy demands that

$$\frac{dU}{dt} = -\oint_S \vec{N} \cdot \overrightarrow{da} \qquad (9.45)$$

This is a special form of *Poynting's theorem*, valid in the case where the volume *V* contains no mobile charges.

We have:

$$\frac{dU(t)}{dt} = \frac{\partial U(t)}{\partial t} = \frac{\partial}{\partial t} \int_V u(\vec{r}, t)\, dv = \int_V \frac{\partial u}{\partial t}\, dv$$

and

$$\oint_S \vec{N} \cdot \overrightarrow{da} = \int_V (\vec{\nabla} \cdot \vec{N})\, dv$$

By substituting these expressions into (9.45) and by demanding that the ensuing integral equality be valid for *any* volume *V*, we find the differential form of Poynting's theorem:

$$\vec{\nabla} \cdot \vec{N} + \frac{\partial u}{\partial t} = 0 \qquad (9.46)$$

As is obvious from the way we derived this equation, (9.46) expresses *conservation of energy* of the e/m field.

By comparing (9.46) with the equation of continuity (6.9) for the electric charge, we note the following analogies between physical quantities:

$$\rho(\vec{r},t) \leftrightarrow u(\vec{r},t)$$
$$\vec{J}(\vec{r},t) \leftrightarrow \vec{N}(\vec{r},t)$$

The current density quantitatively describes the rate of flow of electric charge, while the Poynting vector describes the rate of flow of e/m energy.

## References for Chapter 9

## QUESTIONS

**1.** Consider a region of space in which the e/m field is static. By using the Maxwell equations, show that there can be no time-dependent sources (charges or currents) within that region.

**2.** Show that, in a region of space where the magnetic field is time-dependent, the electric field cannot be the *grad* of some scalar function.

**3.** In a region of space the "displacement current" is everywhere equal to zero. Show that there can be no time-dependent charge distribution within that region.

**4.** The potential difference between the plates of a capacitor is $V$. A compass has been placed between the plates. Describe the expected behavior of the compass if (*a*) $V$ is constant in time; (*b*) $V$ is time varying.

**5.** By using the Maxwell equations, show that the total electric charge in Nature is constant in time. [*Hint:* Imagine that the entire Nature is surrounded by an imaginary closed surface $S$. What will be the total current coming "out" of $S$?]

**6.** Assume that by some experiment we determine the values of the e/m field $(\vec{E},\vec{B})$ at all points of a region of space and at all times. Is it possible to determine the e/m potentials $\Phi(\vec{r},t)$ and $\vec{A}(\vec{r},t)$ in that region *uniquely*? So, can $\Phi$ and $\vec{A}$ be regarded as measurable physical quantities with absolute significance, as the case is with $\vec{E}$ and $\vec{B}$?

**7.** (*a*) What is the physical significance of the Poynting vector? (*b*) A point charge $q$ is firmly positioned at the center of a spherical surface $S$. What is the total e/m energy exiting $S$ per unit time?

# PROBLEMS

**1.** Show that in the interior of a "perfect" conductor ($\sigma \approx \infty$) the electric field is zero and the magnetic field is time-independent. What can you say about the e/m field *just outside* the conductor?

***Solution:*** By Ohm's law, $\vec{J} = \sigma \vec{E} \;\Rightarrow\; \vec{E} = \dfrac{\vec{J}}{\sigma} \simeq 0$ when $\sigma \approx \infty$ .

Then, by the Faraday-Henry law, $\vec{\nabla} \times \vec{E} = -\dfrac{\partial \vec{B}}{\partial t} = 0 \;\Rightarrow\; \vec{B} = \vec{B}(\vec{r})$ .

We regard the surface of the conductor as the interface of the conductor and its environment. From the boundary conditions (9.26) of Sec. 9.5 we know that the component $\vec{E}^{\parallel}$, parallel to the interface, is continuous, having the same value on both sides of the interface. Hence, given that $\vec{E}_{in}^{\parallel} = 0$, it follows that $\vec{E}_{out}^{\parallel} = 0$ also. That is, *the electric field just outside the perfect conductor is normal to the surface of the conductor.* On the other hand, the component $\vec{B}^{\perp}$, normal to the interface, is also continuous at the interface. So, given that $\partial \vec{B}_{in}^{\perp} / \partial t = 0$, it follows that $\partial \vec{B}_{out}^{\perp} / \partial t = 0$ also. That is, *just outside the perfect conductor, the component of the magnetic field normal to the surface of the conductor is time-independent.*

**2.** Show that a conductor cannot sustain net charge different from zero in its interior: any excessive charge is quickly transferred to the surface of the conductor. Thus, non-zero net charge may only exist on the surface of the conductor, while the substance is electrically neutral in its interior.

***Solution:*** Our basic equations are the following:

| | | |
|---|---|---|
| Ohm's law: | $\vec{J} = \sigma \vec{E}$ | (1) |
| Gauss' law: | $\vec{\nabla} \cdot \vec{E} = \dfrac{\rho}{\varepsilon_0}$ | (2) |
| Equation of continuity: | $\vec{\nabla} \cdot \vec{J} + \dfrac{\partial \rho}{\partial t} = 0$ | (3) |

The quantity $\rho$ is the *total* density of free charge in the interior of the conductor, due to both the mobile electrons *and* the stationary positive ions of the metal. On the other hand, the current density $\vec{J}$ is only due to the mobile electrons. We have:

$$(3) \overset{(1)}{\Rightarrow} 0 = \vec{\nabla} \cdot (\sigma \vec{E}) + \frac{\partial \rho}{\partial t} = \sigma (\vec{\nabla} \cdot \vec{E}) + \frac{\partial \rho}{\partial t} \overset{(2)}{=} \frac{\sigma}{\varepsilon_0} \rho + \frac{\partial \rho}{\partial t} \;\Rightarrow\; \frac{\partial \rho}{\partial t} = -\frac{\sigma}{\varepsilon_0} \rho$$

Integrating the differential equation on the right with respect to $t$, we find:

$$\rho(t) = \rho(0)\, e^{-(\sigma/\varepsilon_0)t} = \rho(0)\, e^{-t/\tau} \quad \text{where we have put} \quad \tau = \varepsilon_0 / \sigma$$

We notice that $\tau \to 0$ as $\sigma \to \infty$ (in general, $\tau$ is small for a good conductor). We also notice that $\rho(t) \to 0$ as $t$ increases. Thus, if at some moment there is net charge different from zero in the interior of the conductor, the excess of charge quickly flows toward the surface of the conductor so that the substance stays electrically neutral in the interior.

**3.** An inertial observer *O* records an electric field $\vec{E}$ and a magnetic field $\vec{B}$ in some region of space. Determine the electric field $\vec{E}'$ recorded by another inertial observer *O′* moving with velocity $\vec{\upsilon}$ with respect to *O*. As an application, assume that the observer *O* is inside a *purely magnetic* field (thus, $\vec{E}$ =0). Show that the observer *O′* will additionally record an *electric* field ($\vec{E}'$≠0). Assume that $\upsilon<<c$ or $\upsilon/c<<1$, so that the laws of non-relativistic Mechanics apply.

***Solution:*** We use the following trick: We consider a charge *q* that is momentarily at rest relative to *O′*, thus moves with velocity $\vec{\upsilon}$ relative to *O*. According to *O*, this charge is subject to a force $\vec{F}=q[\vec{E}+(\vec{\upsilon}\times\vec{B})]$ by the e/m field. According to *O′*, however, the force on the (stationary) charge is $\vec{F}'=q\vec{E}'$, due to the electric field alone. Now, since the relative velocity $\upsilon$ of the two observers is small, we can make the non-relativistic approximation $\vec{F}=\vec{F}'$. Thus, by eliminating *q* we find that

$$\vec{E}'=\vec{E}+(\vec{\upsilon}\times\vec{B}) \qquad (1)$$

We emphasize that (1) is only approximately valid, for $\upsilon<<c$, and requires correction in the context of Relativity. Now, in the special case where $\vec{E}$ =0, relation (1) yields $\vec{E}'=\vec{\upsilon}\times\vec{B}$. That is, whereas observer *O* records only a magnetic field, observer *O′* records an electric field as well. We conclude that the separation of the e/m field into an electric and a magnetic component is not absolute but depends on the state of motion of the observer.

**4.** Show that if the e/m field $(\vec{E},\vec{B})$ is static in a region *R* of space, this region cannot contain time-dependent sources $(\rho,\vec{J})$. Is the converse true?

***Solution:*** Static e/m field $\Rightarrow$ $\partial\vec{E}/\partial t=0$, $\partial\vec{B}/\partial t=0$, $\forall\vec{r}\in R$. We then have:

$$\vec{\nabla}\cdot\vec{E}=\frac{\rho}{\varepsilon_0} \;\Rightarrow\; \frac{\partial\rho}{\partial t}=\varepsilon_0\frac{\partial}{\partial t}(\vec{\nabla}\cdot\vec{E})=\varepsilon_0\vec{\nabla}\cdot\frac{\partial\vec{E}}{\partial t}=0 \,,$$

$$\vec{\nabla}\times\vec{B}=\mu_0\vec{J}+\varepsilon_0\mu_0\frac{\partial\vec{E}}{\partial t}=\mu_0\vec{J} \;\Rightarrow\; \frac{\partial\vec{J}}{\partial t}=\frac{1}{\mu_0}\frac{\partial}{\partial t}(\vec{\nabla}\times\vec{B})=\frac{1}{\mu_0}\vec{\nabla}\times\frac{\partial\vec{B}}{\partial t}=0$$

The converse is *not* valid, in general. For example, the region *R* may even contain no sources at all but there may be some time-dependent sources *outside* that region. These sources will generate a time-dependent e/m field everywhere in space, including the region *R*.

**5.** A charge *q* moves with constant velocity $\vec{\upsilon}$ relative to an inertial observer. Show that, relative to this observer, the electric and the magnetic field produced by the charge are everywhere related by

$$\vec{B}=\frac{1}{c^2}(\vec{\upsilon}\times\vec{E}) \qquad (1)$$

where $c=1/\sqrt{\varepsilon_0\mu_0}$ is the speed of light in vacuum. [Although (1) is of general validity, consider for simplicity that the speed $\upsilon$ of the charge is much lower than the speed of light: $\upsilon << c$. This allows us to ignore the finite time needed for the observer to receive the information that the charge passes from a certain point of space at a certain time. In reality, this information – like any e/m signal, in general – travels at a finite speed $c$ and does not reach the observer instantly (see Chap. 10).]

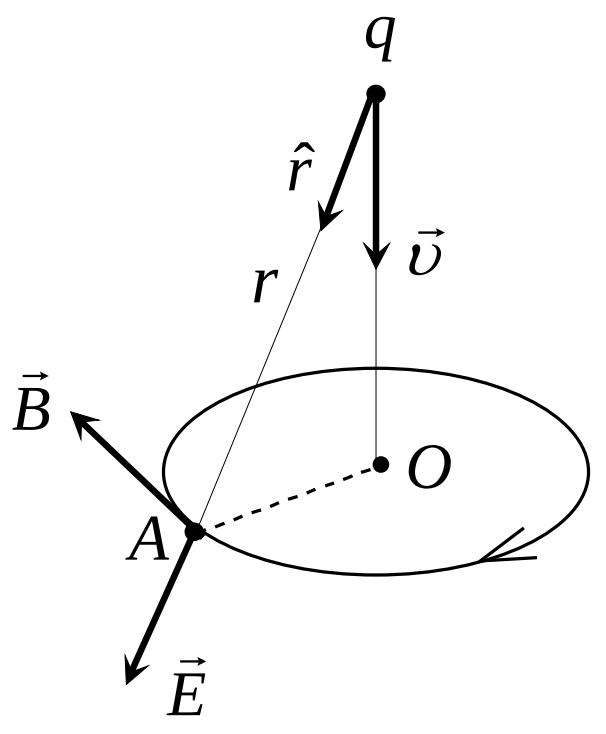

Fig. 9.5. Electric and magnetic field produced by a uniformly moving charge $q$.

***Solution:*** We assume that $q>0$. Let $A$ be the location of the observer (Fig. 9.5). Since $\upsilon << c$, we can consider that the electric field at $A$ is the same as the Coulomb field created by a stationary charge $q$ a distance $r$ from $A$:

$$\vec{E} = \frac{1}{4\pi\varepsilon_0}\frac{q}{r^2}\,\hat{r} \qquad (\upsilon << c) \tag{2}$$

(Note that this is valid for a charge moving with *constant* velocity. The electric field of an *accelerating* charge is no longer radial relative to the charge.) The magnetic field lines are circular, centered on the axis of the velocity of $q$, each circle lying on a plane normal to that axis. The direction of traversing the circle is determined by the right-hand rule and by assuming that the moving positive charge corresponds to an electric current flowing in the direction of motion of the charge. Under these assumptions the Biot-Savart law (Sec. 7.1) is written:

$$\vec{B} = \frac{\mu_0}{4\pi}\frac{q(\vec{\upsilon}\times\hat{r})}{r^2} \qquad (\upsilon << c) \tag{3}$$

By combining (2) and (3) and by taking into account that $\varepsilon_0\mu_0 = 1/c^2$ (see Appendix A) it is not hard to verify (1).

**6.** Compare the magnetic interaction between two electric charges with the electric interaction between them. Show that if the speeds of the charges, relative to an inertial observer, are small compared to the speed $c$ of light, the magnetic force between the charges is negligible compared to the electric force between them, while in the limit of very high speeds the two forces become comparable to each other.

***Solution:*** We consider two charges $q$ and $q'$ moving with corresponding velocities $\vec{\upsilon}$ and $\vec{\upsilon}'$ relative to an inertial observer. We regard $q'$ as the source of an e/m field and $q$ as a test charge within this field. We are interested in the force on $q$ due to the e/m field produced by $q'$. Let $(\vec{E}', \vec{B}')$ be the value of this field at the location of $q$. The electric force on $q$ is $\vec{F}_e = q\vec{E}'$ or, in magnitude, $F_e = qE'$, while the magnetic force on $q$ is $\vec{F}_m = q(\vec{\upsilon}\times\vec{B}')$. Now, according to Problem 5,

$$\vec{B}' = \frac{1}{c^2}(\vec{\upsilon}' \times \vec{E}')$$

Thus, $\vec{F}_m = \frac{q}{c^2}[\vec{\upsilon} \times (\vec{\upsilon}' \times \vec{E}')] \Rightarrow F_m \approx \frac{q}{c^2}\upsilon\upsilon' E' = \frac{\upsilon\upsilon'}{c^2}F_e$ , and so, $\frac{F_m}{F_e} \approx \frac{\upsilon\upsilon'}{c^2}$ .

We notice that $F_m << F_e$ when $\upsilon << c$ and $\upsilon' << c$, while $F_m \sim F_e$ when $\upsilon \sim c$ and $\upsilon' \sim c$. This means that, while in the world of low energies (or low speeds) that we experience in our everyday lives the electric interaction between charged particles seems to be stronger than their magnetic interaction, in a world of high energies the two interactions become comparable to each other. This is natural since, after all, these interactions are the two "faces" of a single *electromagnetic* interaction!

**7.** Consider a circuit consisting of an ideal battery (i.e., one with no internal resistance) connected to an external resistor (Fig. 9.6). Show that the emf of the circuit *in the direction of the current* is equal to the voltage $V$ of the battery. Also show that the emf in this case represents the *work per unit charge* done by the source (battery).

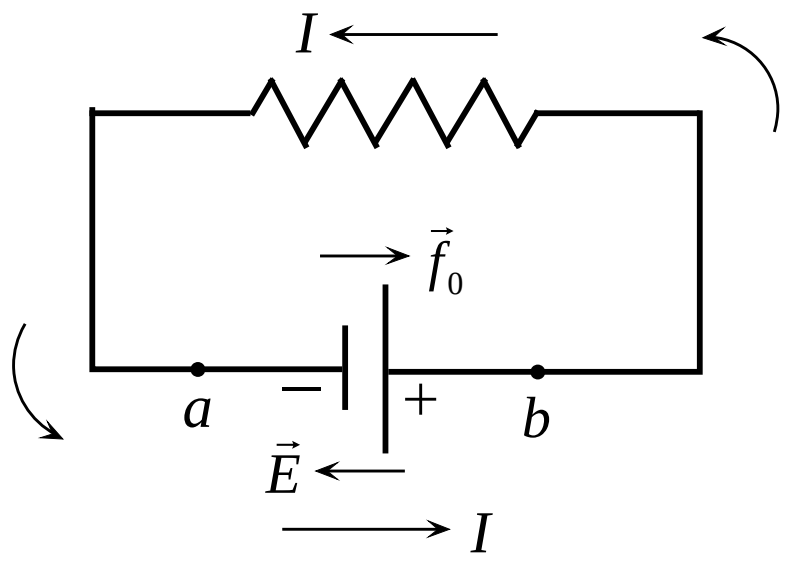

Fig. 9.6. An ideal battery connected to an external resistor. The circuit is positively oriented in the direction of the current.

***Solution:*** We recall that, in general, the emf of a circuit $C$ at time $t$ is equal to the integral

$$\mathcal{E} = \oint_C \vec{f} \cdot \overrightarrow{dl}$$

where $\vec{f} = \vec{F}/q$ is the force per unit charge at the location of the element $\overrightarrow{dl}$ of the circuit, at time $t$. In essence, we assume that in every element $\overrightarrow{dl}$ we have placed a test charge $q$ (this could be, e.g., a free electron in the conducting part of the circuit). The force $\vec{F}$ on each $q$ is then measured *simultaneously* for all charges at time $t$. Since here we are dealing with a *static* (time-independent) situation, however, we can treat the problem somewhat differently: The measurements of the forces $\vec{F}$ on the charges $q$ need not be made at the same instant, given that nothing changes with time, anyway. So, instead of placing several charges $q$ around the circuit and measuring the forces $\vec{F}$ on each of them at a particular instant, we imagine *a single charge q* making a complete tour around the loop $C$. We may assume, e.g., that the charge $q$ is one of the (*conventionally positive*) free electrons taking part in the constant current $I$ flowing in the circuit. We then measure the force $\vec{F}$ on $q$ at each point of $C$.

We thus assume that $q$ is a *positive* charge moving *in the direction of the current I*. We also assume that the direction of circulation of $C$ is the *same as the direction of the current* (counterclockwise in the figure). During its motion, $q$ is subject to two forces: (*a*) the force $\vec{F}_0$ by the source (battery) that carries $q$ from the negative pole, $a$, to the positive pole, $b$, *through the source*; (*b*) the electrostatic force $\vec{F}_e = q\vec{E}$ due to the electrostatic field $\vec{E}$ at each point of the circuit $C$ (both inside and outside the source). The total force on $q$ is

$$\vec{F} = \vec{F}_0 + \vec{F}_e = \vec{F}_0 + q\vec{E} \;\Rightarrow\; \vec{f} = \frac{\vec{F}}{q} = \frac{\vec{F}_0}{q} + \vec{E} \equiv \vec{f}_0 + \vec{E}$$

Then,

$$\mathcal{E} = \oint_C \vec{f} \cdot \overrightarrow{dl} = \oint_C \vec{f}_0 \cdot \overrightarrow{dl} + \oint_C \vec{E} \cdot \overrightarrow{dl} = \oint_C \vec{f}_0 \cdot \overrightarrow{dl} \qquad (1)$$

since $\oint_C \vec{E} \cdot \overrightarrow{dl} = 0$ for an electrostatic field. However, the action of the source on $q$ is limited to the region between the poles of the battery, that is, the section of the circuit from $a$ to $b$. Hence, $\vec{f}_0 = 0$ *outside* the source, so that (1) reduces to

$$\mathcal{E} = \int_a^b \vec{f}_0 \cdot \overrightarrow{dl} \qquad (2)$$

Now, since the current $I$ is constant, the charge $q$ moves at constant speed along the circuit. This means that the *total* force on $q$ in the direction of the path $C$ is zero. In the interior of the resistor, the electrostatic force $\vec{F}_e = q\vec{E}$ is counterbalanced by the force on $q$ due to the collisions of the charge with the positive ions of the metal (this resistive force does *not* contribute to the emf and is *not* counted in its evaluation). In the interior of the (ideal) battery, however, where there is no resistance, the electrostatic force $\vec{F}_e$ must be counterbalanced by the *opposing* force $\vec{F}_0$ exerted by the source. Thus, in the section of the circuit between $a$ and $b$,

$$\vec{F} = \vec{F}_0 + \vec{F}_e = 0 \;\Rightarrow\; \vec{f} = \frac{\vec{F}}{q} = \vec{f}_0 + \vec{E} = 0 \;\Rightarrow\; \vec{f}_0 = -\vec{E}$$

Equation (2) then takes the final form [cf. Eq. (5.17)]:

$$\mathcal{E} = -\int_a^b \vec{E} \cdot \overrightarrow{dl} = V_b - V_a = V$$

where $V_a$ and $V_b$ are the electrostatic potentials at $a$ and $b$, respectively. This is, of course, what every student knows from elementary e/m courses!

The work done by the source on $q$ upon transferring the charge from $a$ to $b$ is

$$W = \int_a^b \vec{F}_0 \cdot \overrightarrow{dl} = q \int_a^b \vec{f}_0 \cdot \overrightarrow{dl} = q\,\mathcal{E}$$

[where we have used (2)]. So, the *work of the source per unit charge* is[4] $W/q = \mathcal{E}$. This work is converted into heat in the resistor, so that the source must again supply energy in order to carry the charges once more from $a$ to $b$. This is similar to the torture of Sisyphus in ancient Greek mythology!

[4] Note that this is *not* a general property of the emf! See [8] for an analysis.

**8.** Examine Poynting's theorem for the case of a metal wire of total resistance *R*, carrying a current *I*. Thus, assume that the power spent as Joule heat within the wire is transferred from the source (battery) to the wire by means of the e/m field surrounding the circuit, and show that this power is precisely equal to $dU/dt = I^2R$.

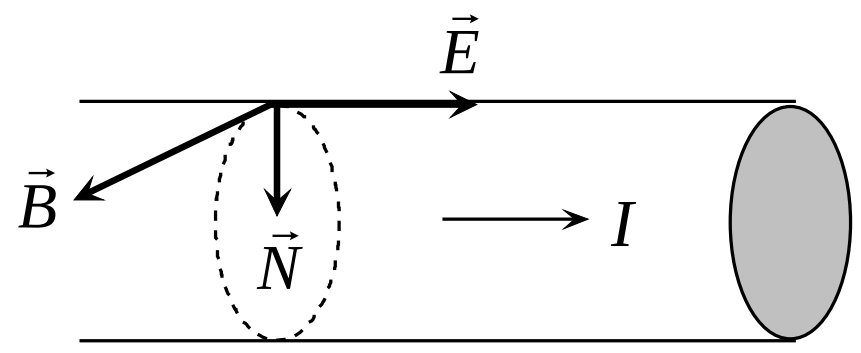

Fig. 9.7. Electromagnetic field and Poynting vector on the surface of a current-carrying wire.

***Solution:*** We consider a long, cylindrical wire of radius *ρ*, carrying a constant current *I* (Fig. 9.7). We want to evaluate the Poynting vector $\vec{N} = \vec{E} \times \vec{H} = (1/\mu_0)\vec{E} \times \vec{B}$ on the surface of the wire and thus find the power entering the wire from outside through the wire's surface. To be specific, we consider a section of the wire of length *l* and of total resistance *R*. Let $\varDelta V$ be the potential difference between the ends of this section. The electric field *inside* the wire is uniform and has the direction of the current *I*, while its magnitude is $E = \varDelta V / l$ (cf. Sec. 6.3). Since the component of the electric field parallel to the interface of two media is continuous upon crossing the interface (Sec. 9.5), we consider that, *just outside* the wire, the component of the electric field parallel to the surface of the wire is also in the direction of *I* and has magnitude $E = \varDelta V / l$. On the other hand, the current *I* produces a magnetic field whose direction *just outside* the wire is tangent to the circular cross-section of the wire and depends on the direction of the current, according to the right-hand rule (see figure). In accordance with Problem 7.3, the magnitude of the magnetic field is $B = \mu_0 I / 2\pi\rho$, where *ρ* is the radius of the cross-section of the wire.

The Poynting vector $\vec{N} = (1/\mu_0)\vec{E} \times \vec{B}$ *just outside* the wire is normal to the surface of the wire and is directed toward the *interior* of the wire. That is, $\vec{N}$ is normal to the axis of the wire and directed toward that axis. The magnitude of $\vec{N}$ is

$$N = \frac{1}{\mu_0} |\vec{E} \times \vec{B}| = \frac{1}{\mu_0} EB = \frac{1}{\mu_0} \frac{\varDelta V}{l} \frac{\mu_0 I}{2\pi\rho} = \frac{I \varDelta V}{2\pi\rho l} \qquad (1)$$

Let now $\overrightarrow{da}$ be a surface element of the wire. This element is normal to the wire's surface and is directed *outward*. The energy per unit time *coming out* of the considered wire section is equal to $\int_S \vec{N} \cdot \overrightarrow{da}$, where *S* denotes the surface of this section. Therefore, the power *entering* the wire section is

$$\frac{dU}{dt} = -\int_S \vec{N} \cdot \overrightarrow{da} = \int_S N da = N \int_S da = N(2\pi\rho l) \qquad (2)$$

(since *N* is constant on *S*). Substituting (1) into (2) and taking into account Ohm's law ($\varDelta V = IR$), we finally get:

$$\frac{dU}{dt} = I \varDelta V = I^2 R$$

**9.** Prove equation (9.5) of Sec. 9.2,

$$\mathcal{E} = -\frac{d\Phi_m}{dt},$$

for the case of a closed plane wire moving inside a *static* magnetic field $\vec{B}(\vec{r})$, where $\mathcal{E}$ is the emf (9.4) along the wire at time *t* and where $\Phi_m$ is the magnetic flux passing through the wire at this time.

***Solution:*** Assume that, at time *t*, the wire describes a closed curve *C* that is the boundary of a plane surface *S* (Fig. 9.8). At time $t' = t + dt$, the wire (which has moved in the meanwhile) describes another curve *C′* that encloses a surface *S′*. Let $\overrightarrow{dl}$ be an element of *C* in the direction of circulation of the curve, and let $\vec{\upsilon}$ be the velocity of this element relative to an inertial observer (the velocity of the elements of *C* may vary along the curve). The direction of the surface elements $\overrightarrow{da}$ and $\overrightarrow{da'}$ is consistent with the chosen direction of $\overrightarrow{dl}$, according to the right-hand rule. The element of the side ("cylindrical") surface *S′′* formed by the motion of *C*, is equal to

$$\overrightarrow{da''} = \overrightarrow{dl} \times (\vec{\upsilon}\, dt) = (\overrightarrow{dl} \times \vec{\upsilon})\, dt$$

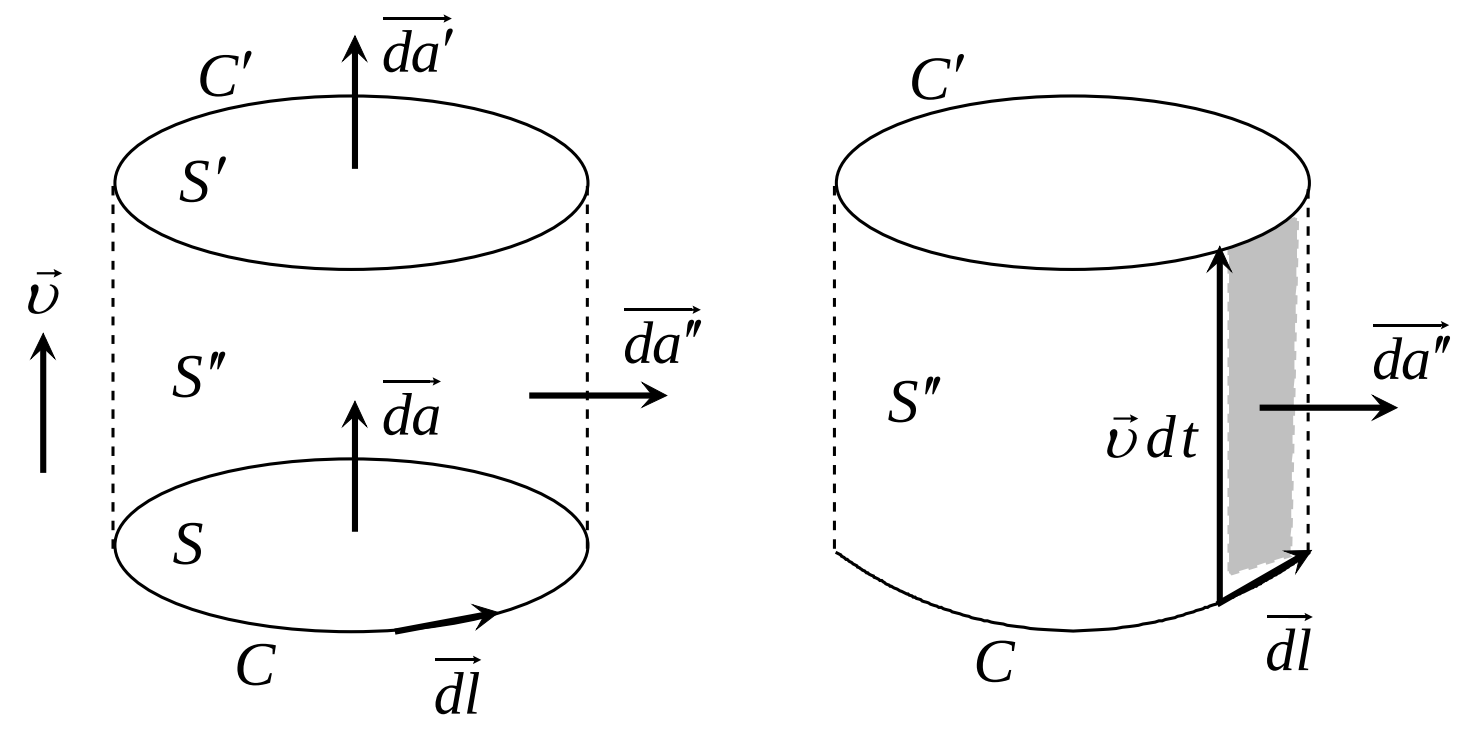

Fig. 9.8. A closed plane wire moving inside a static magnetic field.

Since the magnetic field is static, we can view the situation in a somewhat different way: Rather than assuming that the curve *C* moves within the time interval *dt* so that its points coincide with the points of the curve *C′* at time *t′*, we consider two *constant* curves *C* and *C′ at the same instant t*. In the case of a *static* field $\vec{B}$, the magnetic flux through *C′* at time $t' = t + dt$ (according to our original assumption of a moving curve) is the same as the flux through this same curve at time *t*, given that no change of the magnetic field occurs within the time interval *dt*. Now, we notice that the open surfaces $S_1 = S$ and $S_2 = S' \cup S''$ share a common boundary, namely, the curve *C*. Since the magnetic field is solenoidal, the same magnetic flux $\Phi_m$ passes through $S_1$ and $S_2$ at time *t*. That is,

$$\int_{S_1} \vec{B} \cdot \overrightarrow{da_1} = \int_{S_2} \vec{B} \cdot \overrightarrow{da_2} \;\Rightarrow\; \int_S \vec{B} \cdot \overrightarrow{da} = \int_{S'} \vec{B} \cdot \overrightarrow{da'} + \int_{S''} \vec{B} \cdot \overrightarrow{da''}$$

But, returning to our initial assumption of a moving curve, we notice that

$$\int_S \vec{B} \cdot \overrightarrow{da} = \Phi_m(t) = \textit{magnetic flux through the wire at time } t$$

and

$$\int_{S'} \vec{B}\cdot\overrightarrow{da'} = \Phi_m(t+dt) = \textit{ magnetic flux through the wire at time t+dt}$$

Hence,

$$\Phi_m(t) = \Phi_m(t+dt) + \int_{S''} \vec{B}\cdot\overrightarrow{da''} \;\Rightarrow$$

$$d\Phi_m = \Phi_m(t+dt) - \Phi_m(t) = -\int_{S''} \vec{B}\cdot\overrightarrow{da''} = -\,dt \oint_C \vec{B}\cdot(\overrightarrow{dl}\times\vec{\upsilon}) \;\Rightarrow$$

$$-\frac{d\Phi_m}{dt} = \oint_C \vec{B}\cdot(\overrightarrow{dl}\times\vec{\upsilon}) = \oint_C (\vec{\upsilon}\times\vec{B})\cdot\overrightarrow{dl} = \mathcal{E}$$

in accordance with Eqs. (9.4) and (9.5).

**10.** (*a*) A conductor of capacitance *C* carries a total charge *Q*. Show that the electrostatic energy stored in the conductor is

$$U = \frac{Q^2}{2C}$$

(*b*) Calculate the energy *U* for the case of a spherical conductor of radius *R*. By using this example, verify that the energy density of the electric field is equal to

$$u_e = \frac{1}{2}\varepsilon_0 E^2 \quad (E = |\vec{E}|)$$

***Solution:*** (*a*) Let us follow the charging process of the conductor from zero initial net charge to a final net charge *Q*. We assume that charging is achieved by transferring small amounts of charge *dq* from infinity. Let *q* be the total charge already existing on the conductor at some instant, and let *V* be the electrostatic potential of the conductor at that instant, *where we agree that* $V$=0 when there isn't any net charge on the conductor. (We recall from Sec. 5.6 that nonzero net charge can only exist on the *surface* of the conductor. Moreover, the space occupied by a conductor is a space of constant potential. Therefore, the potential at *every* point of the conductor is *V*.) We have [Eq. (5.29)]:

$$C = \frac{q}{V} \;\Rightarrow\; V = \frac{q}{C} \qquad (1)$$

Now, suppose we transfer an additional charge *dq* to the conductor from infinity (we assume that *dq* has the same sign as *q*). This requires doing some work *dW* in order to overcome the repulsive force on *dq* due to the charge *q* already existing on the conductor. This work results in an increase of the electrostatic energy "stored" in the conductor and is the *opposite* of the work $dW_e$ done on *dq* by the electrostatic field (we assume that *dq* moves with constant velocity, i.e., without acceleration, so that the force we exert on it is always equal and opposite to the force exerted on *dq* by the electrostatic field):

$$dW = -\,dW_e = -dq(V_\infty - V) = dq(V - V_\infty)$$

where use has been made of Eq. (5.26). By making the assumption that $V_\infty=0$ (consistently with the fact that no work is done on *dq* when the conductor is uncharged, i.e., when $q$=0 and $V$=0) we see that the increase $dU$ of the electrostatic energy in the conductor is

$$dU = dW = V\,dq$$

By using (1) and by integrating, we find the final value of the energy $U$ in the conductor when the latter carries a total charge $Q$:

$$dU = \frac{q}{C}dq \;\Rightarrow\; \int_0^U dU = \frac{1}{C}\int_0^Q q\,dq \;\Rightarrow\; U = \frac{Q^2}{2C}$$

(*b*) For a spherical conductor (see Chap. 5, Prob. 6) we have:

$$V = \frac{1}{4\pi\varepsilon_0}\frac{Q}{R} \;\Rightarrow\; C = \frac{Q}{V} = 4\pi\varepsilon_0 R$$

Hence,

$$U = \frac{Q^2}{2C} = \frac{Q^2}{8\pi\varepsilon_0 R}$$

Now, the magnitude of the electric field produced by the charged conductor is

$$\begin{aligned} E &= 0\,, \quad r < R \\ &= \frac{1}{4\pi\varepsilon_0}\frac{Q}{r^2}\,, \quad r \geq R \end{aligned}$$

We define the quantity $u_e = \frac{1}{2}\varepsilon_0 E^2$, so that

$$\begin{aligned} u_e &= 0\,, \quad r < R \\ &= \frac{Q^2}{32\pi^2\varepsilon_0 r^4}\,, \quad r \geq R \end{aligned}$$

We want to evaluate the integral $\int u_e\,dv$ over the entire space. It suffices, of course, to evaluate it at the *exterior* of the sphere ($r \geq R$) since, inside the sphere, $u_e$=0. The volume element is $dv=4\pi R^2 dr$. Hence,

$$\int u_e\,dv = \int_R^\infty \frac{Q^2}{32\pi^2\varepsilon_0 r^4}(4\pi r^2 dr) = \frac{Q^2}{8\pi\varepsilon_0}\int_R^\infty \frac{dr}{r^2} = \frac{Q^2}{8\pi\varepsilon_0 R} = U$$

The fact that the integral $\int u_e\,dv$ over all space equals the total electrostatic energy $U$ suggests that the function $u_e = \frac{1}{2}\varepsilon_0 E^2$ represents some sort of *energy density*. We can consider that the energy $U$ of the conductor (equal to the work done in order to charge it) is stored in the electric field surrounding the conductor. According to this interpretation, the function $u_e$ represents the energy density of the electric field.

**11.** A circuit having inductance $L$ carries a current $I(t)$. Show that the magnetic energy stored in the circuit at time $t$ is given by

$$U_m = \frac{1}{2}LI^2$$

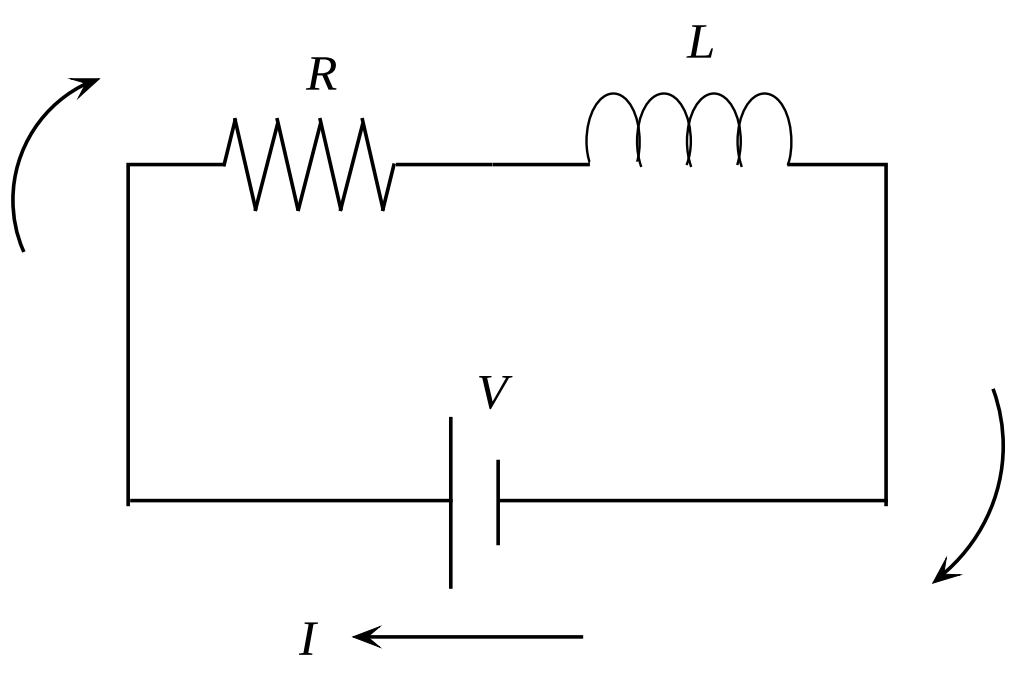

Fig. 9.9. A circuit having total resistance *R* and inductance *L*.

***Solution:*** Let *R* be the total resistance of the circuit and let *V* be the voltage of an ideal battery connected to the circuit and supplying the necessary power for the generation of the current *I* (Fig. 9.9). The moment we connect the battery to the circuit we apply to the latter an emf *V*, which tends to generate a current. The circuit then reacts, producing a self-induced emf $V_L$ that tends to oppose the increase in the current. As soon as the current stabilizes to its final value *V*/*R*, however, $V_L$ vanishes. At an arbitrary time *t*, the total emf of the circuit *in the direction of the current I* is the sum of the emf *V*, due to the battery alone (Prob. 7), and the emf $V_L$ given by the Faraday-Henry law [Eq. (9.14)]:

$$\mathcal{E} = V + V_L = V - L\frac{dI}{dt}$$

By Ohm's law (see [7]),

$$\mathcal{E} = IR \;\Rightarrow\; V - L\frac{dI}{dt} = IR \;\Rightarrow\; V = IR + L\frac{dI}{dt}$$

The total power supplied to the circuit by the battery at time *t*, when the current is *I*(*t*), is equal to

$$P = IV = I^2 R + LI\frac{dI}{dt}$$

The term $I^2R$ represents the power *irreversibly lost* as Joule heat in the resistor. This power must necessarily be supplied back by the battery in order to *maintain* the current against dissipative losses in the resistor. On the other hand, the term *LIdI*/*dt* represents the energy per unit time required to *build up* the current against the self-induced back emf $V_L$. This energy is *retrievable* and is given back to the source when the current decreases. It may be interpreted as the rate of change of the magnetic energy stored in the circuit or, alternatively, as energy per unit time required to establish the magnetic field surrounding the circuit. We thus have:

$$P_m = \frac{dU_m}{dt} = LI\frac{dI}{dt} \quad \text{so that} \quad dU_m = LI\,dI \;\Rightarrow\; \int_0^{U_m} dU_m = L\int_0^I I\,dI \;\Rightarrow$$

$$U_m = \frac{1}{2}LI^2$$

**12.** A long solenoid of length $l$ consists of $N$ turns, each of area $S$. As can be proven[5], the magnetic field in the *interior* of the solenoid is almost uniform and directed parallel to the axis of the solenoid, while its magnitude is

$$B=\mu_0 \frac{N}{l} I=\mu_0 n I \quad \text{where} \quad n=\frac{N}{l}= \textit{number of turns per unit length}$$

In the *exterior* of the solenoid the magnetic field is zero. (*a*) Find the inductance $L$ of the solenoid. (*b*) Verify that the magnetic-energy density in the interior of the solenoid is $u_m=\frac{1}{2\mu_0}B^2$.

***Solution:*** (*a*) The magnetic flux through each turn is

$$\int_S \vec{B}\cdot\overrightarrow{da}=\int_S B\,da=B\int_S da=BS$$

while the total flux through the $N$ turns is $\Phi_m=NBS$. The emf in the circuit *due to the solenoid alone* is, by the Faraday-Henry law,

$$\mathcal{E}_m=-\frac{d\Phi_m}{dt}=-NS\frac{dB}{dt}=-nlS\,(\mu_0 n\frac{dI}{dt})=-\mu_0 n^2(Sl)\frac{dI}{dt}=-\mu_0 n^2 V\frac{dI}{dt}$$

where $V=S\,l$ is the volume occupied by the solenoid. But, in general, $\mathcal{E}_m=-L\frac{dI}{dt}$ [Eq. (9.14)]. Hence, $L=\mu_0 n^2(Sl)=\mu_0 n^2 V$.

(*b*) According to Problem 11, the magnetic energy "stored" in the interior of the solenoid is

$$U_m=\frac{1}{2}LI^2=\frac{1}{2}(\mu_0 n^2 V)\left(\frac{B}{\mu_0 n}\right)^2=\frac{1}{2\mu_0}B^2 V$$

The energy density in the interior is, therefore, $u_m=\frac{U_m}{V}=\frac{1}{2\mu_0}B^2$.

(In the exterior we have $B=0$ and $u_m=0$.)

**13.** In Sec. 9.6 we showed that the Maxwell equations in differential form yield Eq. (9.31), which is the mathematical expression for conservation of charge. Show that the same equation can be obtained by using the *integral* form of the Maxwell equations [in particular, Eqs. (5.5) and (9.16)].

***Solution:*** Relation (5.5) (Gauss' law), combined with (5.7), gives

$$\oint_S \vec{E}\cdot\overrightarrow{da}=\frac{1}{\varepsilon_0}\int_V \rho\,dv \qquad (1)$$

while (9.16) (Ampère-Maxwell law), combined with (9.15), gives

$$\oint_C \vec{B}\cdot\overrightarrow{dl}=\mu_0\int_S \vec{J}\cdot\overrightarrow{da}+\varepsilon_0\mu_0\frac{d}{dt}\int_S \vec{E}\cdot\overrightarrow{da} \qquad (2)$$

[5] See, e.g., [1], Example 5.9.

Note carefully that $S$ in (1) is a closed surface, whereas in (2) the surface $S$ is open (see Fig. 9.10). We may, however, close the latter surface by letting the closed curve $C$ (which constitutes the boundary of $S$) shrink to a point, taking care that the surface element $\overrightarrow{da}$ point toward the *exterior* of the ensuing closed surface $S$ (Fig. 9.11).

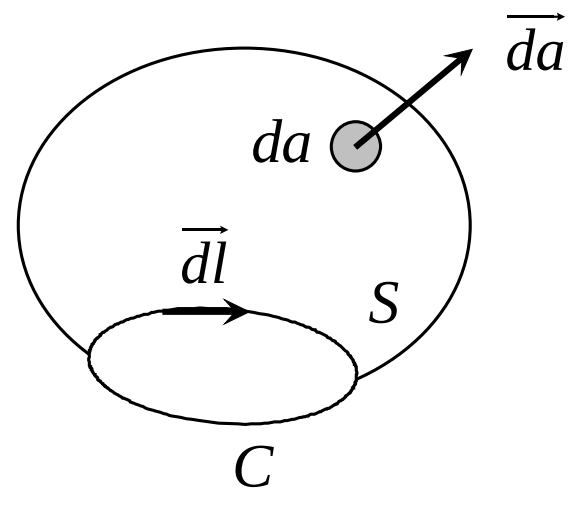

Fig. 9.10. An open surface $S$ bordered by a closed curve $C$.

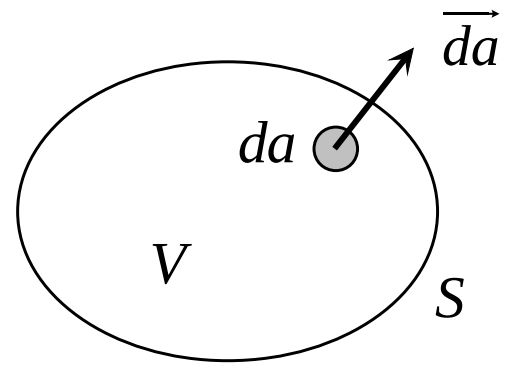

Fig. 9.11. The surface $S$ of Fig. 9.10 becomes closed when its boundary $C$ shrinks to a point.

Now, as $C$ shrinks to a point, the closed line integral on the left-hand side of (2) vanishes, so that this relation takes on the new form (after eliminating $\mu_0$):

$$\oint_S \vec{J} \cdot \overrightarrow{da} + \varepsilon_0 \frac{d}{dt} \oint_S \vec{E} \cdot \overrightarrow{da} = 0 \qquad (3)$$

By assuming that the closed surfaces $S$ in (1) and (3) coincide, and by substituting (1) into (3), we find:

$$\frac{d}{dt} \int_V \rho \, dv = - \oint_S \vec{J} \cdot \overrightarrow{da} \ ,$$

which is precisely Eq. (9.31).

## CHAPTER 10

# ELECTROMAGNETIC WAVES

### 10.1 The Wave Equation

As we know, a *field* is the assignment of a specific value to a certain physical quantity at each point of a region of space. A field can be described mathematically by a function of the form $f(\vec{r})$ (*scalar field*) or $\vec{F}(\vec{r})$ (*vector field*), where $f$ or $\vec{F}$ represents the physical quantity while $\vec{r}$ is the position vector of a point in the considered region. A field is called *static* if it is time-independent: $\partial f/\partial t=0$, $\partial\vec{F}/\partial t=0$.

In a static field nothing changes with time. It looks like a quiet pond in which the water is undisturbed and stays calm over some period of time. Now, you must have observed what happens when we throw a stone into the water of a pond: the local disturbance caused by the stone expands in every direction in the form of a series of ripples, destroying the peacefulness of the static situation even at remote locations relative to the point of incidence of the stone. This disturbance constitutes a *change in time* of a physical condition (in our example, the degree of deviation from the state of absolute calmness of the water). Moreover, the disturbance is not confined to the point where it initially occurred but *propagates* with finite speed, thus creating a time-dependent situation to other points of space (which space here is the surface of the pond). What we have described is perhaps the simplest example of a wave.

Generally speaking, a *wave* is a physical condition that, when produced at some point of space, propagates with finite speed and later becomes perceptible at other points, possibly having suffered some change in the meanwhile. What basically propagates is the *disturbance* (or *perturbation*) of a given field that describes a certain physical quantity, such as, e.g., the deformation of a spring, the pressure in a gas, the vertical displacement of a string, the compression in a solid, the electromagnetic field in a region of space, etc.

We now seek a mathematical description of wave propagation. We assume for simplicity that the disturbance suffers no change as it travels. The simplest kind of field is described by a scalar function of the form $\xi=f(x)$. This field is graphically represented by a curve on the $x\xi$-plane, as seen in Fig. 10.1.

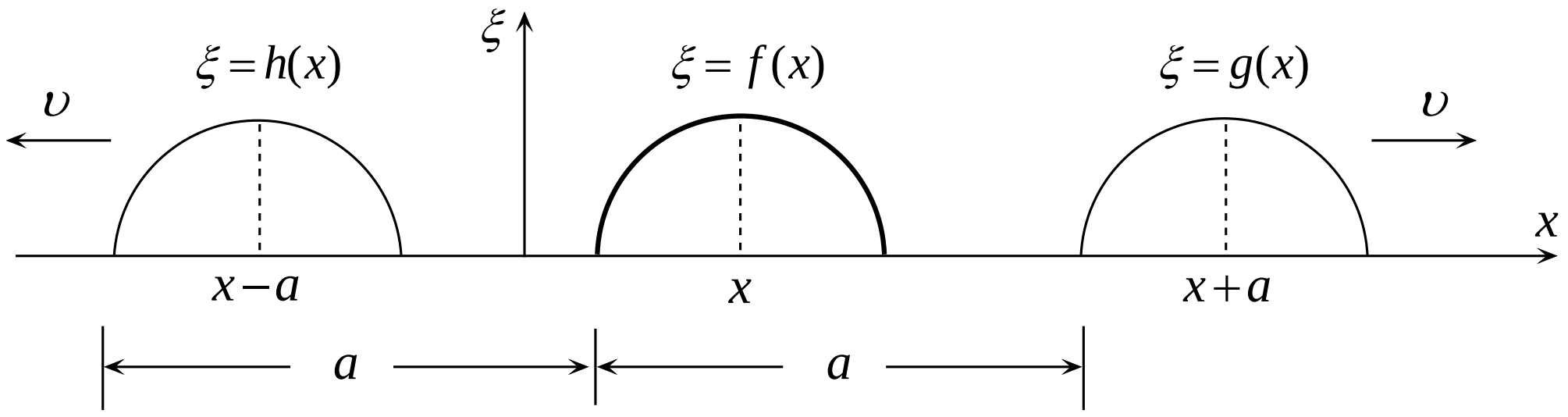

Fig. 10.1. A curve displaced in both directions along the *x*-axis.

If the initial curve $\xi = f(x)$ (the middle one in the figure) is displaced a distance $a$ to the right, we get a new function $\xi = g(x)$, while by displacing the initial curve by $a$ to the left we get a function $\xi = h(x)$. We notice that $f(x) = g(x+a) = h(x-a)$, for all $x$, so that

$$g(x) = f(x-a)\ , \quad h(x) = f(x+a) \qquad (10.1)$$

Suppose now that the distance $a$ is not fixed but changes with time according to the relation $a = \upsilon t$, where $\upsilon$ is a positive constant. We then write:

$$\boxed{\xi = g(x,t) = f(x - \upsilon t)\ , \quad \xi = h(x,t) = f(x + \upsilon t)} \qquad (10.2)$$

The function $g(x,t)$ represents a curve moving to the *right* with speed $\upsilon$, while the function $h(x,t)$ represents a curve moving to the *left* with speed $\upsilon$. At time $t=0$, the two curves are coincident with the constant curve $f(x)$, since $g(x,0)=h(x,0)=f(x)$. We conclude that a mathematical expression of the form

$$\xi = \xi(x,t) = f(x \pm \upsilon t) \qquad (10.3)$$

describes a physical condition that propagates unchanged along the $x$-axis with constant speed equal to $\upsilon$. An effect of this kind constitutes *wave motion*.

We note the following:

1. In the expression (10.3) the positive sign corresponds to a wave traveling in the $-x$ direction, while the negative sign indicates a wave traveling in the $+x$ direction.

2. Equivalent to (10.3) is the expression

$$\xi = \xi(x,t) = F(t \pm x/\upsilon) \qquad (10.4)$$

Indeed, simply notice that $f(x \pm \upsilon t) = f[\pm \upsilon (t \pm x/\upsilon)] \equiv F(t \pm x/\upsilon)$.

3. Consider an observer located at a certain point $x=x_0$ of the $x$-axis. This observer records the values of the field $\xi$ at her location and finds that they change with time according to the relation $\xi=\xi(x_0, t)= f(x_0 \pm \upsilon t)$, where $\upsilon$ is the velocity of displacement of the curve representing the field, along the $x$-axis. If, however, this curve were at rest on the $x$-axis ($\upsilon=0$), the observer at $x_0$ would record a fixed value $\xi = f(x_0)$ while the totality of observations on the $x$-axis would indicate the existence of a *static* field $\xi = f(x)$. We conclude that

*a static field cannot be associated with wave motion.*

Indeed, a wave is the propagation of a *disturbance*, i.e., of a *time change* of some physical quantity. So, if there is no disturbance to begin with, there is nothing to be propagated anyway!

*Proposition:* The functions (10.2): $\xi = f(x \pm \upsilon t)$, as well as their sum, satisfy the *wave equation*

$$\boxed{\frac{\partial^2 \xi}{\partial x^2} - \frac{1}{\upsilon^2}\frac{\partial^2 \xi}{\partial t^2} = 0} \tag{10.5}$$

*Proof:* Let $\xi = f(x \pm \upsilon t)$. We set $w = x \pm \upsilon t$, so that $\xi = f(w)$. We have:

$$\frac{\partial \xi}{\partial x} = \frac{df(w)}{dw}\frac{\partial w}{\partial x} = f'(w)\cdot 1 = f'(w)\,; \text{ similarly, } \frac{\partial^2 \xi}{\partial x^2} = f''(w)$$

$$\frac{\partial \xi}{\partial t} = \frac{df(w)}{dw}\frac{\partial w}{\partial t} = \pm\upsilon f'(w)\,; \text{ similarly, } \frac{\partial^2 \xi}{\partial t^2} = (\pm\upsilon)^2 f''(w) = \upsilon^2 f''(w)$$

Thus,

$$\frac{\partial^2 \xi}{\partial x^2} - \frac{1}{\upsilon^2}\frac{\partial^2 \xi}{\partial t^2} = f''(w) - \frac{1}{\upsilon^2}\upsilon^2 f''(w) = 0$$

That is, the functions $\xi = f_1(x - \upsilon t)$ and $\xi = f_2(x + \upsilon t)$ satisfy the partial differential equation (10.5). Since this equation is *linear* (in the sense that it does not contain products or powers of $\xi$-dependent terms), the sum of any two solutions of the equation also is a solution. The *general solution* of (10.5) is, therefore,

$$\xi = \xi(x,t) = f_1(x - \upsilon t) + f_2(x + \upsilon t) \tag{10.6}$$

for *arbitrary* choices of the functions $f_1$ and $f_2$ [1,2].

## 10.2 Harmonic Wave

Consider a wave $\xi(x, t)$ traveling in the positive $x$-direction with velocity $\upsilon$. At time $t=0$ the field $\xi$ has the form

$$\xi(x,0) = f(x) = A\cos(kx + \alpha) \quad (A > 0,\ -\infty < x < \infty) \tag{10.7}$$

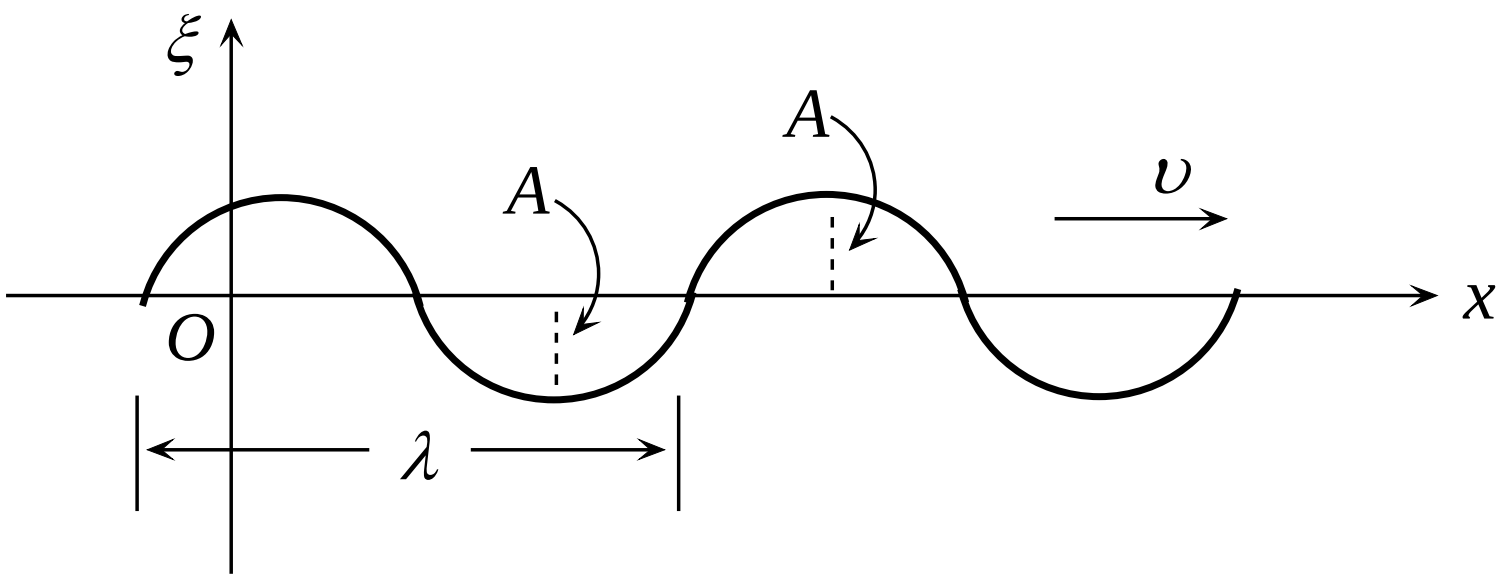

Fig. 10.2. Graph of a harmonic wave at $t$=0.

(only a part of the curve $\xi(x,0)$ is shown in Fig. 10.2, which curve is assumed to extend over the entire $x$-axis). For $t>0$ the wave is described by the equation

$$\xi(x,t) = f(x-\upsilon t) = A\cos[k(x-\upsilon t)+\alpha] \tag{10.8}$$

By putting

$$\omega = k\upsilon \tag{10.9}$$

equation (10.8) is written:

$$\boxed{\xi(x,t) = A\cos(kx-\omega t+\alpha)} \tag{10.10}$$

The positive constant $A$ is called the *amplitude* of the wave, while the constant $\alpha$ may assume any value. For example, for $\alpha=0$, Eq. (10.10) is written: $\xi=A\cos(kx-\omega t)$, while for $\alpha=\pi/2$ it is written: $\xi=A\cos(kx-\omega t+\pi/2)=-A\sin(kx-\omega t)$. We also set

$$\boxed{k=\frac{2\pi}{\lambda}\,, \quad \omega=\frac{2\pi}{T}=2\pi f} \tag{10.11}$$

so that

$$\boxed{\upsilon=\frac{\omega}{k}=\frac{\lambda}{T}=\lambda f} \tag{10.12}$$

We call $k$ the *wave number*, $\lambda$ the *wavelength*, $T$ the *period*, $f=1/T$ the *frequency*, and $\omega$ the *angular frequency*, of the wave. We notice that

$$\xi(x+\lambda,\ t) = \xi(x,t)\,, \quad \xi(x,\ t+T) = \xi(x,t) \tag{10.13}$$

That is, the function $\xi(x,t)$ is periodic in both $x$ and $t$, with corresponding periods[1] $\lambda=2\pi/k$ and $T=2\pi/\omega$ (see also Prob. 1). Furthermore, from the relation $\lambda=\upsilon T$ it follows that, *within the time interval T of a period the wave travels a distance λ equal to one wavelength*.

For a given value of $x$, e.g., at the point $x=0$, we have:

$$\xi(0,t) = A\cos(-\omega t+\alpha) = A\cos(\omega t-\alpha)$$

We notice that, *at every point x, the value ξ of the field varies harmonically with time with period* $T=2\pi/\omega$. For this reason the wave (10.10) is called a *harmonic wave*. Thus, if we set the molecules of an elastic medium into harmonic motion with frequency $f$ at the point $x=0$, the disturbance will be transmitted with finite velocity to other points in the medium, setting the molecules everywhere into harmonic motion with the same frequency $f$.

---

[1] In general, the function $\cos(\omega t+\beta)$ has period $T=2\pi/\omega$. Equivalently, the function $\cos[(2\pi t/T)+\beta]$ has period $T$. Analogous remarks apply to the function $\cos(kx+\gamma)$ with period $\lambda=2\pi/k$.

## 10.3 Plane Waves in Space

Consider a harmonic wave $\xi(x,t)=A\cos(kx-\omega t+\alpha)$ traveling in the positive $x$-direction with velocity $\upsilon=\omega/k$. The argument $\varphi=kx-\omega t+\alpha$ of the cosine is called the *phase* of the wave. At any given time $t$ the phase $\varphi$ takes on the same value at all points of space corresponding to a given value $x=c$ of the variable $x$. The locus of these points for a given $c$ is a *plane* intersecting normally the $x$-axis (different values of $c$ correspond to different planes parallel to one another); see Fig. 10.3.

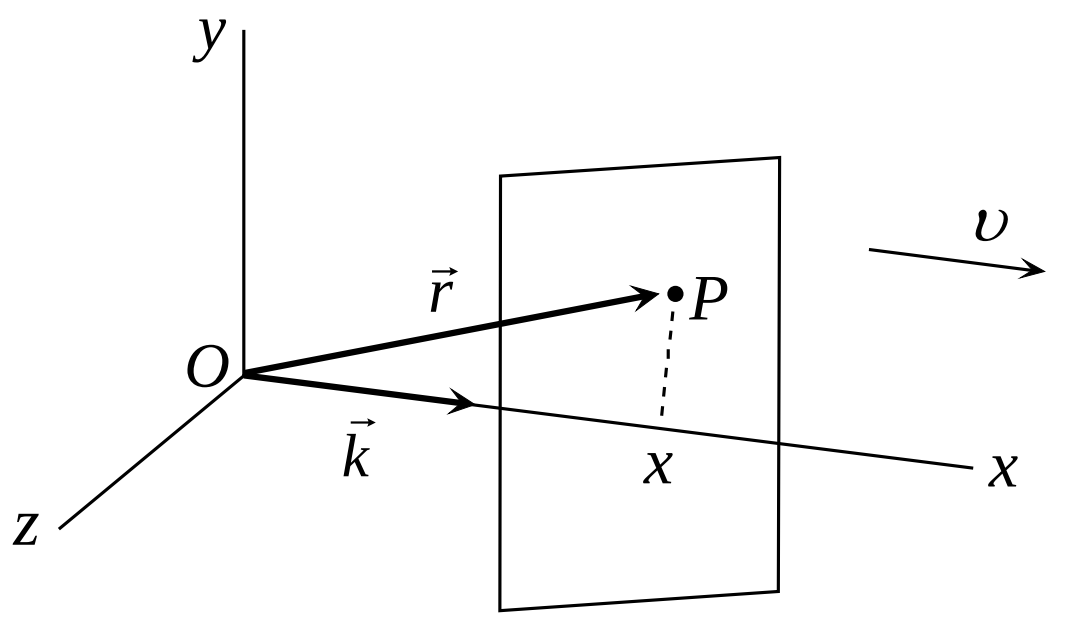

Fig. 10.3. Wave surface of a plane wave traveling in the +$x$-direction.

In general, a surface on which a wave takes on a constant value at any given time is called a *wave surface*. A wave whose wave surfaces are planes is called a *plane wave*. In particular, the harmonic wave $\xi(x,t)=A\cos(kx-\omega t+\alpha)$ is a plane wave whose wave surfaces are surfaces of *constant phase*.

Let $P$ be a point of space having position vector $\vec{r} \equiv (x, y, z)$ relative to the origin $O$ of our coordinate system. It must be emphasized that, although traveling in the direction of the $x$-axis, the harmonic wave considered above is perceptible at *all* points of space (not just those on the $x$-axis!). In particular, the value of $\xi$ at the given point $P\equiv(x,y,z)$ is $\xi(\vec{r}, t)=A\cos(kx-\omega t+\alpha)$. Let now $\vec{k}$ be a vector in the direction of propagation of the wave (here, $+x$), of magnitude $k=\omega/\upsilon$. Then,

$$\vec{k}\cdot\vec{r} = (k\hat{u}_x)\cdot(x\hat{u}_x + y\hat{u}_y + z\hat{u}_z) = kx$$

so that $\xi(\vec{r}, t)=A\cos(\vec{k}\cdot\vec{r}-\omega t+\alpha)$. This expression is not restricted to the case of a wave traveling in the +$x$-direction but is generally valid for *any* direction of propagation of the harmonic wave. Therefore, the function

$$\boxed{\xi(\vec{r}, t) = A\cos(\vec{k}\cdot\vec{r} - \omega t + \alpha)} \qquad (10.14)$$

represents a *harmonic plane wave traveling in the direction of* $\vec{k}$ *with speed* $\upsilon=\omega/k$, where $k=|\vec{k}|$. The vector $\vec{k}$ is called the *wave vector*. At any given time $t_0$ the wave surfaces, which are surfaces of constant phase, are *planes perpendicular to* $\vec{k}$ (Fig. 10.4):

$$\varphi = \vec{k}\cdot\vec{r} - \omega t_0 + \alpha = const. \;\Leftrightarrow\; \vec{k}\cdot\vec{r} = const. \;\Leftrightarrow\; k_x x + k_y y + k_z z = const.$$

(equation of a plane perpendicular to a given vector $\vec{k}$ ).

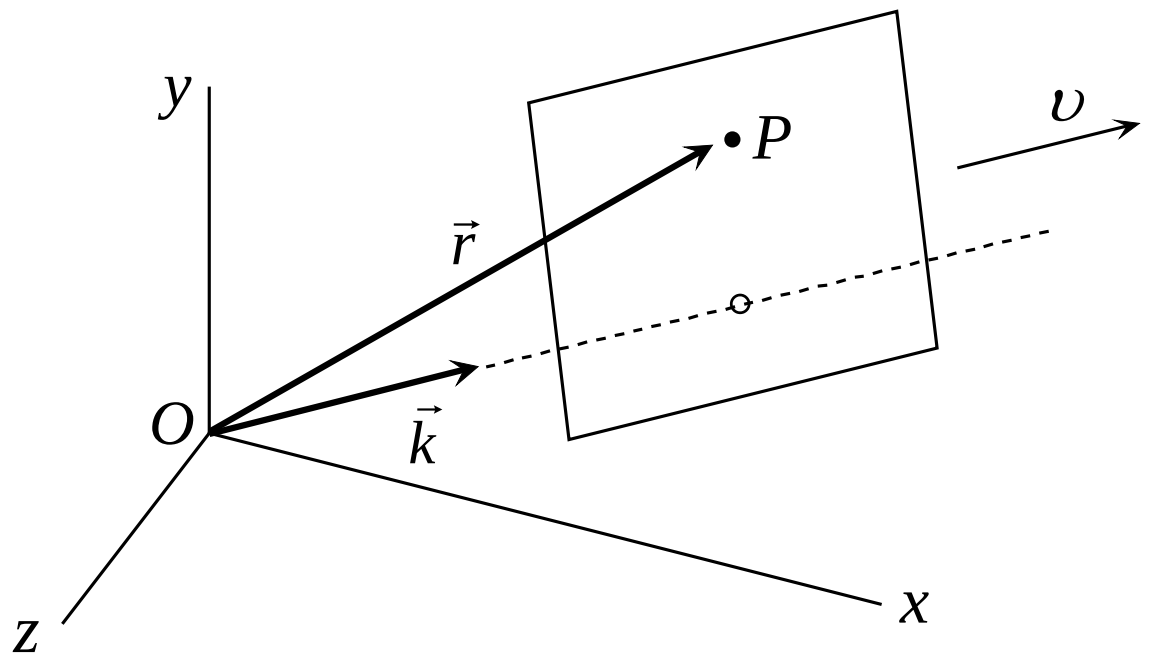

Fig. 10.4. Wave surface of a plane wave traveling in an arbitrary direction.

*Examples:*

1. For a harmonic wave traveling in the *negative* $x$-direction, we have: $\vec{k} = -k\,\hat{u}_x$, $\vec{k}\cdot\vec{r} = -kx$ and $\xi(\vec{r}, t) = A\cos(-kx - \omega t + \alpha) = A\cos(kx + \omega t + \beta)$, where we have put $\beta = -\alpha$.

2. For a harmonic wave in the positive $z$-direction, we have: $\vec{k} = k\hat{u}_z$, $\vec{k}\cdot\vec{r} = kz$ and $\xi(\vec{r}, t) = A\cos(kz - \omega t + \alpha)$ (find the corresponding expression for a wave in the negative $z$-direction).

More generally, now, the expression

$$\boxed{\xi(\vec{r}, t) = f(\vec{k}\cdot\vec{r} - \omega t)} \qquad (10.15)$$

where $f$ is an *arbitrary* function, represents a *plane wave* traveling in the direction of a given vector $\vec{k}$, with speed

$$\upsilon = \frac{\omega}{k} \qquad (10.16)$$

where $k = |\vec{k}|$. The function $\xi(\vec{r}, t) \equiv \xi(x, y, z, t)$ satisfies the *wave equation*

$$\boxed{\nabla^2\xi - \frac{1}{\upsilon^2}\frac{\partial^2\xi}{\partial t^2} = 0 \quad \Leftrightarrow \quad \frac{\partial^2\xi}{\partial x^2} + \frac{\partial^2\xi}{\partial y^2} + \frac{\partial^2\xi}{\partial z^2} - \frac{1}{\upsilon^2}\frac{\partial^2\xi}{\partial t^2} = 0} \qquad (10.17)$$

*Proof:* We set $s = \vec{k}\cdot\vec{r} - \omega t = k_x x + k_y y + k_z z - \omega t$, so that $\xi = f(s)$. We have:

$$\frac{\partial\xi}{\partial x} = \frac{df(s)}{ds}\frac{\partial s}{\partial x} = k_x f'(s), \quad \frac{\partial^2\xi}{\partial x^2} = k_x^{\,2} f''(s), \quad \text{etc.}$$

$$\nabla^2\xi = \frac{\partial^2\xi}{\partial x^2} + \frac{\partial^2\xi}{\partial y^2} + \frac{\partial^2\xi}{\partial z^2} = (k_x^{\,2} + k_y^{\,2} + k_z^{\,2})\, f''(s) = k^2 f''(s)$$

$$\frac{\partial\xi}{\partial t} = \frac{df(s)}{ds}\frac{\partial s}{\partial t} = -\omega f'(s), \quad \frac{\partial^2\xi}{\partial t^2} = (-\omega)^2 f''(s) = \omega^2 f''(s)$$

Therefore,

$$\nabla^2 \xi - \frac{1}{\upsilon^2} \frac{\partial^2 \xi}{\partial t^2} = k^2 f''(s) - \frac{\omega^2}{\upsilon^2} f''(s) = (k^2 - \frac{\omega^2}{\upsilon^2}) f''(s) = 0$$

since $k=\omega/\upsilon$ by (10.16).

*Complex form of a harmonic plane wave*

I hope you feel comfortable with complex numbers [2,3] since we are going to make extensive use of them in this chapter! The reason is simple: instead of working with sines and cosines, it is easier to work with exponential functions – albeit complex ones.

A harmonic plane wave is written in complex form:

$$\boxed{\xi(\vec{r}, t) = A \exp\{i(\vec{k} \cdot \vec{r} - \omega t)\} \equiv A e^{i(\vec{k}\cdot\vec{r} - \omega t)}} \quad (10.18)$$

where $A$ is a *complex* amplitude. Like any complex number, $A$ can be expressed as

$$A = |A| e^{i\alpha} \quad (10.19)$$

where $|A|$ is the *modulus* (absolute value) of $A$ and where $\alpha \in R$. Thus (10.18) is written as

$$\xi(\vec{r}, t) = |A| e^{i\alpha} e^{i(\vec{k}\cdot\vec{r} - \omega t)} = |A| e^{i(\vec{k}\cdot\vec{r} - \omega t + \alpha)} \quad (10.20)$$

Now, in general, $e^{i\theta} = \cos\theta + i \sin\theta$. Hence,

$$\xi(\vec{r}, t) = |A| \cos(\vec{k} \cdot \vec{r} - \omega t + \alpha) + i\, |A| \sin(\vec{k} \cdot \vec{r} - \omega t + \alpha)$$

We observe that the *real part* of $\xi$ is equal to

$$\mathrm{Re}\, [\xi(\vec{r}, t)] = |A| \cos(\vec{k} \cdot \vec{r} - \omega t + \alpha) \quad (10.21)$$

which is our familiar real form (10.14) of the harmonic plane wave.

It should be noted that

> *complex expressions are devoid of physical meaning and cannot themselves represent physical quantities; in every case the actual physical quantity will be the real part of the corresponding complex expression.*

## 10.4 Electromagnetic Waves

Maxwell's discovery that all electromagnetic (e/m) disturbances propagate in space as waves traveling at the speed of light constitutes one of the greatest triumphs of Theo-retical Physics of all time. It could justly be said that this discovery is no less important than the Theory of Relativity! (You should remember this every time you enjoy your favorite music on the radio, watch an interesting soccer game on TV, text messages to your friends on your cell phone, etc.)

We begin with some useful remarks on the Maxwell equations:

1. In *empty space* (*vacuum*) there are no charges or currents, either free or bound (cf. Chap. 8). Hence the *total* charge and current densities are zero: $\rho=0$, $\vec{J}=0$. The general Maxwell equations (9.20) are written:

$$\begin{aligned} &(a) \quad \vec{\nabla}\cdot\vec{E}=0 &\qquad &(c) \quad \vec{\nabla}\times\vec{E}=-\frac{\partial\vec{B}}{\partial t} \\ &(b) \quad \vec{\nabla}\cdot\vec{B}=0 &\qquad &(d) \quad \vec{\nabla}\times\vec{B}=\varepsilon_0\mu_0\frac{\partial\vec{E}}{\partial t} \end{aligned} \tag{10.22}$$

2. Inside matter we *cannot* use relations (10.22), even for insulators where there are no *free* charges or currents. Indeed, inside matter there may exist *bound* charges and currents, so that the total densities $\rho$ and $\vec{J}$ are *not* zero even if the *free* densities $\rho_f$ and $\vec{J}_f$ *are* zero. In the case of a *non-conducting* medium, therefore, we will use the alternative form (9.25) of the Maxwell equations, setting $\rho_f=0$ and $\vec{J}_f=0$:

$$\begin{aligned} &(a) \quad \vec{\nabla}\cdot\vec{D}=0 &\qquad &(c) \quad \vec{\nabla}\times\vec{E}=-\frac{\partial\vec{B}}{\partial t} \\ &(b) \quad \vec{\nabla}\cdot\vec{B}=0 &\qquad &(d) \quad \vec{\nabla}\times\vec{H}=\frac{\partial\vec{D}}{\partial t} \end{aligned} \tag{10.23}$$

3. If the non-conducting medium is a *homogeneous linear dielectric,* then $\vec{D}=\varepsilon\vec{E}$ and $\vec{H}=\vec{B}/\mu$, with constant $\varepsilon$ and $\mu$ (cf. Chap. 8). As is easy to show, Eqs. (10.23) take on the form

$$\begin{aligned} &(a) \quad \vec{\nabla}\cdot\vec{E}=0 &\qquad &(c) \quad \vec{\nabla}\times\vec{E}=-\frac{\partial\vec{B}}{\partial t} \\ &(b) \quad \vec{\nabla}\cdot\vec{B}=0 &\qquad &(d) \quad \vec{\nabla}\times\vec{B}=\varepsilon\mu\frac{\partial\vec{E}}{\partial t} \end{aligned} \tag{10.24}$$

By comparing (10.24) with (10.22) we notice that the Maxwell equations for a homogeneous linear dielectric are just the vacuum Maxwell equations with $\varepsilon_0$ and $\mu_0$ replaced by $\varepsilon$ and $\mu$, respectively. This means that

> *every physical or mathematical conclusion drawn by the Maxwell equations for empty space is also valid for a homogeneous linear dielectric upon substitution of* $\varepsilon_0$ and $\mu_0$ *with* $\varepsilon$ *and* $\mu$, *respectively.*

After these preliminary remarks, we display a useful vector identity that will be needed below:

$$\vec{\nabla}\times(\vec{\nabla}\times\vec{A})=\vec{\nabla}(\vec{\nabla}\cdot\vec{A})-\nabla^2\vec{A} \qquad (10.25)$$

where $\vec{A}=A_x\,\hat{u}_x+A_y\,\hat{u}_y+A_z\,\hat{u}_z$ is a vector field. In Cartesian coordinates, where the unit vectors are constant (i.e., coordinate-independent), we have:

$$\nabla^2\vec{A}=(\nabla^2 A_x)\,\hat{u}_x+(\nabla^2 A_y)\,\hat{u}_y+(\nabla^2 A_z)\,\hat{u}_z \qquad (10.26)$$

*Note:* In curvilinear coordinates [2,4], such as spherical or cylindrical coordinates, a relation of the form (10.26) is *not* valid. In such cases we simply take (10.25) as the *definition* of $\nabla^2\vec{A}$. Alternatively, we apply the operator $\nabla^2$ directly on $\vec{A}$ by differentiating the components of the latter *as well as the unit vectors* associated with the given system of coordinates.

*A. Wave equation in empty space*

Taking the *rot* of (10.22*c*) and using (10.25), (10.22*a*) and (10.22*d*), we have:

$$\vec{\nabla}\times(\vec{\nabla}\times\vec{E})=-\vec{\nabla}\times\frac{\partial\vec{B}}{\partial t} \;\Rightarrow\; \vec{\nabla}(\vec{\nabla}\cdot\vec{E})-\nabla^2\vec{E}=-\frac{\partial}{\partial t}(\vec{\nabla}\times\vec{B}) \;\Rightarrow$$

$$\boxed{\nabla^2\vec{E}-\varepsilon_0\mu_0\frac{\partial^2\vec{E}}{\partial t^2}=0} \qquad (10.27)$$

Similarly, taking the *rot* of (10.22*d*) and using (10.25), (10.22*b*) and (10.22*c*), we find:

$$\boxed{\nabla^2\vec{B}-\varepsilon_0\mu_0\frac{\partial^2\vec{B}}{\partial t^2}=0} \qquad (10.28)$$

The Maxwell system (10.22) with which we started is a system of *first-order* partial differential equations for $\vec{E}$ and $\vec{B}$. What we have achieved is to *uncouple* the system by deriving *separate* equations for $\vec{E}$ and $\vec{B}$. These latter equations, however, are now *second-order* differential equations (this was the price to be paid for uncoupling). Note carefully that, while any two fields $\vec{E}$ and $\vec{B}$ satisfying the system (10.22) separately satisfy (10.27) and (10.28), respectively, the converse is *not* true; that is, *not any two solutions* of (10.27) and (10.28), when taken together, satisfy the Maxwell system (10.22)!

Now, you must have noticed that (10.27) and (10.28) are actually the same differential equation written twice, once for each field $\vec{E}$ and $\vec{B}$. This equation is of the form

$$\nabla^2 \vec{A} - \varepsilon_0 \mu_0 \frac{\partial^2 \vec{A}}{\partial t^2} = 0 \quad \text{with} \quad \vec{A} = \vec{E} \text{ or } \vec{B} \tag{10.29}$$

By setting

$$\boxed{\varepsilon_0 \mu_0 \equiv \frac{1}{c^2} \iff c = \frac{1}{\sqrt{\varepsilon_0 \mu_0}}} \tag{10.30}$$

we write (10.29) as

$$\boxed{\nabla^2 \vec{A} - \frac{1}{c^2} \frac{\partial^2 \vec{A}}{\partial t^2} = 0} \tag{10.31}$$

where $\vec{A} = \vec{E}$ or $\vec{B}$. Explicitly, (10.31) is valid if in place of $\vec{A}$ we put *any component* of $\vec{E}$ or $\vec{B}$ (a total of 6 fields).

We note the following:

1. Relation (10.31) is the equation of a wave traveling with speed *c*. In particular, the plane waves of Sec. 10.3 obey this equation for $\upsilon=c$, although such waves constitute only a special category of solutions. (Solutions of different kinds include, for example, *spherical waves,* which will be briefly examined later on in the context of radiation.)
2. By substituting the values for $\varepsilon_0$ and $\mu_0$ into (10.30), we find that the velocity of propagation of the wave is $c \simeq 3\times 10^8 \, m/s$. This is, of course, the *speed of light in vacuum*.

*Definition:* A pair $(\vec{E}, \vec{B})$ of solutions of the wave equation (10.31), which solutions together satisfy the Maxwell system of equations, constitutes an *electromagnetic (e/m) wave*.

The physical significance of the situation can be stated as follows:

> *Any change (disturbance) of the e/m field at a point of space is not felt instantly at other points but propagates in space in the form of a wave traveling at the speed of light. In particular, light itself is an e/m wave having the special property that our eye can perceive it.*

In the case of a *static* e/m field there is no disturbance (time change) anywhere. Therefore, *static e/m fields do not generate e/m waves*.

*B. Wave equation for a linear non-conducting medium*

By the Maxwell equations (10.24) for a linear non-conducting medium or, equivalently, by Eqs. (10.27) and (10.28) with the replacements $\varepsilon_0 \to \varepsilon$ and $\mu_0 \to \mu$, we obtain the wave equations

$$\boxed{\nabla^2 \vec{E} - \varepsilon\mu \frac{\partial^2 \vec{E}}{\partial t^2} = 0\,, \quad \nabla^2 \vec{B} - \varepsilon\mu \frac{\partial^2 \vec{B}}{\partial t^2} = 0} \tag{10.32}$$

These equations describe an e/m wave traveling within the non-conducting medium with speed

$$\boxed{\upsilon = \frac{1}{\sqrt{\varepsilon\mu}}} \tag{10.33}$$

Visible light is a special kind of an e/m wave. For most transparent media, $\mu \simeq \mu_0$. Given that, in general, $\varepsilon > \varepsilon_0$, we have that $\varepsilon\mu > \varepsilon_0\mu_0$. By comparing (10.30) and (10.33) we then conclude that $\upsilon < c$. That is, the speed of light in matter is less than that in vacuum. The *index of refraction* of a transparent medium is defined as [5,6]

$$n = \frac{c}{\upsilon} = \left( \frac{\varepsilon\mu}{\varepsilon_0\mu_0} \right)^{1/2} \simeq \sqrt{\varepsilon / \varepsilon_0} = \sqrt{\kappa_e} \tag{10.34}$$

where $\kappa_e = 1 + \chi_e$ is the dielectric constant of the medium (Sec. 8.2). We observe that the optical properties of a medium are intimately related to its electromagnetic properties.

## 10.5 Monochromatic Plane E/M Wave in Empty Space

Consider the Maxwell equations in empty space:

$$\begin{aligned} &(a) \quad \vec{\nabla} \cdot \vec{E} = 0 &\qquad &(c) \quad \vec{\nabla} \times \vec{E} = -\frac{\partial \vec{B}}{\partial t} \\ &(b) \quad \vec{\nabla} \cdot \vec{B} = 0 &\qquad &(d) \quad \vec{\nabla} \times \vec{B} = \varepsilon_0 \mu_0 \frac{\partial \vec{E}}{\partial t} \end{aligned} \tag{10.35}$$

As we saw in the previous section, each of the fields $\vec{E}$ and $\vec{B}$ satisfies the wave equation

$$\nabla^2 \vec{A} - \frac{1}{c^2} \frac{\partial^2 \vec{A}}{\partial t^2} = 0 \quad \text{where} \quad c = \frac{1}{\sqrt{\varepsilon_0 \mu_0}} \tag{10.36}$$

In Sec. 10.3 we showed that this equation admits plane-wave solutions of the form

$$\vec{A} = \vec{F}(\vec{k} \cdot \vec{r} - \omega t) \quad \text{with} \quad \omega / k = c \quad (k = |\vec{k}|) \tag{10.37}$$

The simplest such solution is a *monochromatic plane wave* of angular[2] frequency $\omega$, traveling in the direction of the wave vector $\vec{k}$:

---

[2] Sometimes the word *"angular"* will be omitted and $\omega$ will be simply called *"the frequency"*. Literally speaking, of course, this term refers to the quantity $f = \omega / 2\pi$.

$$\boxed{\begin{array}{ll} \vec{E}(\vec{r},t)=\vec{E}_0\, e^{\,i(\vec{k}\cdot\vec{r}-\omega t)} & (a) \\ \vec{B}(\vec{r},t)=\vec{B}_0\, e^{\,i(\vec{k}\cdot\vec{r}-\omega t)} & (b) \end{array}} \qquad (10.38)$$

where the $\vec{E}_0$ and $\vec{B}_0$ are constant *complex* amplitudes and where $\omega/k=c$. (The term *"monochromatic"* is related to the fact that the e/m wave is a harmonic wave that, by definition, contains a single frequency $\omega$.)

The solutions (10.38) seem too simple to be useful in real situations. It can be shown [6,7], however, that *every* solution of the wave equation (10.36) can be expressed as a linear combination of monochromatic solutions with different $\omega$ and $\vec{k}$ such that $\omega/k=c$, where $k=|\vec{k}\,|$. For example, the general solution for $\vec{E}(\vec{r},t)$ can be written as a (triple) integral

$$\vec{E}(\vec{r},t)=\int \vec{E}_0(\vec{k})\, e^{\,i(\vec{k}\cdot\vec{r}-\omega t)}\, d\vec{k} \quad \text{where} \quad d\vec{k}\equiv dk_x dk_y dk_z \qquad (10.39)$$

In the particular case of an e/m wave traveling in the $+x$ direction, we have that $\vec{k}=k\,\hat{u}_x$, $\vec{k}\cdot\vec{r}=kx$ and

$$\vec{E}(x,t)=\int \vec{E}_0(k)\, e^{\,i(kx-\omega t)}\, dk=\int \vec{E}_0(k)\, e^{\,ik(x-ct)}\, dk \equiv \vec{F}(x-ct) \qquad (10.40)$$

As we have already noted, whereas *every* pair $(\vec{E},\vec{B})$ satisfying the Maxwell system (10.35) also satisfies the wave equation (10.36), the converse is *not* true. This means that the solutions (10.38) of the wave equation, in the general form they are written, are not automatically solutions of the Maxwell equations, hence do not *a priori* represent actual e/m waves. We must therefore substitute the general solutions (10.38) into the Maxwell system (10.35) to find the additional constraints that this system imposes on the parameters of the problem.

To this end we need two vector identities. If $\Phi$ is a scalar field and if $\vec{A}$ is a vector field, then

$$\begin{aligned} \vec{\nabla}\cdot(\Phi\vec{A}) &= (\vec{\nabla}\Phi)\cdot\vec{A}+\Phi(\vec{\nabla}\cdot\vec{A}) \\ \vec{\nabla}\times(\Phi\vec{A}) &= (\vec{\nabla}\Phi)\times\vec{A}+\Phi(\vec{\nabla}\times\vec{A}) \end{aligned} \qquad (10.41)$$

In our case we set $\Phi=e^{\,i(\vec{k}\cdot\vec{r}-\omega t)}=e^{\,i\vec{k}\cdot\vec{r}}\,e^{-i\omega t}$ and $\vec{A}=\vec{E}_0$ or $\vec{B}_0$. We also note that

$$\vec{\nabla}\cdot\vec{E}_0=\vec{\nabla}\cdot\vec{B}_0=0\,,\quad \vec{\nabla}\times\vec{E}_0=\vec{\nabla}\times\vec{B}_0=0 \quad \text{(since } \vec{E}_0,\, \vec{B}_0 \text{ are constants) and}$$

$$\vec{\nabla}e^{\,i\vec{k}\cdot\vec{r}}=(\hat{u}_x\frac{\partial}{\partial x}+\hat{u}_y\frac{\partial}{\partial y}+\hat{u}_z\frac{\partial}{\partial z})\,e^{\,i(k_x x+k_y y+k_z z)}=i\,(k_x\hat{u}_x+k_y\hat{u}_y+k_z\hat{u}_z)\,e^{\,i\vec{k}\cdot\vec{r}}=i\,\vec{k}\,e^{\,i\vec{k}\cdot\vec{r}}\,,$$

$$\frac{\partial}{\partial t}e^{-i\omega t}=-i\,\omega\, e^{-i\omega t}$$

Substituting (10.38)(*a*) and (*b*) into (10.35)(*a*) and (*b*), respectively, we have:

$$\begin{aligned}(\vec{E}_0 e^{-i\omega t})\cdot\vec{\nabla} e^{i\vec{k}\cdot\vec{r}} = 0 \;&\Rightarrow\; (\vec{k}\cdot\vec{E}_0)\, e^{i(\vec{k}\cdot\vec{r}-\omega t)} = 0 \\ (\vec{B}_0 e^{-i\omega t})\cdot\vec{\nabla} e^{i\vec{k}\cdot\vec{r}} = 0 \;&\Rightarrow\; (\vec{k}\cdot\vec{B}_0)\, e^{i(\vec{k}\cdot\vec{r}-\omega t)} = 0\end{aligned} \quad\Rightarrow$$

$$\vec{k}\cdot\vec{E}_0 = 0\,, \quad \vec{k}\cdot\vec{B}_0 = 0 \qquad (10.42)$$

Multiplying by $e^{i(\vec{k}\cdot\vec{r}-\omega t)}$ and using (10.38), we find:

$$\vec{k}\cdot\vec{E} = 0\,, \quad \vec{k}\cdot\vec{B} = 0 \qquad (10.43)$$

This indicates that, in a monochromatic plane e/m wave the fields $\vec{E}$ and $\vec{B}$ oscillate perpendicularly to the wave vector $\vec{k}$, thus perpendicularly to the direction of propagation of the wave. In other words,

*the monochromatic plane e/m wave is a transverse wave.*

Also, substituting (10.38)(*a*) and (*b*) into (10.35)(*c*) and (*d*), respectively, we have:

$$\begin{aligned}e^{-i\omega t}(\vec{\nabla} e^{i\vec{k}\cdot\vec{r}})\times\vec{E}_0 = i\,\omega\,\vec{B}_0\, e^{i(\vec{k}\cdot\vec{r}-\omega t)} \;&\Rightarrow\; (\vec{k}\times\vec{E}_0)\, e^{i(\vec{k}\cdot\vec{r}-\omega t)} = \omega\,\vec{B}_0\, e^{i(\vec{k}\cdot\vec{r}-\omega t)} \\ e^{-i\omega t}(\vec{\nabla} e^{i\vec{k}\cdot\vec{r}})\times\vec{B}_0 = -i\,\omega\,\varepsilon_0\mu_0\vec{E}_0\, e^{i(\vec{k}\cdot\vec{r}-\omega t)} \;&\Rightarrow\; (\vec{k}\times\vec{B}_0)\, e^{i(\vec{k}\cdot\vec{r}-\omega t)} = -\frac{\omega}{c^2}\,\vec{E}_0\, e^{i(\vec{k}\cdot\vec{r}-\omega t)}\end{aligned} \quad\Rightarrow$$

$$\vec{k}\times\vec{E}_0 = \omega\,\vec{B}_0\,, \quad \vec{k}\times\vec{B}_0 = -\frac{\omega}{c^2}\,\vec{E}_0 \qquad (10.44)$$

Multiplying by $e^{i(\vec{k}\cdot\vec{r}-\omega t)}$ and using (10.38), we find:

$$\vec{k}\times\vec{E} = \omega\,\vec{B}\,, \quad \vec{k}\times\vec{B} = -\frac{\omega}{c^2}\,\vec{E} \qquad (10.45)$$

We notice that (see Fig. 10.5)

*at each instant, the fields $\vec{E}$ and $\vec{B}$ are mutually perpendicular as well as perpendicular to the direction of propagation $\vec{k}$ of the wave. Specifically, the vectors $(\vec{E},\vec{B},\vec{k})$ define a right-handed rectangular system of axes.*

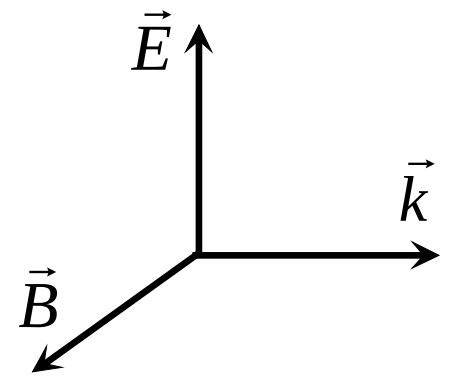

Fig. 10.5. Relative directions of the electromagnetic field and the wave vector. The latter vector determines the direction of propagation of the wave.

Notice also that the direction of propagation $\vec{k}$ is the direction of $\vec{E}\times\vec{B}$. Physically, this can be interpreted as follows: As it travels, the e/m wave transfers the energy of the e/m field in its direction of propagation. On the other hand, the flow of e/m energy is in the direction of the Poynting vector, $\vec{N}=(1/\mu_0)\vec{E}\times\vec{B}$ (cf. Sec. 9.8). Therefore the wave vector $\vec{k}$ must be in the direction of $\vec{E}\times\vec{B}$.

Let us now assume that the complex amplitudes $\vec{E}_0$, $\vec{B}_0$ can be written as

$$\vec{E}_0=\vec{E}_{0,R}\,e^{\,i\alpha}\,,\quad \vec{B}_0=\vec{B}_{0,R}\,e^{\,i\beta}$$

where the $\vec{E}_{0,R}$, $\vec{B}_{0,R}$ are *real* vectors and where $\alpha$, $\beta$ are real numbers. As we can show, relations (10.44) demand that $\alpha=\beta$ and that

$$\vec{k}\times\vec{E}_{0,R}=\omega\,\vec{B}_{0,R}\;,\quad \vec{k}\times\vec{B}_{0,R}=-\frac{\omega}{c^2}\,\vec{E}_{0,R} \tag{10.46}$$

The monochromatic waves (10.38) are now written:

$$\vec{E}=\vec{E}_{0,R}\,e^{\,i(\vec{k}\cdot\vec{r}-\omega t+\alpha)}\;,\quad \vec{B}=\vec{B}_{0,R}\,e^{\,i(\vec{k}\cdot\vec{r}-\omega t+\alpha)} \tag{10.47}$$

Taking the real parts of (10.47), we find the expressions for the *real* fields $\vec{E}$ and $\vec{B}$:

$$\vec{E}=\vec{E}_{0,R}\cos(\vec{k}\cdot\vec{r}-\omega t+\alpha)\;,\quad \vec{B}=\vec{B}_{0,R}\cos(\vec{k}\cdot\vec{r}-\omega t+\alpha) \tag{10.48}$$

We notice that

> *the fields $\vec{E}$ and $\vec{B}$ oscillate in phase, reaching their maximum, minimum and zero values simultaneously.*

Taking the magnitudes of the vector relations (10.46), using the fact that the $\vec{E}_{0,R}$ and $\vec{B}_{0,R}$ are normal to the wave vector $\vec{k}$, and remembering that $\omega/k=c$, we find:

$$E_{0,R}=c\,B_{0,R} \tag{10.49}$$

where $E_{0,R}=|\vec{E}_{0,R}|$ and $B_{0,R}=|\vec{B}_{0,R}|$. Also, taking the magnitudes of Eqs. (10.48) and using (10.49), we find a relation for the *instantaneous values* of the electric and the magnetic field:

$$\boxed{E=c\,B} \tag{10.50}$$

where $E=|\vec{E}|$ and $B=|\vec{B}|$.

Let us summarize:

*In a monochromatic plane e/m wave the fields $\vec{E}$ and $\vec{B}$ are mutually perpendicular as well as perpendicular to the direction of propagation of the wave; they oscillate in phase and their instantaneous values are related by E=cB.*

*Note:* The expressions (10.48) describe a *linearly polarized* e/m wave. We will explain this term in Problem 8, where we will also present plane-wave solutions exhibiting different polarization properties.

## 10.6 Plane E/M Waves of General Form

A plane e/m wave traveling in *empty space* in the direction of the unit vector $\hat{\tau}$ has the general form

$$\vec{A} = \vec{F}(\hat{\tau} \cdot \vec{r} - ct) \quad \text{where} \quad \vec{A} = \vec{E} \text{ or } \vec{B} \tag{10.51}$$

For a wave in the $+x$ direction, we have $\hat{\tau} = \hat{u}_x$, $\hat{\tau} \cdot \vec{r} = \hat{u}_x \cdot \vec{r} = x$, and

$$\vec{A} = \vec{F}(x - ct) \quad \text{where} \quad \vec{A} = \vec{E} \text{ or } \vec{B} \tag{10.52}$$

A wave of this form can be expressed as a linear combination of monochromatic plane waves with different $\omega$ and $k$, as seen in Eq. (10.40).

A plane e/m wave of the form (10.51) has the following physical characteristics (see Problem 5):

1. The fields $\vec{E}$ and $\vec{B}$ are mutually perpendicular as well as perpendicular to the direction of propagation $\hat{\tau}$, so that the $(\vec{E}, \vec{B}, \hat{\tau})$ form a right-handed rectangular system.
2. The magnitudes of the instantaneous values of $\vec{E}$ and $\vec{B}$ are related in *empty* space by $E = cB$.
3. For a wave traveling within a linear *non-conducting* medium, the relation between $E$ and $B$ becomes $E = \upsilon B$, where $\upsilon$ is the speed of propagation of e/m waves in this medium.

As we saw in the previous chapter, the energy density (energy per unit volume) of an e/m field in empty space is given by

$$\begin{gathered} u = \frac{1}{2}\varepsilon_0 E^2 + \frac{1}{2\mu_0} B^2 = u_e + u_m \\ u_e = \frac{1}{2}\varepsilon_0 E^2, \quad u_m = \frac{1}{2\mu_0} B^2 \end{gathered} \tag{10.53}$$

In the case of a plane e/m wave we have that $u_e = u_m$ . Indeed, given that, in vacuum, $E = cB$ and $c^2 = 1/\varepsilon_0 \mu_0$ , we have:

$$u_e = \frac{1}{2}\varepsilon_0 E^2 = \frac{1}{2}\varepsilon_0 c^2 B^2 = \frac{1}{2\mu_0} B^2 = u_m$$

Hence,

$$u = 2u_e = 2u_m = \varepsilon_0 E^2 = \frac{1}{\mu_0} B^2 \tag{10.54}$$

For a non-conducting medium the above relations must be written with $\varepsilon$ and $\mu$ in place of $\varepsilon_0$ and $\mu_0$ , respectively, and with $\upsilon$ in place of $c$. We observe that

> *the electric and the magnetic field contribute equally to the energy of a plane e/m wave.*

The energy per unit time per unit area transferred by a plane e/m wave in empty space is given by the Poynting vector $\vec{N} = (1/\mu_0)\vec{E} \times \vec{B}$ , which is in the direction of propagation $\hat{\tau}$ of the wave. Due to the mutual perpendicularity of $\vec{E}$ and $\vec{B}$ , the magnitude of the Poynting vector is $N=(1/\mu_0)EB$ . By using (10.54) to express $E$ and $B$ as functions of the energy density $u$ , and by taking into account that $c^2 = 1/\varepsilon_0 \mu_0$ , we find that $N = uc$ , so that

$$\vec{N} = u c \hat{\tau} \tag{10.55}$$

The analogous expression for a non-conducting medium is

$$\vec{N} = u \upsilon \hat{\tau} = u \vec{\upsilon} \tag{10.56}$$

where $\upsilon$ is the speed of propagation of e/m waves in that medium. [Compare (10.56) with (6.2), $\vec{J} = \rho_\kappa \vec{\upsilon}$ ; notice the analogies between physical quantities.]

## 10.7 Frequency Dependence of Wave Speed

As we showed in Sec. 10.4, an e/m wave propagates inside a homogeneous linear *non-conducting* medium with speed $\upsilon = 1/\sqrt{\varepsilon\mu}$ . We also showed that the index of refraction of a transparent medium is $n = c/\upsilon \simeq \sqrt{\kappa_e}$ , where $\kappa_e = 1 + \chi_e$ is the dielectric constant of the medium. We thus have:

$$\upsilon = \frac{c}{n} \simeq \frac{c}{\sqrt{\kappa_e}} = \frac{c}{\sqrt{1 + \chi_e}} \tag{10.57}$$

Assume now that a monochromatic e/m wave of angular frequency $\omega = 2\pi f$ enters this medium from empty space. As follows [5,6] from the boundary conditions of Maxwell's equations (Sec. 9.5), *the frequency $\omega$ of a wave is unchanged as the wave passes from one medium to another.*

You may think of it as follows: When an e/m wave of frequency $\omega$ in empty space enters some medium, the electric field of the wave induces forced oscillations to the electrons in the medium, of frequency $\omega$ equal to the frequency of the wave. The oscillating electrons, in turn, emit a secondary e/m wave of the same frequency $\omega$ (this will be explained in Sec. 10.16). Thus the total wave within the medium is the superposition of two waves, a primary and a secondary one, having the same frequency $\omega$. In contrast to the frequency, *the propagation speed $\upsilon$ and the wavelength $\lambda$ do change as the wave passes from one medium to another.*

In empty space all monochromatic waves propagate at the same speed *c*, regardless of their frequency. Inside a material medium, however, *the propagation speed $\upsilon$ may depend on the frequency $\omega$ of the wave*. According to (10.57), this means that the dielectric "constant" $\kappa_e$ of a medium may not be as constant as it sounds but it may depend to some extent on $\omega$. In this case, the index of refraction *n* of the medium will also depend on the frequency of the wave. A medium with these frequency-dependent characteristics is said to be *dispersive*.

White light is a composite e/m wave containing a large number of monochromatic waves of various frequencies. When white light passes through, say, a prism, it undergoes *dispersion*; that is, its various monochromatic components travel at different speeds within the prism, depending on their respective frequencies. Since the index of refraction varies with frequency, the prism deflects differently the various components of light, producing the familiar effect of separation of white light into different colors. We note that the index of refraction increases with frequency. This is why, for example, the red is deflected less than the violet.

If the frequency $\omega$ of the e/m wave passing through the medium is small, the electric field $\vec{E}(x,t)=\vec{E}_0\cos(kx-\omega t)$ changes slowly. In this case the motion of the electrons in the material, as well as the orientation of any molecules or ions possessing permanent electric dipole moment, can easily follow the changes of the field and adjust to them instantly. Thus the polarization vector $\vec{P}$ oscillates in phase with the electric field, with the same frequency $\omega$: $\vec{P}=\vec{P}_0\cos(kx-\omega t)$. (For larger values of $\omega$ the phases of $\vec{P}$ and $\vec{E}$ generally differ.) By Eq. (8.12), $\vec{P}=\varepsilon_0\,\chi_e\vec{E}$. Therefore,

$$\vec{P}_0\cos(kx-\omega t)=\varepsilon_0\,\chi_e\vec{E}_0\cos(kx-\omega t) \;\;\Rightarrow\;\; \vec{P}_0=\varepsilon_0\,\chi_e\vec{E}_0$$

It follows from this that, *at low frequencies*, the electric susceptibility $\chi_e$ of the material is constant, independent of the frequency $\omega$ of the wave. The same is true, therefore, for the dielectric constant $\kappa_e=1+\chi_e$ of the medium.

We conclude that, for small values of $\omega$, the values of the dielectric constant and the index of refraction do not differ appreciably from those when the electric field within the medium is *static*. For this reason the effect of dispersion is not evident at very low frequencies (much lower than the optical frequencies).

## 10.8 Traveling and Standing Waves

A plane wave of the form $\xi(x,t)=f(x \pm \upsilon t)$ or, equivalently, $\xi(x,t)=f(kx \pm \omega t)$ with $\omega/k=\upsilon$, is called a *traveling wave* and is characterized by the presence of terms such as $(x \pm \upsilon t)$ or $(kx \pm \omega t)$ (the lower sign corresponds to wave motion in the positive direction of the $x$-axis, while the upper sign indicates a wave traveling in the negative direction). We define the function

$$\varphi(x,t) = kx \pm \omega t \quad (\text{where } \omega/k=\upsilon) \tag{10.58}$$

so that $\xi(x,t)=f(\varphi)$. The function $\varphi$ is called the *phase* of the wave and its value for given $x$ and $t$ determines the value of $\xi$.

Consider an infinitesimal displacement from $x$ to $x+dx$ on the $x$-axis, where $dx$ is assumed to be positive (negative) for a wave traveling in the positive (negative) direction. At a given time $t$ the phase of the wave at the point $x$ is $\varphi(x,t)$. Assume now that, as the wave advances in the direction from $x$ to $x+dx$, the phase at $x+dx$ at a later time $t+dt$ is the same as the phase at $x$ at time $t$. We then have:

$$\varphi(x+dx,\ t+dt) = \varphi(x,\ t) \ \Rightarrow\ \varphi(x+dx,\ t+dt) - \varphi(x,\ t) \simeq d\varphi = 0 \ \Rightarrow$$

$$(\partial\varphi/\partial x)\,dx + (\partial\varphi/\partial t)\,dt = 0 \ \Rightarrow\ k\,dx \pm \omega\,dt = 0 \ \Rightarrow$$

$$\frac{dx}{dt} = \mp\frac{\omega}{k} = \mp\upsilon \tag{10.59}$$

where (10.58) has been used. Note carefully the correspondence of the signs with the possible directions of propagation of the wave. In particular, the *lower* sign always corresponds to the *positive* direction of propagation.

We conclude that the propagation velocity $\upsilon=\omega/k$ of the wave is actually the *velocity of propagation of phases*. For this reason the quantity $\upsilon$ is called the *phase velocity* (or, more appropriately, the "phase speed") of the traveling wave.

As we know, the general solution of the wave equation for a plane wave traveling along the $x$-axis is of the form

$$\xi(x,t) = f_1(x-\upsilon t) + f_2(x+\upsilon t) = F_1(kx-\omega t) + F_2(kx+\omega t) \quad (\omega/k=\upsilon)$$

In essence, the above expression represents the *superposition* (or *interference*) of two waves traveling in opposite directions along this axis. As an application, let us consider two monochromatic e/m waves of the same frequency $\omega$, traveling in empty space in opposite directions along the $x$-axis (Fig. 10.6). The electric fields of the waves have equal amplitudes $E_0$ and they both oscillate along the $y$-axis.

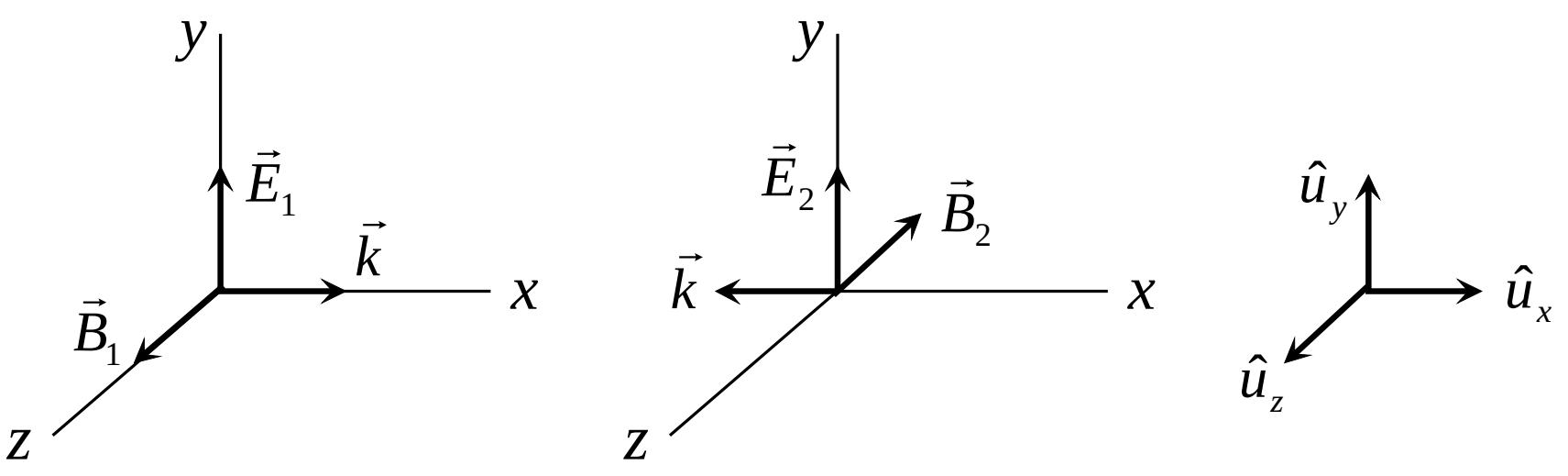

Fig. 10.6. Two monochromatic e/m waves traveling in opposite directions along the *x*-axis.

Wave *1* travels in the $+x$ direction, thus is of the form $F_1(kx-\omega t)$, while wave *2* travels in the $-x$ direction, thus is of the form $F_2(kx+\omega t)$. In every case the wave vector $\vec{k}$, which determines the direction of propagation of the wave, is in the direction of $\vec{E}\times\vec{B}$. Also, the vectors $\vec{E}$ and $\vec{B}$ oscillate in phase and their amplitudes are related by $E_0= cB_0$, where $c=\omega/k$ (see Sec. 10.5). Analytically,

$$\vec{E}_1 = E_0 \cos(kx-\omega t)\,\hat{u}_y \ , \quad \vec{B}_1 = B_0 \cos(kx-\omega t)\,\hat{u}_z = \frac{E_0}{c}\cos(kx-\omega t)\,\hat{u}_z$$

$$\vec{E}_2 = E_0 \cos(kx+\omega t)\,\hat{u}_y \ , \quad \vec{B}_2 = B_0 \cos(kx+\omega t)(-\hat{u}_z) = -\frac{E_0}{c}\cos(kx+\omega t)\,\hat{u}_z$$

Let us now examine the superposition of waves *1* and *2*:

$$\vec{E} = \vec{E}_1 + \vec{E}_2 = E_0\,[\cos(kx-\omega t)+\cos(kx+\omega t)]\,\hat{u}_y$$

$$\vec{B} = \vec{B}_1 + \vec{B}_2 = \frac{E_0}{c}\,[\cos(kx-\omega t)-\cos(kx+\omega t)]\,\hat{u}_z$$

By using the trigonometric identities

$$\cos\alpha + \cos\beta = 2\cos[(\alpha+\beta)/2]\cos[(\alpha-\beta)/2]$$

$$\cos\alpha - \cos\beta = 2\sin[(\alpha+\beta)/2]\sin[(\beta-\alpha)/2]$$

we find:

$$\vec{E}(x,t) = (2E_0 \cos kx \cos\omega t)\,\hat{u}_y$$
$$\vec{B}(x,t) = (\frac{2E_0}{c}\sin kx \sin\omega t)\,\hat{u}_z \qquad (10.60)$$

We note the following:

1. Both $\vec{E}$ and $\vec{B}$ satisfy the wave equation

$$\frac{\partial^2\vec{A}}{\partial x^2} = \frac{k^2}{\omega^2}\frac{\partial^2\vec{A}}{\partial t^2} \quad\Leftrightarrow\quad \frac{\partial^2\vec{A}}{\partial x^2} - \frac{1}{c^2}\frac{\partial^2\vec{A}}{\partial t^2} = 0 \qquad (10.61)$$

(show this!). This was to be expected, given that each of the fields $\vec{E}$ and $\vec{B}$ is the sum of two solutions of the linear differential equation (10.61) for given $\omega$, $k$ and $c=\omega/k$. However, the $\vec{E}$ and $\vec{B}$ in (10.60) do *not* represent traveling waves since they do not contain the characteristic factor $kx \pm \omega t$.

2. Let us write (10.60) in the form

$$\vec{E}(x,t) = [A(x) \cos\omega t]\, \hat{u}_y \quad \text{where} \quad A(x) = 2E_0 \cos kx$$

(similarly for $\vec{B}$). At every point $x$ this expression describes a harmonic oscillation of amplitude equal to $|A(x)|$, which amplitude varies along the $x$-axis according to the factor $\cos kx$. At points where $\cos kx=0$ (called *nodes*) the amplitude $A(x)$, thus also $\vec{E}$ itself, is zero at all times, while at points where $\cos kx= \pm 1$ (called *antinodes*) the field $\vec{E}$ has maximum amplitude. (Analogous observations apply to $\vec{B}$.) On the contrary, in a *traveling* wave of the form $\vec{E} = \vec{E}_0 \cos(kx \pm \omega t)$ the amplitude $E_0$ is constant, the same at all points $x$.

3. At any given time $t$ the phase $\omega t$ of the harmonic oscillation has the same value at all points $x$. This is in contrast to a traveling wave where the phase $kx \pm \omega t$ for a given $t$ changes from one point to another.

Waves such as these in (10.60), with the above-described characteristics, are called *standing waves*. This name is justified as follows: In a *traveling* e/m wave the phases propagate from point to point with speed $c$ (in vacuum). In a *standing* wave, on the other hand, the phase at any time is the same everywhere. It is thus meaningless to speak of "phase propagation" in the latter case. Furthermore, in contrast to a traveling wave, *a standing wave does not transfer energy of the e/m field*. (To be accurate, the *average* value of the Poynting vector over the interval $T=2\pi/\omega$ of a period of oscillation is zero at all points.)

In practice, the standing wave we have described is produced when a traveling e/m wave falls on the surface of a "perfect" conductor. The interference between the incident and the reflected wave then results in the formation of a standing e/m wave.

## 10.9 Propagation of E/M Waves in a Conducting Medium

So far we have studied the propagation of e/m waves inside non-conducting media. For such media it is reasonable to assume that the *free* charge and current densities are zero: $\rho_f = 0$, $\vec{J}_f = 0$. However, this is not true in the case of a *conducting* medium since this medium may possess free charges, such as, e.g., free electrons in a metal, ions in a chemical solution, electrons and ions in a plasma, etc.

We must note at this point that, in the case of a metal the density $\rho_f$ of free charge contains contributions from *both* the free electrons *and* the positive ions, while the free-current density $\vec{J}_f$ is, of course, due to the motion of free electrons. The positive ions count as *free* charges in $\rho_f$ since they do not owe their existence to electric polarization, in contrast to the bound charges of dielectrics.

We consider a homogeneous linear medium that has constant values of $\varepsilon$, $\mu$ and, in addition, possesses a finite conductivity $\sigma$. The fact that, in general, $\varepsilon \neq \varepsilon_0$ suggests that even within a conducting medium there may exist some degree of electric polarization. This polarization may be due to, e.g., the *bound* electrons of the ions in a metal, the molecules of $H_2O$ in saline water (which is conductive due to the ions of $Na^+$ and $Cl^-$ it contains), etc. In a metal the effects of electric polarization due to the bound electrons are not significant; we thus usually set $\varepsilon=\varepsilon_0$ .

Since the considered medium is linear, we write $\vec{D}=\varepsilon\vec{E}$ and $\vec{H}=(1/\mu)\vec{B}$ . The Maxwell equations are written:

$$\vec{\nabla}\cdot\vec{D}=\rho_f \quad \Rightarrow \quad \varepsilon(\vec{\nabla}\cdot\vec{E})=\rho_f$$

$$\vec{\nabla}\cdot\vec{B}=0\,, \qquad \vec{\nabla}\times\vec{E}=-\frac{\partial\vec{B}}{\partial t}$$

$$\vec{\nabla}\times\vec{H}=\vec{J}_f+\frac{\partial\vec{D}}{\partial t} \quad \Rightarrow \quad \frac{1}{\mu}(\vec{\nabla}\times\vec{B})=\vec{J}_f+\varepsilon\frac{\partial\vec{E}}{\partial t}$$

Assuming that the medium is electrically neutral in its interior (see Chap. 9, Prob. 2) we put $\rho_f \simeq 0$. Also, by Ohm's law we have that $\vec{J}_f=\sigma\vec{E}$ . The Maxwell equations then take on the form

$$(a) \quad \vec{\nabla}\cdot\vec{E}=0 \qquad (c) \quad \vec{\nabla}\times\vec{E}=-\frac{\partial\vec{B}}{\partial t}$$

$$(b) \quad \vec{\nabla}\cdot\vec{B}=0 \qquad (d) \quad \vec{\nabla}\times\vec{B}=\mu\sigma\vec{E}+\varepsilon\mu\frac{\partial\vec{E}}{\partial t} \qquad (10.62)$$

Notice that for $\sigma=0$ these equations reduce to the relations (10.24) for a linear insulator.

By taking the *rot* of (*c*) and (*d*), above, and by using the vector identity (10.25), we find that $\vec{E}$ and $\vec{B}$ satisfy the *modified wave equations*

$$\nabla^2\vec{E}-\varepsilon\mu\frac{\partial^2\vec{E}}{\partial t^2}-\mu\sigma\frac{\partial\vec{E}}{\partial t}=0$$

$$\nabla^2\vec{B}-\varepsilon\mu\frac{\partial^2\vec{B}}{\partial t^2}-\mu\sigma\frac{\partial\vec{B}}{\partial t}=0 \qquad (10.63)$$

The system (10.63) admits plane-wave solutions that, for a wave propagating in the $+x$ direction, are of the form

$$\vec{E}(x,t)=\vec{E}_0\,e^{-sx}e^{i(kx-\omega t)}$$

$$\vec{B}(x,t)=\vec{B}_0\,e^{-sx}e^{i(kx-\omega t)} \qquad (10.64)$$

with

$$k = \omega \sqrt{\frac{\varepsilon\mu}{2}} \left[ \sqrt{1 + \left(\frac{\sigma}{\varepsilon\omega}\right)^2} + 1 \right]^{1/2}$$
$$s = \omega \sqrt{\frac{\varepsilon\mu}{2}} \left[ \sqrt{1 + \left(\frac{\sigma}{\varepsilon\omega}\right)^2} - 1 \right]^{1/2} \qquad (10.65)$$

The *propagation speed* of the e/m wave (10.64) inside the conducting medium is

$$\upsilon = \frac{\omega}{k} = \frac{1}{\sqrt{\varepsilon\mu}} \left[ \frac{2}{\sqrt{1 + \left(\frac{\sigma}{\varepsilon\omega}\right)^2} + 1} \right]^{1/2} \qquad (10.66)$$

We note the following:

1. For a *non-conducting* medium, $\sigma$=0 $\Rightarrow$ $s$=0 and $k = \omega\sqrt{\varepsilon\mu}$. Hence, $\upsilon = \omega / k = 1/\sqrt{\varepsilon\mu}$, in agreement with (10.33).

2. By comparing (10.33) and (10.66) we see that, for a given wave frequency $\omega$, the propagation speed inside a conducting medium ($\sigma \neq 0$) is less than the speed inside a non-conducting medium ($\sigma$=0) when the two media have the same values of $\varepsilon$ and $\mu$.

3. The propagation speed in a conducting medium is dependent upon the frequency $\omega$ of the wave. We thus have a *dispersion* effect analogous to that which occurs when light passes through a transparent dielectric (see Sec. 10.7).

4. The presence of the real exponential factor $e^{-sx}$ in the plane waves (10.64) is a consequence of the terms containing first-order time derivatives in the differential equations (10.63). Due to these “damping” terms, the solutions of the system (10.63) represent an *attenuated* e/m wave whose amplitudes decrease exponentially as the wave advances within the conducting medium. We thus conclude that

*an e/m wave suffers attenuation when it passes through a conducting medium; this means that the energy transferred by the wave is absorbed by the medium.*

This effect is due to the dissipation of the energy of the e/m wave into heat. That is, the electric field of the wave induces electric currents inside the conducting medium, which currents then result in resistive heating of the material according to Joule’s law.

We remark that the effect of absorption of e/m waves also occurs in *non-conducting* media, although the associated mechanism there is different. Specifically, in the case of a non-conducting material there is a possibility of *resonance absorption* of the e/m wave by the *bound* electrons of the atoms of the material (we will examine this process in detail in Sec. 10.16).

We define the *skin depth* (or *penetration depth*)

$$\Delta = \frac{1}{s} \tag{10.67}$$

Given that $e^{-sx}$ is a dimensionless quantity, the dimension of $s$ must be that of inverse length; thus $\Delta$ represents length. We notice that when the e/m wave advances a distance $\Delta$ from a point $x$ to the point $x+\Delta$, the above exponential factor becomes $e^{-s(x+\Delta)} = e^{-sx} e^{-s\Delta} = \frac{1}{e} e^{-sx}$. That is, *the skin depth $\Delta$ is the distance the wave must advance in order for its amplitude to reduce to 1/e (approximately 1/3) of its initial value*. For all practical purposes,

> *we consider that the e/m wave penetrates the conducting material up to a depth equal to the skin depth $\Delta$ before it is totally absorbed by the material.*

That is, the wave practically ceases to exist for $x>\Delta$.

It follows from (10.65) that $s$ increases with conductivity $\sigma$. By taking into account (10.67) we conclude that

> *the skin depth $\Delta$ decreases as the conductivity increases.*

In the case of a "perfect" conductor we have $\sigma \approx \infty$, $s \approx \infty$ and $\Delta=0$. Such a conductor therefore does not allow the propagation of e/m waves in its interior. On the other extreme, for a perfect insulator, $\sigma=0$, $s=0$ and $\Delta=\infty$. That is, *theoretically*, the e/m wave should be able to advance up to an unlimited depth within the insulator without suffering absorption. In reality, however, this does *not* happen since, as we have already mentioned, non-conducting materials may also absorb e/m waves.

By definition, $\vec{E}_0$ and $\vec{B}_0$ in (10.64) are complex vectors. Assuming that the e/m wave is linearly polarized (Prob. 8), we set:

$$\vec{E}_0 = \vec{E}_{0,R}\, e^{i\alpha}, \quad \vec{B}_0 = \vec{B}_{0,R}\, e^{i\beta} \tag{10.68}$$

where $\vec{E}_{0,R}$ and $\vec{B}_{0,R}$ are real vectors. Equations (10.64) then take the form

$$\vec{E} = \vec{E}_{0,R}\, e^{-sx} e^{i(kx-\omega t+\alpha)}, \quad \vec{B} = \vec{B}_{0,R}\, e^{-sx} e^{i(kx-\omega t+\beta)} \tag{10.69}$$

In contrast to non-conducting media (Sec. 10.5), here $\alpha \neq \beta$; specifically, $\alpha<\beta$. That is,

> *the fields $\vec{E}$ and $\vec{B}$ do not oscillate in phase inside the conductor but the magnetic field $\vec{B}$ lags behind the electric field $\vec{E}$*

(the phase difference tends to zero in the limit $\sigma \to 0$). Is "*lags behind*" inconsistent with the fact that $\alpha<\beta$? To understand the situation, let us write $\varphi_E(x,t)=kx-\omega t+\alpha$, $\varphi_B(x,t)=kx-\omega t+\beta$. We consider a fixed point $x=x_0$ and we assume that, at this point, the phase $\varphi_E$ at time $t_1$ equals the phase $\varphi_B$ at time $t_2$ . That is, $\varphi_E(x_0, t_1) = \varphi_B(x_0, t_2) \Rightarrow \omega(t_2 - t_1) = \beta - \alpha > 0 \Rightarrow t_1 < t_2$ . This means that the electric field acquires a given phase *before* the magnetic field acquires the same phase. The term *"lags behind"*, therefore, refers to the time order of events, *not* to the sizes of the phases of the two fields!

As in the case of a non-conducting medium (Sec. 10.5), inside a conducting medium

*the fields $\vec{E}$ and $\vec{B}$ of a plane e/m wave are mutually perpendicular as well as perpendicular to the direction of propagation of the wave.*

However,

*the instantaneous values of the electric and the magnetic field are no longer related by $E = \upsilon B$, except in the limit $\sigma \to 0$.*

(For a proof of the above-stated properties of a plane wave in a conducting medium, see Problem 7.)

*Plane e/m wave of a relatively low frequency $\omega$ inside a "good" conductor*

This case is expressed mathematically by the condition

$$\omega << \frac{\sigma}{\varepsilon} \quad \Leftrightarrow \quad \frac{\sigma}{\varepsilon\omega} >> 1 \qquad (10.70)$$

We can then make the approximation $\sqrt{1 + \left(\frac{\sigma}{\varepsilon\omega}\right)^2} \simeq 1 + \frac{\sigma}{\varepsilon\omega}$ , so that by (10.65) we find: $s \simeq \sqrt{\omega\mu\sigma/2} \quad \Rightarrow$

$$\boxed{\varDelta = \frac{1}{s} \simeq \sqrt{\frac{2}{\omega\mu\sigma}}} \qquad (10.71)$$

We observe that the skin depth $\varDelta$ decreases when the frequency $\omega$ of the wave increases. This means that

*when e/m waves of low frequencies enter a good conductor, the waves with the lower frequencies penetrate deeper into the conductor.*

Relation (10.71) is useful for determining the appropriate frequency $f$ in order to achieve penetration of the e/m wave at a given depth $\varDelta$ inside the material:

$$f = \frac{\omega}{2\pi} \simeq \frac{1}{\pi\mu\sigma\varDelta^2} \qquad (10.72)$$

Careful, however: When using (10.72) we must check that the value of $f$ we found does indeed conform to the condition (10.70) for $\omega$!

*Application:* A submarine is submerged at a depth $h=10\,m$. Is it possible to communicate with it by sending an e/m signal in the range of radio frequencies? The following data is given for seawater:

$$\mu \simeq \mu_0 = 4\pi \times 10^{-7} N/A^2 , \quad \varepsilon \simeq 70\varepsilon_0 \simeq 6\times 10^{-10} C^2 / N\cdot m^2 , \quad \sigma \simeq 5\,(\Omega\cdot m)^{-1} .$$

*Solution:* The conductivity of seawater is mostly due to $Na^+$ and $Cl^-$ ions. By the given data we find $\sigma/\varepsilon \simeq 10^{10} Hz$, which is much higher than radio frequencies. In order for the submarine to receive an e/m signal, the submarine must be at a depth that does not exceed the skin depth $\Delta$ for the frequency of that signal: $h \le \Delta$. According to (10.72),

$$\Delta^2 = \frac{1}{\pi\mu\sigma f} \ge h^2 \quad \Rightarrow \quad f \le \frac{1}{\pi\mu\sigma h^2} \simeq 500\,Hz$$

which is indeed in the region of radio frequencies (see Sec. 10.15). The result is admissible since condition (10.70) is satisfied. Let us now recall that the frequency of a wave is unchanged when the wave passes from one medium to another, in contrast to the wavelength and the velocity of propagation. Therefore the frequency of the signal we must emit *from the air* is equal to the frequency we wish to be received by the submarine, that is, 500 *Hz*. The wavelength of our signal *in the air* for that frequency is $\lambda = c/f = 600\,km$! (We take the propagation speed of e/m waves in the air to be approximately equal to the speed in empty space, i.e., $c=3\times10^8$ *m/s*.) The emission of such a signal would require an antenna of gigantic size (a conventional antenna would deliver a very weak signal to the submarine).

In conclusion,

> *the conductivity of seawater makes it very difficult to communicate with submarines by means of e/m signals.*

For similar reasons,

> *the radar is not a useful instrument for submarine detection*

(the *sonar* is used instead for this task).

## 10.10 Reflection of an E/M Wave on the Surface of a Conductor

When an e/m wave falls on the surface of a conductor, a part of its energy goes through the surface and enters the conductor where it is finally *absorbed*, while another part of the wave energy is *reflected* back to the medium (e.g., air) of the incident wave. *A good conductor is also a good reflector*, in the sense that the proportion of energy reflected increases with the conductivity of the material [5,6]. Ideally, on the surface of a "perfect" conductor ($\sigma = \infty$) the incident e/m wave will suffer *total reflec-* *tion*. For this reason excellent conductors such as silver are used to make high-quality mirrors.

On the other hand, as we saw previously, *a good conductor is also a good absorber of e/m waves*, in the sense that, for a given wave frequency $\omega$, the skin depth $\Delta$ decreases as the conductivity of the medium increases.

In conclusion, a good conductor exhibits the following characteristic properties:

- The major part of the power carried by an incident e/m wave is reflected on the surface of the material.
- The small part of the wave that crosses the surface of the conductor (*transmitted wave*) dies out quickly, being absorbed by the material.

We notice that

*a good reflector of e/m waves is also a good absorber of these waves*.

As mentioned above, both these properties increase with conductivity. It should be noted, however, that these conclusions are valid on the condition that the frequency $\omega$ of the incident e/m wave is *relatively small*, specifically, smaller than a limit frequency characteristic of the material, called *plasma frequency* and denoted $\omega_p$ . Our conclusions will have to be revised for e/m waves with frequencies $\omega>\omega_p$ , as we will learn in Sec. 10.17.

## 10.11 Electromagnetic Radiation

By *electromagnetic (e/m) radiation* we mean the *transport of energy by means of e/m waves*. Accordingly, any physical system emitting energy in the form of e/m waves is said to *emit e/m radiation* or, simply, to *radiate*. Examples of radiating systems include atoms, molecules, nuclei, hot bodies, radio and TV broadcasting antennas, etc.

As we know, the sources of an electric field are electric charges (regardless of their state of motion) while the sources of a magnetic field are *moving* charges and electric currents. But, what are the sources of e/m radiation? Let us see how far our logic can take us.

A *static* charge and current distribution ($\partial\rho/\partial t=0$, $\partial\vec{J}/\partial t=0$), such as a system of stationary charges and time-independent currents localized in some finite region of space, *cannot* generate e/m radiation. Indeed, a distribution of this sort produces a *static* e/m field ($\partial\vec{E}/\partial t=0$, $\partial\vec{B}/\partial t=0$) and, as we know, a field that does not change with time does not exhibit wave behavior. It is necessary, therefore, that the densities $\rho(\vec{r},t)$ and $\vec{J}(\vec{r},t)$ be *time-dependent* in order for the produced e/m field to be time-dependent as well. The *disturbance* (time change) of this field then propagates in space in the form of an e/m wave.

A more careful analysis by using the Maxwell equations [5,7] reveals that *e/m radiation can be produced*

- *by accelerated electric charges, or*
- *by time-varying electric currents*

(a current may vary with time in magnitude, direction, or both).

From the moment they are produced, e/m waves travel to "infinity" with speed *c* (in vacuum) transporting part of the energy of the source that produced them.

*The hallmark of e/m radiation is precisely this flow of energy away from the source; an energy that is lost forever, never to be regained by the source.*

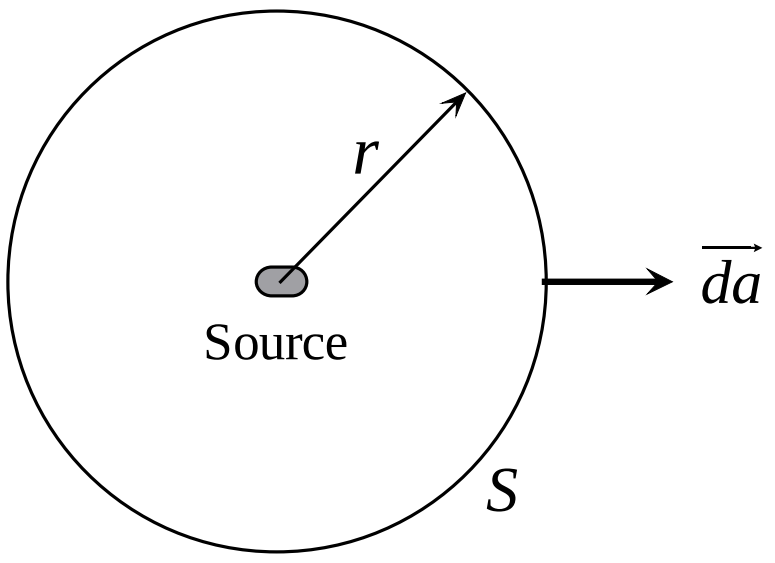

Fig. 10.7. A source of e/m radiation at the center of a spherical surface *S* of radius *r*.

We assume that the source of e/m radiation is localized within a small region of space. Imagine now a huge spherical surface *S* of radius *r*, centered at the location of the source (Fig. 10.7). Let $U(t)$ be the total e/m energy in the interior of *S* at time *t*. The total power passing outward through *S* at this time is given by Eq. (9.45):

$$P(r,t) = -\frac{dU(t)}{dt} = \oint_S \vec{N} \cdot \overrightarrow{da} = \frac{1}{\mu_0} \oint_S (\vec{E} \times \vec{B}) \cdot \overrightarrow{da} \qquad (10.73)$$

where $\vec{N}$ is the Poynting vector. The radiated power at time *t* is

$$P_{rad}(t) = \lim_{r \to \infty} P(r,t) \qquad (10.74)$$

This quantity represents energy per unit time radiated by the source and finally reaching infinity. If $P_{rad}=0$, the system does *not* radiate.

By using the Coulomb and Biot-Savart laws it can be shown [5] that, for *static* e/m fields, $P_{rad}=0$. Thus, as mentioned previously, *static (i.e., time-independent) sources do not radiate*; the energy of their (static) e/m field is confined within some finite volume and never reaches infinity. To produce radiation we therefore need *time-changing* sources and, correspondingly, time-dependent e/m fields, so that $P_{rad} \neq 0$.

As *intensity* of e/m radiation at a point $\Sigma$ of space we define the *average* power transported by the e/m wave per unit surface normal to the direction of propagation at $\Sigma$. The intensity *I* is equal to the average value $<N>$ of the magnitude of the Poynting vector. In the case of a harmonic wave of frequency $\omega=2\pi/T$, the average value of *N* is evaluated over the time interval of a period *T* (or arbitrarily many periods):

$$< N > = \frac{1}{T} \int_0^T N(\vec{r},t)\, dt \qquad (10.75)$$

where $\vec{r}$ is the position vector of point $\Sigma$. But, by (10.55), (10.54) and (10.48) we have that

$$N(\vec{r},t) = u\,c = \varepsilon_0\, c\, E^2 = \varepsilon_0\, c\, E_0^2 \cos^2(\vec{k}\cdot\vec{r} - \omega t + \alpha) \qquad (10.76)$$

The intensity $I$ is, therefore,

$$I = < N > = \varepsilon_0 c E_0^2 \frac{1}{T} \int_0^T \cos^2(\vec{k}\cdot\vec{r} - \omega t + \alpha)\, dt$$

Taking into account that $\omega = 2\pi/T$, we can show that

$$\frac{1}{T} \int_0^T \cos^2(\vec{k}\cdot\vec{r} - \omega t + \alpha)\, dt = \frac{1}{2\pi} \int_0^{2\pi} \cos^2(\varphi - \vec{k}\cdot\vec{r} - \alpha)\, d\varphi = \frac{1}{2}$$

Thus in the case of a harmonic e/m wave the intensity of radiation is

$$\boxed{I = < N > = \frac{1}{2}\, \varepsilon_0 c E_0^2} \qquad (10.77)$$

In the following sections we will see some examples of physical systems that radiate. Due to the mathematical complexity of the subject, we will confine ourselves to the main physical conclusions avoiding mathematical details as much as possible.

## 10.12 Radiation from an Accelerating Point Charge

According to a stationary inertial observer, a moving point charge $q$ produces both an electric and a magnetic field. The charge transfers along with it the energy of its own e/m field. If this energy is constant in time, the charge obviously does not radiate. If, however, the total energy of the e/m field of the charge decreases with time, the loss of energy is due to e/m radiation. Let us examine the situation analytically:

If the charge moves with constant velocity (in magnitude *and* direction) the total energy of its e/m field is constant and is simply being transferred in the direction of motion of $q$. This happens because, as can be proven [5], the integral of the Poynting vector over a spherical surface $S$ of radius $r$, centered at $q$, tends to zero in the limit $r \to \infty$. Hence $P_{rad}=0$, as explained in Sec. 10.11. This means that the energy of the e/m field remains "attached" to $q$ rather than "escaping" to infinity in the form of e/m radiation. We conclude that

*a point charge in uniform rectilinear motion does not emit e/m radiation.*

When the charge *accelerates*, part of the energy of its e/m field is detached, in a sense, and flies away to infinity in the form of an e/m wave traveling at the speed of light [5,7,8]. In this case, $P_{rad} \neq 0$. This is precisely the main characteristic of e/m radiation. Therefore,

*an accelerating point charge emits e/m radiation.*

We consider now a point charge $q$ moving with acceleration $\vec{a}$. We will confine ourselves to the case where $q$ is either *instantaneously* at rest ($\vec{\upsilon} = 0$) or is moving at much lower speed than the speed of light ($\upsilon << c$). Let $\Sigma$ be a point of observation a distance $r$ from the instantaneous position of $q$ (Fig. 10.8). We call $\theta$ the angle between the axis $q\Sigma$ and the acceleration of $q$.

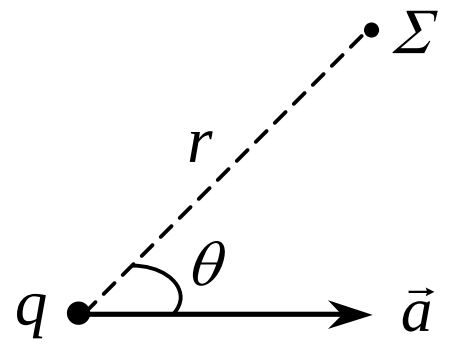

Fig. 10.8. An accelerating charge $q$ a distance $r$ from the observation point $\Sigma$.

The *intensity* of the radiation emitted by $q$ due to its acceleration is, at point $\Sigma$,

$$I(r,\theta) = (\text{constant}) \cdot q^2 a^2 \frac{\sin^2\theta}{r^2} \qquad (a = |\vec{a}|) \tag{10.78}$$

The *total power emitted* is given by *Larmor's formula*:

$$\boxed{P = \frac{q^2 a^2}{6\pi\varepsilon_0 c^3}} \tag{10.79}$$

We note the following:

1. To maintain a charge in accelerated motion over a long period we must constantly supply it with energy to make up for radiation losses. These losses, however, are significant only for very large values of the acceleration $a$, while for relatively small values of $a$ the energy losses are minimal and are often ignored in applications.

2. A charged particle moving along a circular path executes accelerated motion *even if it moves with constant speed*, because of the constant change of its direction of motion (the particle has at least a centripetal acceleration). Thus this particle emits e/m radiation. For example, when an ion is accelerated in a circular accelerator (e.g., a cyclotron) part of its energy is lost in the form of radiation, which is called *synchrotron radiation* and becomes more noticeable as the mass of the ion decreases.

3. In connection with our previous comment we must emphasize the following: The case of a *point* charge in uniform circular motion is different from that of a steady circular *current*. In the latter case the produced e/m field is simply a *static* magnetic field (the electric field outside the current-carrying wire is practically zero since the wire is overall electrically neutral). Thus *a steady circular current does not produce e/m radiation*. On the contrary, the circularly moving point charge produces a *time-varying* e/m field due to the fact that its position and velocity are constantly changing. As we have said, the time dependence of the e/m field is a necessary condition for radiation. We stress again that

> *the e/m radiation is produced either by isolated electric charges in accelerated motion or by continuous currents whose magnitude or/and direction vary with time. Systems of charges or currents that are sources of static e/m fields do not radiate.*

4. The Larmor formula (10.79) is valid regardless of whether the charge $q$ is "accelerating" (its speed $\upsilon$ increases) or "decelerating" ($\upsilon$ decreases). Indeed, the radiated power $P$ depends only on the magnitude $a$ of the acceleration of $q$ (note also that $P$ is independent of the sign of $q$). In particular, a *decelerating* charged particle emits e/m radiation that is called *deceleration radiation* or *bremsstrahlung*. Such a radiation is emitted, for example, when a fast-moving electron falls on a target, losing part or all of its kinetic energy. By conservation of energy, the kinetic energy that is lost is converted into e/m radiation. This is the principal mechanism by which radiation is produced in X-ray tubes.

## 10.13 Electric Dipole Radiation

An example of a radiation-emitting system of charges is an *oscillating electric dipole*. Such an emitting device is, e.g., the linear antenna of a radio station.

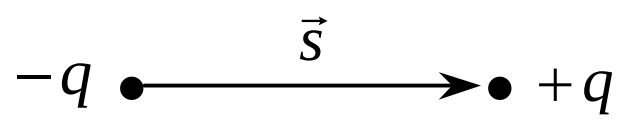

Fig. 10.9. An electric dipole.

Consider an electric dipole of dipole moment $\vec{p} = q\vec{s}$ (Fig. 10.9). Assume that the charge oscillates from one end of the dipole to the other so that, at time $t$, the charge on the right end (as seen in the figure) is

$$q(t) = q_0 \cos \omega t \quad (q_0 > 0) \tag{10.80}$$

We assume that $\vec{s}$ is a *fixed* vector independent of the sign of the instantaneous value $q(t)$ of the charge. This means that, at a given time $t$, the dipole moment $\vec{p} = q\vec{s}$, which is always directed *from the negative to the positive end* of the dipole, is in the direction of $\vec{s}$ or in the opposite direction, depending on whether $q(t)>0$ or $q(t)<0$, respectively. This dipole moment oscillates with frequency $\omega$, changing direction periodically. If $\hat{u} = \vec{s}/s$ $(s = |\vec{s}\,|)$ is the unit vector in the direction of $\vec{s}$ (Fig. 10.10) then

$$\vec{p}(t) = q(t)\,\vec{s} = (q_0\, s \cos\omega t)\,\hat{u} = (p_0 \cos\omega t)\,\hat{u} \quad \text{where} \quad p_0 = q_0\, s \tag{10.81}$$

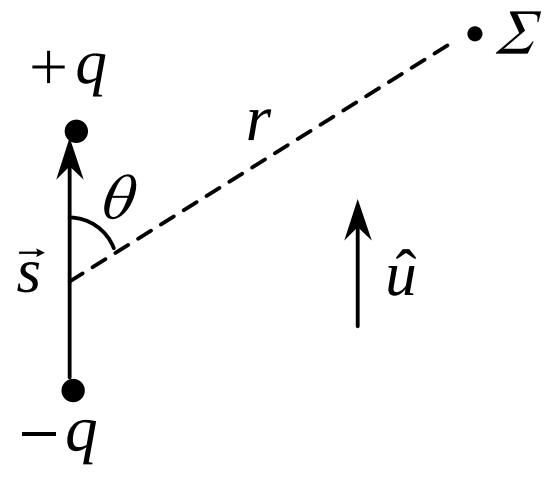

Fig. 10.10. An electric dipole a distance *r* from the observation point *Σ*.

This system emits e/m radiation [5] in the form of e/m waves of frequency $\omega=2\pi f$. The radiation advances radially in all directions at the speed of light. At points far away from the dipole, the fields $\vec{E}$ and $\vec{B}$ oscillate in phase being mutually perpendicular as well as perpendicular to the local direction of propagation of the wave, while their instantaneous values are related by $E=cB$. Moreover, the amplitudes of $\vec{E}$ and $\vec{B}$ decrease by a factor $1/r$ (where $r$ is the distance of the point of observation $\Sigma$ from the dipole) as the wave moves away from the source (dipole). Also, the *wave surfaces* (i.e., the surfaces of constant phase) are spheres centered at the dipole. Waves of this kind are called *spherical waves*.

The *intensity* of the e/m radiation at the observation point $\Sigma$ is

$$I(r,\theta) = (\text{constant}) \cdot p_0^{\,2}\,\omega^4\,\frac{\sin^2\theta}{r^2} \tag{10.82}$$

while the *total power radiated* is

$$P = (\text{constant}) \cdot p_0^{\,2}\,\omega^4 \tag{10.83}$$

An alternative form of an oscillating electric dipole is that in which the end charges are constant in time, equal to $\pm q_0$, but the vector $\vec{s}$ varies harmonically with time:

$$\vec{s}(t) = \vec{s}_0 \cos\omega t = (s_0 \cos\omega t)\,\hat{u}$$

Then,

$$\vec{p}(t) = q_0\,\vec{s}(t) = (q_0\, s_0 \cos\omega t)\,\hat{u} = (p_0 \cos\omega t)\,\hat{u}$$

as before. (This time the vector $\vec{s}$ is not constant; it is always directed, however, from $-q_0$ to $+q_0$, which is also the direction of $\vec{p}$. The charges $+q_0$ and $-q_0$ exchange positions at the ends of the dipole in accordance with the direction of $\vec{s}$.) This oscillator model is used to describe the absorption and emission of radiation by the atoms or molecules of a dielectric (see Sec. 10.16).

To better understand the situation with regard to a time-varying electric dipole, in general, let us see a simple example. Imagine that we connect two small metal spheres with the poles of a source of voltage $V$, as seen in Fig. 10.11. An observer is stationed at a point $\Sigma$ a distance $r$ from the system. Suppose that, initially, the voltage of the source is constant in time ($V=V_0=$*const.*). Then the charges $+q$ and $-q$ on the two spheres are constant, the electric field $\vec{E}$ over all space is static, the magnetic filed $\vec{B}$ is zero, and no change over time (i.e., no disturbance) is recorded by the observer at $\Sigma$.

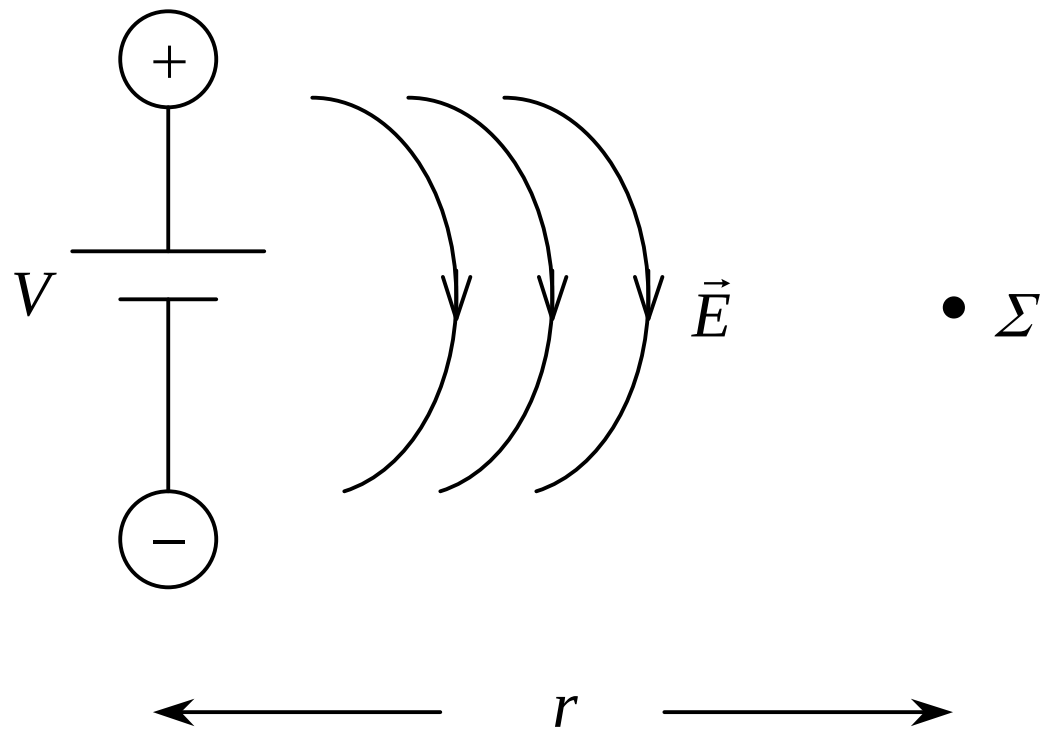

Fig. 10.11. An example of an oscillating electric dipole; the observation point $\Sigma$ is a distance $r$ from the dipole.

Assume now that the voltage $V$ of the source is allowed to vary with time. Any change of $V$ causes a change of the value of the charges $\pm q$ on the spheres and, accordingly, a change of the $\vec{E}$-field in the vicinity of the dipole. The observer at $\Sigma$, however, does not perceive this change immediately but only after a time interval $\Delta t=r/c$ needed in order for the "information" to travel a distance $r$ from the system. Now, the time-change of $q$ implies a flow of charge along the dipole, hence the emergence of a magnetic field $\vec{B}$ near the dipole. Again, the observer at $\Sigma$ will perceive this field after a time interval $\Delta t=r/c$ from the moment of its creation. Notice that the time-change of the electric field is accompanied by the presence of a magnetic field, in accordance with the Ampère-Maxwell law.

To create an oscillating dipole, we must make the voltage $V$ of the source vary harmonically with time ($V=V_0 \cos \omega t$). Notice that, since the charges are oscillating between the two spheres, they are constantly *accelerating*. Moreover, the electric current along the dipole is continuously changing. As we know, these physical conditions are associated with the production of e/m radiation.

## 10.14 Magnetic Dipole Radiation

A special case of a time-dependent current emitting e/m radiation is an *oscillating magnetic dipole*. Consider a circular current loop of radius $R$ (hence of area $a=\pi R^2$) on the $xy$-plane (Fig. 10.12) carrying a harmonically varying current of the form

$$I(t) = I_0 \cos \omega t \quad (I_0 > 0) \qquad (10.84)$$

By convention, a current on the $xy$-plane will be considered positive (negative) if it is directed counterclockwise (clockwise). Thus, at time $t$=0 the initial current $I_0$ is counterclockwise; the direction of the current $I$, however, changes periodically with time.

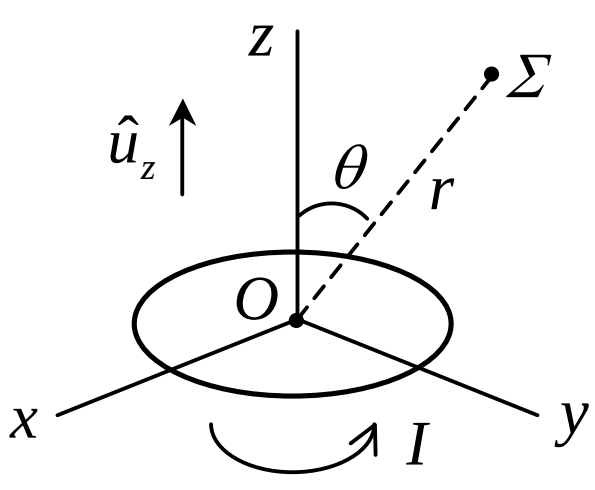

Fig. 10.12. A magnetic dipole a distance *r* from the observation point *Σ*.

The magnetic dipole moment of the loop at time *t* is

$$\vec{m}(t) = I(t)\,\vec{a} = I(t)\,a\,\hat{u}_z = (I_0 \pi R^2 \cos\omega t)\,\hat{u}_z \;\Rightarrow$$

$$\vec{m}(t) = (m_0 \cos\omega t)\,\hat{u}_z \quad \text{where} \quad m_0 = I_0\, a = I_0 \pi R^2 \qquad (10.85)$$

Notice that $\vec{m}$ is in the $+z$ direction when $I>0$ (counterclockwise current) while it is in the $-z$ direction when $I<0$ (clockwise current). Therefore the magnetic dipole moment changes direction periodically in accordance with the direction of *I*.

The above system, like the ones examined in the previous section, emits spherical e/m waves [5] of frequency $\omega=2\pi f$. The *intensity* of the e/m radiation at the observation point *Σ* is

$$I(r,\theta) = (\text{constant}) \cdot m_0^{\,2}\, \omega^4\, \frac{\sin^2\theta}{r^2} \qquad (10.86)$$

while the *total power radiated* is

$$P = (\text{constant}) \cdot m_0^{\,2}\, \omega^4 \qquad (10.87)$$

Notice the similarities of these equations with those for electric dipole radiation.

## 10.15 The Spectrum of E/M Radiation

The e/m radiation spans a very broad range of frequencies that is called the *spectrum of e/m radiation* (or *e/m spectrum*, for short). For practical purposes the e/m spectrum is partitioned into certain regions, depending on the frequency and the way of production of the e/m waves. Ordered according to increasing frequency *f*, these regions are named as follows:

1. *Radio-frequency waves* (a few *Hz* up to $10^9$ *Hz*). They are produced by oscillating electric circuits and antennas and are used in radio and TV broadcasting systems. They are subdivided into frequency bands, a partial list of which is the following: *medium-frequency (MF) waves* (medium-wave AM radio broadcasting; 500-1600 *kHz*), *high-frequency (HF) waves* (shortwave AM radio broadcasting; 6-25 *MHz*) and *very-high-frequency (VHF) waves* (FM radio and TV broadcasting; 30-1000 *MHz*).

2. *Microwaves* (UHF TV broadcasting, mobile phones, satellite communications, etc; $10^9$ - $3\times10^{11}$ *Hz*); also produced by oscillating electric circuits and antennas.

3. *Infrared spectrum* ($3\times10^{11}$ - $4\times10^{14}$ *Hz*); emitted by molecules and hot bodies.

4. *Visible spectrum* or *light* ($4\times10^{14}$ - $8\times10^{14}$ *Hz*); emitted by excited atoms and molecules. In order of increasing frequency: red, orange, yellow, green, blue, violet.

5. *Ultraviolet rays* ($8\times10^{14}$ - $3\times10^{17}$ *Hz*); also emitted by excited atoms and molecules.

6. *X-rays* ($3\times10^{17}$ - $5\times10^{19}$ *Hz*); emitted by excited atoms or produced by the *bremsstrahlung* effect (see Sec. 10.12).

7. *γ-rays* ($3\times10^{18}$ - $3\times10^{22}$ *Hz*); emitted by excited nuclei.

We note that the frequency ranges in the above classification do not have very sharp boundaries, a fact which allows for some degree of overlapping between adjacent regions of the spectrum.

## 10.16 Absorption of E/M Radiation by Non-Conducting Media

When an e/m wave falls on an atom (or a molecule) of a dielectric medium, the fields $\vec{E}$ and $\vec{B}$ in the wave interact with the *bound* electrons of the atom (or molecule). The action of $\vec{B}$ is negligible for small values of the speed $\upsilon$ of an electron, as seen by comparing the magnetic force with the electric force exerted on the electron, taking into account that $E \simeq cB$:

$$F_{magn} \simeq q_e \upsilon B \simeq \frac{\upsilon}{c}(q_e E) = \frac{\upsilon}{c} F_{el} \quad \Rightarrow \quad \frac{F_{magn}}{F_{el}} \simeq \frac{\upsilon}{c} \simeq 0 \quad \text{when} \quad \upsilon << c$$

(where $q_e$ is the absolute value of the charge of the electron). Notice that the two forces would become comparable in the limit $\upsilon \to c$, that is, in a high-energy process (see also Chap. 9, Prob. 6).

Let $\vec{E} = \vec{E}_0 \cos(kx - \omega t)$ be the electric field in a monochromatic wave that falls on a bound electron of an atom of the dielectric. Assuming that $x$ varies very little in the limited space where the electron moves, we set $x$=0, so that $\vec{E} \simeq \vec{E}_0 \cos\omega t$. Under the action of the electric force $\vec{F} = -q_e \vec{E} = -q_e \vec{E}_0 \cos\omega t$, the electron is set into *forced oscillations* of frequency $\omega$ equal to that of the incident wave. The rate at which the electron absorbs energy from the wave (that is, the average power transferred from the wave to the oscillator) acquires a maximum value when *energy resonance* occurs, which takes place when the frequency $\omega=2\pi f$ of the wave equals the natural frequency of oscillation of the electron. The latter frequency can be any frequency in the *absorption spectrum* $\omega_1$, $\omega_2$, $\omega_3$,..., of the atom (or molecule) to which the electron belongs.

The atom now behaves like an *oscillating electric dipole* which is forced to oscillate with frequency $\omega$. (The charges $\pm q_e$ at the ends of the dipole are constant but their

distance – i.e., the distance between the oscillating electron and the positive ion constituting the remaining part of the atom – changes harmonically.) The atom will thus emit e/m radiation of the same frequency $\omega$ as that of the incident wave. The radiated energy is energy that was *absorbed* from the incident e/m wave by one of the bound electrons of the atom and is now reemitted in random directions. This energy, therefore, is lost forever for the incident wave. The above-described process is called *scattering* of e/m radiation and its effect is to reduce the energy content (hence, the intensity) of the incident e/m wave. Notice that this mechanism of energy absorption by *bound* electrons is different from that due to *free* charges in a conducting medium (see Sec. 10.9).

The absorption effect becomes more pronounced as the frequency $\omega$ of the incident wave comes close to any one of the frequencies $\omega_1$ , $\omega_2$ , $\omega_3$ ,…, of the absorption spectrum of the dielectric. If the absorption is minimal for some frequency $\omega$, we say that the material is *transparent* to this frequency. On the other hand, the material is *opaque* for frequencies near the resonance frequencies $\omega_1$ , $\omega_2$ ,…, of its atoms. For example, the lowest frequency $\omega_1$ for glass is in the ultraviolet region of the e/m spectrum. Hence glass is transparent to visible light but relatively opaque to ultraviolet rays.

The observed color of the sky [5,6,9] is the result of scattering of e/m radiation. The solar radiation that falls on the atoms of the atmosphere of the Earth spans a wide range of frequencies (white light). However, the energy absorbed and reemitted (scattered) by these atoms corresponds predominantly to the higher frequencies. Thus scattering is more intense in the blue than in the red. So, when we look at the sky in daytime, what we observe is the scattered blue light – unless, of course, we look directly at the Sun, in which case we see the yellowish color of the latter (which is what remains from white light after a relatively small portion of the blue has been removed). On the other hand, at sunrise and at sunset the solar rays are coming to us from the depths of the horizon, having traversed much greater distances in comparison to those in midday. As a result, a large part of the blue component of solar light has been removed because of scattering in the atmosphere, which explains why the Sun looks reddish at these hours.

## 10.17 Plasma Frequency of a Conducting Medium

The physical phenomena following the incidence of an e/m wave onto a conducting medium were studied in Sections 10.9 and 10.10 by using the Maxwell equations and Ohm's law, $\vec{J}_f = \sigma \vec{E}$ . In our analysis we silently assumed that the conductivity $\sigma$ of the medium is a constant quantity. This is strictly true, however, for relatively low frequencies $\omega$ of the incident wave. For such frequencies we can make the approximation $\sigma \approx \sigma_0$ , where $\sigma_0$ is the *static* value of the conductivity (the one used in Ohm's law when the $\vec{E}$ -field is static). For a good conductor at room temperature, the conductivity can be considered constant when $\omega$ does not exceed microwave frequencies.

So, when e/m radiation of relatively low frequency $\omega$ falls on the surface of a conductor, the previously described phenomena are observed; that is, the major part of the energy of the incident wave is *reflected* on the surface of the conductor, while the small part that manages to enter the material is quickly *absorbed* by it. Moreover, as we showed in Sec. 10.9, an increase of $\omega$ reduces the skin depth $\Delta$, thus the penetrating ability of the e/m wave.

However, as the frequency $\omega$ of the incident wave increases, the conductivity $\sigma$ begins to depend on $\omega$ in a way similar to the dependence of the dielectric "constant" of an insulator on $\omega$ (Sec. 10.7). When $\omega$ exceeds a limit value $\omega_p$ , characteristic of the conducting material, the situation changes dramatically [9-12]: the skin depth ceases to decrease with $\omega$ and increases indefinitely. The material then allows the radiation to go through it without any appreciable absorption and with only minor reflection on its surface. The frequency $\omega_p$ is called the *plasma frequency* of the material (strictly speaking, the term refers to the quantity $f_p=\omega_p/2\pi$). We conclude that

> *a conducting material is "transparent" to e/m radiation of frequency that exceeds the plasma frequency* $f_p$ *of the material*.

For most metals, $f_p \approx 3\times10^{16}\ Hz$. Thus, although metals are opaque to visible light, they are transparent to ultraviolet radiation above $f_p$ . Metals are not the only conducting substances, however. Of particular interest is a state of matter called *plasma*. By this term we mean a gaseous substance in which a great number of atoms or molecules have been ionized. Thus the plasma is a mixture of free electrons and positive ions. Its conductivity is mostly due to the free electrons, given that the mobility of the positive ions is relatively small because of their much larger mass.

A physical example of plasma is the *ionosphere*, a highly ionized layer in the atmosphere extending from about 60 *km* up to 200-300 *km* above the surface of the Earth. Ionization is produced when solar ultraviolet radiation falls on the atoms of the atmosphere. The plasma frequency $f_p$ of the ionosphere is in the region of FM radio waves. Any radiation of frequency $f < f_p$ (such as, e.g., AM radio waves) falling on that layer suffers reflection (for the most part) and partial absorption. On the other hand, radiation having frequency $f > f_p$ (e.g., FM radio waves, microwaves, infrared radiation, visible light, etc.) passes through the ionosphere with only minor losses due to reflection or absorption.[3]

Let us examine these processes in more detail:

When an AM radio wave, say, reaches the ionosphere from the ground, the electric field in the wave induces forced oscillations on the free electrons of the ionosphere, of frequency equal to that of the wave. A small fraction of the wave's energy is given up to this oscillation and is finally absorbed in the interior of the plasma, while the major part of the incident radiation is reflected back to the ground.

The frequency of an FM radio wave, on the other hand, is higher than the plasma frequency of the ionosphere and, as a result of this, the free electrons cannot respond fast enough to be set into forced oscillations in a similar way. The ionosphere thus simply lets the FM wave pass through and reach the outer space, having suffered only minimal absorption and reflection. The same occurs, of course, for every radiation of even higher frequency, such as microwaves, infrared radiation, visible light, etc.

---

[3] For the values of the related frequencies, see Sec. 10.15.

The effect of reflection for $f < f_p$ is used in AM radio broadcasting to transmit short-wave AM signals[4] around the Earth. Upon reaching the ionosphere, the signal is bounced back to the ground, as seen in Fig. 10.13. In this way, communication is possible between two points $A$ and $B$ separated by a large distance on the surface of the Earth (a straight-line transmission from $A$ to $B$ is impossible due to the curvature and the conductivity of the Earth). This effect is intensified during the night as the height at which the ionosphere begins increases due to the lack of solar ultraviolet radiation, and the reflected wave is thus able to reach remote locations that cannot be reached in daytime.

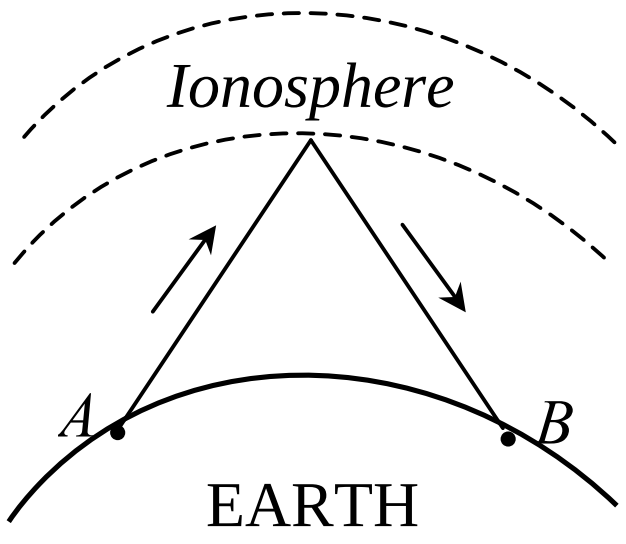

Fig. 10.13. Transmission of a radio wave by reflection on the Ionosphere.

Now, in order to communicate with spaceships or satellites we must use signals with frequencies *exceeding* the plasma frequency of the ionosphere. Such communications are achieved by using *microwaves*. As an application, the transmission of high-frequency signals at great distances on the surface of the Earth is done with the aid of *telecommunications satellites*, which play a role analogous to the ionosphere for AM radio waves.

The fact that both the infrared and the visible solar radiation can penetrate the ionosphere (since $f >> f_p$ ) and finally reach the surface of the Earth is of enormous importance for life on Earth. On the other hand, the plasma frequency of seawater is much above the visible spectrum ( $f_p \geq 10^{15}\ Hz$), making it impossible for light to reach great depths into the sea.

[4] The process is not as effective with medium-wave AM signals since the ground wave of such a signal, because of its lower frequency, can reach much greater distances and finally meet with the reflected wave out of phase. The interference of the direct and the reflected wave then produces a distorted sound effect on the receiver.

## References for Chapter 10

# QUESTIONS

**1.** Give the general mathematical expression for a harmonic plane wave $\xi(\vec{r},t)$ traveling ($a$) in the $+y$ direction; ($b$) in the $-z$ direction.

**2.** A plane e/m wave is traveling in the $-x$ direction. The magnetic field at some point of space is instantaneously oriented in the $-z$ direction. What is the corresponding instantaneous orientation of the electric field at that point?

**3.** Show that the standing waves (10.60) satisfy the wave equation (10.61).

**4.** Two media, a conducting and a non-conducting one, share the same constant values of ε and $\mu$. Compare the propagation speeds of plane e/m waves of a certain frequency $\omega$ in these media.

**5.** A conducting medium has constant values of $\varepsilon$, $\mu$, $\sigma$. Compare the propagation speeds $\upsilon_1$ and $\upsilon_2$ in this medium, of two plane e/m waves of frequencies $\omega_1$ and $\omega_2$ where $\omega_1 < \omega_2$.

**6.** Two plane waves of low frequencies $\omega_1$ and $\omega_2 = 2\omega_1$ fall on the surface of a conductor. Which wave will penetrate deeper into the conductor?

**7.** Consider two metals $M_1$ and $M_2$. The skin depths in these metals for the visible part of the spectrum of e/m radiation are such that $\Delta_1 << \Delta_2$. Which of the two metals would you use to make a mirror?

**8.** Examine which of the following two systems emits e/m radiation: ($a$) an electron in uniform circular motion; ($b$) a steady circular current $I$.

**9.** At the center of a very large sphere there is a small electric dipole oscillating along a vertical axis passing through the center of the sphere. At which points of the spherical surface is the intensity of the emitted e/m radiation a maximum and at which points is it a minimum? Repeat the problem, this time considering an oscillating magnetic dipole at the center of the sphere.

**10.** Where is it preferable to listen to the radio, inside a metal room or inside a glass room? Explain.

**11.** On a sunny summer day we lock ourselves inside a dark metal room. Do we need to put on our sunscreen?

**12.** A sailor in a submarine (which is submerged at a very low depth) and an astronaut in a spacecraft are both interested in the radio broadcasting of a football game. The game is being broadcasted at two radio stations: SPORT FM (107.4 *MHz*) and SPORT AM (560 *kHz*). Which station should each of these persons choose? (*Hint:* The plasma frequency of the sea is much higher than radio frequencies while the plasma frequency of the ionosphere is in the region of FM radio waves. How does the penetrating ability of an e/m wave vary with the frequency $\omega$ of the wave for $\omega << \omega_p$? What happens when $\omega > \omega_p$?)

**13.** Maxwell is assigned the post of commander of a warship. The ship possesses (*a*) a radar, (*b*) a sonar, (*c*) an AM radio transmitter, and (*d*) a microwave emitter. His multitask mission is (1) to detect enemy submarines; (2) to check for enemy warships; (3) to send a radio signal to a friendly submarine diving at a depth of 5 *m*; (4) to send a signal to a spacecraft in orbit around the Earth. Which apparatus must he use for each task?

**14.** How do you explain the fact that in nighttime we can hear remote AM radio stations that we are not able to hear in daytime? Why do we receive these stations in the shortwave frequency band?

**15.** Imagine our world if the plasma frequency of the ionosphere happened to be as high as that of the sea. How many hours would you study Physics during daytime and how many hours during nighttime? (*Hint:* The plasma frequency of the sea is much above the visible spectrum.)

# PROBLEMS

**1.** Consider a plane wave of the form $\xi(x,t)=f(x-\upsilon t)$. Show that, if the function $\xi$ is periodic in $t$, then it is necessarily periodic in $x$ as well, and vice versa.

***Solution:*** Assume that $\xi$ is periodic in $t$ with period $T$:

$$\xi(x,t) = \xi(x,\ t+T) \ \Rightarrow\ f(x-\upsilon t) = f[x-\upsilon(t+T)] = f[(x-\upsilon T)-\upsilon t] = f[(x-\lambda)-\upsilon t]$$

where we have set $\lambda=\upsilon T$. Hence, $\xi(x,t) = \xi(x-\lambda,\ t)$, which means that $\xi$ is periodic in $x$, with period $\lambda$. Conversely, by assuming that $\xi$ is periodic in $x$ with period $\lambda$, it is easy to show that it will also be periodic in $t$, with period $T=\lambda/\upsilon$ . If $\xi$ represents a harmonic plane wave, what is the physical significance of $\lambda$ and $T$ ?

**2.** According to the *principle of relativity*, the laws of Physics must have the same form in all *inertial* frames of reference. In particular, the Maxwell equations must be of the same form (i.e., must "look the same") for all inertial observers. Moreover, as found by experiment, the speed of light in empty space has the same value $c$ in every inertial reference frame (indeed, from the Maxwell equations it follows that the value of $c$ depends only on the constants $\varepsilon_0$ and $\mu_0$). Consider now a charge $q$ in uniform rectilinear motion relative to an inertial observer $O$. Show that, by the principle of relativity, this charge cannot emit e/m radiation.

***Solution:*** Since $q$ moves with constant velocity relative to the inertial observer $O$, it itself defines the origin of an inertial frame of reference. With respect to an observer $O'$ of this frame, $q$ is at rest. Thus the only thing recorded by $O'$ is a *static* electric field with no presence of any e/m wave. Let us now assume that, according to $O$, the charge $q$ emits e/m radiation. This means that $O$ records the presence of an e/m wave in her own frame, which wave travels at speed $c$. But, since the propagation speed of an e/m wave (not just of light!) is the same in all inertial frames, it follows that, relative to $O'$, the wave observed by $O$ also propagates at speed $c$. This, however, contradicts the fact that $O'$ does not perceive any e/m wave whatsoever! We conclude that neither $O$ can perceive an emission of e/m radiation from $q$.

In the case of an *accelerating* charge $q$ the above rationale breaks down since $q$ no longer defines the origin of an inertial frame of reference (indeed, there is no inertial frame relative to which $q$ is at rest at all moments). Therefore the principle of relativity does not apply in this case.

**3.** Consider two harmonic e/m waves of frequencies $\omega_1$ and $\omega_2$ , propagating in the $+x$ direction in empty space. The electric fields in the waves are

$$\vec{E}_1(x,t) = E_0\cos(k_1 x-\omega_1 t)\,\hat{u}_y\ ,\ \ \vec{E}_2(x,t) = E_0\cos(k_2 x-\omega_2 t)\,\hat{u}_y\ ;\quad \frac{\omega_1}{k_1}=\frac{\omega_2}{k_2}=c$$

($a$) Find the corresponding magnetic fields in the waves. ($b$) Find the expression for the e/m wave that results from the superposition of these waves. Is this a standing or a traveling wave?

***Solution:*** In both waves the fields $\vec{E}$ and $\vec{B}$ oscillate in phase and their amplitudes are related by $B_0=E_0/c$ . Furthermore, the cross product $\vec{E}\times\vec{B}$ is in the direction of propagation, i.e., in the $+x$ direction. Therefore we must have:

$$\vec{B}_1(x,t) = \frac{E_0}{c}\cos(k_1 x - \omega_1 t)\,\hat{u}_z \ , \quad \vec{B}_2(x,t) = \frac{E_0}{c}\cos(k_2 x - \omega_2 t)\,\hat{u}_z$$

By using the trigonometric identity

$$\cos\alpha + \cos\beta = 2\cos[(\alpha+\beta)/2]\cos[(\alpha-\beta)/2]$$

and by noticing that $\omega_1 \pm \omega_2 = c\,(k_1 \pm k_2)$, show that

$$\vec{E} = \vec{E}_1 + \vec{E}_2 = 2E_0 \cos\left[\frac{k_1 - k_2}{2}\,(x - ct)\right]\cos\left[\frac{k_1 + k_2}{2}\,(x - ct)\right]\hat{u}_y$$

$$\vec{B} = \vec{B}_1 + \vec{B}_2 = \frac{2E_0}{c}\cos\left[\frac{k_1 - k_2}{2}\,(x - ct)\right]\cos\left[\frac{k_1 + k_2}{2}\,(x - ct)\right]\hat{u}_z$$

The wave is a *traveling* one since it is of the form $F(x-ct)$ (compare with the example of Sec. 10.8). We can view it as a harmonic wave of frequency $(\omega_1+\omega_2)/2$, the amplitude of which constitutes a *separate* harmonic wave of frequency $|\omega_1-\omega_2|/2$. This is the simplest example of *amplitude modulation*, a method on which AM radio broadcasting is based.

**4.** A submarine submerged at depth $20\,m$ can barely receive an e/m signal of very long wavelength $\lambda$ (in the air). To what depth must the submarine shift in order for it to receive a signal of wavelength $\lambda/4$? (Assume that, approximately, $c=\lambda f$ in the air, where $f$ is the frequency of the original signal.)

***Solution:*** According to Eq. (10.71), the skin depth for a wave of (angular) frequency $\omega=2\pi f=2\pi c/\lambda$ is

$$\Delta \simeq \sqrt{\frac{2}{\omega\mu\sigma}} = \sqrt{\frac{\lambda}{\pi c\mu\sigma}}$$

Hence,

$$\Delta_2 / \Delta_1 = \sqrt{\lambda_2 / \lambda_1}$$

By substituting $\Delta_1=20\,m$ and $\lambda_2=\lambda_1/4$, we get: $\Delta_2 = 10\,m$.

**5.** We will study the characteristics of a general (non-monochromatic) plane e/m wave in empty space. Such a wave can be expressed as a linear combination of monochromatic waves of various frequencies $\omega$, which waves are traveling in the same direction, say, $+x$. If $\omega$ varies continuously, this linear combination has the general form

$$\vec{E}(x,t) = \int \vec{E}_0(k)\, e^{i(kx-\omega t)}\, dk \ , \quad \vec{B}(x,t) = \int \vec{B}_0(k)\, e^{i(kx-\omega t)}\, dk$$

where $\omega/k = c \Leftrightarrow \omega = ck$, with $c = (\varepsilon_0\,\mu_0)^{-1/2}$ (note that the wave vector is written $\vec{k} = |\vec{k}\,|\,\hat{u}_x = k\,\hat{u}_x$). As we know (Sec. 10.5), the monochromatic waves

$$\vec{E}_0(k)\, e^{i(kx-\omega t)} \ , \quad \vec{B}_0(k)\, e^{i(kx-\omega t)}$$

are solutions of the Maxwell equations in empty space, provided that

$$\vec{k}\cdot\vec{E}_0(k) = k\,\hat{u}_x\cdot\vec{E}_0(k) = 0\,,\quad \vec{k}\cdot\vec{B}_0(k) = k\,\hat{u}_x\cdot\vec{B}_0(k) = 0$$

$$\vec{k}\times\vec{E}_0(k) = k\,\hat{u}_x\times\vec{E}_0(k) = \omega\vec{B}_0(k)\,,\quad \vec{k}\times\vec{B}_0(k) = k\,\hat{u}_x\times\vec{B}_0(k) = -\frac{\omega}{c^2}\vec{E}_0(k)$$

(*a*) Show that $\vec{E}(x,t)$ and $\vec{B}(x,t)$ are plane waves traveling in the $+x$ direction with speed $c=\omega/k$. [*Hint:* Such waves are expressed as functions of the form $F(x-ct)$.]

(*b*) Show that $\vec{E}(x,t)$ and $\vec{B}(x,t)$ satisfy the Maxwell equations in empty space. (Remember that $\varepsilon_0\mu_0 = 1/c^2$.)

(*c*) Show that $\hat{u}_x\cdot\vec{E} = 0$, $\hat{u}_x\cdot\vec{B} = 0$. That is, the $\vec{E}(x,t)$ and $\vec{B}(x,t)$ are normal to the direction of propagation of the wave (*transverse wave*).

(*d*) Show that $\hat{u}_x\times\vec{E} = c\vec{B}$, $\hat{u}_x\times\vec{B} = -(1/c)\vec{E}$. That is, the $(\vec{E},\vec{B},\hat{u}_x)$ form a right-handed rectangular system.

(*e*) Show that the instantaneous values of $\vec{E}$ and $\vec{B}$ are related by $E = cB$.

***Solution:***

(*a*) $\vec{E}(x,t) = \int \vec{E}_0(k)\,e^{ik(x-ct)}\,dk \equiv \vec{F}(x-ct)$ ; similarly for $\vec{B}$

(*b*) The Maxwell equations in empty space are

$$\vec{\nabla}\cdot\vec{E} = 0 \qquad \vec{\nabla}\times\vec{E} = -\frac{\partial\vec{B}}{\partial t}$$

$$\vec{\nabla}\cdot\vec{B} = 0 \qquad \vec{\nabla}\times\vec{B} = \varepsilon_0\mu_0\frac{\partial\vec{E}}{\partial t} = \frac{1}{c^2}\frac{\partial\vec{E}}{\partial t}$$

By using Eqs. (10.41) and by taking into account that

$$\vec{\nabla}e^{i(kx-\omega t)} = ik\,e^{i(kx-\omega t)}\,\hat{u}_x$$

we have:

$$\vec{\nabla}\cdot\vec{E} = \int \vec{E}_0(k)\cdot\vec{\nabla}e^{i(kx-\omega t)}\,dk = \int ik\,\hat{u}_x\cdot\vec{E}_0(k)e^{i(kx-\omega t)}\,dk = 0\ ;\ \text{similarly, } \vec{\nabla}\cdot\vec{B} = 0$$

$$\vec{\nabla}\times\vec{E} + \frac{\partial\vec{B}}{\partial t} = \int[\vec{\nabla}e^{i(kx-\omega t)}]\times\vec{E}_0(k)\,dk - \int i\omega\vec{B}_0(k)e^{i(kx-\omega t)}\,dk$$

$$= i\int dk\; e^{i(kx-\omega t)}\,[k\,\hat{u}_x\times\vec{E}_0(k) - \omega\vec{B}_0(k)] = 0$$

$$\vec{\nabla}\times\vec{B} - \frac{1}{c^2}\frac{\partial\vec{E}}{\partial t} = i\int dk\; e^{i(kx-\omega t)}\,[k\,\hat{u}_x\times\vec{B}_0(k) + \frac{\omega}{c^2}\vec{E}_0(k)] = 0$$

(*c*) $\hat{u}_x\cdot\vec{E} = \int \hat{u}_x\cdot\vec{E}_0(k)\,e^{i(kx-\omega t)}\,dk = 0\,,\quad \hat{u}_x\cdot\vec{B} = \int \hat{u}_x\cdot\vec{B}_0(k)\,e^{i(kx-\omega t)}\,dk = 0$

(*d*) $\hat{u}_x\times\vec{E} = \int \hat{u}_x\times\vec{E}_0(k)\,e^{i(kx-\omega t)}\,dk = \int \frac{\omega}{k}\vec{B}_0(k)e^{i(kx-\omega t)}\,dk = c\vec{B}$

$$\hat{u}_x \times \vec{B} = \int \hat{u}_x \times \vec{B}_0(k)\, e^{i(kx-\omega t)}\, dk = -\int \frac{\omega}{kc^2} \vec{E}_0(k) e^{i(kx-\omega t)}\, dk = -\frac{1}{c}\vec{E}$$

(*e*) We now consider the relations $\hat{u}_x \times \vec{E} = c\vec{B}$, $\hat{u}_x \times \vec{B} = -(1/c)\vec{E}$ for *real* values of $\vec{E}$ and $\vec{B}$. By taking the results of part (*c*) into account, the first relation yields:

$$|\hat{u}_x \times \vec{E}| = c|\vec{B}| \;\Rightarrow\; |\vec{E}| = c|\vec{B}| \;\Rightarrow\; E = cB$$

We get the same result by using the second relation.

**6.** A lamp emits a huge number of monochromatic e/m waves of various frequencies $\omega$ and wave vectors $\vec{k}$. The e/m field in the surrounding empty space is the result of superposition of these waves and, for continuously varying $\omega$ and $\vec{k}$, it is written as

$$\vec{E}(\vec{r},t) = \iiint \vec{E}_0(\vec{k}) e^{i(\vec{k}\cdot\vec{r}-\omega t)}\, dk_x dk_y dk_z \equiv \int \vec{E}_0(\vec{k})\, e^{i(\vec{k}\cdot\vec{r}-\omega t)}\, d\vec{k}$$
$$\vec{B}(\vec{r},t) = \iiint \vec{B}_0(\vec{k}) e^{i(\vec{k}\cdot\vec{r}-\omega t)}\, dk_x dk_y dk_z \equiv \int \vec{B}_0(\vec{k})\, e^{i(\vec{k}\cdot\vec{r}-\omega t)}\, d\vec{k}$$

where $\omega = ck \Leftrightarrow \omega/k = c$ ( $k = |\vec{k}|$ ) . As we know (Sec. 10.5), the monochromatic waves

$$\vec{E}_0(\vec{k}) e^{i(\vec{k}\cdot\vec{r}-\omega t)}\ , \quad \vec{B}_0(\vec{k}) e^{i(\vec{k}\cdot\vec{r}-\omega t)}$$

are solutions of the Maxwell equations in empty space if

$$\vec{k}\cdot\vec{E}_0(\vec{k}) = 0, \quad \vec{k}\cdot\vec{B}_0(\vec{k}) = 0$$
$$\vec{k}\times\vec{E}_0(\vec{k}) = \omega\vec{B}_0(\vec{k}), \quad \vec{k}\times\vec{B}_0(\vec{k}) = -\frac{\omega}{c^2}\vec{E}_0(\vec{k})$$

(*a*) Show that the $\vec{E}(\vec{r},t)$ and $\vec{B}(\vec{r},t)$ satisfy the wave equation in empty space.

(*b*) Show that the $\vec{E}(\vec{r},t)$ and $\vec{B}(\vec{r},t)$ satisfy the Maxwell equations in empty space. (Remember that $\varepsilon_0\mu_0 = 1/c^2$ .)

(*c*) Explain why the e/m radiation emitted by the lamp does not, in general, correspond to a plane e/m wave, thus does not exhibit the properties stated in Problem 5. What if in place of the ordinary lamp we had a laser-beam emitter?

***Solution:***

(*a*)

$$\nabla^2\vec{E} = \frac{\partial^2\vec{E}}{\partial x^2} + \frac{\partial^2\vec{E}}{\partial y^2} + \frac{\partial^2\vec{E}}{\partial z^2} = -\int (k_x{}^2 + k_y{}^2 + k_z{}^2)\vec{E}_0(\vec{k}) e^{i(\vec{k}\cdot\vec{r}-\omega t)}\, d\vec{k}$$
$$= -\int k^2 \vec{E}_0(\vec{k}) e^{i(\vec{k}\cdot\vec{r}-\omega t)}\, d\vec{k}$$

$$\frac{\partial^2\vec{E}}{\partial t^2} = -\int \omega^2 \vec{E}_0(\vec{k}) e^{i(\vec{k}\cdot\vec{r}-\omega t)}\, d\vec{k}$$

(Note that the $k^2$ and $\omega^2$ *cannot* be taken out of the corresponding integrals since they are *not* constant quantities!) Hence,

$$\nabla^2\vec{E} - \frac{1}{c^2}\frac{\partial^2\vec{E}}{\partial t^2} = \int \left(\frac{\omega^2}{c^2} - k^2\right)\vec{E}_0(\vec{k}) e^{i(\vec{k}\cdot\vec{r}-\omega t)}\, d\vec{k} = 0 \;; \quad \text{similarly for } \vec{B}$$

(*b*) The Maxwell equations in empty space are

$$\vec{\nabla}\cdot\vec{E}=0 \qquad \vec{\nabla}\times\vec{E}=-\frac{\partial\vec{B}}{\partial t}$$

$$\vec{\nabla}\cdot\vec{B}=0 \qquad \vec{\nabla}\times\vec{B}=\varepsilon_0\mu_0\frac{\partial\vec{E}}{\partial t}=\frac{1}{c^2}\frac{\partial\vec{E}}{\partial t}$$

By using Eqs. (10.41) and the relation

$$\vec{\nabla}e^{i(\vec{k}\cdot\vec{r}-\omega t)}=i\vec{k}\,e^{i(\vec{k}\cdot\vec{r}-\omega t)}$$

we have:

$$\vec{\nabla}\cdot\vec{E}=\int\vec{E}_0(\vec{k})\cdot\vec{\nabla}e^{i(\vec{k}\cdot\vec{r}-\omega t)}\,d\vec{k}=\int i\vec{k}\cdot\vec{E}_0(\vec{k})e^{i(\vec{k}\cdot\vec{r}-\omega t)}\,d\vec{k}=0\ ;\ \text{similarly, } \vec{\nabla}\cdot\vec{B}=0$$

$$\vec{\nabla}\times\vec{E}+\frac{\partial\vec{B}}{\partial t}=\int[\vec{\nabla}e^{i(\vec{k}\cdot\vec{r}-\omega t)}]\times\vec{E}_0(\vec{k})\,d\vec{k}-\int i\omega\vec{B}_0(\vec{k})e^{i(\vec{k}\cdot\vec{r}-\omega t)}\,d\vec{k}$$

$$=i\int d\vec{k}\,e^{i(\vec{k}\cdot\vec{r}-\omega t)}[\vec{k}\times\vec{E}_0(\vec{k})-\omega\vec{B}_0(\vec{k})]=0$$

$$\vec{\nabla}\times\vec{B}-\frac{1}{c^2}\frac{\partial\vec{E}}{\partial t}=i\int d\vec{k}\,e^{i(\vec{k}\cdot\vec{r}-\omega t)}[\vec{k}\times\vec{B}_0(\vec{k})+\frac{\omega}{c^2}\vec{E}_0(\vec{k})]=0$$

(*c*) The emitted radiation does not have the form of a plane e/m wave since the e/m field is not of the form $\vec{F}(\hat{\tau}\cdot\vec{r}-ct)$ with a *fixed* direction of $\hat{\tau}$ (this is in contrast to Prob. 5, where we had that $\hat{\tau}=\hat{u}_x=const.$). A laser beam, on the other hand, is characterized by a high degree of directivity; therefore the corresponding e/m wave may be considered as an almost plane wave.

**7.** We will study the propagation of a plane e/m wave inside a conducting medium having constant values of $\varepsilon$, $\mu$, $\sigma$.

(*a*) By using the Maxwell equations for this medium,

$$\vec{\nabla}\cdot\vec{E}=0 \quad (1) \qquad \vec{\nabla}\times\vec{E}=-\frac{\partial\vec{B}}{\partial t} \quad (3)$$

$$\vec{\nabla}\cdot\vec{B}=0 \quad (2) \qquad \vec{\nabla}\times\vec{B}=\mu\sigma\vec{E}+\varepsilon\mu\frac{\partial\vec{E}}{\partial t} \quad (4)$$

show that the fields $\vec{E}$ and $\vec{B}$ satisfy the modified wave equations:

$$\nabla^2\vec{E}-\varepsilon\mu\frac{\partial^2\vec{E}}{\partial t^2}-\mu\sigma\frac{\partial\vec{E}}{\partial t}=0 \quad (5)$$

$$\nabla^2\vec{B}-\varepsilon\mu\frac{\partial^2\vec{B}}{\partial t^2}-\mu\sigma\frac{\partial\vec{B}}{\partial t}=0 \quad (5')$$

(*b*) Show that the above wave equations admit solutions of the form

$$\vec{E}(x,t)=\vec{E}_0e^{-sx}e^{i(kx-\omega t)}, \quad \vec{B}(x,t)=\vec{B}_0e^{-sx}e^{i(kx-\omega t)} \qquad (6)$$

(where $\vec{E}_0$, $\vec{B}_0$ are constant *complex* vectors) on the condition that

$$s^2 - k^2 + \varepsilon\mu\omega^2 = 0 \quad \text{and} \quad \mu\sigma\omega - 2sk = 0 \qquad (7)$$

By solving the system (7) for *k* and *s*, verify Eqs. (10.65). The solutions (6) describe a plane e/m wave traveling in the $+x$ direction.

(*c*) Show that the $\vec{E}$ and $\vec{B}$ are normal to the direction of propagation of the wave (transverse wave). [*Hint:* Substitute Eqs. (6) into (1) and (2).]

(*d*) Show that the $\vec{E}$ and $\vec{B}$ are mutually perpendicular. [*Hint:* Write $\vec{E}_0 = \widetilde{E_0}\,\hat{u}_y$ where $\widetilde{E_0}$ is a complex constant. By substituting the first of Eqs. (6) into (3) and by integrating for *t*, show that $\vec{B}_0 = \dfrac{k+is}{\omega}\,\widetilde{E_0}\,\hat{u}_z$.]

(*e*) Show that the *real* fields are written:

$$\vec{E}(x,t) = E_0\, e^{-sx} \cos(kx - \omega t + \alpha)\,\hat{u}_y$$
$$\vec{B}(x,t) = \frac{\sqrt{k^2+s^2}}{\omega}\, E_0\, e^{-sx} \cos(kx - \omega t + \alpha + \varphi)\,\hat{u}_z \qquad (8)$$

where $E_0$ is real and where $\varphi = \arctan(s/k)$. Check your result at the limit $\sigma \to 0$ (non-conducting medium) where, at this limit, $s \to 0$. [*Hint:* Put $\widetilde{E_0} = |\widetilde{E_0}|\, e^{i\alpha} = E_0\, e^{i\alpha}$ and $k + is = |k+is|\, e^{i\varphi} = \sqrt{k^2+s^2}\; e^{i\varphi}$, where $\tan\varphi = s/k$.]

***Solution:***

(*a*) We take the *rot* of (3) and (4) and we work as in Sec. 10.4.

(*b*) By substituting the first of Eqs. (6) into (5) and by noticing that $\nabla^2\vec{E} = \partial^2\vec{E}/\partial x^2$, we are led to a complex equation whose real and imaginary parts correspond to the first and the second of Eqs. (7), respectively. [Alternatively, we may substitute the second of Eqs. (6) into (5′).] Show that the solution of the system (7) for *k* and *s* yields Eqs. (10.65).

(*c*) We substitute Eqs. (6) into (1) and (2), noticing that

$$\vec{\nabla}\cdot\vec{E} = \vec{E}_0\cdot\vec{\nabla}e^{-sx+i(kx-\omega t)} = (-s+ik)\,\hat{u}_x\cdot\vec{E}, \quad \vec{\nabla}\cdot\vec{B} = (-s+ik)\,\hat{u}_x\cdot\vec{B}$$

Hence, $\hat{u}_x\cdot\vec{E} = 0$, $\hat{u}_x\cdot\vec{B} = 0$.

(*d*) We set $\vec{E}_0 = \widetilde{E_0}\,\hat{u}_y$, so that the first of Eqs. (6) is written as

$$\vec{E} = \widetilde{E_0}\, e^{-sx} e^{i(kx-\omega t)}\,\hat{u}_y \qquad (9)$$

or, briefly, $\vec{E} = \widetilde{E}\,\hat{u}_y$. Substituting (9) into (3), we have:

$$-\frac{\partial\vec{B}}{\partial t} = \vec{\nabla}\times\vec{E} = \begin{vmatrix} \hat{u}_x & \hat{u}_y & \hat{u}_z \\ \dfrac{\partial}{\partial x} & \dfrac{\partial}{\partial y} & \dfrac{\partial}{\partial z} \\ 0 & \widetilde{E} & 0 \end{vmatrix} = \frac{\partial\widetilde{E}}{\partial x}\,\hat{u}_z = (-s+ik)\,\widetilde{E_0}\, e^{-sx} e^{i(kx-\omega t)}\,\hat{u}_z$$

Integrating with respect to $t \Rightarrow$

$$\vec{B} = \frac{k+is}{\omega} \widetilde{E_0}\, e^{-sx} e^{i(kx-\omega t)}\, \hat{u}_z \qquad (10)$$

Then, by (6), $\vec{B}_0 = \dfrac{k+is}{\omega} \widetilde{E_0}\, \hat{u}_z$ .

(*e*) We set $\widetilde{E_0} = |\widetilde{E_0}|e^{i\alpha} = E_0 e^{i\alpha}$ and $k+is = |k+is|e^{i\varphi} = \sqrt{k^2+s^2}\, e^{i\varphi}$, where $\tan\varphi = s/k \Leftrightarrow \varphi = \arctan(s/k)$. Substituting into (9) and (10), and then taking the *real* parts of $\vec{E}$ and $\vec{B}$, we get Eqs. (8).

In the limit $\sigma \to 0$, the second of Eqs. (7) yields $s \to 0$ while the first one reduces to $\omega/k = \upsilon = (1/\varepsilon\mu)^{1/2}$. Moreover, the phase difference tends to $\varphi \to 0$ (that is, the fields $\vec{E}$ and $\vec{B}$ oscillate in phase) while the instantaneous values of the fields are related by $E = \upsilon B$. Therefore, in the limit of vanishing conductivity our results essentially reduce to those of Sec. 10.5 (this time for an arbitrary non-conducting medium instead of empty space).

**8.** We will study the *polarization* of a monochromatic plane e/m wave traveling in the $+x$ direction. The electric field in this wave is written as

$$\vec{E}(x,t) = \vec{E}_0\, e^{i(kx-\omega t)} \qquad (1)$$

where $\vec{E}_0$ is a constant *complex* vector. In the simplest case we set, as in Sec. 10.5,

$$\vec{E}_0 = \vec{E}_{0,R}\, e^{i\theta} \qquad (2)$$

where $\vec{E}_{0,R}$ is a *real* vector, so that

$$\vec{E}(x,t) = \vec{E}_{0,R}\, e^{i(kx-\omega t+\theta)} \qquad (3)$$

Taking the real part of $\vec{E}$, we find the *real* field

$$\vec{E}(x,t) = \vec{E}_{0,R} \cos(kx - \omega t + \theta) \qquad (4)$$

We notice that, at a given position $x = const.$, the tip of the vector $\vec{E}$ oscillates along the axis defined by the constant vector $\vec{E}_{0,R}$. We thus say that this wave is *linearly polarized*.[1]

The above situation, however, is a special one. To make the problem more general, let us return to the complex expression (1). As we know, the $\vec{E}_0$ and $\vec{E}$ are normal to the direction of propagation (i.e., normal to $\hat{u}_x$); these vectors, therefore, belong to the *yz*-plane. In place of (2) we now write:

$$\vec{E}_0 = E_1 e^{i\theta_1}\, \hat{u}_y + E_2 e^{i\theta_2}\, \hat{u}_z \qquad (5)$$

where the $E_1$ and $E_2$ are *real* scalar quantities.

[1] The term "plane polarized" is also often used. We will not use this term here, however, to avoid confusion with the distinct concept of a plane wave.

(*a*) Show that the *real* electric field is written:

$$\vec{E} = E_y\,\hat{u}_y + E_z\,\hat{u}_z \;; \quad E_y = E_1 \cos(kx - \omega t + \theta_1)\;, \quad E_z = E_2 \cos(kx - \omega t + \theta_2) \tag{6}$$

That is, $\vec{E}$ is the superposition of two linearly polarized waves with mutually perpendicular polarizations and with a phase difference equal to $(\theta_1 - \theta_2)$.

(*b*) Show that, for $\theta_1 = \theta_2$ as well as for $|\theta_1 - \theta_2| = \pi$, the field $\vec{E}$ is *linearly* polarized.

(*c*) Show that, for $|\theta_1 - \theta_2| = \pi/2$,

$$\left(\frac{E_y}{E_1}\right)^2 + \left(\frac{E_z}{E_2}\right)^2 = 1 \tag{7}$$

We say that $\vec{E}$ is *elliptically polarized*, since the tip of the vector $\vec{E}$ describes an ellipse on the *yz*-plane. In particular, if $E_1 = E_2 = E_0$, relation (7) is written

$$E_y^{\,2} + E_z^{\,2} = E_0^{\,2} \tag{8}$$

and the e/m wave is said to be *circularly polarized*.

(*d*) For an arbitrary phase difference $(\theta_1 - \theta_2)$, show that

$$\left(\frac{E_y}{E_1}\right)^2 + \left(\frac{E_z}{E_2}\right)^2 - 2\left(\frac{E_y}{E_1}\right)\left(\frac{E_z}{E_2}\right)\cos(\theta_1 - \theta_2) = \sin^2(\theta_1 - \theta_2) \tag{9}$$

Demonstrate that (9) reduces to our previous special results by allowing the phase difference to take on the values $\theta_1 - \theta_2 = 0,\ \pm\pi$ and $\pm\pi/2$.

(*e*) Consider a linearly polarized wave in which the electric field is

$$\vec{E}(x,t) = (E_1\,\hat{u}_y + E_2\,\hat{u}_z)\cos(kx - \omega t + \theta) \tag{10}$$

where $E_1$, $E_2$, $\theta$ are given real constants. Show that the corresponding magnetic field in the wave is

$$\vec{B}(x,t) = \frac{1}{c}(E_1\,\hat{u}_z - E_2\,\hat{u}_y)\cos(kx - \omega t + \theta) \tag{11}$$

where $c = \omega/k$. [*Hint:* The $\vec{E}$ and $\vec{B}$ are normal to the direction of propagation +*x*, they oscillate in phase and they satisfy the relations (cf. Prob. 5)

$$\hat{u}_x \times \vec{E} = c\vec{B}, \quad \hat{u}_x \times \vec{B} = -(1/c)\vec{E} \tag{12}$$

Assume that $\vec{B} = (B_1\,\hat{u}_y + B_2\,\hat{u}_z)\cos(kx - \omega t + \theta)$ and determine the $B_1$, $B_2$.]

***Solution:***

(*a*) We substitute (5) into (1) and then take the real part of $\vec{E}$. We thus find (6).

(*b*) For $\theta_1 = \theta_2 = \theta$, Eq. (6) reduces to (4) with $\vec{E}_{0,R} = E_1\,\hat{u}_y + E_2\,\hat{u}_z$; similarly for $\theta_1 = \theta$ and $\theta_2 = \theta + \pi$, with $\vec{E}_{0,R} = E_1\,\hat{u}_y - E_2\,\hat{u}_z$.

(*c*) We set $\theta_1=\theta$ and $\theta_2=\theta+\pi/2$. Then,

$$E_y = E_1\cos(kx-\omega t+\theta)\ ,\quad E_z = -E_2\sin(kx-\omega t+\theta)$$

By using the identity $\cos^2 A+\sin^2 A=1$, we obtain (7).

(*d*) We solve the system

$$E_y = E_1\cos(kx-\omega t+\theta_1)\ ,\quad E_z = E_2\cos(kx-\omega t+\theta_2)$$

for $\cos(kx-\omega t)$ and $\sin(kx-\omega t)$ (with a little help from Trigonometry!) and we use the identity $\cos^2 A+\sin^2 A=1$. For $\theta_1-\theta_2 = 0$ or $\pi$, Eq. (9) yields:

$$\left(\frac{E_y}{E_1}\mp\frac{E_z}{E_2}\right)^2=0 \;\Rightarrow\; \frac{E_y}{E_z}=\pm\frac{E_1}{E_2}$$

This is precisely what we get from (6) when $\theta_1-\theta_2 = 0$ or $\pi$ (linear polarization). For $\theta_1-\theta_2 = \pi/2$, Eq. (9) reduces to (7) (elliptical polarization).

(*e*) In the first or the second of Eqs. (12), we substitute $\vec{E}$ from (10) and $\vec{B}$ from the expression given in the hint. By equating coefficients of similar unit vectors, we find that $B_1= -E_2/c$ and $B_2= E_1/c$.

**9.** In Problem 8 we saw that the superposition of two linearly polarized e/m waves of equal amplitudes, the electric fields in which waves are mutually perpendicular and differ in phase by $\pm\pi/2$, yields a circularly polarized wave. Show now that, conversely, by a suitable superposition of two circularly polarized waves we can construct a linearly polarized wave.

***Solution:*** We consider the general expression

$$\vec{E}=E_y\,\hat{u}_y+E_z\,\hat{u}_z\ ;$$

$$E_y = E_1\cos(kx-\omega t+\theta_1)\ ,\quad E_z = E_2\cos(kx-\omega t+\theta_2)$$

We let $E_1=E_2=E_0$ and we call $\theta_1=\theta$. For $\theta_2=\theta+\pi/2$ we have the circularly polarized wave

$$\begin{aligned}\vec{E}_L &= E_0\cos(kx-\omega t+\theta)\,\hat{u}_y - E_0\sin(kx-\omega t+\theta)\,\hat{u}_z\\ &= E_0\cos[\omega t-(kx+\theta)]\,\hat{u}_y + E_0\sin[\omega t-(kx+\theta)]\,\hat{u}_z\end{aligned}$$

while for $\theta_2=\theta-\pi/2$ we have the circularly polarized wave

$$\begin{aligned}\vec{E}_R &= E_0\cos(kx-\omega t+\theta)\,\hat{u}_y + E_0\sin(kx-\omega t+\theta)\,\hat{u}_z\\ &= E_0\cos[(kx+\theta)-\omega t]\,\hat{u}_y + E_0\sin[(kx+\theta)-\omega t]\,\hat{u}_z\end{aligned}$$

By adding the electric fields of these two waves, we find:

$$\vec{E}=\vec{E}_L+\vec{E}_R=2E_0\cos(kx-\omega t+\theta)\,\hat{u}_y$$

which represents a linearly polarized wave. The wave $\vec{E}_L$ is called *left-hand polarized* since, at each point *x*, it can be pictured as a vector of magnitude $E_0$, rotating counterclockwise on the *yz*-plane (Fig. 10.14). Similarly, the wave $\vec{E}_R$ is *right-hand polarized* since it can be pictured as a vector rotating clockwise. [Notice that the angle $\omega t-(kx+\theta)$ of the vector $\vec{E}_L$ with the $y$-axis increases with time, while the angle $(kx+\theta)-\omega t$ of $\vec{E}_R$ with this axis decreases with time.]

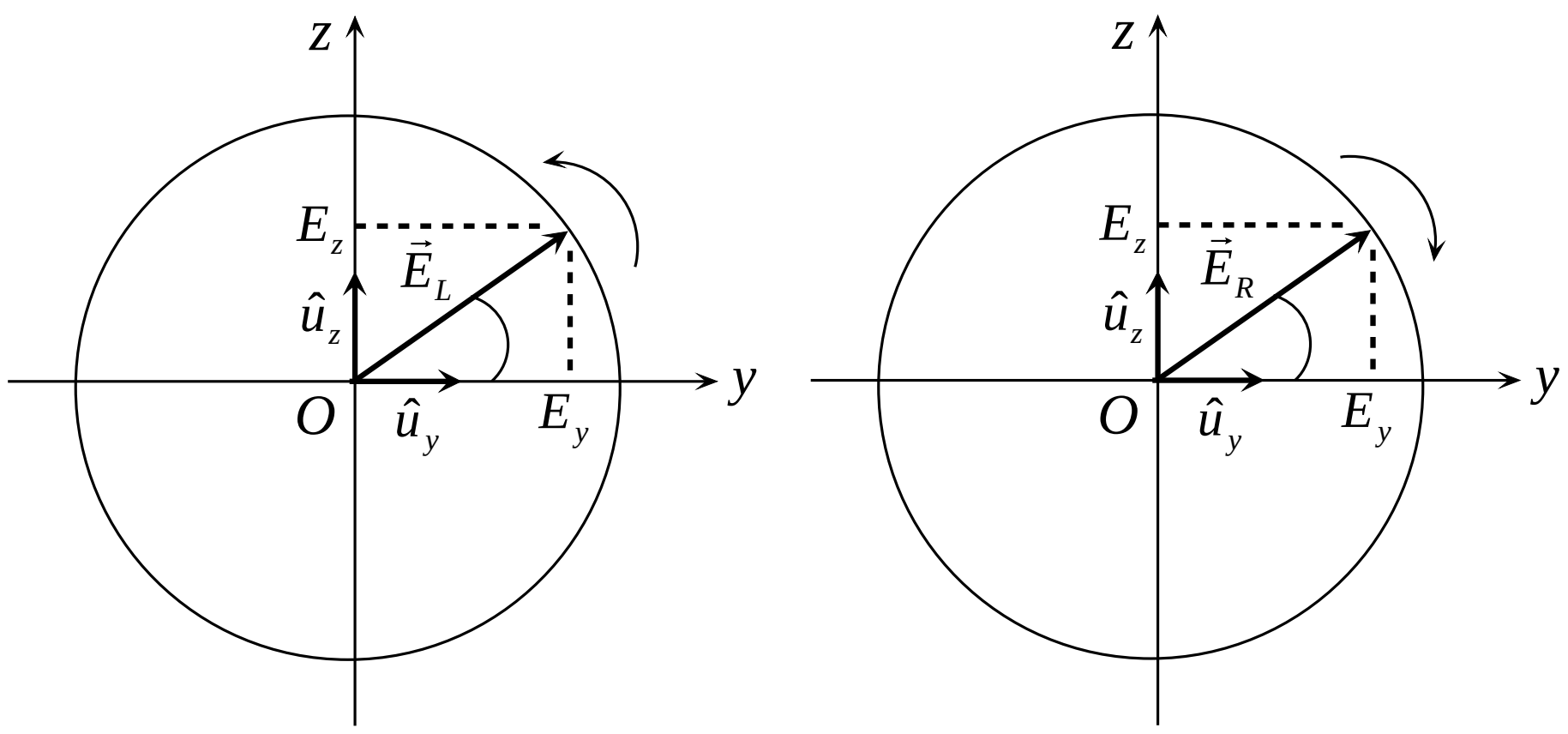

Fig. 10.14. Left-handed and right-handed polarized e/m wave.

# APPENDIX A: A NOTE ON THE SYSTEM OF UNITS [1]

Consider first the familiar Newton's law of gravity:

$$F = G \frac{m_1 m_2}{r^2}$$

Given that the units of mass, distance, and force have already been defined, one may determine the value of the constant $G$ by experiment. In simple terms, take two known masses $m_1$ and $m_2$ a distance $r$ apart, and measure the gravitational force $F$ between them; substitute the corresponding values into the above equation and solve for $G$.

Now, consider the Coulomb force between two charges at rest, a distance $r$ apart:

$$F = K_e \frac{q_1 q_2}{r^2} \qquad \text{(A.1)}$$

Assume that we want to evaluate the constant $K_e$. Although the units of distance and force have already been defined in the context of classical Mechanics, we still need to define a unit for charge. This means that no experimental determination of $K_e$ is possible by using (A.1) before we somehow fix the aforementioned unit.

Consider also the force per unit length between two parallel rectilinear currents a perpendicular distance $R$ apart. As can be shown [1],

$$\frac{dF}{dl} = K_m \frac{2 I_1 I_2}{R} \qquad \text{(A.2)}$$

where the factor 2 on the right-hand side is introduced for future convenience. Again, an experimental determination of the constant $K_m$ is possible *after* we have defined some unit for the current.

In any system of units the charge and the current are related by

$$I = \frac{dq}{dt} \qquad \text{(A.3)}$$

This means that the unit of charge uniquely determines the unit of current, and vice versa.

We may now use the following strategy in order to determine the constants $K_e$ and $K_m$ in Eqs. (A.1) and (A.2), respectively, and simultaneously define units for charge and current: We *arbitrarily* assign a value to $K_e$ and use (A.1) to define the unit of charge. For example, for $K_e=1$ we have that $F=q_1 q_2 / r^2$, so that

$$\text{(unit of charge)}^2 = \text{(unit of force)} \times \text{(unit of distance)}^2$$

---

[1] For a more detailed discussion, see, e.g., [1], Chap. 23.

Then, by (A.3),

(unit of current) = (unit of charge) / (unit of time)

Having fixed the unit of current, we may now use (A.2) to *experimentally* determine the value of $K_m$ .

An alternative strategy could be the following: We arbitrarily define the unit of current by making some definite statement regarding the force between two parallel unit currents a unit distance apart. We then use (A.2) to determine the value of $K_m$ . Now, by (A.3),

(unit of charge) = (unit of current) × (unit of time)

Having fixed the unit of charge, we use (A.1) to *experimentally* determine $K_e$ . This is the process employed in the S.I. system of units (meter, kilogram, second, ampere).

As experiment shows, in *all* systems of units (and regardless of the associated process for determining $K_e$ and $K_m$) the following relation is satisfied:

$$\frac{K_e}{K_m} = c^2 \quad \text{where} \quad c = 3\times 10^8 \, m/s \qquad \text{(A.4)}$$

Let us now concentrate on the S.I. system. We first define the unit of current, called *ampere* ($A$), as follows: If $I_1=I_2=1A$ and $R=1m$, then $dF/dl=2\times 10^{-7}\, N/m$. Relation (A.2) then yields: $K_m=10^{-7}\, N/A^2$. The unit of charge, called *coulomb* ($C$), is equal to $1C=1A.s$. Having defined this unit, we can now experimentally determine $K_e$ from (A.1). The result is: $K_e=9\times 10^9\, N.m^2/C^2$. You may check that $K_e/K_m=c^2$, in agreement with (A.4).

For reasons of convenience, it is customary to set $K_e=1/4\pi\varepsilon_0$ and $K_m=\mu_0/4\pi$. Then (A.4) takes on the form

$$c = \frac{1}{\sqrt{\varepsilon_0 \mu_0}} = 3\times 10^8 \, m/s$$

The quantity $c$ is, of course, the speed of light in vacuum. Notice that this speed is, in a sense, “built into” the fundamental laws of Electrodynamics. Notice also that $c$ is a constant independent of the state of motion of the observer, in agreement with the fundamental postulates of Special Relativity.

## Reference for Appendix A

# APPENDIX B: A REMARK ON THE CHARGING CAPACITOR

The Ampère-Maxwell law in integral form (Sec. 9.4) involves the concept of the total current through a loop *C*, where by "loop" we mean a closed curve in space. As we learned in Problem 5 of Chap. 7, in the static-field case this is a well-defined quantity given by the expression

$$I_{in} = \int_S \vec{J} \cdot \overrightarrow{da} \qquad \text{(B.1)}$$

where *S* is *any* open surface bordered by *C*. That is, the quantity $I_{in}$ has a uniquely defined value that is independent of the particular choice of *S*, i.e., is the same for all open surfaces *S* sharing a common border *C*. This is related to the fact that, in the static case the charge density is constant in time at all points in the region of interest. This condition, however, is not always satisfied for time-varying e/m fields.

As an example, consider a circuit carrying a time-dependent current $I(t)$. If the circuit does not contain a capacitor, no charge is piling up at any point and the charge density at any elementary segment of the circuit is constant in time. Moreover, at each instant *t* the current *I* is constant along the circuit, its value changing only with time. Now, if *C* is a loop encircling some section of the circuit, as shown in Fig. B.1, then at each instant *t* the same current $I(t)$ will pass through any open surface *S* bordered by *C*. Thus, the integral in Eq. (B.1) is a well-defined quantity having the value $I_{in}=I(t)$ for all *S*.

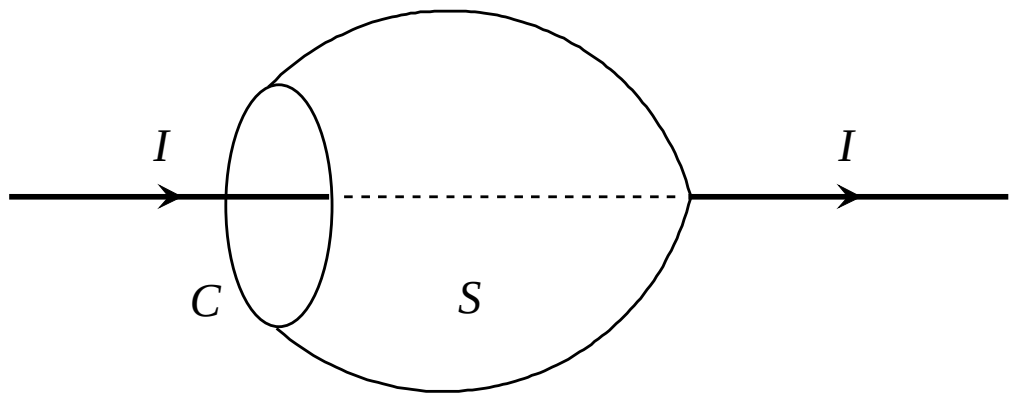

Fig. B.1. A loop *C* encircling the current *I*; the open surface *S* is bordered by *C*.

Things are not the same if the circuit contains a capacitor that is charging or discharging. It is then no longer true that the current $I(t)$ is constant along the circuit, given that $I(t)$ is nonzero outside the capacitor and zero inside. Thus, the value of the integral in (B.1) depends on whether the surface *S* does or does not contain points belonging to the interior of the capacitor.

Figure B.2 shows a simple circuit containing a capacitor that is being charged by a time-dependent current $I(t)$. At time *t* the plates of the capacitor carry charges $\pm Q(t)$. Assume now that we encircle the current *I* by an imaginary plane loop *C* parallel to the plates. The "current through *C*" is here an ill-defined notion since the value of the integral in Eq. (B.1) is $I_{in}=I$ for the flat surface $S_1$ and $I_{in}=0$ for the curved surface $S_2$. This, in turn, implies that Ampère's law of magnetostatics (Sec. 7.3) cannot be valid in this case, given that, according to this law, the integral of the magnetic field $\vec{B}$ along the loop *C*, equal to $\mu_0 I_{in}$, would not be uniquely defined but would depend on the surface *S* bounded by *C*.

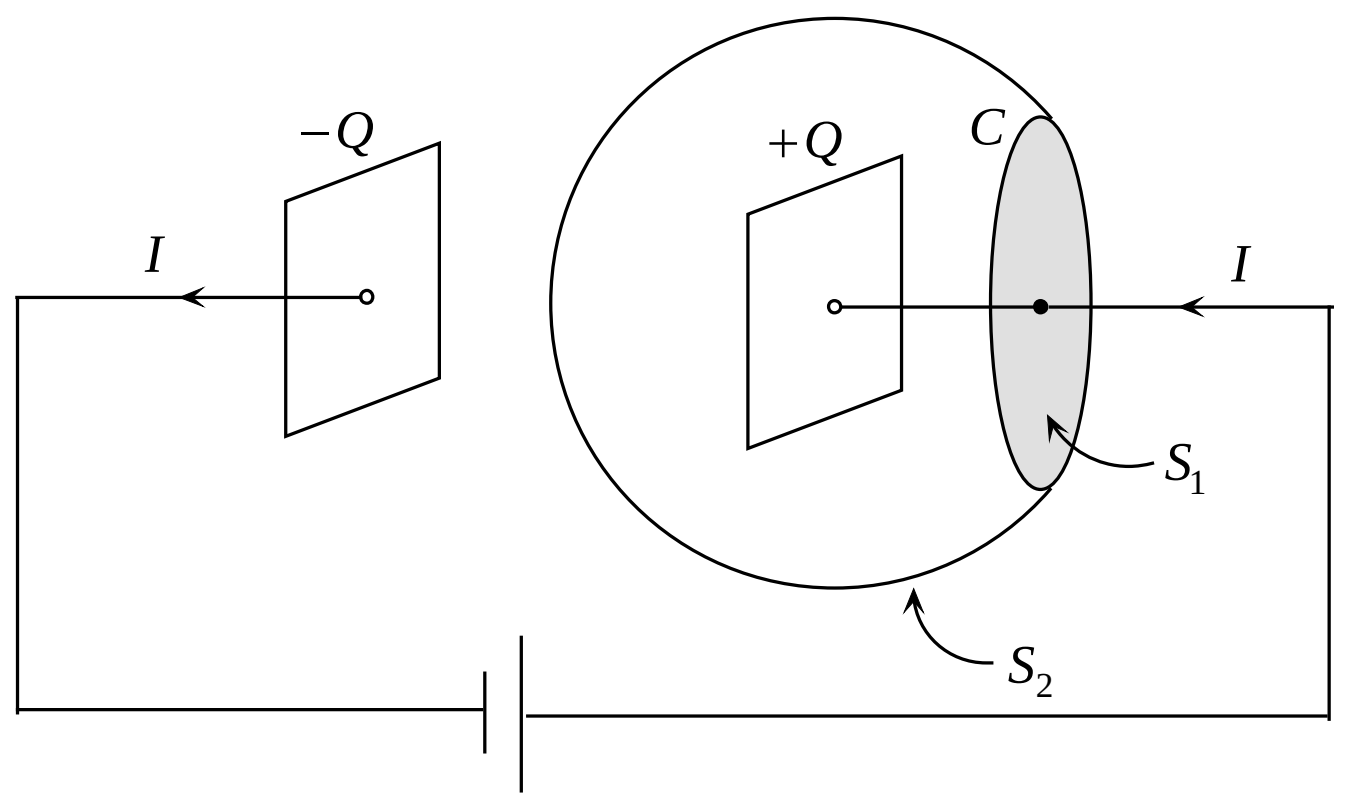

Fig. B.2. A flat surface $S_1$ and a curved surface $S_2$ with common border $C$. Part of $S_2$ lies in the interior of a capacitor.

To restore the single-valuedness of the closed line integral of $\vec{B}$, one has to introduce the so-called *displacement current*, which actually expresses the rate of change of a time-dependent electric field (Sec. 9.4). The Ampère-Maxwell law then guarantees that the integral of $\vec{B}$ along $C$ is uniquely defined, equal to $\mu_0 I(t)$ for both surfaces $S_1$ and $S_2$ (see, e.g., [1], Chap. 21).

Although everything works fine with regard to the Ampère-Maxwell law, there is still one law we have to take into account; namely, the *Faraday-Henry law* (Sec. 9.3). According to that law, a time-changing magnetic field is always accompanied by an electric field (or, as is often said, "induces" an electric field). So, the electric field outside the capacitor cannot be zero, given that the time-dependent current $I(t)$ is expected to generate a time-dependent magnetic field in that region. For a similar reason, the electric field inside the capacitor cannot be of the same form as in the static-field case (there must also be a contribution from the rate of change of the magnetic field between the plates). These corrections to the electric field are usually disregarded in standard textbook approaches to the charging capacitor problem (see [2]).

## References for Appendix B

# APPENDIX C: CLASSICAL AND QUANTUM THEORY OF CONDUCTION

The classical theory of electrical conduction of metals, which leads to Ohm's law (2.7) and the expression $\sigma = qn\mu$ for the conductivity, was proposed by P. Drude in 1900. Drude's model makes the following assumptions:

- The electron is a classical particle obeying Newton's laws and the laws of classical electromagnetism.
- The motion of free electrons in a metal is subject to laws similar to those obeyed by the molecules of an ideal gas.
- The totality of free electrons in the crystal (the concentration of which electrons is given by the electronic density $n$) contributes to the conductivity of the metal.
- Electrical resistance is due to collisions of free electrons with the positive ions of the metal, which ions are assumed to be *stationary*.

This classical model of electrical conduction is not entirely successful. Some of its failures are the following:

1. The model assumes that the conductivity $\sigma$ is proportional to the electronic density $n$. This, however, is not in accord with experimental observations, given that metals with high values of $n$ are found to be less conductive than others having lower values of $n$ [1].

2. The model predicts that the resistivity $\rho$ of a metal must be proportional to $T^{1/2}$, where $T$ is the absolute temperature [1]. Experiments, however, indicate that $\rho$ is in fact proportional to $T$ over a wide range of temperatures. The model also predicts that $\rho \to 0$ as $T \to 0$, which does not agree with experimental results (the resistivity tends to a nonzero value at very low temperatures).

3. The model also makes inaccurate predictions for the molar heat capacity of metals (the heat absorbed by one mole per unit change of temperature) [1].

The main reasons for the failure of the classical Drude model are the following: (*a*) The model treats the electron – an exceedingly microscopic elementary particle – like a classical particle subject to the usual laws of Newton and Maxwell and behaving more or less similarly to the molecules of an ideal gas. However, the behavior of an electron in an atom, a molecule, or a crystal cannot be interpreted within the framework of the classical theory. An approach to the problem is necessary which will take into account the principles of Quantum Mechanics (in particular, the uncertainty principle as well as the Pauli exclusion principle). (*b*) The classical model assumes that the ions of the metal are stationary, thus ignoring their thermal vibrations. (*c*) All free electrons are assumed to contribute equally to the conductivity of the metal, an assumption that is not true in reality.

In 1925, F. Bloch proposed a quantum theory of electrical conduction of metals. According to this theory, the electrical resistance is due to various kinds of imperfections in the crystal lattice, such as impurities or structural defects, as well as due to thermal vibrations of the ions about their equilibrium positions. Bloch's theory ex-plains why the resistivity $\rho$ of an ordinary (i.e., non-superconducting) metal does not vanish for $T\to 0$.

Now, in the classical Drude theory all valence electrons of the atoms of a metal (that is, the totality of free electrons in the crystal) are assumed to contribute to the conductivity of the metal. This theory, however, ignores a purely quantum principle, namely, the Pauli exclusion principle, according to which even at absolute temperature $T$=0 it is impossible for the entire set of free electrons in the metal to occupy the (kinetic) energy level $E$=0 (as the case is with the molecules of an ideal gas). As discussed in Sec. 3.5, at $T$=0 the free electrons occupy all available energy levels in the conduction band from $E$=0 up to the Fermi level $E_F$.

By taking into account the Pauli exclusion principle, one is led to the conclusion that the only electrons which may participate in the conduction process are those that occupy energy levels in the vicinity of the Fermi level. Hence, the number of conduction electrons is some 100 times smaller than the one assumed in the classical theory. The classical expression $\sigma=qn\mu$ for the conductivity, therefore, cannot be correct as it is based on the assumption that the conduction is due to the totality of free electrons in the metal, the concentration of which electrons is $n$. The above expression for $\sigma$, however, can still be used as an approximate one in problems where a rough estimate of physical quantities related to conductivity is needed.

## Reference for Appendix C

# INDEX